\documentclass[dvips,12pt,twoside]{report}
\usepackage{epsfig,a4}

\newcounter{com}
\newsavebox{\comname}

\RequirePackage[hyperindex,breaklinks=true]{hyperref}

\begin{document}

 \pagenumbering{roman}
 \pagestyle{empty}

\null\vskip0.5in

\begin{center}

\large\expandafter{DIELECTRON PRODUCTION IN HEAVY ION COLLISIONS
AT 158~GeV/c PER NUCLEON}

\end{center}

\vfill

\begin{center}

 \rm Vom Fachbereich Physik\\

 der Technischen Universit\"{a}t Darmstadt

\end{center}

\vfill

\begin{center}

 \rm zur Erlangung des Grades\\

 eines Doktors der Naturwissenschaften\\

 (Dr.~rer.~nat.)

\end{center}

\vfill

\begin{center}

 \rm genehmigte Dissertation von\\

 \vspace{2.5mm}

 {\large Dipl.-Phys.~M.A. Gunar Hering}\\

 \vspace{2mm}

  aus Chemnitz

\end{center}

\vfill \vskip0.75in

\begin{center}

 \rm Darmstadt 2001 \\

 D 17

\end{center}

\clearpage

\null\vskip4.5in
 \begin{center}
  \rm Referent: Prof.~Dr.~P.~Braun-Munzinger\\
  \rm Koreferent: Prof.~Dr.~W.~N\"{o}renberg\\
\end{center}

\vfill
\begin{center}
 \rm Tag der Einreichung: 27. November 2001 \\
     Tag der Pr\"{u}fung: 21. Januar 2002 \\
\end{center}
\vfill

\cleardoublepage

\begin{abstract}
In this paper, the low-mass electron pair production in
158~A\,GeV/c Pb-Au collisions is investigated with the Cherenkov
Ring Electron Spectrometer (CERES) at the Super Proton Synchrotron
accelerator (SPS) at CERN\@. The main goal is to search for
modifications of hadron properties in hot and dense nuclear
matter. The presented re-analysis of the 1996 data set is focused
on a detailed study of the combinatorial-background subtraction by
means of the mixed-event technique. The results confirm previous
findings of CERES. The dielectron production in the mass range of
0.25\,$<$\,$m_{\rm ee}$\,$<$\,2\,GeV/c$^2$ is enhanced by a factor
of $3.0\pm1.3({\rm stat.})\pm 1.2({\rm syst.})$ over the
expectation from neutral meson decays. The data is compared to
transport model calculations and seem to favor the version
including in-medium effects.

Furthermore, the development of a new technology to manufacture
ultralightweight mirrors for Ring Imaging Cherenkov detectors
(RICH) is described. Replacement of the RICH-2 glass mirror by a
mirror almost transparent to electrons would considerably improve
the performance of the upgraded CERES detector system including a
radial Time Projection Chamber (TPC).\\

\begin{center}
{\bf Zusammenfassung}
\end{center}

\noindent In dieser Arbeit wird die Produktion von
Elektronenpaaren kleiner Masse in Pb-Au Kollisionen bei
158~A\,GeV/c mit Hilfe des Cherenkov Ring Elektron Spektrometers
(CERES) am Super Proton Synchrotron Beschleuniger (SPS) des CERN
untersucht. Ziel des Experimentes ist der Nachweis von
Ver\"anderungen der Eigenschaften von Hadronen in einem dichten
und hei{\ss}en Medium aus Kernmaterie. Der Schwerpunkt der hier
pr\"asentierten Neuanalyse des Datensets von 1996 ist die
detailierte Untersuchung der kombinatorischen
Untergrundsubtraktion mit der Methode der gemischten Ereignisse.
Die Ergebnisse der Untersuchung best\"atigen die vorhergehenden
Resultate von CERES. Die im Massenbereich 0.25\,$<$\,$m_{\rm
ee}$\,$<$\,2\,GeV/c$^2$ gemessene Rate von Elektronenpaaren
\"ubersteigt die Erwartung von den Zerf\"allen neutraler Mesonen
um einen Faktor $3.0\pm1.3({\rm stat.})\pm 1.2({\rm syst.})$. Die
Daten werden mit theoretischen Vorhersagen im Rahmen eines
Transportmodells verglichen. Die beste \"Ubereinstimmung ergibt
sich mit einem Szenario, welches Mediumeffekte einschlie{\ss}t.

In einem weiteren Teil der Arbeit wird die Entwicklung einer neuen
Technologie zur Herstellung ultra-d\"unner Spiegel f\"ur
Ringabbildende Cherenkov Detektoren (RICH) beschrieben. Der
Austausch des gegenw\"artigen RICH-2 Glas\-spiegels mit einem
f\"ur Elektronen fast transparenten Spiegels w\"urde die Leistung
des mit einer TPC nachger\"usteten CERES Detektorsystems
betr\"achtlich steigern.

\end{abstract}

\cleardoublepage
\pagestyle{plain}

\tableofcontents

\cleardoublepage

\listoffigures

\addcontentsline{toc}{chapter}{List of Figures}

\cleardoublepage

\listoftables

\addcontentsline{toc}{chapter}{List of Tables}

\cleardoublepage

\chapter*{Acknowledgements}

Completing this thesis gives me an opportunity to express my
gratitude to many people who have been important to me during my
time at GSI.\\

\noindent First of all, I would like to thank my advisor,
Prof.~Peter Braun-Munzinger. I have learnt much from him. His
personal amiability and professional enthusiasm have helped make
the research enjoyable and interesting. I wish to thank  Dariousz
Miskowisz, Hiroyuki Sako, Jana Slivova, Peter Gl\"{a}ssel, and
Chilo Garabatos for many inspiring discussions and their
contributions to this work. I am indebted to Ralf Rapp for sending
me his recently computed dielectron spectra.\\

\noindent The people of Composite Optics Inc., Edward Lettner,
Audrey Clark, Randy Clark, and Helmuth Dorth, who were
instrumental in manufacturing the carbon-fiber-composite mirrors,
I thank for the successful cooperation.\\

\noindent I want to show my appreciation of the scientists and
staff of GSI, whose support and intellectual challenges I greatly
value. Prof. Johanna Stachel deserves my gratitude for providing
me with a stimulating work environment and access to the
facilities of the Physics Institute of the University of
Heidelberg. Of all the people who made my stays at CERN such a
pleasant one I would like to specially mention Wolfgang Schmitz,
Thomas Wienhold, Geydar Agakishiev, Harry Appelsh\"auser, Ingrid
Heppe, Heinz Tilsner, Hannes Wessels, and all collaborators of the
CERES experiment.\\

\noindent I am indebted to the German Scholarship Foundation and
The Melton Foundation for supporting me.\\

\noindent Furthermore, I thank my girlfriend Anne K\"ohler for
encouraging me throughout the work and for her invaluable help in
proofreading this thesis. Finally, I would like to thank my
parents for all the support and encouragement they have given me
throughout my education.

\addcontentsline{toc}{chapter}{Acknowledgements}

\cleardoublepage

\bibliographystyle{unsrt}

\chapter{Preface}
 \pagenumbering{arabic}
 \setcounter{page}{1}

This thesis is about the study of dielectron production in Pb-Au
collisions at 158\,GeV/c per nucleon measured with the Cherenkov
Ring Electron Spectrometer (CERES) at the Super Proton Synchrotron
(SPS) accelerator at CERN\@. Part of the heavy ion research
program at CERN, CERES is committed to the exploration of nuclear
matter under extreme conditions of high temperature and high
density.

The interdisciplinary field of ultrarelativistic heavy ion
collisions combines the elementary interaction aspect of
high-energy particle physics with the macroscopic matter aspects
of nuclear physics. It is focused on the investigation of the
properties of nuclear bulk matter made up of strongly interacting
particles, i.e.~hadrons, quarks, and gluons. The prediction is
that nuclear matter would undergo a phase transition to a
quark-gluon plasma (QGP)~\cite{Shuryak:1978ij}, a gas of freely
moving quarks and gluons. This triggered not only a strong
theoretical interest in this field but also initiated a huge
experimental effort to verify the prediction.

First evidence for this new state of matter was found at CERN in
the beginning of the year 2000~\cite{Heinz:2000bk}. CERES was one
of the experiments contributing to this fundamental discovery.

During the phase transition, the quarks are expected to lose their
constituent mass which leads to the restoration of chiral
symmetry. Left- and right-handed quarks decouple and hadronic
states of opposite parity become degenerate.

The study of deconfined or chiral matter is not only relevant for
the understanding of heavy ion collisions but also for
astrophysics and cosmology. The environmental conditions of a
158~A\,GeV/c Pb-Au collision at the SPS accelerator resemble those
encountered in the evolution of the early universe, where a few
tens of microseconds after the {\em big bang} a transient stage of
strongly interacting matter persisted at temperatures of about
$10^{12}$\,K and low baryon density. Another extreme of high
densities and low temperatures created in heavy ion collisions at
the SIS accelerator is close to the conditions occurring in the
interior of neutron
stars~\cite{Alford:1999pb,Heiselberg:1997mx,Maruyama:1999td},
where mass densities are likely to exceed $10^{18}$\,kg/m$^3$ -
about four times the density of nuclei.

\chapter{Relativistic heavy ion physics}
\section{Hot and dense nuclear matter}
\label{sec:theo intro}

The observations of particle production in heavy ion collisions
are related to the evolution of hot and dense matter - a general
question of relativistic heavy ion physics - and in particular to
the transition of hadronic matter to a quark-gluon plasma and the
simultaneous restoration of chiral symmetry.

First, the fundamental physics properties to describe a strongly
interacting system of nucleons in vacuum shall be introduced.
Next, the modifications in the presence of a hot and dense medium
are discussed with special emphasis on theoretical concepts
applied to explain experimental data. Finally, dielectron
production is related to the properties of hadronic processes
occurring in nuclear collisions. In this section, the velocity of
light $c$ and Planck's constant $\hbar$ are set to
$c$\,=\,$\hbar$\,=\,1.

In the {\em Standard Model} of particle physics, nucleons are
constituted of quarks and gluons. Each flavor of quark comes in
three colors: red, blue, and green. The quark color wavefunction
can be written as a vector of Dirac spinors:
\begin{equation}
\psi = \left(\begin{array}{c} \psi_{\rm r} \\ \psi_{\rm b}
\\ \psi_{\rm g} \end{array} \right) \,.
 \label{equ:color wavefunction}
\end{equation}
The strong interactions of quarks and gluons are described by
Quantum Chromodynamics (QCD), a local $SU(3)$ gauge theory. The
dynamics are governed by the Lagrangian of QCD:
\begin{equation}
 {\cal L}_{\rm QCD} =
 \underbrace{\overline{\psi}_{q}(i\gamma^{\mu}\mathbf{D}_{\mu}
 -\widehat{\mathbf{M}})\psi_q}_{\rm quarks + interaction}-\underbrace{\mathbf{G}^{a}_{\mu\nu}\mathbf{G}^{\mu\nu
 a}/4}_{\rm gluons+interaction} \,.
 \label{equ:QCD Lagrangian}
\end{equation}
Considering the relevant light quark flavors, the spinor
$\psi_{q}$ is represented by u, d, and s quark
(i.e.~$\overline{\psi}=(\overline{{\rm u}},\overline{{\rm
d}},\overline{{\rm s}})$). The matrix $\widehat{\mathbf{M}}$ in
flavor space is composed of the bare quark masses, i.e.~$m^0_{\rm
u}$, $m^0_{\rm d}$, and $m^0_{\rm s}$, on the diagonal.

The gluonic part of the Lagrangian is determined by the gluonic
field strength tensor:
\begin{equation}
 \mathbf{G}^{a}_{\mu\nu} =
  \partial_{\mu}\mathbf{A}^a_{\nu}-\partial_{\nu}\mathbf{A}^a_{\mu}+
  igf^{abc}\mathbf{A}^b_{\mu}\mathbf{A}^c_{\nu} \,,
 \label{equ:gluons}
\end{equation}
where $g$ and $f^{abc}$ denote the strong coupling constant and
the structure constants of the group $SU(3)$~\cite{Zuber},
respectively.

To achieve invariance of QCD under local SU(3) gauge
transformation the derivative takes the form:
\begin{equation}
 \mathbf{D}_{\mu} = \partial_{\mu} - {\rm
 i}g\,\mathbf{t}^a\cdot\mathbf{A}^a_{\mu} \,.
 \label{equ:gauge derivative}
\end{equation}
It induces a coupling between the spin-1/2 colored quarks fields
and the gluonic spin-1 gauge fields. The coupling strength $g$ -
commonly expressed in terms of the strong ``fine-structure''
constant $\alpha_s$\,$=$\,$g^2/4\pi$ - increases with space-time
distance or equivalently decreases with the momentum transfer $Q$
of a given strong process~\cite{Schmelling:1996wm}:
\begin{equation}
  \alpha_s(Q)=\frac{\alpha_s(\Lambda)}{1+
  9\alpha_s(\Lambda)/4\pi\ln(|Q^2|/\Lambda^2)}\;.
\end{equation}
This particular behaviour is a consequence of the self-interaction
of gluonic fields leading to an {\em antiscreening} of the strong
interaction which dominates the screening of the quark color by
quark-antiquark bubbles. The reference $\alpha_s(\Lambda)$ is
fixed by measurements at a certain scale $\Lambda$ where
$\alpha_s(\Lambda)$ is small enough to justify a perturbation
expansion, e.g. $\alpha_s(m_Z)=0.118$ at the Z boson mass
$m_Z$\,$=$\,91\,GeV/c$^2$~\cite{PDBook}.

At large distance scales, quarks and gluons are confined in
colorless mesons and baryons. Only at short distances can
perturbation theory be applied, as quarks and gluons are
quasi-free ({\em asymptotic freedom}). The breakdown of
perturbation theory for $\alpha_s\geq1$ at momentum transfer of
about $Q\simeq1$\,GeV/c, encountered in heavy ion collisions,
poses the most challenging problem in theory. Both the effective
couplings and the relevant degrees of freedom change rapidly with
scale.

In the limit of vanishing quark mass, the QCD Lagrangian exhibits
additional symmetries that can be explored. It becomes invariant
under global vector $\lambda^b$ and axial-vector
$\lambda^b\gamma^5$ transformations in flavor space:
\begin{equation}
 \psi_{\rm q} \rightarrow \exp(-({\rm
 i}/2)\lambda^{b}\alpha^b_{V})\psi_{\rm q}\quad\mbox{and}\quad\psi_{\rm q} \rightarrow \exp(-({\rm
 i}/2)\lambda^{b}\alpha^b_{A}\gamma^5)\psi_{\rm q} \;,
 \label{equ:flavor transform}
\end{equation}
with parameters $\alpha^b_{V}$ and $\alpha^b_{A}$ being arbitrary
vectors in flavor space. This results in conserved vector and
axial-vector Noether currents:
\begin{equation}
 j^{\mu}_{\rm V,b}=\overline{\psi}\gamma^{\mu}\lambda^b/2\psi
 \qquad \mbox{and}\qquad j^{\mu}_{\rm
A,b}=\overline{\psi}\gamma^{\mu}\gamma_5\lambda^b/2 \psi \;.
 \label{equ:currents}
\end{equation}
Introducing the quark-spinor projections of right- and left-handed
components:
\begin{equation}
 \psi_{\rm R/L}=\frac{1}{2}\,(1\pm\gamma^5)\psi_{\rm q}\;,
 \label{equ:left/right}
\end{equation}
the transformation~\ref{equ:flavor transform} can be rewritten as:
\begin{eqnarray}
 \psi_{\rm R} & \rightarrow & \exp(-({\rm
 i}/2)\lambda^{b}\alpha^b_{\rm R})\psi_{\rm R}\;,\quad \psi_{\rm L} \rightarrow \psi_{\rm
 L}\;,\\
 \psi_{\rm L} & \rightarrow & \exp(-({\rm
 i}/2)\lambda^{b}\alpha^b_{\rm L})\psi_{\rm L}\;,\quad \psi_{\rm R} \rightarrow \psi_{\rm
 R}\;,
 \label{equ:chiral}
\end{eqnarray}
which describes a global SU(3)$_{\rm R}\times$SU(3)$_{\rm L}$ {\em
chiral symmetry} in flavor space. This symmetry has two important
implications. First, left- and right-handed quarks are not mixed
dynamically and their {\em handedness}, i.e.~the sign of the
projection of spin on its momentum direction, is conserved.
Second, corresponding vector and axial-vector resonances are
degenerate, as the respective current-current correlation
functions $\Pi^{\mu\nu}_{\rm V/A}$\@:
\begin{equation}
 \Pi^{\mu\nu}_{\rm V/A}(q)=i\int d^4x\,e^{iq\do\cdot x}
 \langle0|\mathcal{T} j^{\mu}_{\rm V/A}(x)j^{\nu}_{\rm
 V/A}(0)|0\rangle\;,
 \label{equ:cc correlator}
\end{equation}
which determine the spectral shape of unstable resonances, are
identical~\cite{Klingl:1998qy}.

In the physical world, chiral symmetry is apparently spontaneously
broken because chiral partners such as $\rho$(770)- and
$a_1$(1260)-meson show a large mass splitting ($\Delta
m$\,=\,500\,MeV/c$^2$). Therefore, the ground state, i.e.~the QCD
vacuum, is not invariant under chiral transformation. In
particular, the vacuum state $|0\rangle$ only respects vector
symmetries ($j_V$\,=\,$j_R$+$j_L$)~\cite{Vafa:1984tf}, while the
axial-vector symmetry ($j_A$\,=\,$j_R$-$j_L$) is spontaneously
broken:
\begin{equation}
 Q_{A,b}|0\rangle \equiv |PS_a\rangle \neq 0\qquad
 \mbox{with}
 \qquad Q_{A,b}=\int d^3x\;\psi^\dag\frac{\lambda^b}{2}\gamma_5\psi  \;,
 \label{equ:axial-anomaly}
\end{equation}
where $Q_{A,b}$ is the axial-vector charge corresponding to the
axial-vector current $j_{A,b}$ (see Eq.~\ref{equ:currents}).

The strength of the symmetry breaking can be characterized by the
vacuum expectation value of the Goldstone
boson~\cite{Goldstone:1961eq}, which is the (nearly) massless pion
$\pi$:
\begin{equation}
 \langle0|j^{\mu}_{A,k}|\pi_j(p)\rangle =
 -i\delta_{jk}f_{\pi}p^{\mu}e^{-ipx}
 \label{equ:pi-decay}
\end{equation}
where {\em f}$_{\pi}$ is the measured pion decay constant of {\em
f}$_{\pi}=93$\,MeV. It is expected that a transition from the
asymmetric phase observed (see Eq.~\ref{equ:axial-anomaly}) to a
phase where the symmetry of the vacuum is restored
($Q_{A,b}|0\rangle$\,=\,$Q_{V,b}|0\rangle$\,=\,0) can be triggered
by external parameters such as temperature and/or pressure. The
expectation value of the so-called chiral condensate
$\langle\overline{\psi}\psi\rangle$:
\begin{equation}
\langle\overline{\psi}\psi\rangle = \frac{1}{2}\,
\langle0|\overline{u}u+\overline{d}d|0\rangle =
\langle0|\overline{\psi}_{\rm L}\psi_{\rm R}+\overline{\psi}_{\rm
R}\psi_{\rm L}|0\rangle\;,
 \label{equ:<qq>}
\end{equation}
is the lowest-dimensional order parameter characterizing the
chiral phase transition. The quark condensate respects all
unbroken symmetries of the Lagrangian, as it is a scalar density,
diagonal in flavor space, and carrying a baryon number of zero.

The chiral condensate vanishes in the chiral symmetric phase but
becomes finite in the asymmetric phase corresponding to a mixing
of left- and right-handed quarks in the ground state. In other
words, there is a finite expectation value to create a light
quark-antiquark pair from a zero-point energy fluctuation of the
physical vacuum.

The mixing strength of left- and right-handed quarks in vacuum
$\langle\overline{\psi}\psi\rangle$ is connected to the pion decay
constant according to the Gell-Mann-Oakes-Renner
relation~\cite{Gell-Mann:1968rz}:
\begin{equation}
 m^2_{\pi}f^2_{\pi}=-2\overline{m}\langle\overline{\psi}\psi\rangle
 \qquad (\overline{m}\approx 6\,{\rm MeV})\,.
 \label{equ:GOR}
\end{equation}
A value of $f_{\pi}$\,=\,93\,MeV from pion decay measurements
leads to a vacuum expectation value of
$\langle\overline{\psi}\psi\rangle$\,$\simeq$\,$-$(240\,MeV)$^3$\,=\,$-$1.8\,fm$^{-3}$,
which is large compared to the normal nuclear density of about
0.17\,fm$^{-3}$~\cite{PDBook}, indicating a strong dynamical
breaking of chiral symmetry.

Until now what has been discussed is the properties of hadronic
matter in vacuum but what is of more interest is the dynamics in
the presence of a hot and dense medium.

When hadronic matter is heated and/or compressed, initially
confined quarks and gluons start to percolate between hadrons to
finally be liberated. This phase transition to a plasma of quarks
and gluons is accompanied by a melting of the quark condensate
indicating chiral symmetry restoration. Even before the critical
region is approached, the chiral symmetry is partially restored by
the presence of hadrons. The valence quarks and the pionic cloud
of a hadron produce a positive scalar density inside the hadron,
thus effectively decreasing the (negative) quark condensate.

The expected modification of the condensate are derived for the
case of high temperature {\em T} and low density. The equilibrium
properties of a hadron gas in contact with a heat bath are
described by the grand canonical partition function:
\begin{equation}
 \mathcal{Z}(V,T,\mu_q)={\rm
 Tr} \left( e^{-(\mathbf{\hat{H}}-\mu_q\mathbf{\hat{N}})/T}
 \right)\;,
 \label{equ:partition function}
\end{equation}
where $\mathbf{\hat{H}}$ is the Hamiltonian of the system,
$\mathbf{\hat{N}}$ is the quark number operator, and $\mu_q$
denotes the quark chemical potential. The expectation value of the
quark condensate is then given by the thermal average:
\begin{equation}
 \langle\langle\overline{\psi}\psi\rangle\rangle =
 \mathcal{Z}^{-1} \sum_n\langle n|\overline{\psi}\psi|n\rangle
 e^{-(E_n-\mu_q)/T}\;,
 \label{equ:qq thermal}
\end{equation}
where the sum is carried out over all eigenstates of the QCD
Hamiltonian. Equation~\ref{equ:qq thermal} can be solved for the
simplified case of a non-interacting hadron gas. The resulting
correction reduces the quark condensate for increasing
temperature:
\begin{equation}
 \frac{\langle\langle\overline{\psi}\psi\rangle\rangle}{\langle\overline{\psi}\psi\rangle}
 \simeq 1- \sum_{\rm hadr.}
 \frac{\sum_{h}\rho^s_{h}(T)}{f^2_{\pi}m^2_{\pi}}\;.
 \label{equ:pion gas}
\end{equation}
Each hadron species present with scalar density $\rho^s_{h}$
contributes to the reduction of the condensate according to its
sigma commutator $\sum_{h}$. The latter quantity divided by quark
mass is a measure for the integrated scalar quark density inside a
hadron $h$:
\begin{equation}
 \frac{\sum_h}{\overline{m}}=\int_h d\mathbf{r}\,\langle
 h|\overline{\psi}\psi|h\rangle\;.
\end{equation}
The particular temperature dependence of Eq.~\ref{equ:pion gas}
can be explained as follows. In the physical vacuum, the color
fields are squeezed into hadrons by the repulsion of the quark
condensate which fills the vacuum. With increasing temperature
this mechanism becomes inefficient, as thermal pions are produced
as excitations of the quark condensate, locally changing the
expectation value of the quark condensate.

Increasing density also reduces the quark condensate because more
and more space is occupied by baryons (equivalent to turning on a
baryon chemical potential). In this case, the nucleons give the
dominant correction leading to a formula similar to
Eq.~\ref{equ:pion gas}:
\begin{equation}
 \frac{\langle\langle\overline{\psi}\psi\rangle\rangle}{\langle\overline{\psi}\psi\rangle}
 \simeq 1- \frac{\sum_{N}\rho^s_{N}(\mu_N)}{f^2_{\pi}m^2_{\pi}}\;.
 \label{equ:nucleon gas}
\end{equation}
where $\sum_{N}$ is the nucleon sigma commutator and $\rho^s_{N}$
denotes the nucleon scalar density at a given nucleon chemical
potential $\mu_N$. In nuclear matter at normal density, the quark
condensate is already quenched by 30\% according to a value of
about 45\,MeV for the nucleon sigma
commutator~\cite{Jameson:1992ep}.

\begin{figure}[tb]
    \begin{minipage}[t]{.65\textwidth}
        \vspace{0pt}
        \epsfig{file=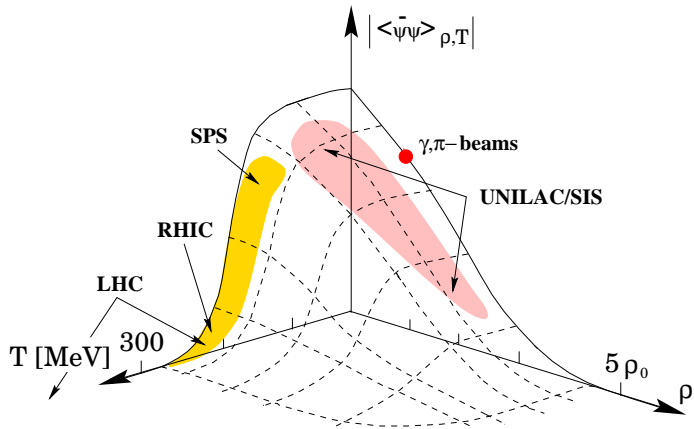,width=\textwidth}
    \end{minipage}%
    \begin{minipage}[t]{.35\textwidth}
      \vspace{0.5cm}
      \caption[Expectation value of the quark condensate]
      {\newline
       Expectation value of the quark condensate as described by the Nambu-Jona-Lasinio
       model~\cite{Lutz:1992dv,Friman:1997sv}. The regions accessible by various
       accelerators are highlighted.
      }
      \label{fig:qq NJL}
    \end{minipage}
\end{figure}

The sigma commutator and the dropping of the chiral condensate can
be estimated in the framework of effective theories, e.g. the
$\sigma$-model~\cite{Gell-Mann:1960np}, the Nambu-Jona-Lasinio
model~\cite{Nambu:1961tp}, or the Walecka
model~\cite{Walecka:1974qa}. Figure~\ref{fig:qq NJL} illustrates
the melting of the condensate for the example of the
Nambu-Jona-Lasinio model~\cite{Lutz:1992dv}.
Several effects preceding the phase transition towards a
restoration of chiral symmetry were predicted by effective
mean-field models. The most important effects are dropping hadron
masses (BR-scaling)~\cite{Brown:1991kk} and mixing of vector and
axial-vector currents~\cite{Dey:1990ba}, both leading to
modifications of the hadron spectral functions~\cite{Rapp:1999ej}.
The competing models shall be discussed in detail in connection
with the experimental results presented in Sec.~\ref{sec:theory}.

Model-independent results are obtained by perturbative low-density
expansion. However, this procedure is restricted to temperatures
below 120\,MeV and cannot address the nature of the phase
transition. Of the non-perturbative approaches, only numerical
lattice QCD calculations provide a stringent framework even though
many-body theory~\cite{Aouissat:1996va} and renormalization-group
techniques~\cite{Schaefer:1999em,Papp:1999he} are promising
developments.

\begin{figure}[bt]
  \centering
  \mbox{
   \epsfig{file=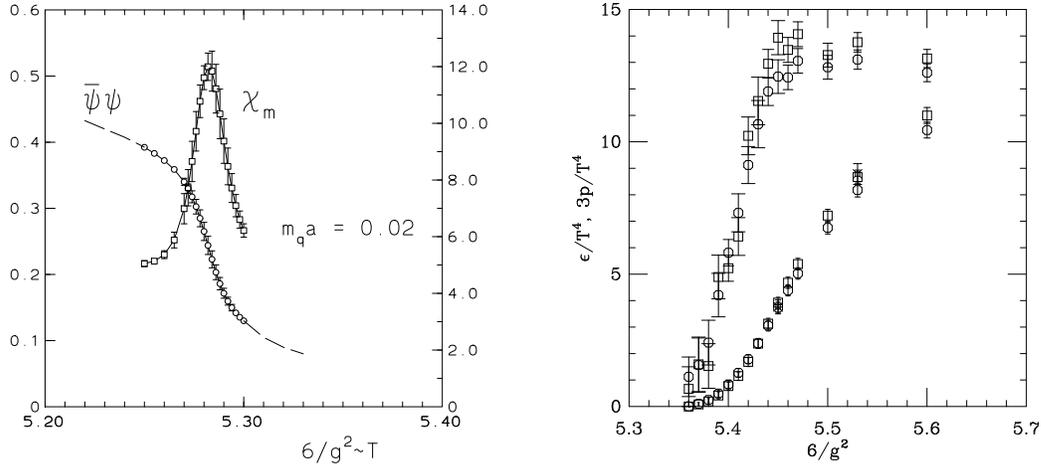,width=.5\textwidth}
   \hfill
   \epsfxsize=.46\textwidth \epsfbox[300 325 4396 4121]{eos0125syserr.eps}
  }
  \caption[Lattice calculation of the temperature dependence of the quark condensate and the energy density]
  {Lattice calculation including two quark flavors of the temperature dependence $T\sim6/g^2$ of the quark
   condensate $\langle\overline{\psi}\psi\rangle$ and the associated susceptibility $\chi_m\sim\delta\langle\overline{\psi}\psi\rangle/\delta
   m_q$~\cite{Karsch:1994hm}(left panel) and the energy density $\epsilon/T^4$ and the pressure
   $3p/T^4$~\cite{Bernard:1997cs} (right panel). In the limit of an ideal gas of quarks and gluons,
   the energy density should approach a value of
$\epsilon/T^4$\,=\,$40\pi^2/30=17.5$ according to the number
degrees
   of freedom in the plasma phase of 16 and 24 for gluons and quarks, respectively.}
   \label{fig:lattice}
\end{figure}

The results of a lattice calculation including two quark flavors
are depicted in Fig.~\ref{fig:lattice}.
The expectation value of the quark condensate plotted in the left
panel, as expected, shows the sudden drop at the critical
temperature indicating a phase transition restoring chiral
symmetry. This transition is accompanied by a jump in energy
density (see upper symbols in the right panel of
Fig.~\ref{fig:lattice}) from a low hadronic value to nearly the
level expected for an ideal gas of quarks and gluons
(QGP)~\cite{Blum:1995zf}. Up to now, lattice calculation do not
allow to conclude whether deconfinement is a phase transition of
first order, second order, or just a rapid crossover.

Lattice theory including non-vanishing baryon density has been
impeded by technical difficulties. Other methods applicable at
finite densities include random matrix
theory~\cite{Shuryak:1993pi,Verbaarschot:2000dy}, random phase
approximation~\cite{Bertrand:2001fk,Rabhi:2001ta}, instanton
models~\cite{Shuryak:1999fe,Rapp:1998zu,Schafer:1998wv},
percolation~\cite{Fortunato:2000ge,Fortunato:2000fa}, and
supersymmetric models~\cite{Randall:1998ra}. But their predictive
power has been rather limited. A general overview and evaluation
of the available theoretical models is presented
in~\cite{Rapp:1999ej}.

\begin{figure}[tb]
  \centering
  \epsfig{file=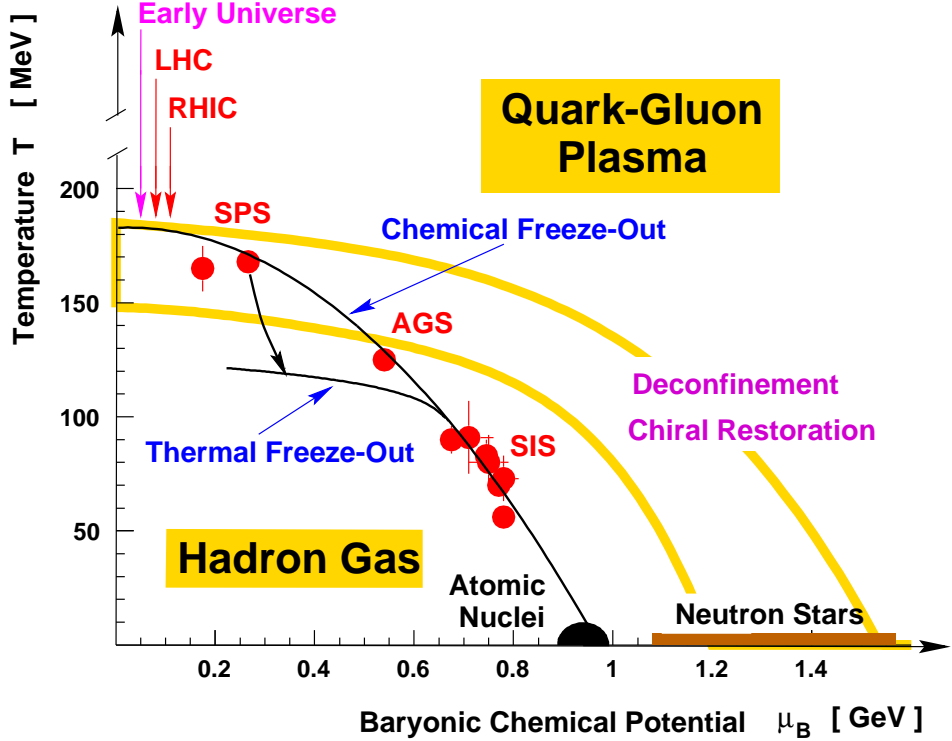}
  \caption[QCD phase diagram]
  {QCD phase diagram for the transition of hadronic to quark matter.
  An overview of the status of the experimental mapping of the QCD
  phase diagram can be found in~\cite{Stachel:1998rc,Braun-Munzinger:1999vr}.
  The theoretical aspects are summarized in~\cite{Rajagopal:2000wf}.}
  \label{fig:qcd phase}
\end{figure}

The emerging picture of the transition from hadronic to quark
matter is illustrated in a schematic phase diagram in
Fig.~\ref{fig:qcd phase}.

Despite the tremendous progress in recent years, dispute remains
about the exact circumstances, at high temperatures and/or high
densities, of restoration of broken symmetries in the medium under
extreme conditions. Whether deconfinement and chiral symmetry
restoration are two distinct phase transitions, or only one, is a
matter of current debate. Lattice calculation indicate the
critical temperatures of chiral restoration and deconfinement to
coincide in the low-density scenario. In fact,~\cite{Satz:2000bn}
argues deconfinement in the light-quark sector to be triggered by
the chiral transition.

Many
probes~\cite{Matsui:1986eu,Satz:1995sa,Lourenco:1996wn,Wang:1997qa}
have been proposed to map out the behavior of hot and dense
hadronic matter and also to highlight its eventual transition to a
quark-gluon plasma. Among those probes dileptons stand out for two
reasons.

First, they couple directly to vector mesons. Therefore, hadronic
processes are expected to reveal their properties in dilepton
spectra. In particular, the dilepton rate allows for direct
measurement of the imaginary part of the current-current
correlation function in the medium (see Eq.~\ref{equ:cc
correlator})~\cite{Feinberg:1976ua,McLerran:1985ay}:
\begin{equation}
 E_+E_-\frac{d^6N_{e^+e^-}}{d^3p_+d^3p_-} = \frac{2e^2}{(2\pi)^6}\frac{1}{k^4}[p^{\mu}_+p^{\nu}_-+p^{\nu}_+p^{\nu}_-
 g^{\mu\nu}p_+p_+]\times {\rm Im}(\Pi_{\mu\nu}(k))\frac{1}{e^{\omega/T}-1}
 \label{equ:thermal ee prod}
\end{equation}
which determines in part the vector meson
resonance~\cite{Shuryak:1993kg,Steele:1997tv}. Apart from the
kinematical constants $p$ describing the meson decay, the
influence of the medium enters in the current-current correlation
function $\Pi_{\mu\nu}(T,\mu_B)$ which can be calculated by
theoretical models discussed in Sec.~\ref{sec:theory}.

Second, dileptons suffer minimal final-state interaction because
they interact only electromagnetically and are therefore likely to
bring information about the innermost zones of high-density and
high-temperature matter, formed in the early stages of nuclear
collisions, to the detector essentially unscathed. On the other
hand, hadrons are rescattered and carry little information about
the time prior to the freeze-out stage of the collision.

\begin{figure}[tb]
    \centering
    \begin{minipage}[t]{.65\textwidth}
        \vspace{0pt}
        \epsfig{file=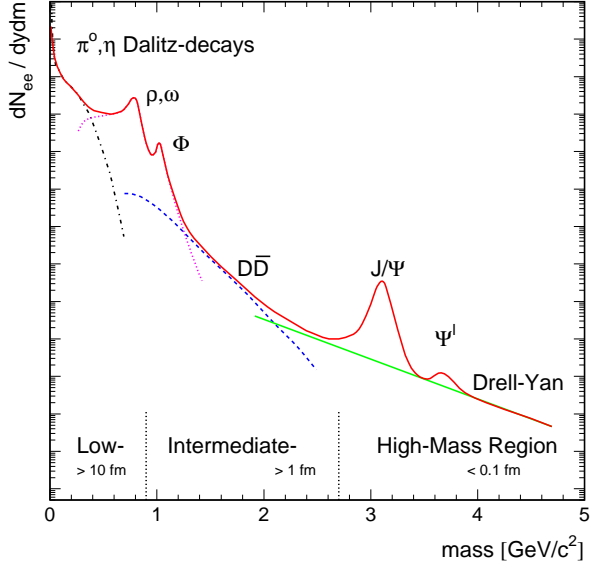,width=\textwidth}
    \end{minipage}%
    \begin{minipage}[t]{.35\textwidth}
      \vspace{.5cm}
      \caption[Schematic dilepton mass spectrum of ultrarelativistic heavy ion collisions]
      {\newline
      Schematic dilepton mass spectrum of ultrarelativistic heavy ion collisions.}
      \label{fig:ll sources}
    \end{minipage}
\end{figure}

The schematic dilepton mass spectrum in Fig.~\ref{fig:ll sources}
indicates the major dilepton sources in ultrarelativistic heavy
ion collisions.
In the high-mass region, dileptons stem from hard processes
(mostly Drell-Yan annihilation) occurring in the early
pre-equilibrium stage of the collision. Furthermore, a suppression
of the $J/\psi$ and $\psi'$ abundance has the potential to signal
the onset of deconfinement, as the heavy-quark bound states are
dissolved due to freely moving color charges (Debye screening). If
not masked by an enhanced open-charm production, a thermal signal
from plasma could be revealed by the observation of associated
$D\overline{D}$ production in the intermediate-mass
region~\cite{Shuryak:1978ij,McLerran:1985ay}. The low-mass region
is the exclusive domain of the CERES experiment. It is dominated
by soft processes involving the light quark sector. The dilepton
continuum originates from Dalitz decays of neutral mesons such as
$\pi^0,\eta,\eta'~\rightarrow~e^+e^-\gamma$ and $\omega
\rightarrow e^+e^-\pi^0$. The resonance peaks are due to direct
decays $\rho$,
 $\omega$, $\phi\rightarrow e^+e^-$. This region is particularly
sensitive to in-medium modifications of the light hadrons which
can signal the restoration of chiral symmetry. The $\rho$-meson is
of special importance because, once produced in a dense and hot
hadronic environment, it will decay predominantly within the
fireball due to its short lifetime. Compared to the other sources,
the relative contribution of thermal dielectron radiation from a
quark-gluon plasma is expected to be negligible at SPS
energies~\cite{Rapp:2000pe}.

\section{The CERES physics program}
 \label{sec:ceres prog}

The production of dileptons in hadronic collisions has been of
great experimental and theoretical interest for more than 30
years. The early dilepton measurements were motivated by the
search for the vector mesons in pp, pA, and $\pi^-$A collisions.
The unexpected observation of a continuous dilepton spectrum for
mass below 600\,MeV/c$^2$ motivated the development of several
theoretical models which are still relevant. Most notably was the
prediction of the quark-gluon plasma - a new phase of matter - in
1978~\cite{Shuryak:1978ij}. The thermal radiation of the plasma
comprises low-mass dileptons and direct photons.

\begin{figure}[hbt]
  \centering
  \mbox{
   \epsfig{file=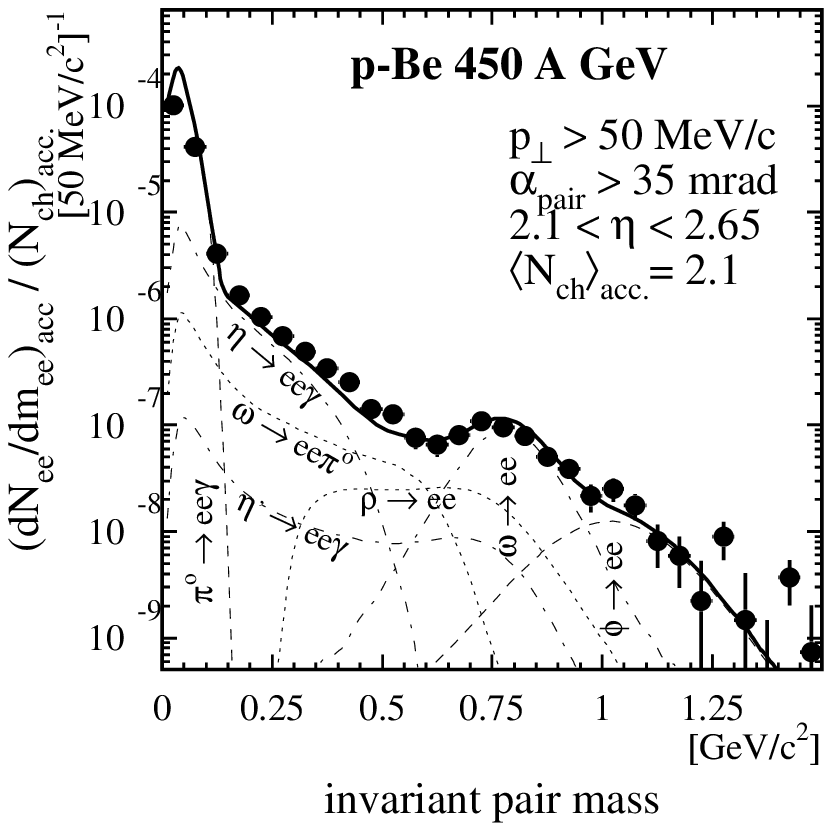,width=.49\textwidth}
   \hfill
   \epsfig{file=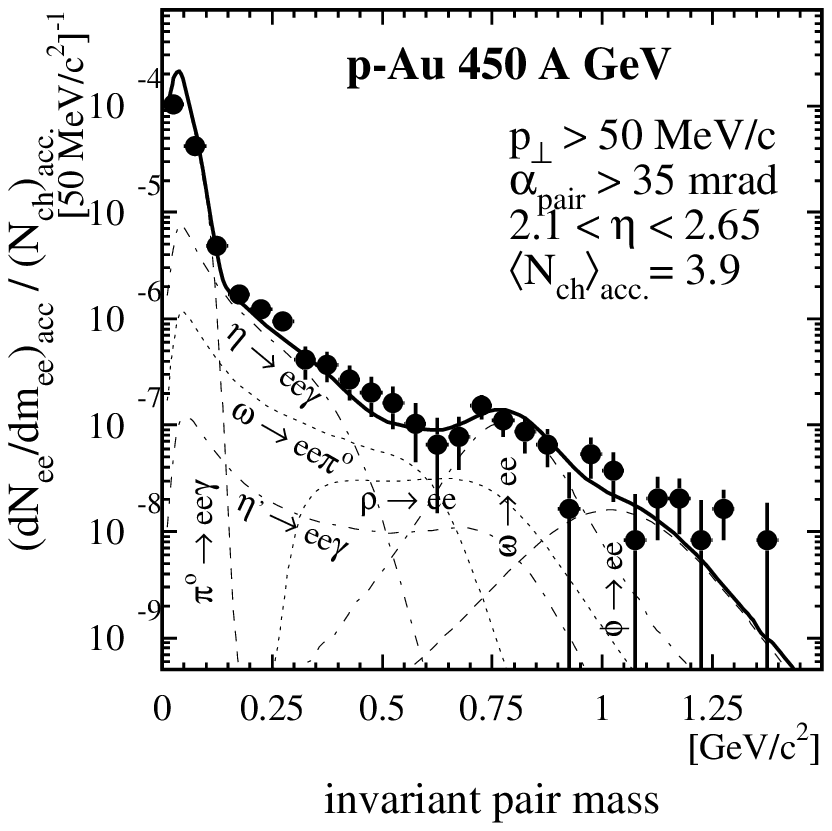,width=.49\textwidth}
  }
  \caption[Dielectron invariant mass spectrum of p-Be and p-Au collisions]
  {Dielectron invariant mass spectrum of p-Be and
  p-Au collisions at 450\,GeV/c~\cite{Agakishiev:1998mv}. The yield observed is saturated
  by the expected cocktail of hadronic sources.}
  \label{fig:pA data}
\end{figure}

CERN was the first laboratory worldwide to systematically
investigate dielectron production in ultrarelativistic
hadron-nucleus and nucleus-nucleus collisions. The Helios-1/NA34
collaboration was the first to measure $e^+e^-$ and $\mu^+\mu^-$
pair production in p-Be collisions~\cite{Goerlach:1992sh}. The
CERES/TAPS collaboration reproduced them with much greater
precision. Figure~\ref{fig:pA data} shows the measured dielectron
invariant mass spectrum of p-Be and p-Au collisions at 450~GeV/c
in comparison to the expected contributions of hadron decays.
The simulation of the so-called {\em hadronic cocktail} is based
on the knowledge of the branching ratios of all relevant leptonic
and semi-leptonic decays and the total production cross sections
of neutral mesons measured in pp collisions (see
App.~\ref{app:genesis}).

As evident from Fig.~\ref{fig:pA data}, the hadronic cocktail
accounts for the measured dielectron yield. Previous speculations
about an anomalous source of dileptons with mass below
600~MeV/c$^2$ were found to have originated from an
underestimation of $\eta$ Dalitz yield. This was proven by the
exclusive measurement of the $\eta \rightarrow e^+e^- \gamma$
decay~\cite{Agakishiev:1998mw}.

Most important, a reference based on pp and pA data was
established to be used in nucleus-nucleus collisions for
distinguishing between new in-medium effects and {\em trivial}
dielectron sources.

Recent measurements of the low-mass dilepton yield in 200\,GeV/c
p-U collisions by the NA38/50 collaboration~\cite{Abreu:1999av}
and in 12\,GeV/c p-C(Cu) collisions at KEK~\cite{Ozawa:2000iw}
could not be explained exclusively by the decay of the known
hadronic sources. The NA38/NA50 collaboration found a significant
excess in the mass window $0.4$--$0.6$\,GeV/c$^2$. In-medium
modification of the $\rho$-meson could not explain the observed
enhancement. NA38/50 conjectured that it may be due to
$q\overline{q}$ annihilations (Drell-Yan process). This production
mechanism is considered to be negligible in the CERES acceptance,
i.e.~for low transverse pair momentum $q_{\rm t}$\,$<$1\,GeV/c,
but may become important for lepton pairs with large transverse
momentum predominating in the NA50 acceptance. Therefore, this
measurement is not necessarily contradicting the CERES results. At
KEK, the mass spectra of p-C and p-Cu collisions were found to
differ significantly below the $\omega$-meson peak (i.e.~mass
window $0.4$--$0.6$\,GeV/c$^2$). This difference was interpreted
as an in-medium modification of the $\rho$-meson spectral shape at
normal nuclear density. Since no such effect was found by CERES,
further studies are necessary to settle the dispute about the role
of in-medium modifications of vector mesons in pA collisions.

\begin{figure}[hbt]
  \centering
  \mbox{
   \epsfig{file=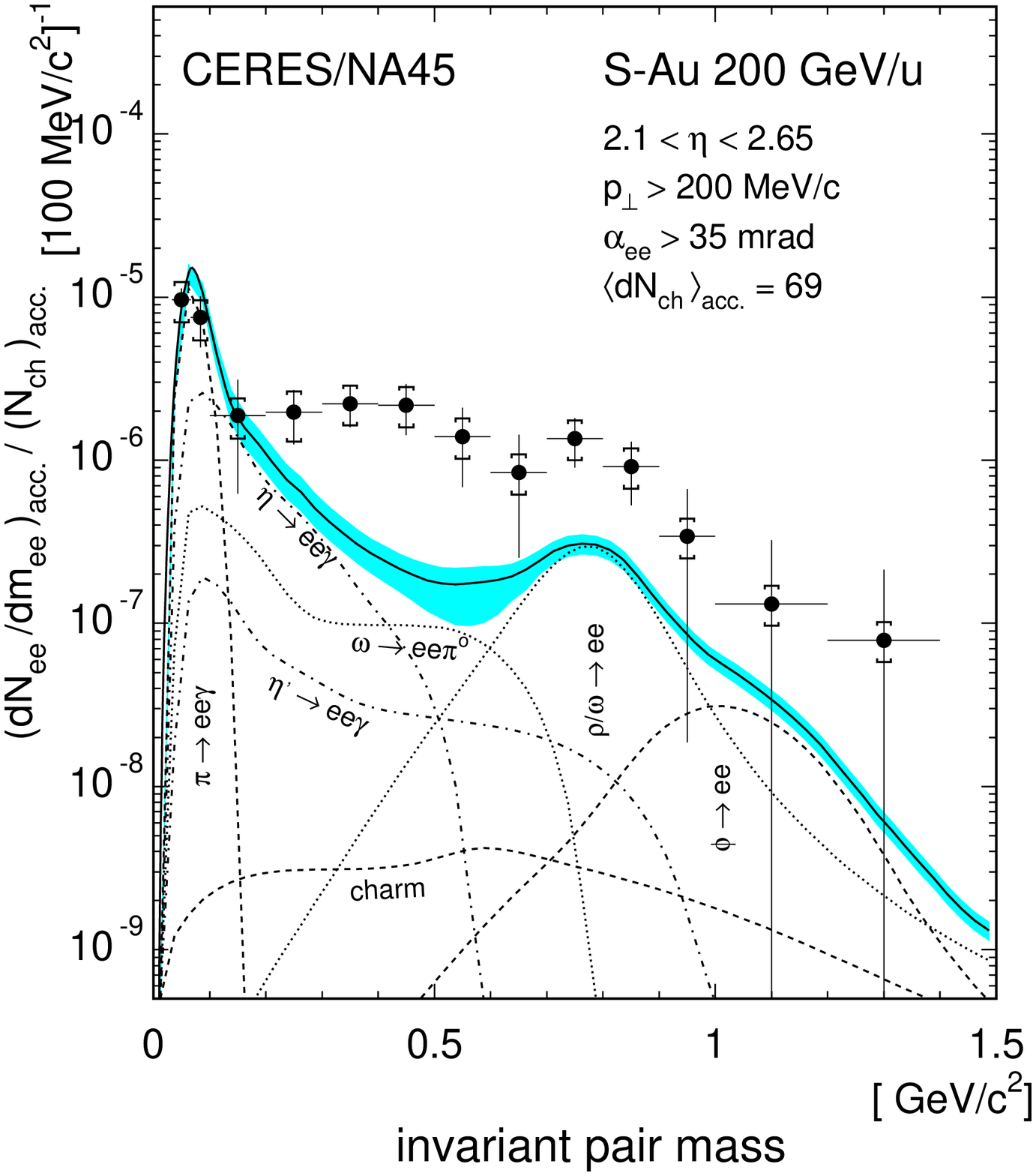,width=.465\textwidth,height=171pt}
   \hfill
   \epsfig{file=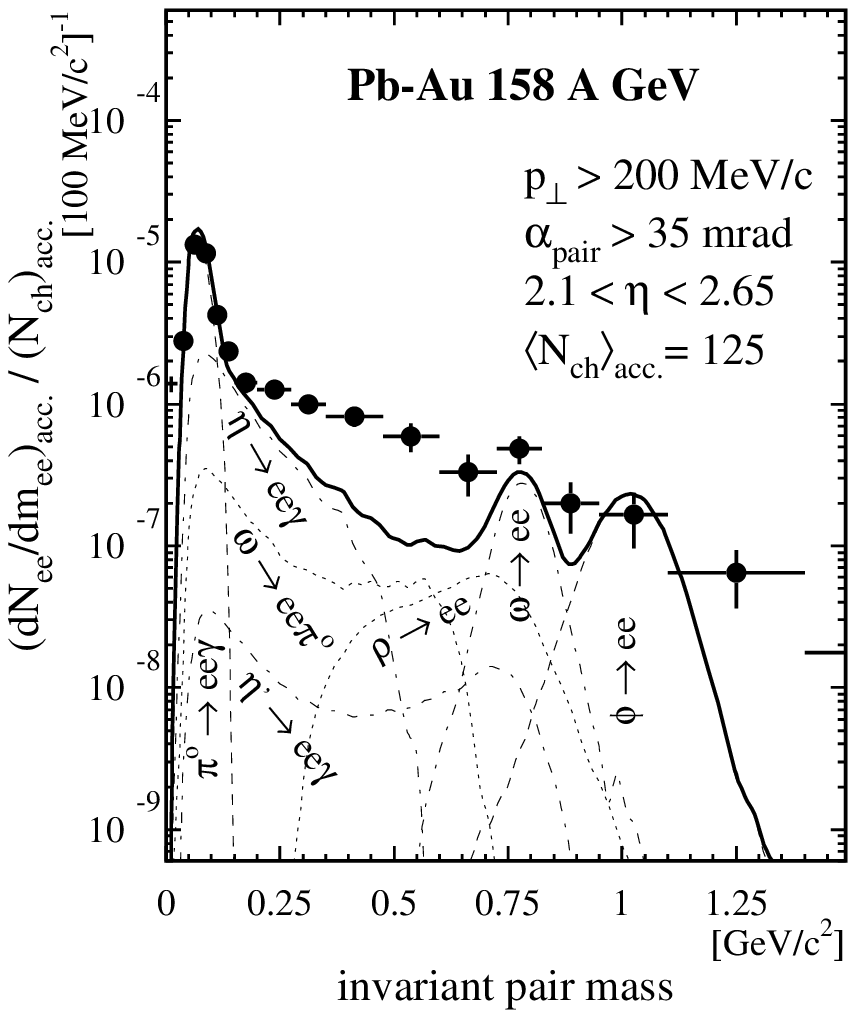,width=.49\textwidth}
  }
  \caption[Dielectron invariant mass spectrum of S-Au and Pb-Au collisions]
  {Dielectron invariant mass spectrum of S-Au and Pb-Au
  collisions~\cite{Agakishiev:1995xb,Agakishiev:1998vt}. The
  Pb-Au data plotted is a weighted average of the 1995 and the 1996 data
  sample~\cite{Agakishiev:2001xx}. The data is compared to the expected cocktail
  of hadronic sources. The Pb-Au cocktail plotted was corrected and extended
  compared to previous CERES publications (see
  App.~\ref{app:genesis}). The integrated yield for invariant mass above 200\,MeV/c$^2$ of
  $(1.13\pm0.16)\cdot 10^{-5}$ (S-Au) and $(5.4\pm0.9)\cdot 10^{-5}$ (Pb-Au) corresponds to
  an enhancement factor of $5.0\pm0.7$(stat.) and $2.7\pm0.4$(stat.),
  respectively.
  }
  \label{fig:AA data}
\end{figure}

The situation changes dramatically for nucleus-nucleus collision.
The dielectron yield observed in S-Au and Pb-Au collisions at
200~A\,GeV/c and 158~A\,GeV/c, respectively, significantly exceeds
the expectations extrapolated from p-p
collisions~\cite{Agakishiev:1995xb,Agakishiev:1998vt,Agakishiev:2001xx}.
While the $\pi^0$ Dalitz peak is well reproduced by the hadronic
cocktail, the local minimum expected between the $\eta$-Dalitz
component and the $\rho/\omega$-resonance peak at around
500\,MeV/c$^2$ is entirely filled up, as apparent from
Fig.~\ref{fig:AA data}.
The integrated yield of pairs with mass above 200\,MeV/c$^2$
exceeds the hadronic cocktail by a factor of $5.0\pm0.7$(stat.)
and $2.7\pm0.4$(stat.) for S-Au and Pb-Au collisions,
respectively.

This result was confirmed by the observation of an enhanced
$\mu^+\mu^-$ production in 200~A\,GeV/c S-W collisions compared to
200\,GeV/c p-W collisions by the HELIOS/3
collaboration~\cite{Angelis:1998pw}.

The comparison of nucleus-nucleus collision at different
bombarding energies, i.e.~different initial conditions, should
allow for independent interpretation of temperature and
baryon-density driven changes of the dielectron spectrum.

\begin{figure}[htb]
    \begin{minipage}[t]{.65\textwidth}
        \vspace{0pt}
        \epsfig{file=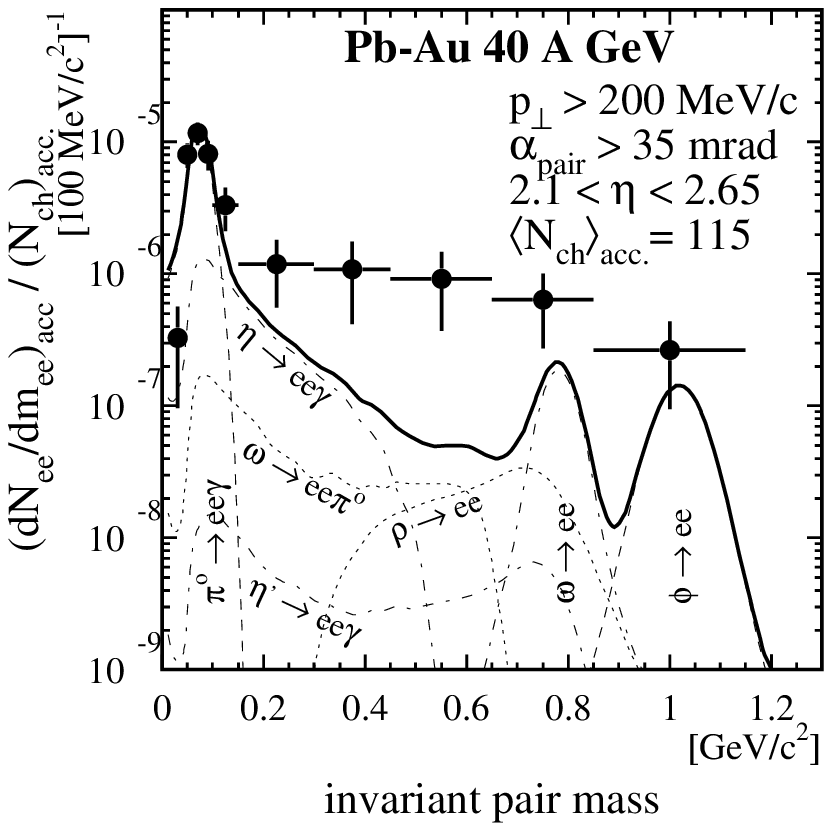,width=\textwidth}
    \end{minipage}%
    \begin{minipage}[t]{.35\textwidth}
      \vspace{0.5cm}
      \caption[Dielectron invariant mass spectrum of Pb-Au collisions at 40\,GeV/c per nucleon]
      {\newline
      Dielectron invariant mass spectrum of Pb-Au collisions at 40\,GeV/c per nucleon.
      The data in~\cite{Damjanovic:2002} were normalized to the
      expected low-mass yield simulated with GENESIS (see App.~\ref{app:genesis}).
      The integrated yield of pairs with invariant mass above 200\,MeV/c$^2$ of
      $(4.2\pm1.1)\cdot 10^{-6}$ corresponds to
      an enhancement factor of $4.5\pm1.2$(stat.).}
      \label{fig:40GeV data}
    \end{minipage}
\end{figure}

The most recent result for dileptons is the measurement of the
invariant mass spectrum of Pb-Au collisions at 40\,GeV/c per
nucleon (see Fig.~\ref{fig:40GeV data}). A recently resolved
problem in GENESIS (see App.~\ref{app:genesis}) resulted in a 30\%
increase of the predicted low-mass yield compared
to~\cite{Appelshauser:2002,Damjanovic:2002}. The data plotted in
Fig.~\ref{fig:40GeV data} were taken from~\cite{Damjanovic:2002}
and normalized to the expected yield of pairs with mass below
200\,MeV/c$^2$ according to the procedure described
in~\cite{Appelshauser:2002}.

 An enhancement of the dielectron yield, larger even than in the
158~A\,GeV/c data, is observed, relative to the expected yield of
hadronic sources. Since the detector system upgraded in 1998 was
not yet fully operational, the data set is limited in terms of
statistics and momentum resolution.

\begin{figure}[!b]
    \begin{minipage}[t]{.65\textwidth}
        \vspace{0pt}
        \centering
        \epsfig{file=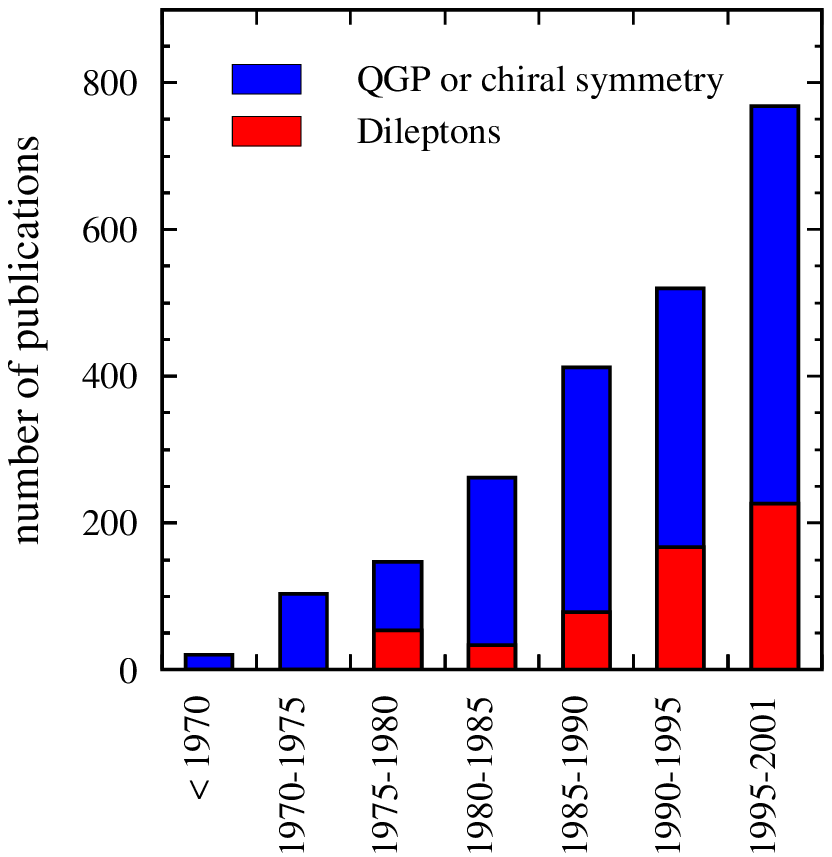,width=.96\textwidth}
    \end{minipage}%
    \begin{minipage}[t]{.35\textwidth}
      \vspace{0.5cm}
      \caption[Number of publications related to dileptons, chiral symmetry, and quark-gluon plasma]
      {\newline
      Number of publications related to dileptons, chiral symmetry, and quark-gluon plasma
      in the SLAC-SPIRES-HEP database~\cite{connell:2001}.}
      \label{fig:publix}
     \end{minipage}
\end{figure}

The experimental results on dilepton production in nucleus-nucleus
collisions have experienced many responses from theoretical
physicists. These were mainly stimulated by the prospects of
chiral symmetry restoration and deconfinement (see
fig.~\ref{fig:publix}).
Aside from the focus on dielectrons, the CERES collaboration has
also extensively studied charge particle
production~\cite{Ceretto:1998}, high momentum
pions~\cite{Ceretto:1998,Agakishiev:2000bt}, direct photon
production~\cite{Irmscher:1994rg,Messer:1998}, and azimuthal
correlations of charged particles
(i.e.~flow)~\cite{Agakishiev:1998wu}. The upgrade of the
spectrometer with the TPC allowed to greatly extend the scope of
CERES towards hadronic observables. Recent
results~\cite{Appelshauser:2002,Schmitz:2001} include the
measurement of Bose-Einstein correlations, mean transverse
momentum fluctuations, and hadronic production of $\Lambda$,
$\overline{\Lambda}$, and K$_0$.

\chapter{The CERES detector}

\section{Setup}

The CERES experiment was designed for the detection of dielectrons
with invariant mass up to 2.0\,GeV/c$^2$, covering a range in
pseudo-rapidity of
$2.1<\eta<2.65$~\cite{Specht:1992ht,Tserruya:1993sa}. The initial
setup consisted of two ring imaging Cherenkov detectors (RICH),
placed before and after an azimuthally deflecting magnetic field.
The RICH detectors provide particle identification and a
measurement of the trajectory. The azimuthal deflection in the
magnetic field determines the momentum. The measurements of p-Be,
p-Au, and S-Au
collisions~\cite{Agakishiev:1998mv,Agakishiev:1998mw,Agakishiev:1995xb}
were performed with this configuration.

\begin{figure}
  \epsfig{file=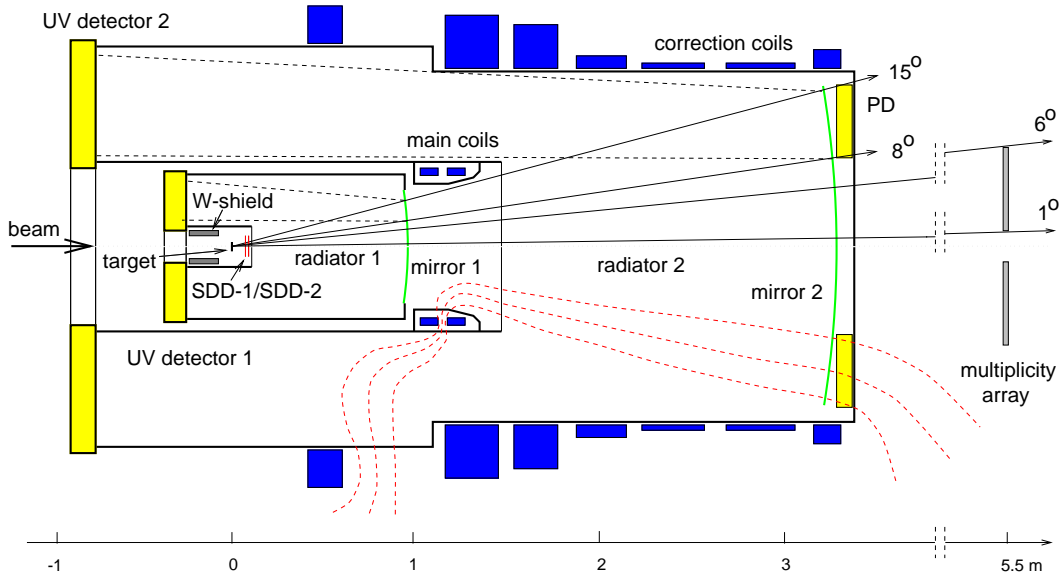,width=14cm}
  \caption[Schematic view of CERES experiment]
  {Schematic view of the CERES setup used for the measurement of Pb-Au collisions in 1995 and 1996.}
  \label{fig:setup 96}
\end{figure}

In 1994 and 1995, the spectrometer was substantially upgraded in
order to cope with the high multiplicity environment encountered
in ultrarelativistic Pb-Au collisions. The original setup was
extended by two silicon drift detectors (SDD) and a multiwire
proportional counter with pad readout (PD) as illustrated in
Fig.~\ref{fig:setup 96}.
The SDDs sample each track on two additional points. This allows
for a precision reconstruction of the event vertex, a measurement
of the energy deposition $dE/dx$, and a reliable determination of
the charged-particle multiplicity. The main purpose of the PD is
to help the ring recognition in the RICH detectors. The enhanced
track reconstruction and electron recognition capabilities were
demonstrated with the study of high-momentum pion and dielectron
production in Pb-Au collisions at 158~GeV/c per nucleon recorded
in 1995~\cite{Ceretto:1998,Voigt:1998}.

The 158~A\,GeV/c Pb-Au collision data analyzed in this work was
recorded in 1996 with the setup shown in Fig.~\ref{fig:setup 96}.
The following section gives a brief description of the individual
components of the experimental setup.

\section{Target region}

\begin{figure}[htb]
    \begin{minipage}[t]{.65\textwidth}
        \vspace{0pt}
        \epsfig{file=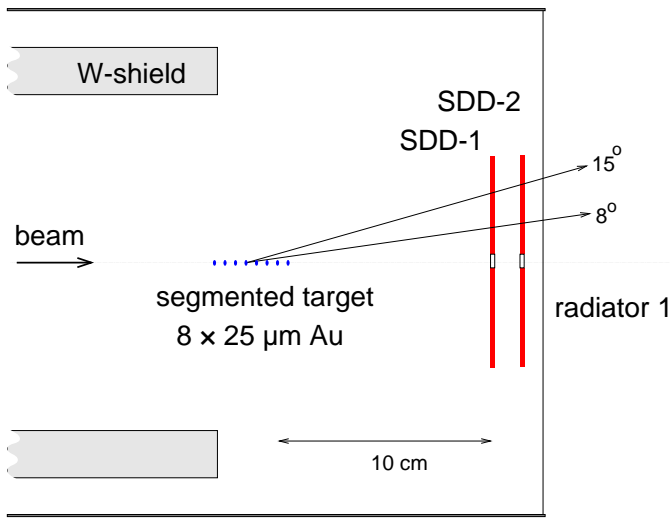,width=\textwidth}
    \end{minipage}%
    \begin{minipage}[t]{.35\textwidth}
      \vspace{0.5cm}
      \caption[Target area]
      {\newline
      Schematic view of the CERES target region.
      }
      \label{fig:target}
    \end{minipage}
\end{figure}

The setup of the target region is shown in Fig.~\ref{fig:target}.
The target consists of 8 gold foils of $600\,{\rm \mu m}$ diameter
and $25\,{\rm \mu m}$ thickness. For the particles to hit just one
of the consecutive targets disks, a space of 3\,mm between targets
is chosen, minimizing the probability of secondary interactions.
The target is surrounded by a tungsten shield to protect the
readout of the RICH UV-detectors from highly ionizing particles
scattered backwards.

\section{Silicon drift detector}
 \label{sec:sdd set up}

The first detector system of the CERES apparatus, located
$\sim12\,{\rm cm}$ downstream of the target, is a doublet of
silicon drift detectors (SDD) of $4''$
diameter~\cite{Rehak:1992hq}. The CERES experiment was first to
successfully employ this type of radially symmetric
position-sensitive detector~\cite{Chen:1993sh}. A detailed
description of these detectors is given
elsewhere~\cite{Weber:1997}.
\begin{figure}
  \centering
  \epsfig{file=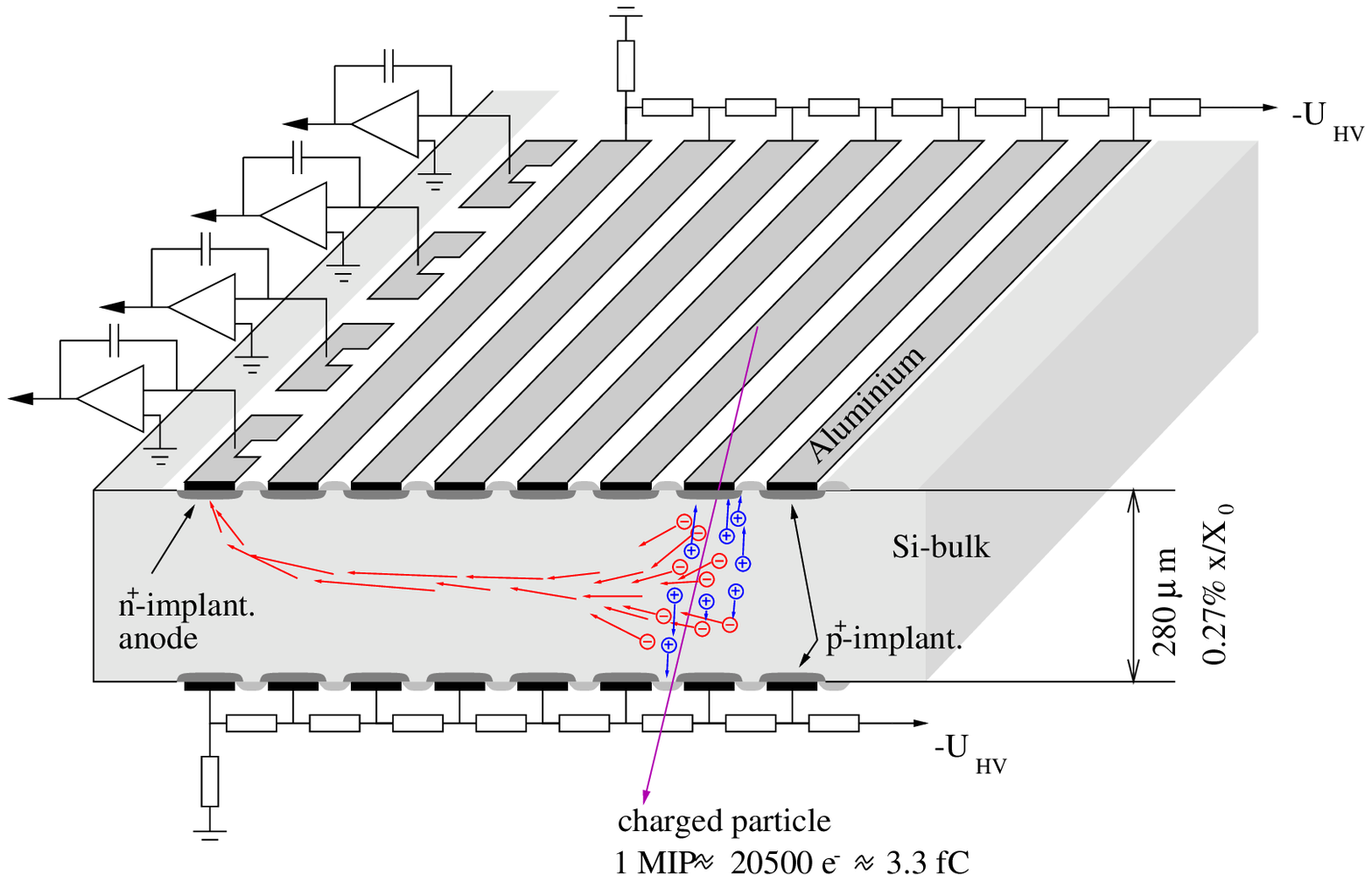,width=12cm}

  \vspace*{.6cm}
  \epsfig{file=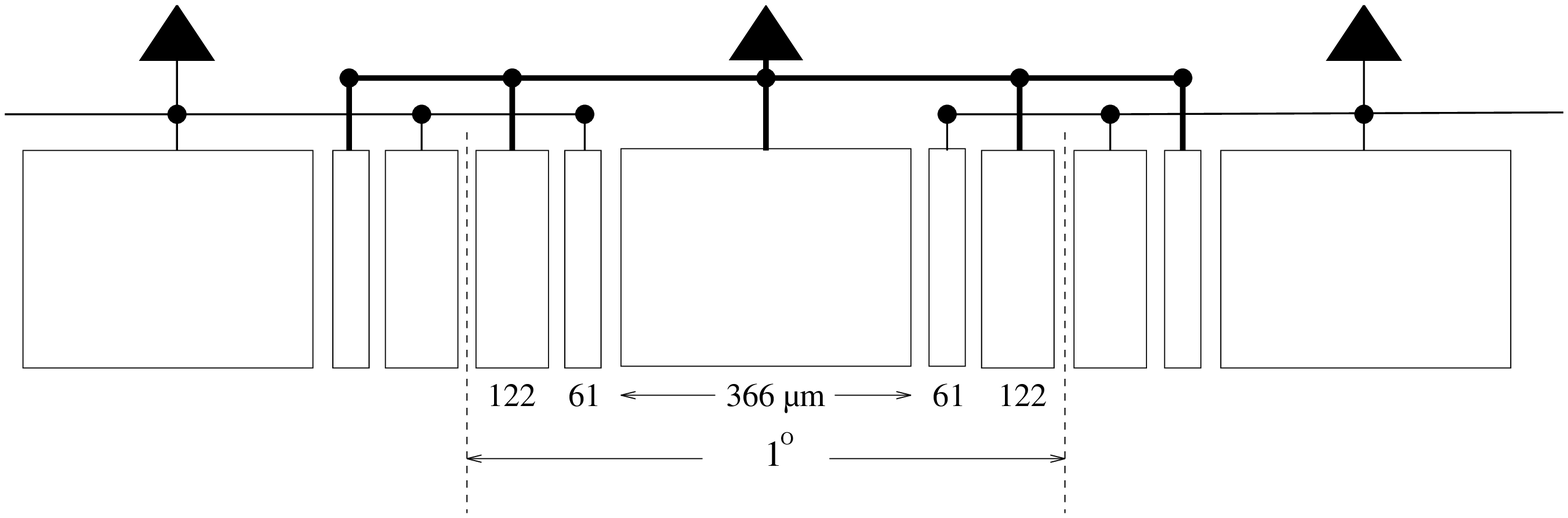,width=12cm}
  \caption[Schematic view of SDD]
 {Schematic view of a (radial) slice of the SDD (upper panel). The interlaced anode
 structure improves the single-hit resolution by charge sharing (lower panel).}
 \label{fig:schematic-sidc}
\end{figure}
The basic principle of operation is illustrated in the upper panel
of Fig.~\ref{fig:schematic-sidc}. A charged particle traversing
the detector produces a cloud of electron-hole pairs inside the
depleted region of the semiconductor. The particle energy required
to create an electron-hole pair is about 3.6\,eV~\cite{Ashcroft}.
The charge deposited by a minimum ionizing particle in a
280\,$\mu$m thick Si-detector is about 3.3\,fC (20500 electrons)
corresponding to an average energy loss of 74\,keV.

In radial electric field generated by a set of concentric
implanted voltage dividers, the electrons drift radially towards a
segmented anode at the outer circumference of the silicon wafer.
The segmented anode, shown in the lower panel of
Fig.~\ref{fig:schematic-sidc}, consists of 360 partially
interlaced pixels to provide a precise azimuthal position
measurement. The charge signal collected for each anode is
digitized by a fast FADC with a sampling frequency of 50\,MHz.
Given the drift velocity known, the radial position of a
charged-particle hit can be determined by a measurement of the
drift time with respect to the first-level trigger.

\section{Ring Imaging Cherenkov detector}

\begin{figure}[htb]
    \begin{minipage}[t]{.65\textwidth}
        \vspace{0pt}
        \centering
        \epsfig{file=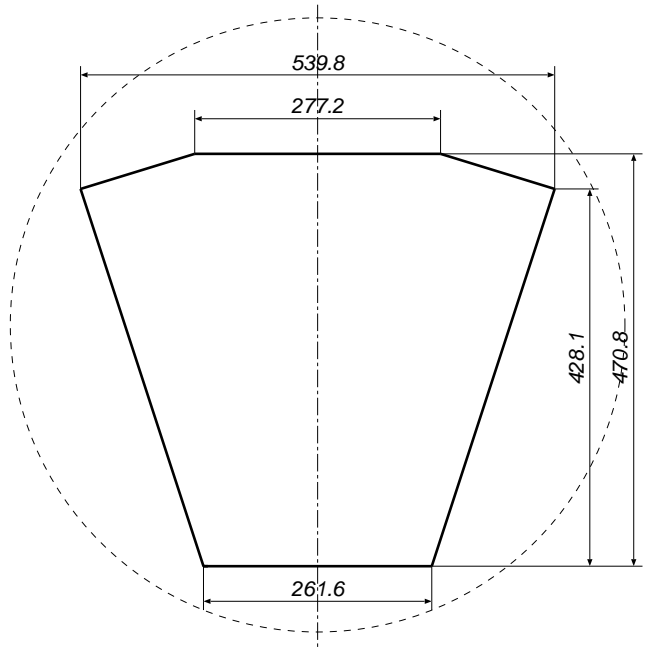,width=0.9\textwidth}
    \end{minipage}%
    \begin{minipage}[t]{.35\textwidth}
      \vspace{0.5cm}
      \caption[Geometry of RICH-2 mirror panel]
      {\newline
      Geometry of RICH-2 mirror panel.
      }
      \label{fig:panel}
    \end{minipage}
\end{figure}

The essential components of the 1996 CERES apparatus are two Ring
Imaging Cherenkov detectors (RICH). The first of these is situated
between the SDDs and a short super-conducting double solenoid, and
the second is behind the solenoid. An electron produced in a
collision emits Cherenkov photons while traversing the Methane
filled radiator volume. A spherical mirror reflects the Cherenkov
light to form a ring image at the mirrors focal plane. In case of
RICH-2, the mirror is split up in 8 smaller panels of equal size
for manufacturing reasons. The geometry of one such panel is shown
in Fig.~\ref{fig:panel}.
The imaging properties are described in detail in
Sec.~\ref{sec:rich intro}. The photons are detected in a position
sensitive UV-detector located at the image plane. This gas
detector is filled with helium, methane, and
Tetrakis-dimethylamino-ethylene vapor (TMAE)~\cite{Rewick:1988bu}.
The high quantum efficiency of TMAE in the UV region made it the
preferable detector gas at the time when CERES was designed. The
UV-detectors of RICH-1 and RICH-2 are separated from the radiator
volume by a CaF$_2$ and a quartz window, respectively.

Electrons emitted after photoabsorption in TMAE are amplified in
three steps: two parallel-plate avalanche stages and a subsequent
Multi-Wire Proportional Counter (MWPC). After an amplification by
a factor of $2$--$5\cdot 10^{5}$, signals are read out on 50000
individual pads covering the geometric acceptance. A complete
description is given
in~\cite{Tserruya:1993sa,Baur:1994mw,Baur:1995ux}.

\section{Magnetic field}
 \label{sec:B field}

To determine particle momentum, a magnetic field of 7\,T is
produced between RICH-1 and RICH-2 detector by a pair of
super-conducting coils with alternating currents. The currents in
the additional correction coils are adjusted for the field lines
in the RICH-2 radiator to point back to the target and the RICH-1
detector to become almost magnetic field free. This particular
field shape is illustrated in Fig.~\ref{fig:setup 96}.

\begin{figure}[htb]
    \begin{minipage}[t]{.65\textwidth}
        \vspace{0pt}
        \epsfig{file=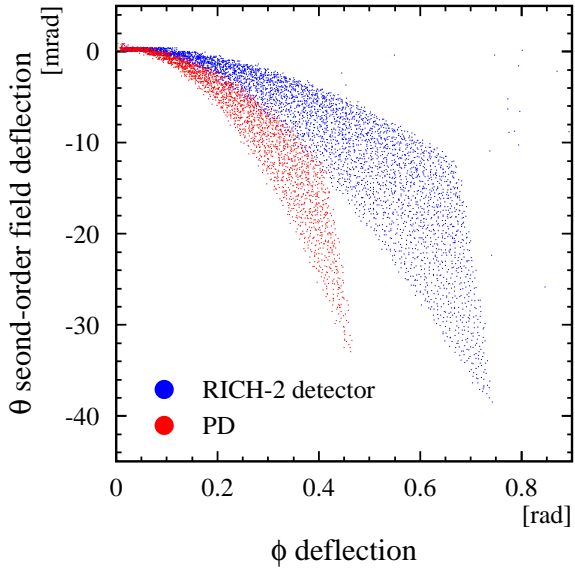,width=\textwidth}
    \end{minipage}%
    \begin{minipage}[t]{.35\textwidth}
      \vspace{0.5cm}
      \caption[$\theta$-deflection in RICH-2 and PD because of the second-order-field effect]
      {\newline
      Second-order-field effect modification of track polar angle as
      a function of the deflection in the magnetic field. Small
      deviations from the $1/r$-radial-field gradient result in a variation of
      the second-order-field deflection depending on the polar angle of the particle.
      }
      \label{fig:second order B}
    \end{minipage}
\end{figure}

Particles crossing the radially increasing magnetic field
($B$\,$\sim$\,$1/r$) between RICH-1 and RICH-2 are azimuthally
deflected by an angle $\Delta\phi$\,:
\begin{equation}
 \Delta\phi \approx \frac{144~{\rm mrad}}{p}\,{\rm GeV/c}\;.
 \label{equ:deflect}
\end{equation}
The polar angle of the particles is approximately conserved
because particle trajectories and magnetic field lines in RICH-2
do not cross in this direction. However, all particles traversing
the RICH-2 detector are slightly bend towards the beam axis
according to their initial azimuthal deflection. This so-called
{\em second-order-field effect} is illustrated in
Fig.~\ref{fig:second order B}.

While RICH-2 measures the local space direction of the trajectory
in the radiator, the Pad Chamber samples a point on the particle
trajectory with respect to the vertex. Therefore, the azimuthal
deflection observed in RICH-2 is about 1.5 times larger than in
the pad chamber.

To simulate the particle trajectories in the presence of the
B-field, the {\em Poisson} program package~\cite{Poisson:1993b}
was used to calculate a field map for the CERES geometry because
the magnetic field distribution has not been measured directly.
The accuracy of the map was estimated to be better than $\Delta
B/B$\,$\approx$\,0.5\%.

\section{Pad Chamber}

The Pad Chamber (PD) is located downstream of the RICH-2 radiator.
It consists of a MWPC with pad readout. Each of the 20000 pads has
a dimension of 7$\times$7\,mm$^2$, which results in an angular
resolution of better than 0.6\,mrad. The main purpose of the PD is
to limit the number of possible electron track candidates, given
by the combination of rings in the RICH-1 and RICH-2 detectors, by
measuring an additional point on the particle trajectory. This
background reduction is vital in the high charged-particle
multiplicity environment of Pb-Au collisions. Furthermore, the PD
provides an independent reference frame for the geometry
calibration of the other detectors~\cite{Ceretto:1998}. An
exhaustive description is given in~\cite{Sokol:1999}.

\section{Trigger system}
 \label{sec:trigger}

 The CERES trigger system consists of three Cherenkov
counters (BC1, BC2, BC3) and two plastic scintillator detectors:
veto counter (VC) and multiplicity detector (MD). BC1 is located
60\,m upstream of the experiment, BC2 and BC3 are directly before
and after the target, respectively. MD is at the downstream end of
the spectrometer.

The minimum-bias collision trigger requires a lead signal in BC1
and BC2, and no signal in BC3. The central trigger additionally
demands a signal in MD, with the threshold set at a level
corresponding to 100 charged particles. This is equivalent to the
most central 35\% of the geometrical cross section. To avoid beam
pile-up, the particle which triggers the reaction must not be
followed or preceded by any other lead particle going through BC1
for several microseconds.

\section{1998 detector upgrade}
\label{sec:upgrade}

\begin{figure}
  \centering
  \epsfig{file=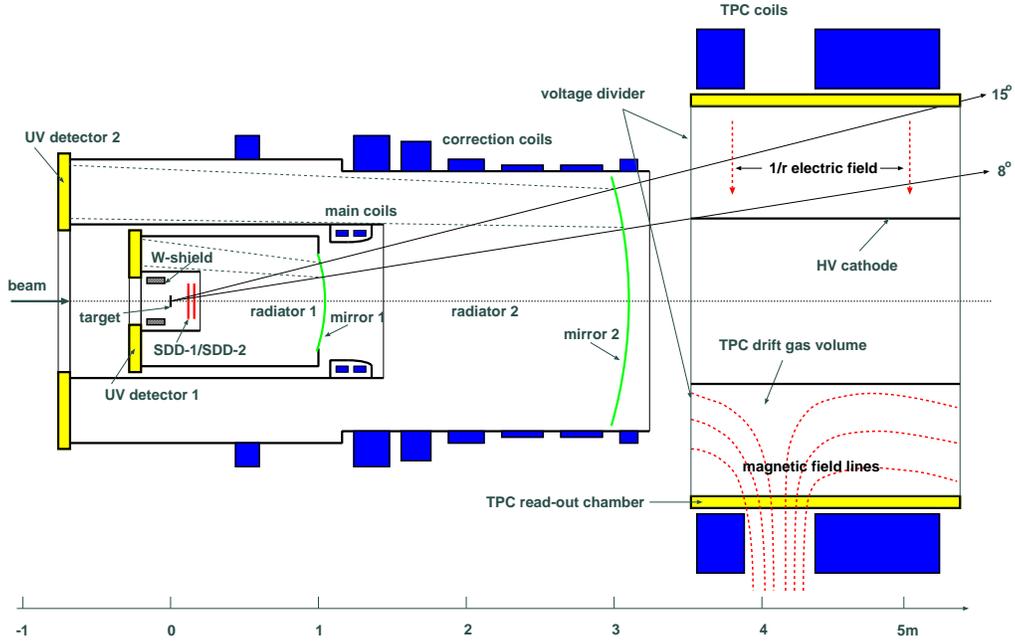,angle=270,width=13.4cm}
  \caption[Schematic view of the upgraded CERES setup]
  {Schematic view of the upgraded CERES setup used for the
  measurement of Pb-Au collisions in 1998,1999, and 2000.}
  \label{fig:setup 98}
\end{figure}

The CERES detector system was upgraded in 1998 by the addition of
a new magnet system and a radial Time Projection Chamber
(TPC)~\cite{Agakishiev:1999ka}. The PD and the multiplicity
detector were removed. This new setup is illustrated in
Fig.~\ref{fig:setup 98}.

There are three major benefits of the upgrade. First, the
high-precision momentum measurement of the TPC is expected to
improve the mass resolution to $dp/p=\sqrt{(0.0105)^2+(0.0103\cdot
p\cdot{\rm GeV}^{-1}{\rm c})^2}$~\cite{Schmitz:2001}. Second,
between the RICH detectors no magnetic field is applied, allowing
to operate both detectors in a combined way with doubled photon
statistics for the RICH rings and increasing
efficiency~\cite{Appelshauser:2002}. Third, considerable reduction
of the combinatorial background can be achieved due to higher
photon statistics in the RICH detector and the additional
d$E$/d$x$ measurement in the TPC.~Furthermore, the spectrometer
capabilities for the study of hadronic observables are
significantly extended.


\chapter{Development of an ultra\-light\-weight mirror for RICH detectors}

\section{Reasons for the replacement of the RICH-2 mirror}

After the upgrade of the CERES experiment with a TPC downstream of
the existing detector as described in Sec.~\ref{sec:upgrade}, a
new tracking scheme has been developed in order to improve
background rejection. According to this scheme, both RICH
detectors are operated without magnetic field and allow a combined
use for electron identification and tracking while the momentum is
measured separately in the TPC\@. All particles must traverse the
RICH-2 mirror before entering the TPC\@. Multiple scattering in
the mirror material changes the particle direction and, thus,
deteriorates the invariant mass resolution. Additionally,
electrons lose energy by Bremsstrahlung. The resulting low energy
tail impedes the spectroscopy of vector resonances. Therefore, the
replacement of the thick RICH-2 glass mirror by an
ultralightweight mirror almost transparent to dielectrons would
considerably improve the performance of the new detector system as
will be discussed in detail in the following sections.

\section{Impact of the RICH-2 mirror on the spectrometer performance}

\subsection{Interaction of electrons in matter}

\begin{figure}[tb]
  \centering
  \epsfig{file=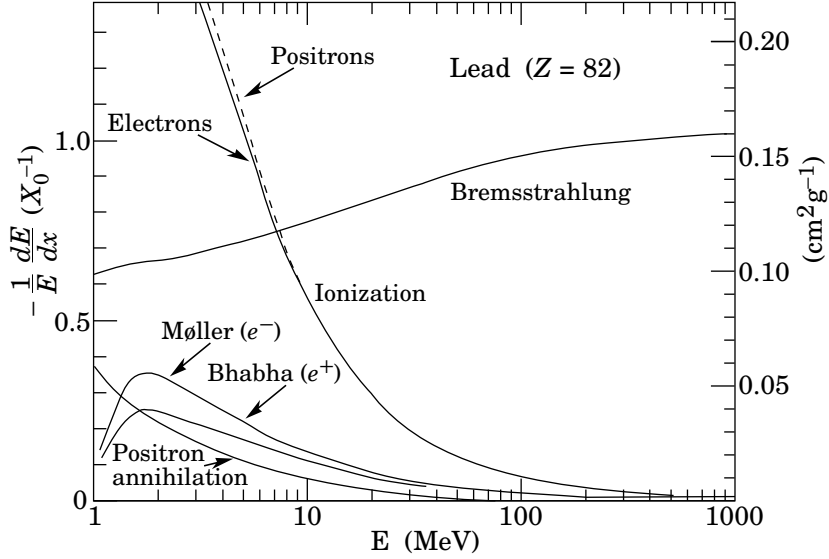}
  \caption[Fractional energy loss per radiation length]
  {Fractional energy loss per radiation length in lead as a
  function of electron/positron energy~\cite{PDBook}.}
  \label{fig:energy loss}
\end{figure}

High-energy electrons traversing the matter of a mirror are
affected in two ways. First, all electrons with momentum {\em p}
are deflected due to multiple Coulomb scattering from nuclei with
a probability depending on the thickness~{\em x} and the radiation
length~$X_0$ of the material. The Coulomb scattering distribution
is roughly Gaussian for small deflection angles, with a width
given by:
\begin{equation}
 \theta^{\rm rms}_{\rm plane}=\frac{13.6~{\rm MeV}}{\beta\,c\,p}
 \sqrt{\frac{x}{X_0}}\left(1+0.038\ln\left(\frac{x}{X_0}\right)\right)\;.
 \label{equ:scatter}
\end{equation}
Second, electrons lose energy by bremsstrahlung at a rate nearly
proportional to their energy. The cross section can be
approximated in the ''complete screening case'' as~\cite{PDBook}:
\begin{equation}
 \frac{d\sigma}{dk}=\frac{A}{X_0\,N_A\,k}\,\left(\frac{4}{3}-\frac{4}{3}y+y^2\right)\;,
 \label{equ:brems}
\end{equation}
with {\em y=k/E} being the fraction of the electron's energy
transferred to the radiated photon. The energy loss due to
scattering and ionization is negligible compared to bremsstrahlung
for electrons with $E
> 50$\,MeV/c$^2$ as illustrated in Fig.~\ref{fig:energy loss}~\cite{PDBook}.
The thickness of matter in terms of electron energy loss can be
conveniently measured in units of the radiation length $X_0$ which
is the mean distance over which a high-energy electron loses all
but $1/e$ of its initial energy by brems\-strahlung. The radiation
length can be approximated for a material with a charge number
{\em Z} and a mass number {\em A} as~\cite{PDBook}:
\begin{equation}
\label{equ:rad-length} X_0=\frac{716.4~{\rm g\,cm}^{-2}
A}{Z(Z+1)~\ln(287/\sqrt{Z})}\;.
\end{equation}
For compounds or mixtures each material contributes with a
fraction {\em w$_j$} proportional to its molecular weight {\em
A$_j$}:
\begin{equation}
 \label{equ:mixture}
 \frac{1}{{\rm X}_{\rm 0}}=\sum_{j}\,\frac{w_j}{{\rm X}_{\rm j}}
  = \sum_j\,\frac{n_j\,A_j}{A_{\rm comp}\,X_j}\;.
\end{equation}
The material thickness, which corresponds to $0.01 X_0$, is given
in Table~\ref{tab:thick} for several materials.
\begin{table}[h]
\begin{tabular}{|c|c|c|c|}
  \hline material & density & thickness  & comment \\
               & in g/cm${^2}$ &  ($x/X_0 = 0.01$) & \\
  \hline soda lime float glass & $2.5$ &{\em $1.4$~mm } & CERES RICH-2 mirror \\
  \hline carbon fiber (MAN)  & 1.7 & {\em $2.6$~mm}   &  CERES RICH-1 mirror\\
  \hline carbon fiber (HEXCEL)& $1.6$ &{\em $2.7$~mm} & COI prototype\\
  \hline Poly-Carbon ceramic& $1.54$ &{\em $2.8$~mm}  & HADES mirror~\cite{Muntz:1999td}\\
  \hline
\end{tabular}
 \caption{Equivalent thickness of mirror materials}
 \label{tab:thick}
\end{table}

\subsection{Imaging properties of the RICH detector}
 \label{sec:rich intro}

A brief review on the fundamentals of RICH detectors with special
emphasis on the CERES RICH detector will precede more extensive
discussion of the impact of the RICH-2 mirror on the performance
of the CERES spectrometer.

\begin{figure}[thb]
 \centering
 \epsfig{file=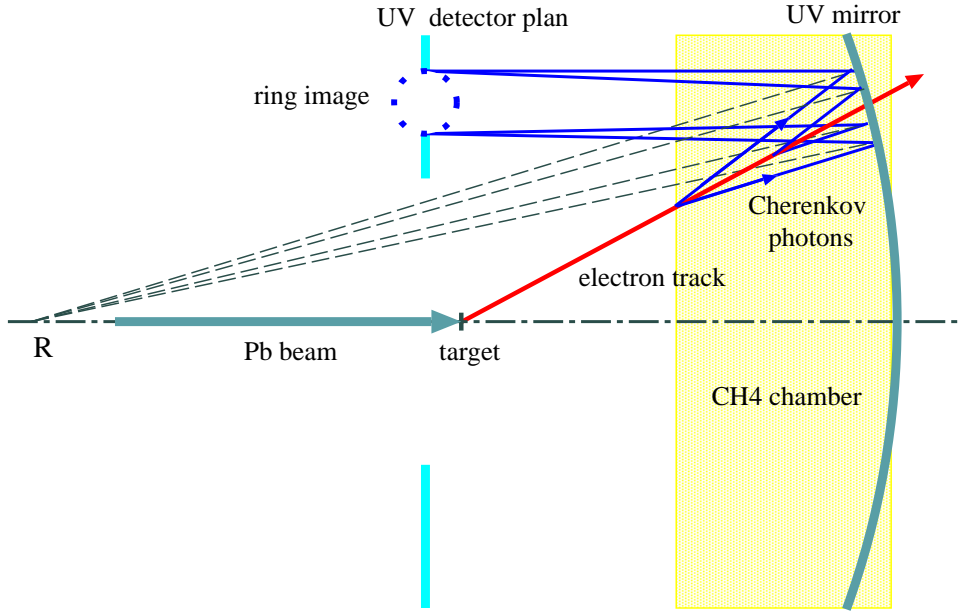}
 \caption[Schematic view of the CERES RICH detector]
 {Schematic view of the CERES RICH detector illustrating the origin of ring images.}
 \label{fig:rich}
\end{figure}

\begin{figure}[th]
    \begin{minipage}[t]{.65\textwidth}
        \vspace{0pt}
        \centering
        \epsfig{file=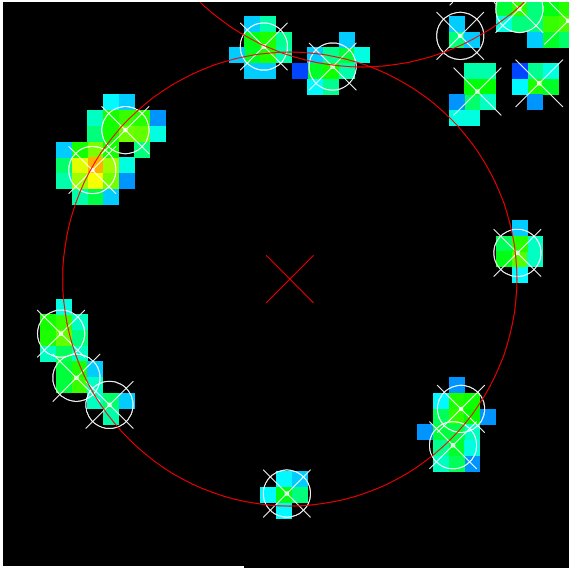,width=0.8\textwidth}
    \end{minipage}%
    \begin{minipage}[t]{.35\textwidth}
      \vspace{0.5cm}
      \caption[Ring reconstruction in RICH-1]
      {\newline
      Asymptotic electron ring reconstructed from 11
      photon hits in RICH-1~\cite{Agakishiev:1996ey}.
      }
      \label{fig:ring hits}
    \end{minipage}
\end{figure}

A Ring Imaging Cherenkov detector measures the photons radiated by
a charged particle traversing a transparent medium (radiator) with
a velocity higher than that of light in the medium. The emission
angle $\theta$ of the so-called Cherenkov photons is then
determined by the index of refraction {\em n} of the radiator
medium and the velocity $\beta$ of the charged particle:
\begin{equation}
 \label{equ:cher_angle}
 \cos{\theta_{\rm photon}}=\frac{1}{n\,\beta}\;.
\end{equation}
At atmospheric pressure gases have refraction indices close to
one. The threshold velocity is best expressed in terms of the
Lorentz factor $\gamma_{\rm th}$:
\begin{equation}
 \label{equ:gamma}
 \gamma_{\rm th}=\left(1-\frac{1}{n^2}\right)^{-\frac{1}{2}}\;.
\end{equation}
Methane~({\rm CH$_4$}) with $\gamma_{\rm{th}}=32$ was chosen for
the radiator gas of the CERES RICH~\cite{Baur:1994mw}. It makes
the detector almost blind to hadrons, except to pions with a
momentum of more than $4.5$\,GeV/c. The Cherenkov photons emitted
along the trajectory of the particle in the radiator and reflected
by a spherical mirror create a ring image at the focal plane. The
particular optical scheme is illustrated in Fig.~\ref{fig:rich}.
A photon detector, located at the focal plane, allows to determine
the ring position and radius, as well as the number of photons
(see Fig.~\ref{fig:ring hits}).

The ring-center position is a measure of $\theta$- and
$\phi$-coordinates of the original charged-particle track. The
ring radius is related to the photon emission angle and, thus, to
the velocity of the particle.  The Cherenkov angle approaches its
asymptotic value $\theta_{max}$ for electrons with a velocity
close to the speed of light ($\beta\!\approx\!1$):
\begin{equation}
 \sin{\theta_{\rm max}}=\frac{1}{\gamma_{\rm th}}\;.
 \label{equ:theta_max}
\end{equation}
Formula~\ref{equ:theta_max} applied to CERES geometry results in
an electron ring radius of R$_{\infty}\approx$\,$31.25$\,mrad,
slightly depending on radiator temperature and atmospheric
pressure.

The efficiency of the ring reconstruction depends strongly on the
number of detected photon per charged particle. The total number
of photons $N_{\rm ph}$ emitted per unit path length $x$ and unit
photon energy interval $E_{\rm ph}$ is related to the half-angle
$\theta_{\rm ph}$ of the Cherenkov cone:
\begin{equation}
 \frac{d^2N_{\rm ph}}{dE_{\rm ph}\,dx}=\frac{\alpha}{\hbar c}\,\sin^2{\theta_{\rm ph}}=\frac{\alpha}{\hbar
 c}\,\left(1-n_{\rm CH_{4}}(E_{\rm
 ph},T)^{-2}\right)\;\;(\beta_{e}\approx1)\;.
\label{equ:dn/de ceres}
\end{equation}
and, thus, to the index of refraction of the radiator gas $n_{\rm
CH_{4}}$ which is a function of photon energy.

The energy range of the detected photons is limited by the
photo-sensitivity of TAME ($E>5.4\,$eV)~\cite{Coyne:1993} and the
ultraviolet transparency of the quartz entrance window
($E<7.4\,$eV)~\cite{CERES:29,Braem:1994pf} which separates the UV
detector from the radiator. In this energy range, the index of
refraction is very close to one and nearly constant. The
temperature gradient in the radiator is small due the continuous
gas flow. Therefore, the expected total number of photons for a
radiator of length {\em L} is given by:
\begin{equation}
 \langle{N}\rangle \approx
 L\,N_{0}\,\langle{\sin^2{\theta_{\rm ph}}}\rangle\;.
\label{equ:n photon}
\end{equation}
The factor $N_0$, called the {\em figure of merit}, is defined by
the product of quantum efficiency of the UV-detector {\em Q},
total transmission of radiator gas and quartz entrance window {\em
T}, and mirror reflectivity {\em R}:
\begin{equation}
  \label{equ:n_0}
  N_0=\frac{\alpha}{\hbar\,c} \int{Q(E)\,T(E)\,R(E)\,dE}\;.
\end{equation}
The emission probability for {\em k} Cherenkov photons is
described by a Poisson distribution:
\begin{equation}
  \label{equ:poisson}
  P(N=k) =
  \frac{\langle{N}\rangle^k}{k!}\,\exp(-\langle{N}\rangle)\;\;(k=0,1,2,...)\;,
\end{equation}
where $\langle{N}\rangle$, the mean number of photons, is given by
Eq.~\ref{equ:n photon}. Successful detection of an electron
requires the reconstruction of a ring image composed of {\em k}
single photon hits. The reconstruction efficiency depends strongly
on the mean number of photon hits per ring excluding background
contributions~\cite{Glassel:1999rf}. Thus, it is limited by all
contributions in equation~\ref{equ:n_0}, in particular by the UV
reflectivity of the mirror coating. Additionally, the UV detector
spokes and small gaps between adjacent mirror segments lead to a
local reduction of the number of detected photons.

The right assignment of a RICH ring to the external track
information is of great importance for efficient particle
tracking. It is determined by the ring-center resolution
$\sigma_{{\rm Ring}}$ and, thus, by the mean number of hits per
ring $\langle N \rangle$ and the single-hit resolution
$\sigma_{{\rm hit}}$ according to:
\begin{equation}
 \label{equ:sigma_center}
 \sigma_{{\rm Ring}}=\sqrt{\frac{2}{\langle N \rangle -2}}\,\sigma_{{\rm
 hit}}\;.
\end{equation}
There are four major contributions to the single photon hit
resolution. First, multiple scattering of the charged particle
within the radiator contributes on average $\sigma_{\rm
mult}\approx 0.26$\,mrad. Second, the chromatic dispersion of the
radiator gas results effectively in a smearing of Cherenkov angle
according to:
\begin{equation}\label{equ:chrom}
\sigma_{{\rm disp}}=\frac{1}{2}\frac{\sigma_{n}}{n-1}\theta
\;,\;\;(\beta\approx 1;n\approx 1)\;\cite{Glassel:1999rf}\;,
\end{equation}
$\sigma_n$ being the rms width of the index of refraction averaged
over the bandwidth and weighted with the probability to detect a
photon {\em QTR} (see Eq.~\ref{equ:n photon}). With $\sigma_{\rm
disp}\approx 0.53$\,mrad, chromatic
dispersion~\cite{Glassel:1999rf} is by far the dominating
contribution to the single-hit resolution. Third, mirror shape
irregularities that occur on a scale of less then the radius of a
Cherenkov light cone distort the ring image and, thus, contribute
to the single-hit resolution. This is one of the main issues of
this chapter. Large scale mirror deformations shifting the entire
ring can be corrected to first order by local adjustment of the
focal length, provided the deformations are continuous for
adjacent mirror segments. Finally, the granularity of the
UV-detector as determined by a pad size of~($2.7$\,mm)$^2$ for
RICH-1 and ($7.6$\,mm)$^2$ for RICH-2 (equivalent to about
$2$\,mrad per pad in both cases) results in an expected single-hit
resolution of $\sigma_{\rm pad}=1.8(1.4)$\,mrad for the
RICH-1(RICH-2) detector.

To protect the UV-detectors from particles produced in the
collision, the target is placed at $0.8\cdot{\rm f}_{\rm mirror}$
which leads to a small deviation from the ideally flat focal
plane. The contribution thereof to the single-hit resolution is
negligible.

All contributions are independent and, hence, add in quadrature to
the single photon resolution:
\begin{equation}
 \sigma_{hit}=\sqrt{\sigma_{mult}^2+\sigma_{disp}^2+\sigma_{mirror}^2+\sigma_{pad}^2}\;.
\end{equation}


\subsection{Simulation of energy loss in the RICH-2 mirror}
\label{sec:energy loss}

The effect of bremsstrahlung was studied using a GEANT detector
simulation~\cite{GEANT:1993a} including contributions of multiple
photon radiation in the inhomogeneous material distribution of the
CERES setup according to Eq.~\ref{equ:brems}.

\begin{figure}[htb]
    \begin{minipage}[t]{.64\textwidth}
        \vspace{0pt}
        \epsfig{file=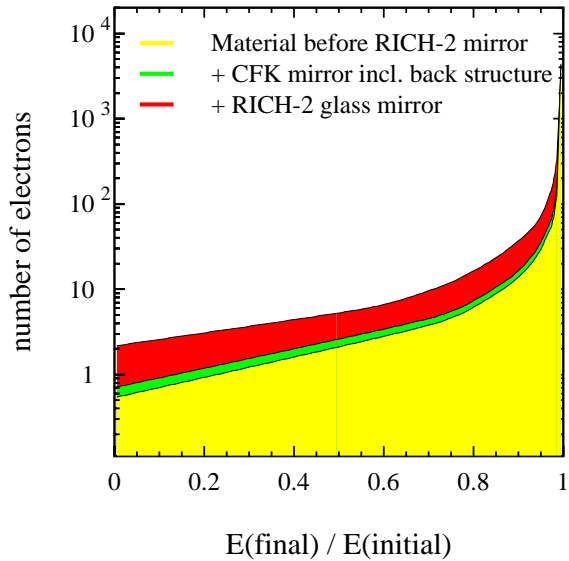,width=\textwidth}
    \end{minipage}%
    \begin{minipage}[t]{.36\textwidth}
      \vspace{0.5cm}
      \caption[Relative energy loss of electrons because of bremsstrahlung]
      {\newline
      Relative energy loss of electrons because of bremsstrahlung.
      }
      \label{fig:brems}
    \end{minipage}
\end{figure}

Figure~\ref{fig:brems} shows the relative energy loss of electrons
due to bremsstrahlung in the CERES setup upstream of RICH-2
($\approx 1.3\%$~of a radiation length), which is dominated by the
target and the RICH-1 mirror, and the additional contribution of
an ultrathin RICH-2 mirror ($0.5\%$~of a radiation length) in
comparison to the present thick mirror ($4.5\%$~of a radiation
length).

\begin{figure}[tb]
    \begin{minipage}[t]{.65\textwidth}
        \vspace{0pt}
        \epsfig{file=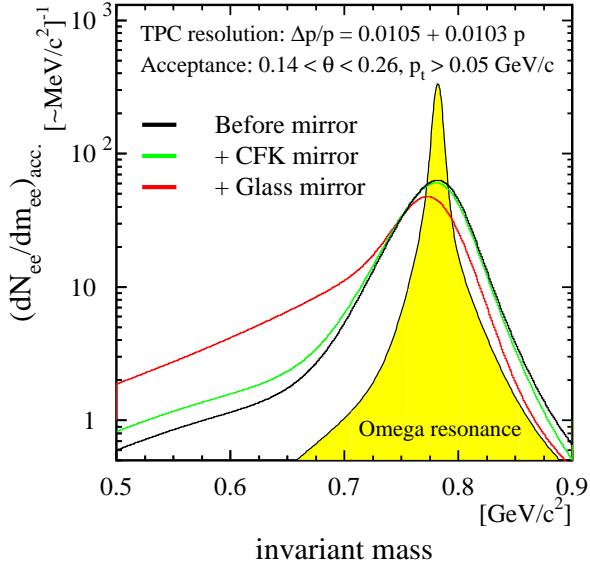,width=\textwidth}
    \end{minipage}%
    \begin{minipage}[t]{.35\textwidth}
      \vspace{0.5cm}
      \caption[Broadening of the $\omega$ resonance because of the energy loss in the RICH-2 mirror]
      {\newline
      Impact of the RICH-2 mirror on the spectrum of the $\omega$
      resonance. A Breit-Wigner function was assumed for the spectral shape of
      the $\omega$ resonance.
      }
      \label{fig:omega}
    \end{minipage}
\end{figure}

Considering the logarithmic scale in Fig.~\ref{fig:brems}, it
becomes obvious that the RICH-2 glass mirror is presently the
dominating source of bremsstrahlung leading to a significant low
energy tail for every resonance in the dielectron pair mass
spectrum. To verify this for the mass region of interest, the
dielectron decay distributions of $\phi$- and $\omega$-mesons were
folded with the bremsstrahlung spectrum. Figure~\ref{fig:omega}
shows the result for the case of the $\omega$~resonance.

The number of dielectrons in the peak drops significantly due to
the brems\-strahlung tail. Discriminating the resonance peak from
the background of other sources becomes increasingly difficult. An
ultrathin mirror, in comparison, would significantly reduce the
dielectron loss almost to the minimum level determined by the
upstream material. Although not shown here, the situation of the
$\phi$-meson is qualitatively comparable.

\subsection{Quantitative estimate of the influence of the RICH-2 mirror on the invariant mass
spectrum}

A dielectron invariant mass spectrum was produced for quantitative
study using the Monte Carlo detector simulation. The input
distributions of various dielectron sources were obtained from the
GENESIS $e^-e^+$ event generator (see App.~\ref{app:genesis}). The
appropriate momentum resolution and the CERES acceptance cuts were
applied to the simulated mass spectrum.

\begin{figure}[tb]
    \begin{minipage}[t]{.65\textwidth}
        \vspace{0pt}
        \centering
        \epsfig{file=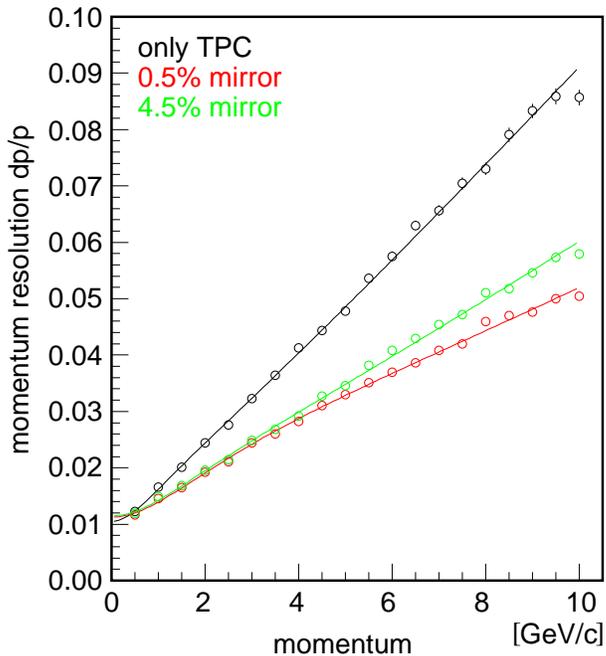,width=0.9\textwidth}
    \end{minipage}%
    \begin{minipage}[t]{.35\textwidth}
      \vspace{0.5cm}
      \caption[Simulation of the CERES momentum resolution]
      {\newline
      Momentum resolution of the CERES detector as simulated with the GEANT package~\cite{Esumi:50}.
      }
      \label{fig:p-resolution}
    \end{minipage}
\end{figure}

\begin{figure}
  \epsfig{file=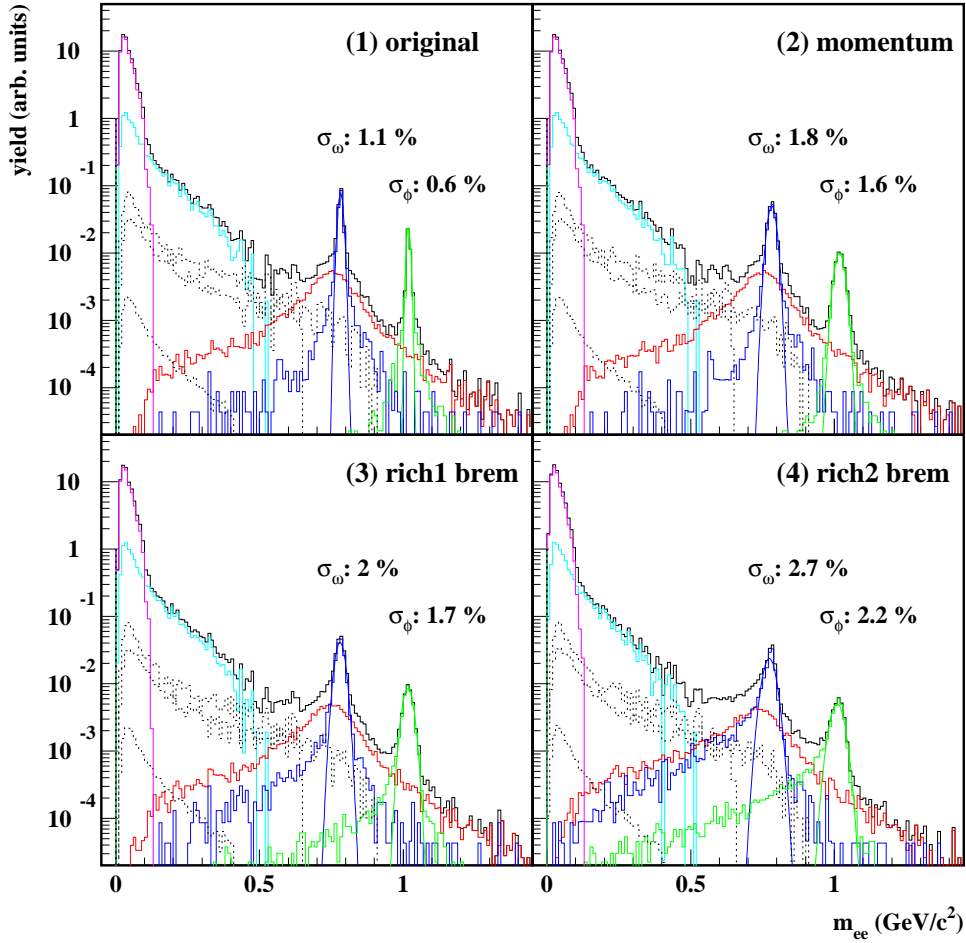,width=14.0cm}
  \caption[Impact of RICH-2 mirror on the dielectron mass spectrum]
  {Impact of RICH-2 mirror on the dielectron mass spectrum.
  \newline (1) cocktail of expected hadronic
  sources~\cite{Sako:2000}
  \newline (2) including the limited momentum resolution
  \newline (3) including the modifications by bremsstrahlung in RICH-1 mirror
  \newline (4) including the modifications by bremsstrahlung in RICH-2 mirror
  }
  \label{fig:pair_spectrum}
\end{figure}

To perform meson spectroscopy, the dielectron signal $S_{\rm
unlike}$ has to be obtained by subtraction of the uncorrelated
background pairs $B_{\rm unlike}$ according to Eq.~\ref{equ:pair
signal} in Sec.~\ref{sec:mix-intro}. Next, the resonance signal
S$_{\rm meson}$ can be extracted from the background of other
dielectron sources $B_{\rm meson}$:
\begin{equation}
 \label{equ:meson_signal}
 S_{\rm meson}=S_{\rm unlike}-B_{\rm meson}\;.
\end{equation}
For an expected signal-to-background ratio of $S_{\rm
unlike}/B_{\rm unlike}$\,$\approx$\,1/10 in the mass region of
$\omega$- and $\phi$-resonance, the relative error of the
signal~$\Delta$S$_{\rm meson}$/S$_{\rm meson}$ is dominated by the
statistical error of the uncorrelated background of dielectron
pairs $B_{\rm unlike}$:
\begin{eqnarray}
 \label{equ:meson_error}
 \frac{\Delta {\rm S}_{\rm meson}}{{\rm S}_{\rm meson}} & = & \frac{\Delta {\rm S}_{\rm unlike}+\Delta {\rm B}_{\rm meson}}{{\rm S}_{\rm
 meson}}\\
& = & \frac{\sqrt{{\rm S}_{\rm unlike}+2{\rm B}_{\rm unlike}} +
\sqrt{{\rm B}_{\rm meson}+2{\rm B}_{\rm unlike}}}{{\rm S}_{\rm
meson}}\approx\frac{\sqrt{8{\rm B}_{\rm unlike}}}{{\rm S}_{\rm
meson}}\;.\nonumber
\end{eqnarray}
In the next step, the meson signal is determined for each mirror
version integrating the pair yield within the 3$\sigma$-width of
the original resonance peak (see Fig.~\ref{fig:omega}). The
results are summarized in Fig.~\ref{fig:pair_spectrum} and
Table~\ref{tab:peak}.
\begin{table}[b]
\begin{tabular}{|c|c|c|c|c|}
\hline Option & Total number & $\%$ in  &Total number & $\%$ in \\
       & in $\omega$ peak & $3\sigma$~width & in $\phi$ peak & $3\sigma$ width \\
\hline\hline $1.3\%$ RICH-1 & 2300 & 89.8 & 500 & 89.2 \\ \hline
$+0.5\%$ RICH-2 (thin) & 2300 & 83.0 & 500 & 81.9 \\ \hline
$+4.5\%$ RICH-2 (thick) & 2300 & 64.6 & 500 & 62.9 \\ \hline
\end{tabular}
 \caption[Impact of bremsstrahlung on the number of dielectrons in the
$\omega$- and the $\phi$-resonance peak] {Impact of bremsstrahlung
on the number of dielectrons in the $\omega$- and the
$\phi$-resonance peak.}
 \label{tab:peak}
\end{table}
It turns out that the number of dielectrons in the peak would
increase by as much as $30\%$ if the present thick mirror was
replaced by an ultrathin mirror. The impact of the 30\% difference
is most profound in a low-statistics and high-background scenario
because large width of mass bins  and large statistical errors of
the background subtraction (see Eq.~\ref{equ:meson_error})
strongly reduce the statistical significance of the meson signal.
The reduction of the multiple scattering in the thin mirror
results in an improvement of the total momentum resolution by
approximately $0.5\%$ at $p=6$\,GeV/c and $1\%$ at $p=10$\,GeV/c
as seen in Fig.~\ref{fig:p-resolution}.

It should be noted that in case of the thick mirror, shown in
panel (4) of Fig.~\ref{fig:pair_spectrum}, the region between
$\omega$- and $\phi$-meson peak, which is particularly sensitive
to possible shape changes of the $\rho$-meson, would be masked by
the bremsstrahlung tail of the $\phi$-meson resonance.

To summarize, spectroscopy of $\omega$- and $\phi$-mesons would
greatly benefit from replacement of the present RICH-2 mirror by
an ultrathin mirror. An increase in signal-to-background ratio of
the meson resonances and better access to the spectral shape of
the $\rho$-meson peak are the two main prospects.


\section{Ultralightweight RICH-2 mirror}

In the past, manufacturing of ultralightweight mirrors has proven
a technologically very challenging endeavor for two
reasons~\cite{Kasl:2000}. First, reduction of the mirror thickness
results in a quadratic decrease in bending stiffness making it
increasingly difficult to maintain sufficient optical imaging
quality. Second, achieving a high reflectivity for photon energies
in the UV range requires not only a surface micro-roughness below
$3$\,nm~\cite{Braem:1993pd} but also very sophisticated coating
technology~\cite{Hass:1982,CERES:41,Zeitelhack:1999ry,Maier-Komor:1999ug}.

\subsection{Mirror distortions}

\begin{figure}[!tb]
    \begin{minipage}[t]{.65\textwidth}
        \vspace{0pt}
        \centering
        \epsfig{file=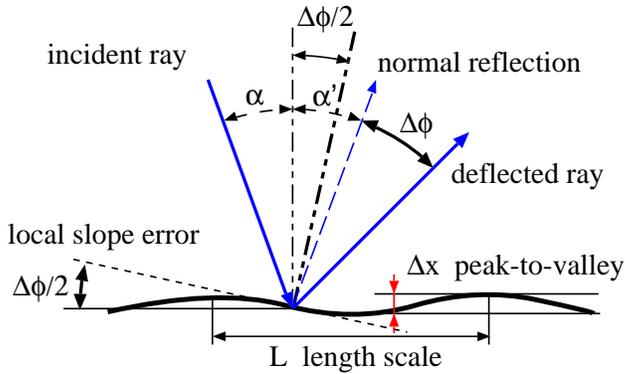,width=0.9\textwidth}
    \end{minipage}%
    \begin{minipage}[t]{.35\textwidth}
      \vspace{0.5cm}
      \caption[Definition of the slope error]
      {\newline
      Definition of the slope error $\Delta\Phi$.
      }
      \label{fig:slope_error_plot}
    \end{minipage}
\end{figure}

Deviations of the mirror surface from an ideal spherical shape
will result in a blurring of the point image produced by parallel
radiated Cherenkov rays. The deformation of the mirror can be
measured in terms of the slope error. It can be specified by a
twofold of the angle between the actual slope of the mirror
surface and the nominal value which corresponds statistically to a
rms-width of distribution of the reflected light (see
Fig.~\ref{fig:slope_error_plot}).

In a Monte Carlo simulation including real background events of
the 1995 data set, the ring reconstruction efficiency (defined as
the probability to reconstruct a ring for a given Monte Carlo
track) and the ring-center resolution in RICH-2 were studied as a
function of the slope error of the mirror. Local distortions
smaller than the mirror area illuminated by a Cherenkov light cone
were assumed to be randomly distributed. This is roughly
equivalent to a random deviation of the reflected light from its
nominal direction with a Gaussian probability distribution.
Therefore, the hit position of each Cherenkov photon in the
UV-detector is smeared by the convolutions of the probability
distributions describing slope error, chromatic dispersion,
electron drift, and finite pad resolution.
\begin{figure}[!tb]
   \centering
   \mbox{
   \epsfig{file=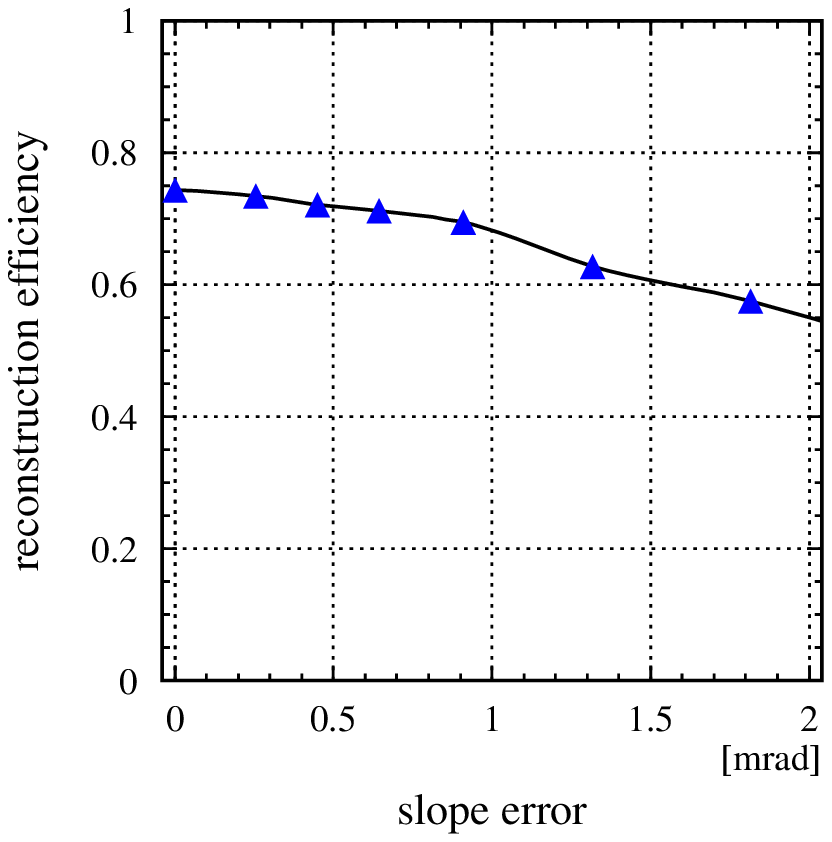,width=.49\textwidth}
   \hfill
   \epsfig{file=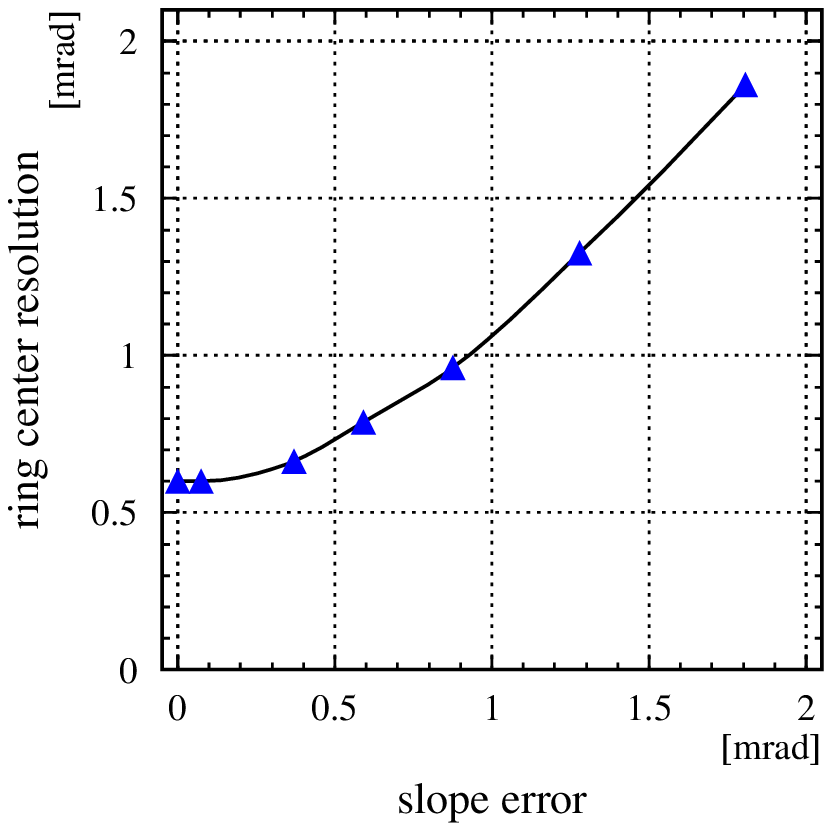,width=.49\textwidth}
   }
   \caption[Ring reconstruction efficiency and ring-center resolution of RICH-2]
   {Ring reconstruction efficiency (left panel) and ring-center resolution (right panel) of RICH-2 as a function
   of the slope error of the RICH-2 mirror.}
   \label{fig:ring-reconstr}
\end{figure}

Figure~\ref{fig:ring-reconstr} shows the decrease of the ring
reconstruction efficiency with increasing slope error. This
reflects the decrease of the number of true hits per ring due to
the fixed size of the ring search mask in the reconstruction
algorithm and the relative increase of misidentified hits on
rings. This analysis allowed for setting of the limits of
tolerance for slope errors at a maximum of about 0.7\,mrad. It is
worth stressing that improving the mirror quality beyond a slope
error of $0.25$\,mrad will not lead to a gain in reconstruction
efficiency. Furthermore, it should to be noted that the
ring-center resolution deteriorates quickly with increasing mirror
deformation as seen in Fig.~\ref{fig:ring-reconstr}. It is clear
that as soon as the errors induced by a certain slope error become
larger than the contribution of the chromatic dispersion ($0.53$
mrad) the ring-center resolution drops significantly. Optimizing
the tracking strategy and fine tuning of the ring fitting
algorithms could slightly improve the result. Combining both
studies, an upper limit of $0.7$\,mrad\,(rms) for the slope error
was estimated. The present glass mirror, in comparison, has an
overall slope error of $0.35$\,mrad. Additionally, the actual
focal length of the mirror must not deviate by more than $1.0\%$
from the nominal value of $f$\,$=$\,4000\,mm. The gap between
adjacent mirror segments must not exceed $2$\,mm which corresponds
to an average loss of about 1.8\% of the photons for 50\% of all
rings. A small gap would allow for an easy adjustment of the
mounted mirror segments.

\subsection{Reflectivity for UV photons}
\label{sec:reflect}

 The only metal known to provide high reflectivity in the UV
region is aluminum (Al). The optical properties of metals are
characterized by the index of refraction {\em n} and the
extinction coefficient {\em k}. The normal-incidence reflectance
is given by:
\begin{equation}
 R_0=\frac{(n-1)^2+k^2}{(n+1)^2+k^2}\;,
 \label{equ:refl}
\end{equation}
which holds true for Cherenkov photons because of the large focal
length of the mirror and the small Cherenkov angle. A maximum
reflectance of $95\%$ can be achieved for aluminum at a wavelength
of $200$\,nm~\cite{Hass:1982}. In case of the size of the surface
roughness exceeding $10\%$ of the photon wavelength $\lambda$, the
diffusely reflected component of the beam becomes large due to
scattering from surface structures according to:
\begin{equation}
 \frac{R}{R_0} \approx
 \exp{\left(-\frac{4\pi\sigma}{\lambda}\right)}\;,
 \label{equ:micro rough}
\end{equation}
with $\sigma$ being the width of the autocorrelation function
$\langle z(r)z(r') \rangle$ of the surface profile. The latter
describes the spatial correlation between the height of the
surface at points $z(r)$ and $z(r')$ which can be approximated by
a Gaussian distribution:
\begin{equation}
 K(r-r')= \langle z(r)z(r') \rangle \approx \sigma
 \exp{\left(-(r-r')/L\right)}\;,
 \label{equ:auto corr}
\end{equation}
where $\sigma$ and {\em L} denote the micro-roughness and the
average correlation length, respectively. Thus for UV photons, a
micro-roughness of less than $3$\,nm is required, comparable to
excellent polished float glass.

The most widely used technique for depositing Al coatings is
evaporation in high vacuum~\cite{Hass:1982}. All coatings for this
study were performed at the CERN coating facility. Purest grade of
Al($99.9999$\%) was evaporated from a tungsten coil. It is then
deposited on the rotating substrate to ensure uniform thickness of
the coating. In order to produce highly reflecting films, extreme
care must be taken to ensure that the evaporated coatings are not
contaminated by residual gases present during deposition. In
principle, the metal should be deposited at the highest possible
rate and not thicker than needed to be just opaquely reflecting.
Otherwise, the surface roughness will increase. The optimal
thickness for Al was found to be about 7\,\AA~\cite{Bream:1999}.
The small thickness of the film means a rather slow deposition
rate and, therefore, an ultrahigh vacuum of $10^{-7}$\,mbar was of
utmost importance. The thickness of the coating was controlled by
measuring the oscillation frequency change of a crystal induced by
the material deposited onto it.

A natural oxide film grows on the evaporated Al surface to an
ultimate $40$\,\AA~in thickness as soon as it is exposed to air.
While this oxide layer prevents aluminum from tarnishing, it also
causes a drastic decrease in reflectance in the UV region.
Therefore, the Al film needs to be protected with a magnesium
fluoride (MgF$_2$) overcoating preventing oxidation. The thickness
of the MgF$_2$ film is chosen such that the reflectivity of the
combined layer is enhanced by destructive interference of the
light reflected on both boundaries of the film. A MgF$_2$ layer of
$3.1$\,\AA~in thickness was used for all coatings produced for
this study.\\
 It is well known that evaporated MgF$_2$ coatings absorb water when
exposed to air. This decreases the reflectance in the UV region
significantly. Therefore, once coated, a mirror must always be
kept free from moisture, preferably in a protective nitrogen
atmosphere. Additionally, outgassing or diffusion of a component
of the mirror substrate can deteriorate the coating. Previous
experience in the coating of the RICH-1 mirror showed that a
replicated gold surface needs to be covered by a blocking layer
such as SiO/Cr to prevent the gold atoms from diffusing into the
aluminum coating~\cite{CERES:41}.

\clearpage

\subsection{Other important aspects}

The mirror is operated in a methane (CH$_4$) environment at
atmospheric pressure in a temperature range of 35 to
45\,$^{\circ}$C. All specifications must be fulfilled at least for
this operating temperature interval. Furthermore, the mounted
mirror should withstand any thermal stresses induced in a
temperature interval of 10 to 45\,$^{\circ}$C due to shut down of
the heating during the off-line period.

 Any additional contamination of the radiator gas with $1$\,ppm
of water or oxygen will reduce the UV photon efficiency by about
$1\%$  due to absorption. The present gas system can achieve an
equilibrium concentration of O$_2$\,$<$\,1\,ppm and
H$_2$O\,$<$\,2\,ppm. If a new mirror introduces a larger oxygen
source an upgrade of the RICH-2 gas system would be required to
maintain the present low level of oxygen contamination. Of further
importance is the fact that cleaning of the CH$_4$ radiator gas
after opening the RICH-2 takes at least 3 to 6 weeks.


\section{Manufacturing technologies}

The manufacturing technology depends on the choice of the mirror
substrate such as glass, metal, or composite materials.
Traditionally, mirrors for optical applications such as the old
RICH-2 mirror were made of glass for the following reasons:
technological control over wide range of physical properties,
excellent surface quality due to polishing, and very low
production cost. For the purpose of reducing the thickness below
$1\%$ of a radiation length, a float glass mirror must be thinner
than $1.4$\,mm according to Table~\ref{tab:thick}. This, however,
is not feasible because glass as a quasi-fluid rapidly loses its
long term shape stability for large-size mirror panels with a
thickness of less than $3$\,mm.

In the scope of this study, three alternative approaches were
considered. First, to construct a mirror from coated Mylar foil
keeping a spherical shape by applying a pressure gradient between
front and back side. A two Mylar foil mirror setup of a
100\,$\mu$m thickness for each would correspond to $0.1\%$ of a
radiation length. This option was rejected because any spatial
anisotropy in elasticity or substrate thickness will lead to large
shape distortions. Second, a novel mirror substrate based on
carbon ceramic was developed by DASA/IAGB~\cite{DASA} and the
Technical University of Munich for the HADES RICH
mirror~\cite{Muntz:1999td}. The extremely high stiffness of the
ceramic allows for minimum mirror thickness of about $2.0$\,mm.
This corresponds to $0.75\%$ of a radiation length for a substrate
density of $1.54$\,g/cm$^2$. Additionally, a very high local
surface quality and, hence, reflectivity can be achieved because
substrate shells are individually polished to a micro-roughness of
about $2$\,nm. The major disadvantage of this type of substrate is
the high risk of residual stresses in the material. At the time of
this evaluation the mirror panels regularly broke during polishing
or trimming indicating insufficient control of the manufacturing
process. In combination with the prohibitive high cost of such a
mirror, this led to discarding of the option.

Third, mirror substrates made of carbon-fiber-composite (CFK)
materials were considered. These materials had been used for many
high-tech applications in the past decade. Many times, it was
successfully shown that large ultralightweight mirrors can be
manufactured using CFK substrates with a replica technology for
coating~\cite{Kasl:2000,CERES:41,Akiba:1999}. The main advantages
of CFK materials are very low area density
($\rho_A$\,$\approx$\,1.7\,g/cm$^2$), high stiffness, and
industrial mass production guaranteeing reasonable prices. The
significant technological advances of recent years led to a
considerable reduction of inherent disadvantages such as fiber
printing, inhomogeneous module of thermal expansion, and moisture
absorption. In view of these promising developments and the lack
of alternatives, all subsequent studies were focused on CFK
materials.

\subsection{Carbon-fiber-composite mirrors}

\begin{figure}[tb]
  \centering
  \epsfig{file=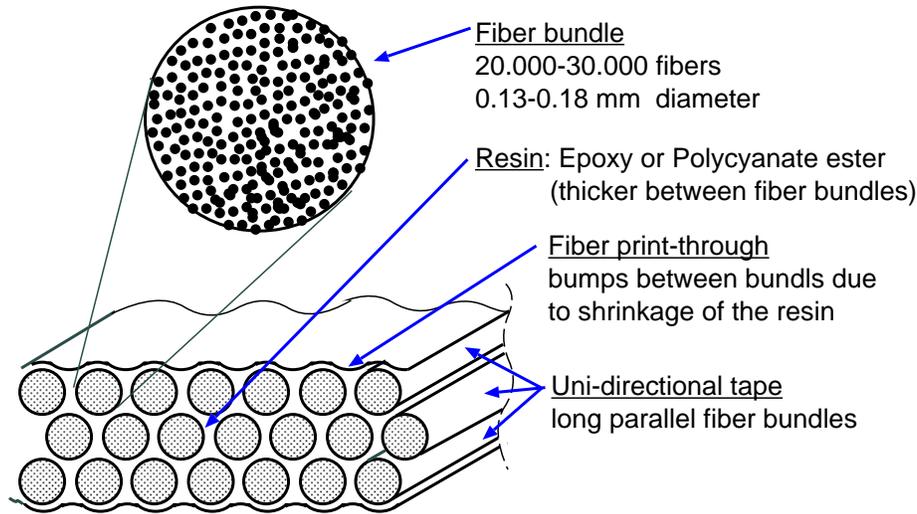}
  \caption[Structure of the carbon-fiber substrate]
  {Structure of a uni-directional carbon-fiber material.}
  \label{fig:cfk_substrate}
\end{figure}

Carbon-fiber-composite mirrors consist of bundles of carbon fibers
which are glued together with a special resin in a baking process.
Three types of substrates can be distinguished depending on the
arrangement of the carbon-fiber bundles: uni-directional materials
with irregular orientation of chopped bundles, materials with long
parallel aligned bundles, and woven fabricates consisting of
interlaced texture of carbon-fiber bundles.
Figure~\ref{fig:cfk_substrate} illustrates the composition of a
typical uni-directional material commonly used as mirror
substrate.
These days, the properties of carbon-fiber substrates can be
controlled within a broad range to fit specific requirements. This
can be done by choosing the proper fiber material and resin system
and by adjusting the parameters in the curing process.

Most difficulties of CFK substrate arise from the spatial
inhomogeneity of the material due to anisotropic fiber structure
and the thermal inhomogeneity due to the difference of the
coefficient of thermal expansion\,(CTE) between resin and carbon
fibers. The lower CTE of the resin system creates valleys between
adjacent carbon-fiber bundles when the substrate cools down after
the high-temperature curing process as shown in
Fig.~\ref{fig:cfk_substrate}.

\begin{figure}
  \centering
  \epsfig{file=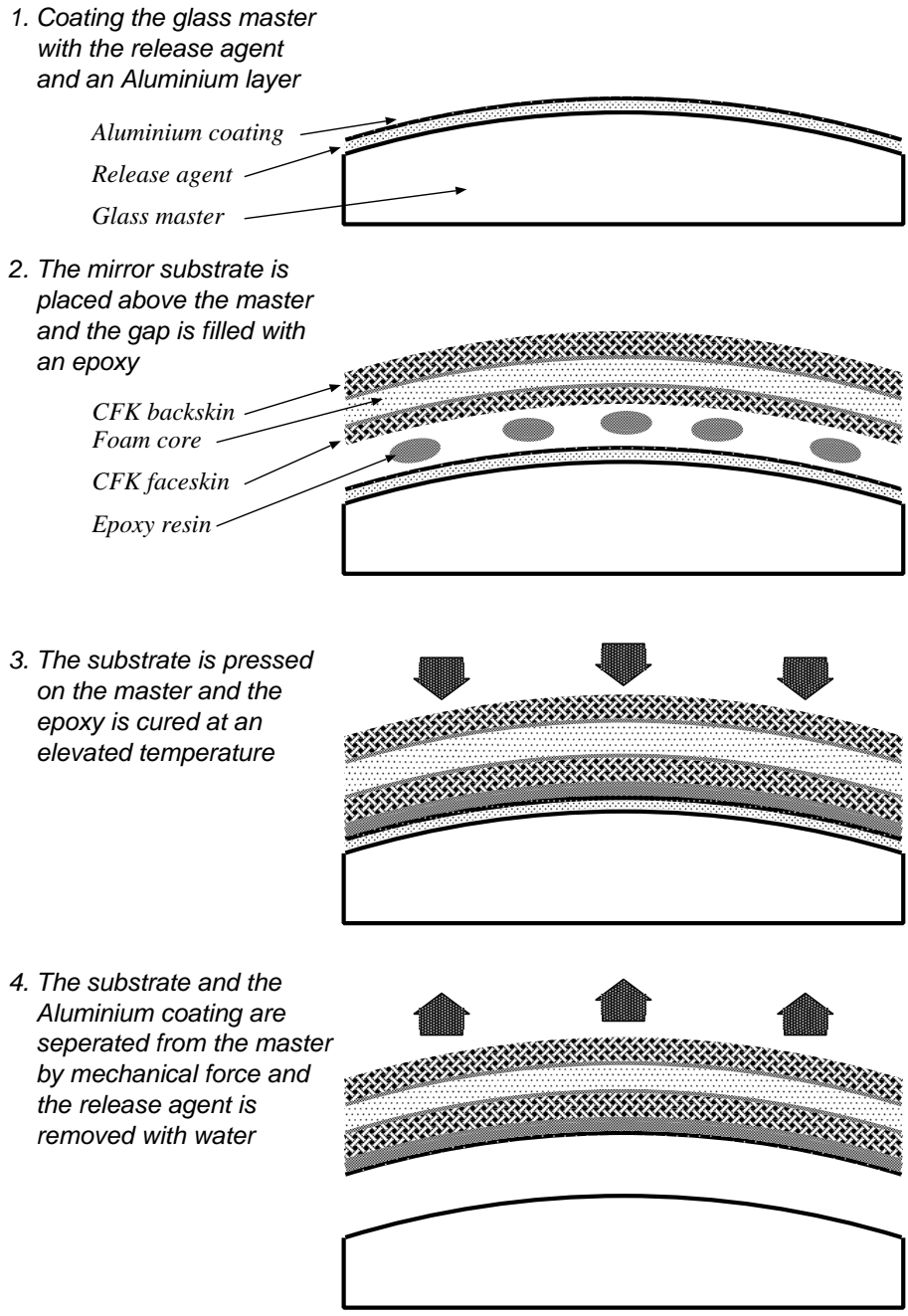, width=12.1cm}
  \caption[Schematic view of the replication process]
  {Replication process for a sandwich-design CFK mirror developed by OPTICON~\cite{OPTICON}.}
  \label{fig:replica}
\end{figure}

This effect, known as fiber printing, can be reduced by applying
an additional resin layer to the substrate surface in a
replication process as illustrated in Fig.~\ref{fig:replica}. This
process can also be utilized to transfer a reflective metal
coating such as aluminum and silver to a mirror surface.

 This technology was successfully applied in the manufacturing
of the CERES RICH-1 mirror by MAN Technologie
AG~\cite{CERES:41,MAN}. However, it cannot be utilized for UV
coatings, as the protective MgF$_2$ coating layer is very fragile
and, therefore, cannot be transferred from master to substrate.

Another drawback of CFK materials is their spatial inhomogeneity.
This can be controlled by combining several uni-directional
carbon-fiber sheets with appropriate fiber orientations.
Approximate spatial homogeneity to thermal, mechanical, or bending
stresses can be achieved through arrangement of 2, 4, or 8 layers
of CFK, respectively. Adherence to absorption of moisture causing
the material to expand and to act as a source of water in a dry
atmosphere such as the CH$_4$ radiator gas is a further
disadvantageous property of CFK materials~\cite{Whitaker}.

\subsection{Evaluation of CFK manufacturing concepts}
\label{sec:cfk-meas}

A survey of CKF mirrors from all major vendors was carried out to
assess the impact of recent improvements in manufacturing
technologies on the inherent critical properties: surface
micro-roughness, fiber print-through, and substrate stiffness.
Mirror samples were obtained from the following companies:
ARDCO~\cite{ARDCO}, SESO~\cite{SESO}, and Composite Optics
Inc.\,(COI)~\cite{COI}. MAN Technologie AG~\cite{MAN}, that had
previously built several 1-m-diameter CFK mirrors for CERES
RICH-1, had no longer the facilities to do the gold coating of the
replication master and could not offer any alternative solution.
Before assessing technologies with respect to the specifications,
experimental methods to measure the samples will be described.

\begin{figure}[t]
  \centering
  \epsfig{file=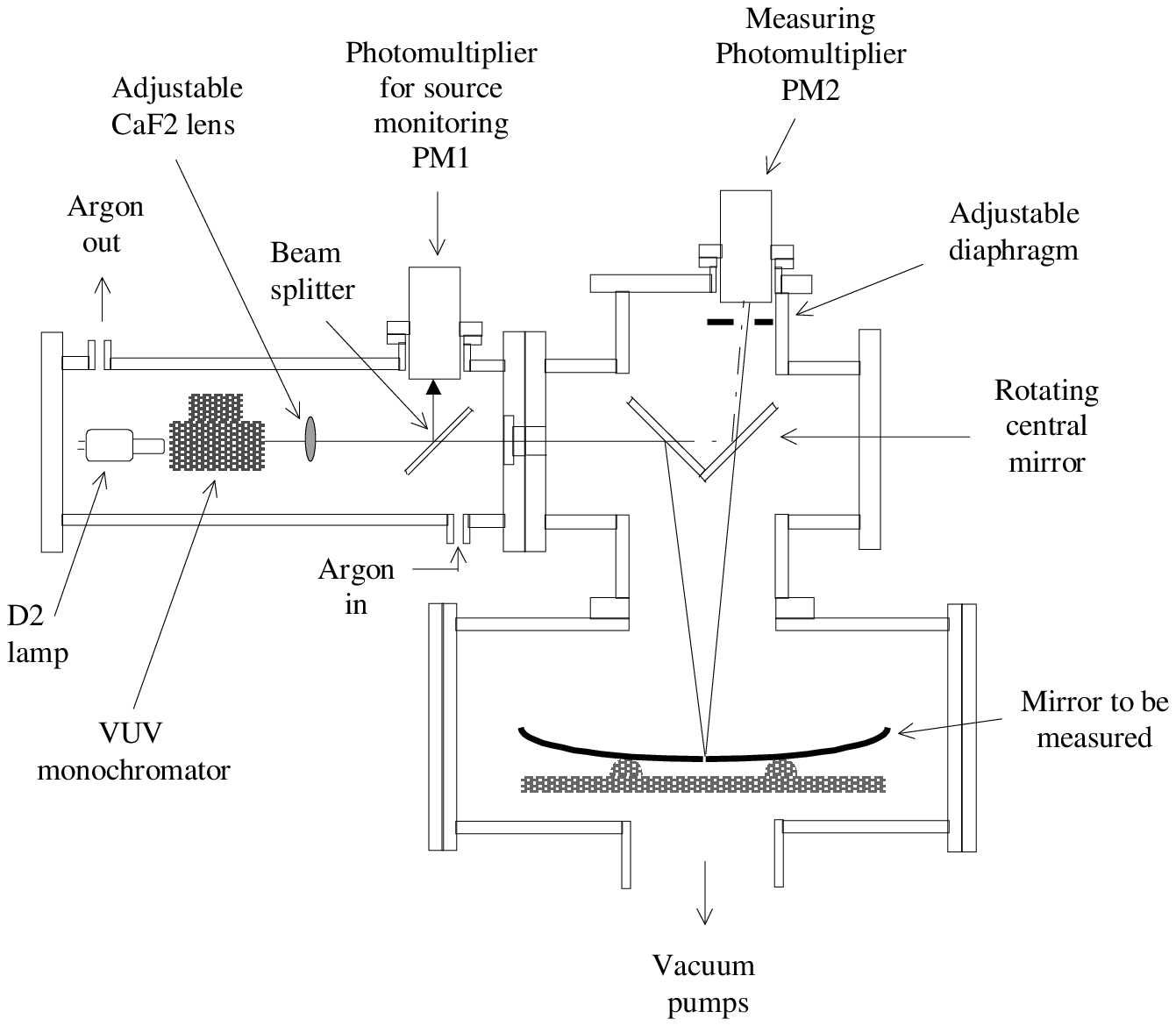}
  \caption[Setup for UV reflectivity measurement at CERN]
  {Setup for UV reflectivity measurement at
  CERN~\cite{Bream:1999}.}
  \label{fig:UV_setup}
\end{figure}

UV reflectivity can be measured directly with a reflectometer or
indirectly by determining the micro-roughness with a surface
interferometer. The samples were evaluated in the CERN
reflectometer depicted in Fig.~\ref{fig:UV_setup}.
A grating selected a monochromatic light beam with a wavelength of
160 to 230\,nm with a bandwidth of 2\,nm from the continuum of a
deuterium lamp. The intensity of the incident beam $I_0$ was
measured with the photo multiplier PM1. After rotation of the
central mirror the beam was sent to the center of the measured
sample. The ratio of the intensity of reflected beam $I_{\rm
refl}$ measured with photo multiplier PM2 to the intensity of the
incident beam $I_{0}$ gave the reflectance of the sample. All
measurements were performed in ultrahigh vacuum to avoid photon
absorption.

\begin{figure}[tb]
  \centering
  \epsfig{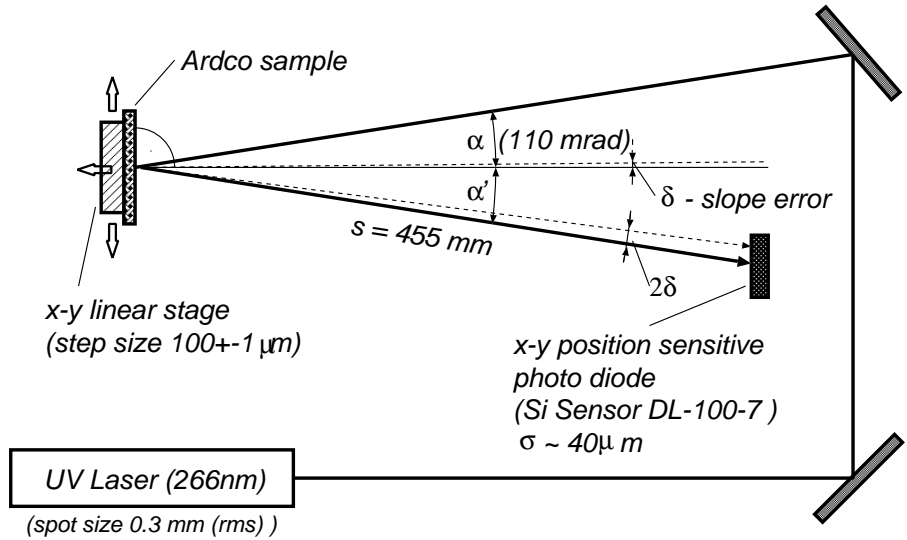}
  \caption[Setup for the mirror surface scan]
  {Setup for the measurement of the local slope error of flat mirror samples.}
 \label{fig:scan-surf setup}
\end{figure}

To evaluate the CFK material specific fiber print-through a
surface profile needs to be measured with a resolution of less
than $0.1$\,mm as determined by the diameter of the carbon-fiber
bundles. Figure~\ref{fig:scan-surf setup} shows the experimental
setup used to scan flat reflective samples with a laser beam. The
UV laser excels because of a small beam diameter of 300\,$\mu$m
and a very high intensity that allows to measure the reflection
from uncoated samples.
A surface profile was obtained by moving the sample on a linear
stage relative to a fixed laser beam and recording the image of
the reflected laser beam on a position sensitive diode. Any
surface deformation with a length scale larger than the beam
diameter results in a shift of the image position on the diode.
The average slope error $\Delta \phi$ can be computed from the
rms-width of the distribution of image positions:
\begin{equation}
 \Delta\phi =
 \arctan{\left(\tan{\alpha}+\overline{\langle(x-\overline{x})^2\rangle}/s\right)}\;,
 \label{equ:slope_scan}
\end{equation}
where {\em s} denotes the distance between mirror and diode and
$\alpha$ is explained in Fig.~\ref{fig:scan-surf setup}. The
orientation of the carbon-fiber bundles needs to be taken into
account for highly anisotropic uni-directional CFK fabric.

The new substrate from ARDCO is a sandwich of CKF layers and
ROHACELL foam. The ROHACELL foam stiffens the mirror considerably
and increases the total thickness only by $0.3\%$ of a radiation
length. The mirror samples manufactured by ARDCO~\cite{ARDCO}
represented a novel sandwich design developed for the RICH mirrors
of the PHENIX experiment~\cite{Akiba:1999}. Two uni-directional
plane CFK shells of $1$\,mm in thickness are stabilized and
stiffened by a $1$\,cm thick ROHACELL foam core. A detailed
analysis of the sandwich design concept will be presented in
Sec.~\ref{sec:isogrid}.

The composite substrates were coated by OPTICON
Inc.~\cite{OPTICON} with an Al/Au layer by a replication process
(see Fig.~\ref{fig:replica}). To achieve high UV reflectance, both
samples were coated with a high quality Al and MgF$_2$-protective
film (by A. Braem, CERN). This required the development of a novel
technology involving differential pumping on the sandwich
structure to prevent outgassing and collapse of the ROHACELL foam
in the ultrahigh vacuum during coating.

\begin{figure}[tb]
    \begin{minipage}[t]{.65\textwidth}
        \vspace{0pt}
        \centering
        \epsfig{file=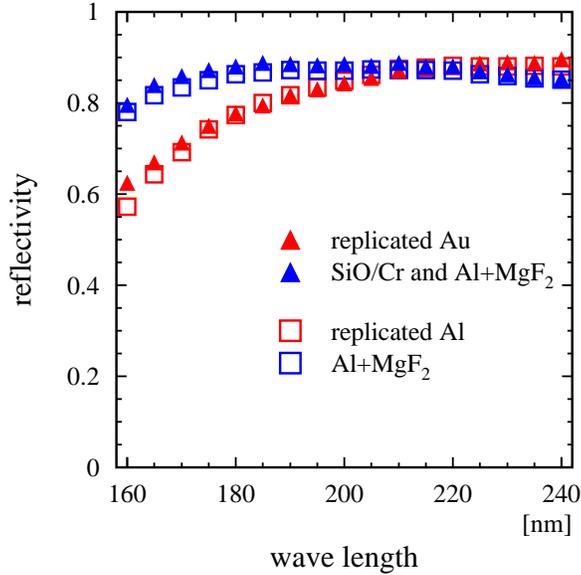,width=\textwidth}
    \end{minipage}%
    \begin{minipage}[t]{.35\textwidth}
      \vspace{0.5cm}
      \caption[Reflectivity measurement of the UV coating of replicated CFK mirrors]
      {\newline
        Reflectivity of the aluminum coating with MgF$_2$ protective layer applied
        to a CFK mirror with aluminum or gold replicated coating.
      }
      \label{fig:ardco refl}
    \end{minipage}
\end{figure}

A summary of the measured reflectance before and after UV coating
is shown in Fig.~\ref{fig:ardco refl}.
A reflectance of 87\% for a wavelength above $180$\,nm is close to
the expected value of 90\%~\cite{Hass:1982}. Furthermore, it is
comparable to the float glass witness sample indicating that the
release agent used by OPTICON during replication did not
deteriorate the micro-roughness of the replicated Al or Au
surface. The strong decrease of reflectance of the replicated
samples without UV coating for a wavelength below $200$\,nm, as
apparent in Fig.~\ref{fig:ardco refl}, is caused by the natural
oxide surface layer as explained in Sec.~\ref{sec:reflect}.
Inasmuch as Au and Al coating exhibit a comparable reflectance,
the replication of Al is easier because Au coating requires an
additional SiO/Cr blocking film to avoid diffusion into the
surface UV coating. It can be concluded that the UV reflectance of
replicated CFK mirrors is not limited by the surface
micro-roughness and is comparable to a glass mirror.

\begin{figure}[tb]
  \centering
  \mbox{
   \epsfig{file=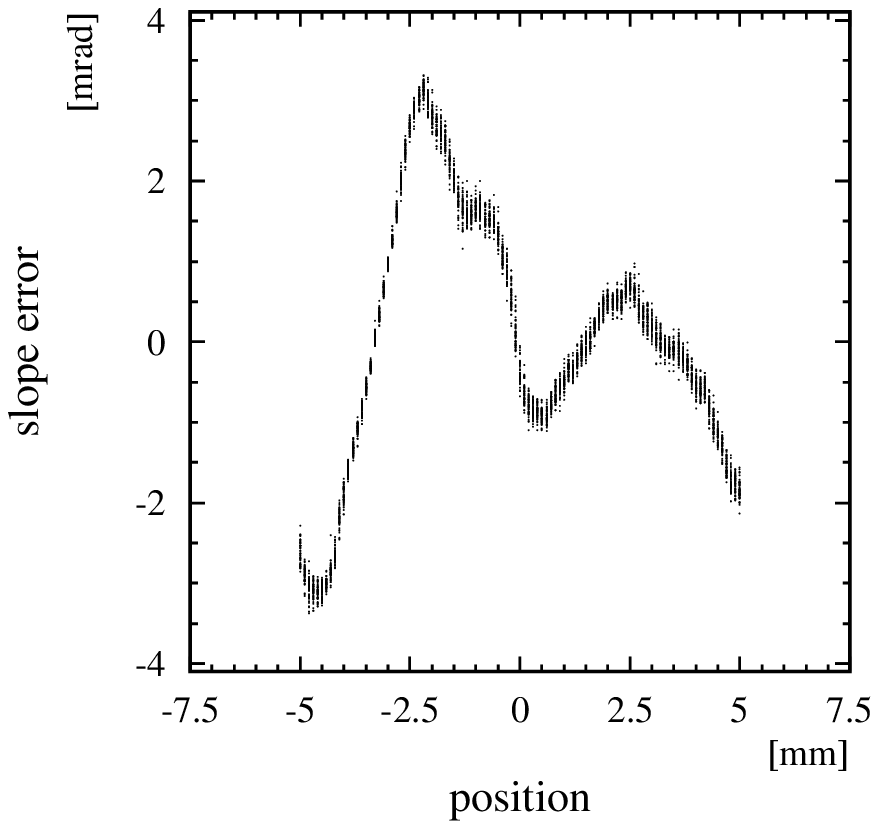,width=.49\textwidth}
   \hfill
   \epsfig{file=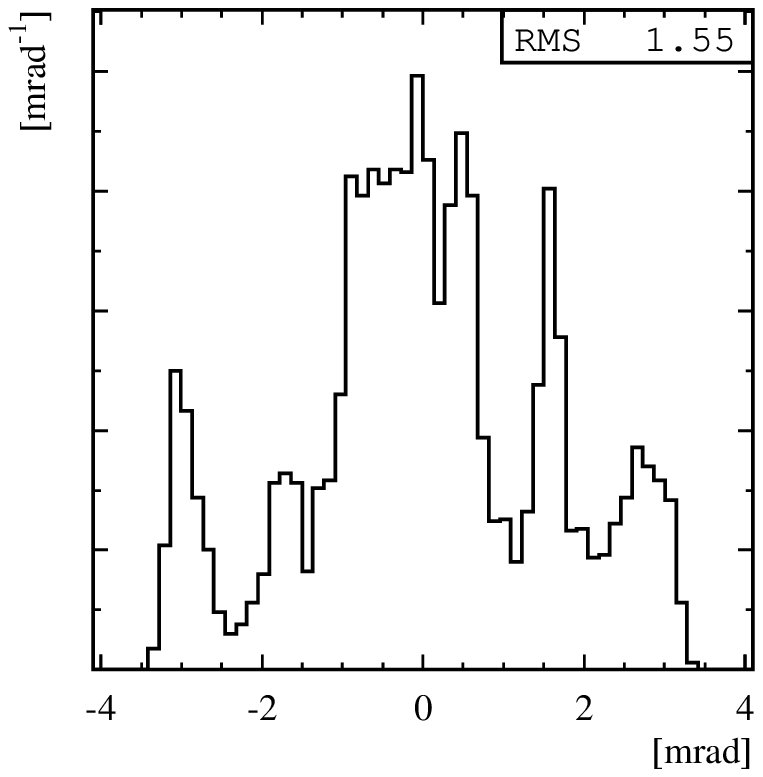,width=.49\textwidth}}
  \caption[Surface profile of the ARDCO mirror sample]
  {Spatial variation the local slope error (left panel) and average slope error
  assuming spatial invariance (right panel) of a
  surface profile of the ARDCO mirror sample.
  }
 \label{fig:ardco slope}
\end{figure}

Figure~\ref{fig:ardco slope} shows the local slope error along a
$12$\,mm surface profile measured using the experimental setup of
Fig.~\ref{fig:scan-surf setup}.
The surface quality of the ARDCO/OPTICON sample is surprisingly
poor. The print-through of individual fiber bundles was already
evident from visual inspection. The deformations have a maximum
amplitude of the order of $3.5$\,mrad resulting in a slope error
of approximately $1.6$\,mrad. OPTICON claimed that the large fiber
printing was solely due to the poor quality of the ARDCO CFK
substrate. Rough surface structures result in an uneven gap width
between substrate and replication master and hence in variations
of the thickness of the epoxy resin layer. These variations induce
mechanical stresses when the resin shrinks during the curing
process.

\begin{figure}
  \centering
  \mbox{
   \epsfig{file=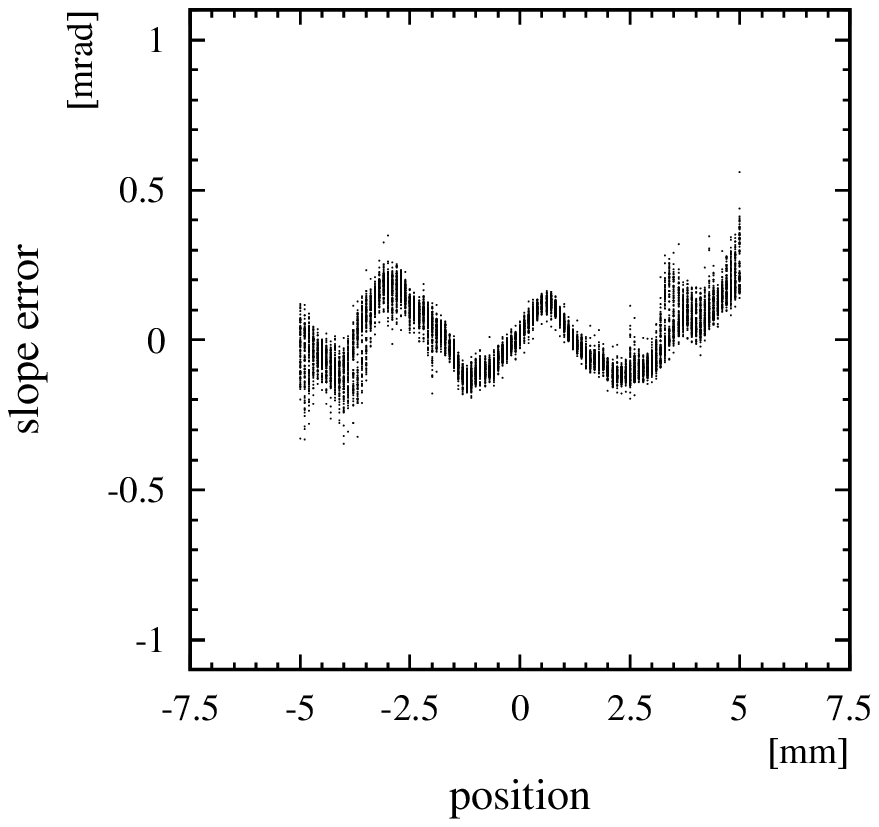,width=.49\textwidth}
   \hfill
   \epsfig{file=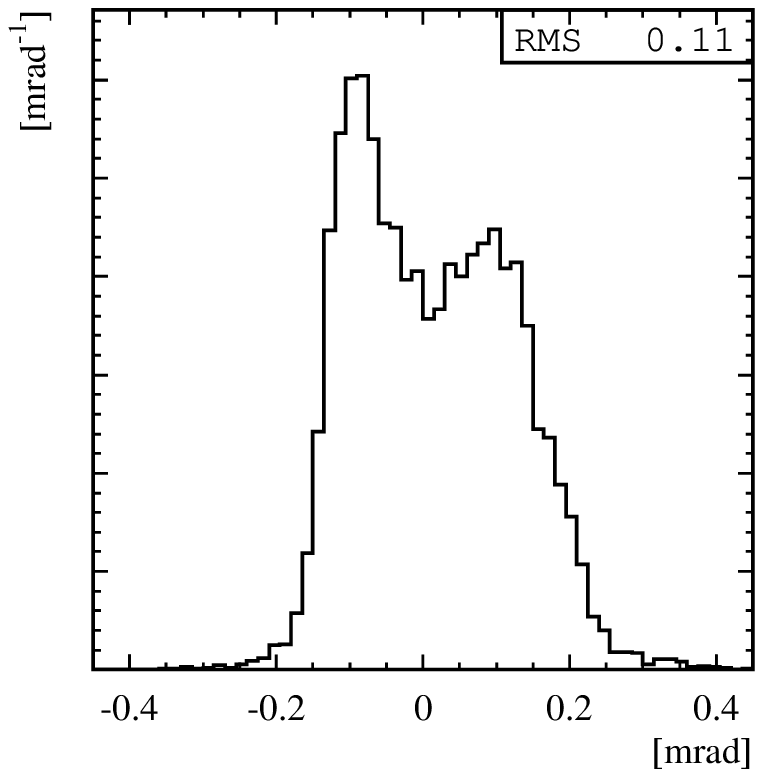,width=.49\textwidth}}
  \caption[Surface profile of the SESO mirror sample]
  {Spatial variation the local slope error (left panel) and average slope error
  assuming spatial invariance (right panel) of a
  surface profile of the SESO mirror sample.}
  \label{fig:myseso}
  \centering
  \mbox{
   \epsfig{file=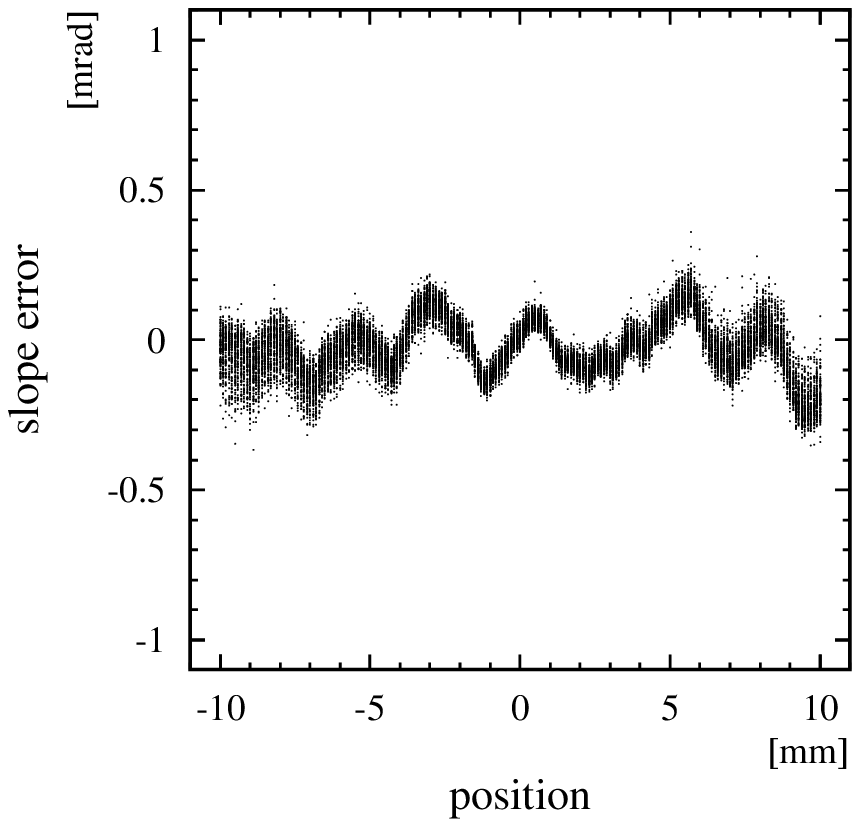,width=.49\textwidth}
   \hfill
   \epsfig{file=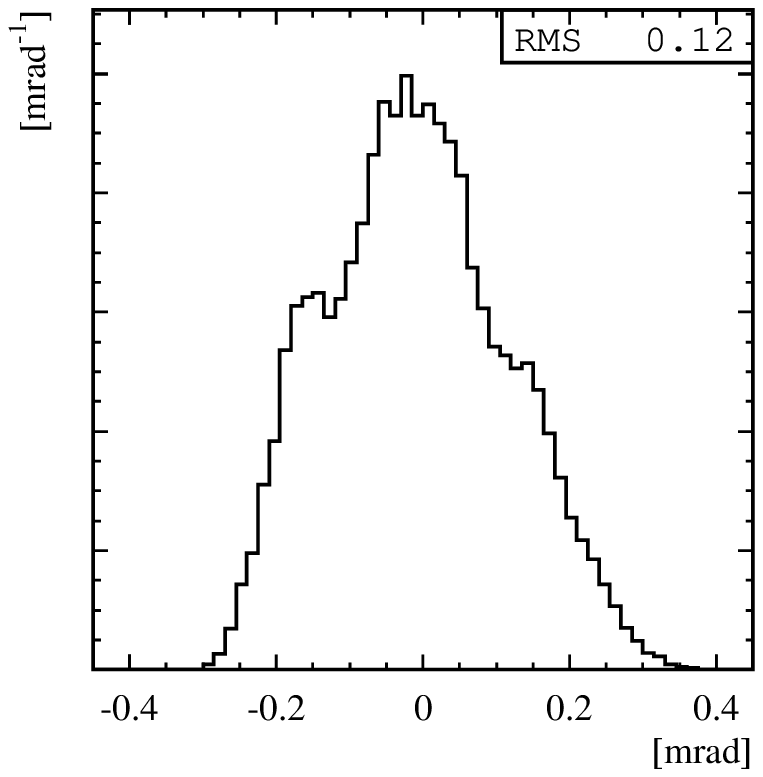,width=.49\textwidth}}
  \caption[Surface profile of the COI mirror sample]
  {Spatial variation the local slope error (left panel) and average slope error
  assuming spatial invariance (right panel) of a
  surface profile of the COI mirror sample.}
\label{fig:coi slope}
\end{figure}

The $1$\,m diameter SESO mirror~\cite{SESO}, originally
manufactured for the HADES experiment~\cite{Muntz:1999td}, was
made of woven CFK fabric which is very cheap but gives inferior
stiffness. Its thickness of $2.0$\,mm corresponds to $0.8\%$ of a
radiation length. The surface profile shown in
Fig.~\ref{fig:myseso} is much smoother and exhibits a slope error
of only 0.11\,mrad\,(rms), one order of magnitude better than the
ARDCO sample. Detailed study of the mirror shape by
HADES~\cite{Friese:2000} revealed a large astigmatism due to the
lack of stiffness. According to SESO, only an increase of
substrate thickness up to 4\,mm would allow to achieve a
sufficiently high stiffness.

The sample provided by COI consisted of a uni-directional pan
fiber substrate (M55J/954-3 made by HEXCEL~\cite{HEXCEL}) of
$0.4$\,mm in thickness. A novel polycyanate ester resin was used
to glue the laminate layers for minimal difference in CTE between
laminate and resin system. An Al surface coating was replicated
onto the substrate by OPTICON using the replication technique
described in Fig.~\ref{fig:replica}. The measurement of the slope
error along a surface profile is summarized in Fig.~\ref{fig:coi
slope}.
The slope error of $0.1$\,mrad is remarkably small and comparable
to that of the SESO sample.

In conclusion, an average slope error of less than $0.15$\,mrad
can be achieved by combining best available CFK substrates with
novel efficient manufacturing approaches. This result invalidates
the widespread opinion that CFK mirrors have poor quality due to
fiber print-through.

COI Inc.~was chosen for further mirror development based on the
measurement and its excellent reputation as the largest
manufacturer for advanced carbon-fiber substrates for all kinds of
optical and spacecraft high-tech applications~\cite{Kasl:2000}.

\subsection{Mechanical stability of the mirror}
\label{sec:isogrid}

Generally, very thin large-size CFK mirrors possess insufficient
strength and stiffness to hold proper shape under their mass
unless mounted or otherwise supported. Fixing the mirror at its
outer perimeter to a mount, as done for the RICH-1 mirror, is
clearly not possible due to the segmentation of the RICH-2 mirror.

\begin{figure}[tb]
  \centering
  \epsfig{file=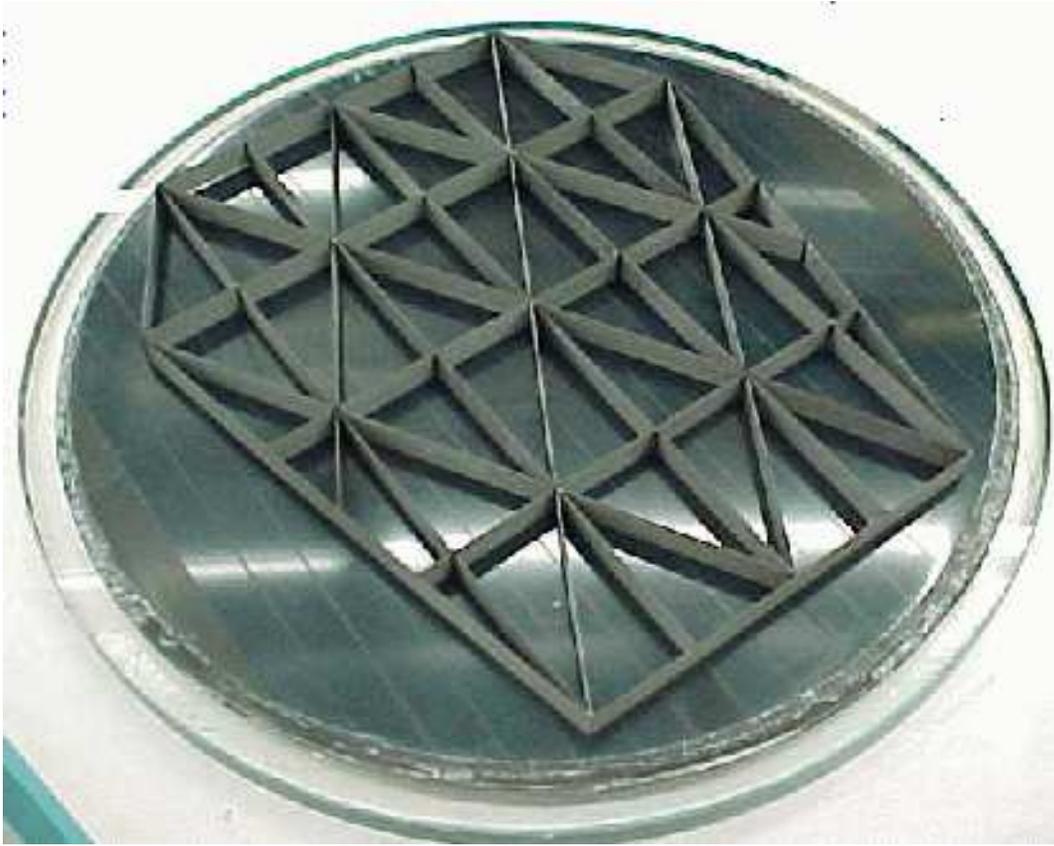}
  \caption[Isogrid design support structure]
  {Isogrid design support structure of the second COI prototype mirror.}
  \label{fig:isogrid}
\end{figure}

There are two concepts for stabilizing a thin mirror shell. First,
the sandwich design: the CFK face skin bonded to a lightweight
core material such as ROHACELL foam and CFK back skin to
counterbalance thermal stresses. Second, the isogrid
design~\cite{Djobadze}: the thin mirror shell is supported with an
isotropic core structure achieved by assembling strips of CFK
flatstock in an egg-crate fashion forming equilateral triangle
core cells (see Fig.~\ref{fig:isogrid}).

The sandwich approach was first adapted to RICH mirrors by the
PHENIX collaboration~\cite{Akiba:1999}. In
Sec.~\ref{sec:cfk-meas}, measurements of a test sample were
described. The main advantages for RICH mirror application are:
very low density foam allowing a high core thickness and resulting
in an excellent mirror stiffness, low material and assembly cost,
and technological feasibility for mounting at the back skin. An
unsealed ROHACELL foam core, used for the PHENIX detector, was
ruled out because ROHACELL has a strong affinity to take up
moisture also causing the foam to expand. As a result, an
additional sealing of the panel edges is necessary to prevent
outgassing into the radiator. Replacing ROHACELL by a reticulated
vitreous carbon core (RVC) with similar properties but less
affinity to moisture would reduce the outgassing rate but sealing
of the core would still be necessary.

The development of the isogrid design was driven by the need for
more thermally stable satellite reflectors for use in
telecommunication and in remote sensing of the atmosphere. From
the fabrication point of view, isogrid mirrors provide higher
bending stiffness and are less prone to thermal distortion because
the entire structure is constructed from a single material. An
exhaustive comparison of sandwich and isogrid design can be found
in~\cite{Djobadze}.
 For the application to RICH mirrors the main advantage of this
solution is that the CFK face skin can be as thin as $0.76$\,mm
assuming 12 layers of pan fiber tape or about $0.5\%$ in terms of
a radiation length. Additionally, 1\% to 5\% of the electrons
would traverse the material of the support structure made of ribs
with $2.5$\,cm height and $0.76$\,mm thickness. This would
correspond to a thickness of 0.6\% to 3.8\% of a radiation length
depending on the incident angle of the electrons. From general
considerations it becomes clear that this solution is favorable
compared to an equivalent (thicker) homogeneous mirror because
most electrons traverse the thinner face skin and remain within
the narrow line width of the $\omega$- or $\phi$-meson whereas the
electrons traversing the much thicker back structure are anyway
''lost'' in the bremsstrahlung tail. The higher number of almost
unaffected electrons improves the background discrimination
considerably. In this design without a foam core, there is no need
to worry about outgassing or special coating arrangements which
leads to substantial time and cost savings. The critical issue of
this design is the possible print-through of the support structure
caused by the gluing of the back structure to the face sheet.

 The apparent advantages of the isogrid design led to the
decision to further pursue this novel solution to ultralightweight
UV mirrors.


\section{Prototype measurements}

The first full size prototype was fabricated by COI and delivered
in October 1999. The substrate consists of 12 layers of
uni-directional pan fiber tape (M55J/954-3 by
HEXCEL~\cite{HEXCEL}) with a combined thickness of $0.76$\,mm. The
specific substrate was selected to provide a near zero CTE to
insure dimensional stability over a wide temperature range and for
its exceptionally high stiffness. The unitape layers are rotated
consecutively by 30 degrees each to obtain best thermal isotropy.
To bond the layers, a polycyanate ester based resin system was
chosen to match the CTE of the substrate. The face skin restrained
to a glass master was cured in an autoclave at a temperature of
approximately 140\,$^{\circ}$C\@. In the same step, the mirror
surface was replicated by use of a newly developed release agent
of COI making any additional replication with a metal coating
unnecessary. This improvement resulted in considerable time and
cost savings. After curing, the rib structure was bonded to the
back side of the face skin which was still restrained to the glass
master. The isogrid structure of the second prototype including
five additional vertical ribs is illustrated in
Fig.~\ref{fig:isogrid}. After the second curing at room
temperature to bond the rib structure, the composite mirror is
released from the glass master. From now on the back structure is
responsible for preserving the shape of the mirror.

\begin{figure}[tb]
  \centering
  \epsfig{file=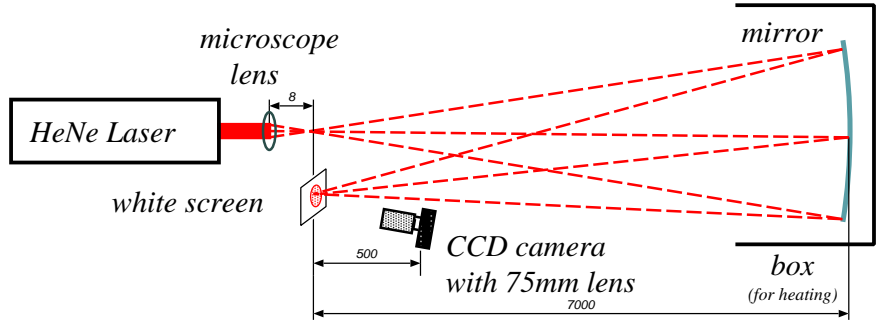,width=0.8\textwidth}
  \caption[Setup for the optical mirror quality measurement]
  {Setup for the optical mirror quality measurement with a HeNe laser (632\,nm).}
  \label{fig:setup-mirror}
\end{figure}

The reflectance of the replicated mirror surface was sufficiently
high even without an additional metal coating to allow for laser
measurement of the mirror. Figure~\ref{fig:setup-mirror} shows a
schematic view of experimental setup used for mirror shape
measurements.
The mirror was illuminated by a point source of monochromatic
light (632\,nm HeNe laser beam focused by a microscope lens)
placed at the center of the mirror sphere. The radius of curvature
of the mirror panel was measured as $7880\pm50$\,mm.

\begin{figure}
      \centering
      \epsfig{file=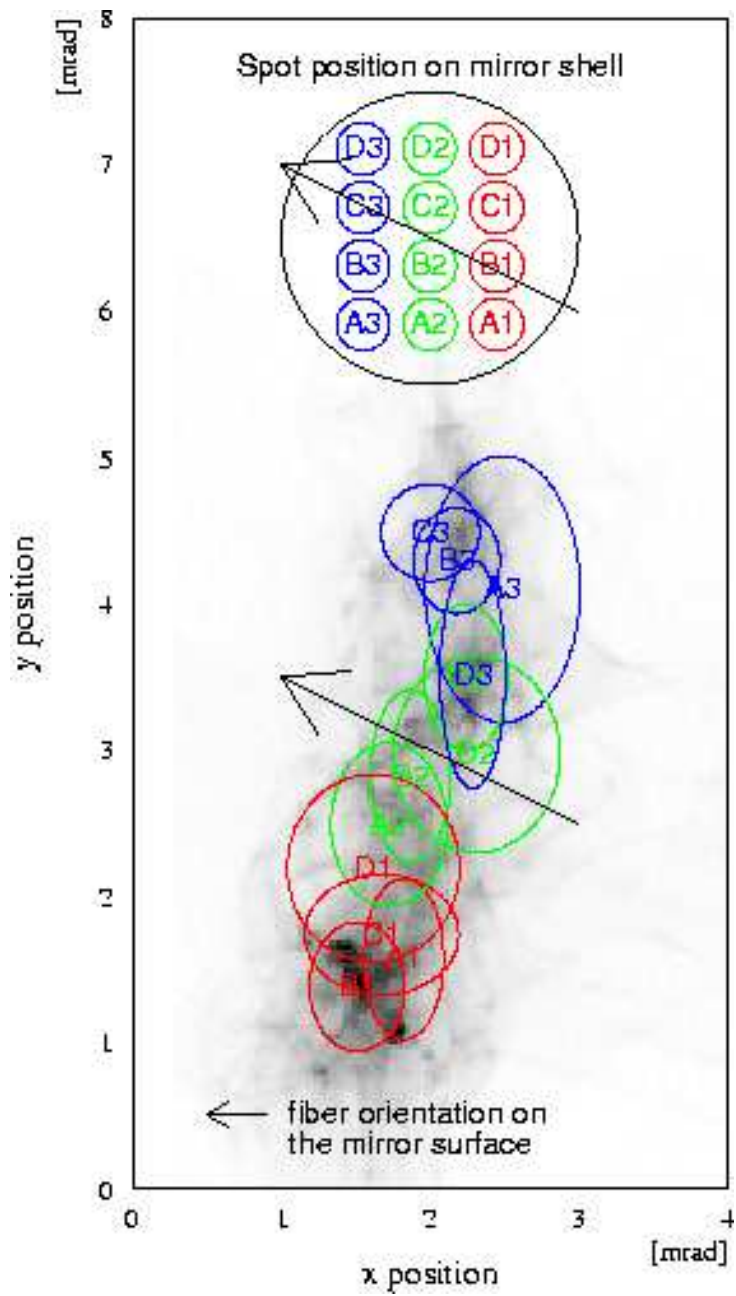}
      \caption[Focal image of the first COI prototype]
      {Variation of position and width of the focal image
       of the first COI prototype overlayed with the
       focal image of the full mirror.}
      \label{fig:coi1999-focal}
\end{figure}

A summary of the optical measurements is presented in
Fig.~\ref{fig:coi1999-focal}.
It shows the focal image of the point source when the whole panel
is illuminated. The apparent structures indicate rather large
surface distortions. The slope error is about 1.4\,mrad~(rms).
This is equivalent to an image size of about 2.5\,cm (diameter).
Heating the mirror up to the operating temperature of
$35^\circ$\,C did not improve the slope error. A closer
examination by illuminating individual surface regions of 10\,cm
diameter, the size of a Cherenkov light cone, reveals a large
spatial variation of the focal point as evident by the position of
the individual focal images. This astigmatism is produced by a
larger radius of curvature along the fiber direction of the front
CFK layers and can be directly attributed to the higher bending
stiffness along the fiber direction.

The average slope error of the spots, shown as the diameter of the
circles, varies between 0.40\,mrad and 0.91\,mrad\,(rms). It
turned out that the surface area within each back structure
triangle is especially distorted. Most likely it is due to
stresses induced by the shrinking of the adhesive used to bond the
isogrid. The direct print-through of the back structure was
visible but fairly small compared to those other distortions.
Although the optical quality for small areas was encouraging,
deficits in manufacturing and design were apparent.

\begin{figure}
  \centering
  \epsfig{file=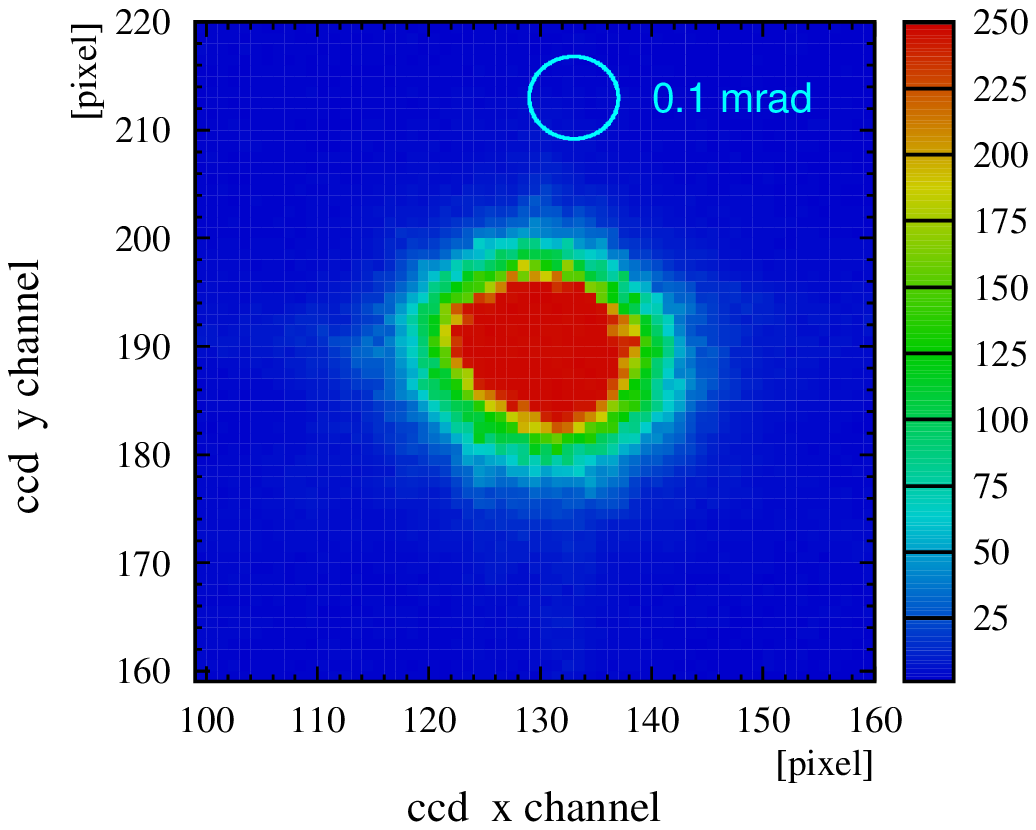}
  \caption[Focal image of the second COI prototype]
  {Focal image of the second COI prototype mirror.}
  \label{fig:mirror2000-peak}

  \vspace*{.8cm}
  \mbox{
   \epsfig{file=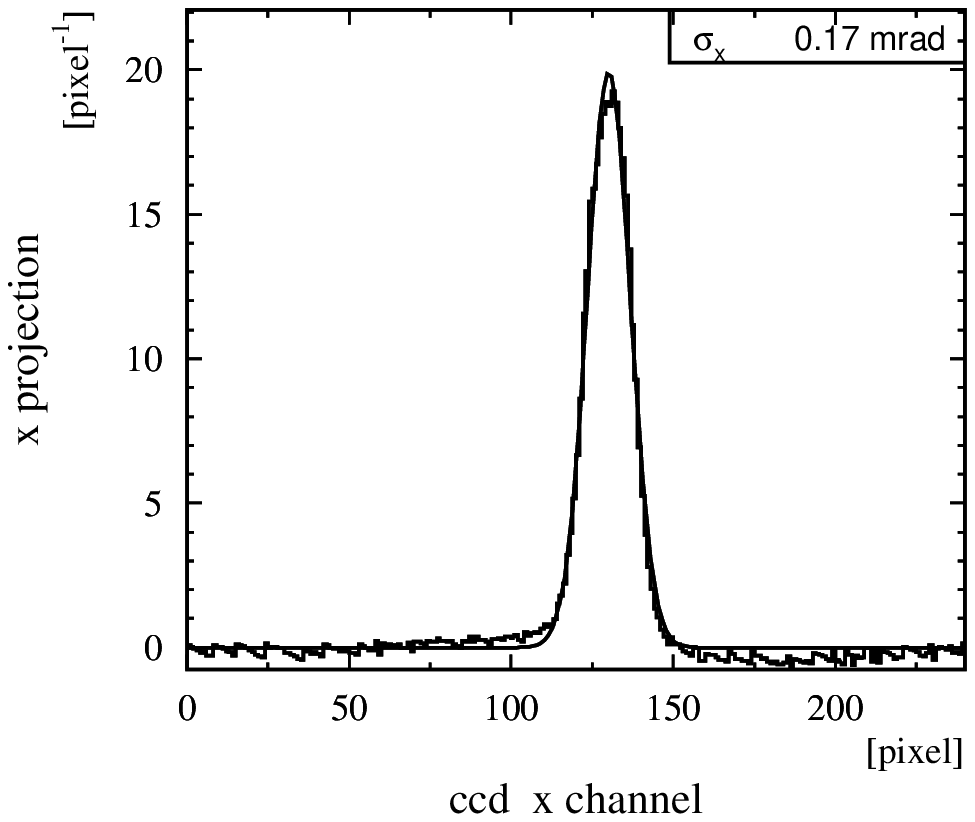,width=.49\textwidth}
   \hfill
   \epsfig{file=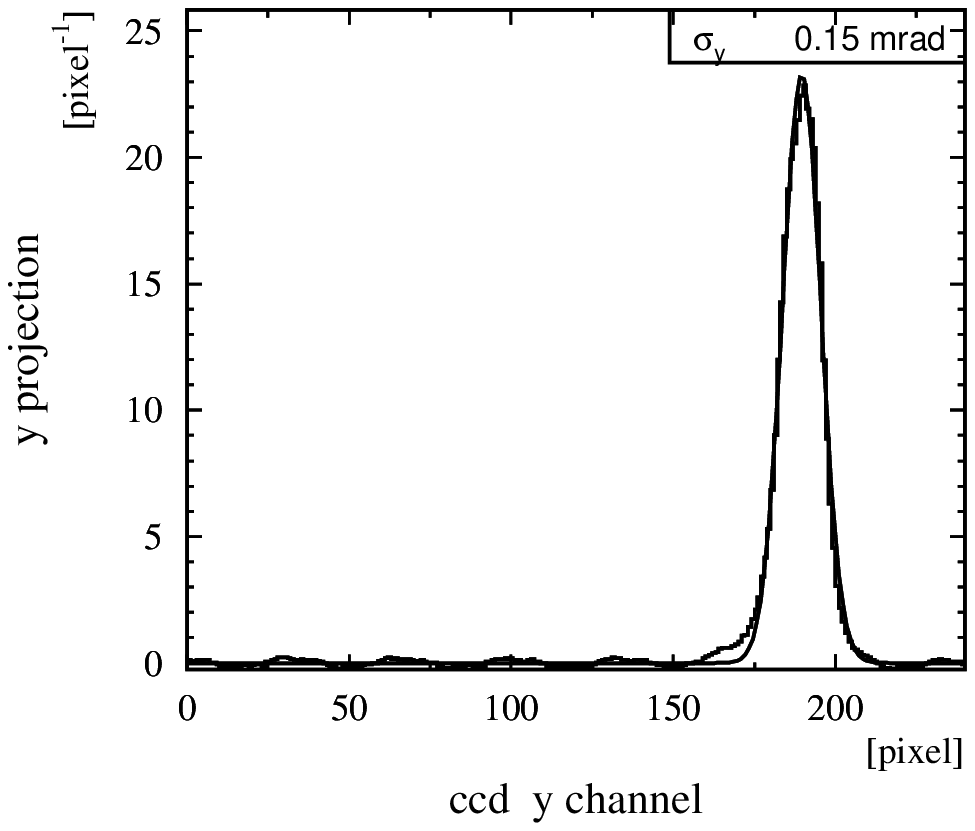,width=.49\textwidth}}
  \caption[Projections of the focal image of the second COI prototype]
  {Projections of the focal image of the second COI prototype.}
 \label{fig:mirror2000-peak-proj}
\end{figure}

Based on this experience the following steps were taken to improve
the local and the overall mirror shape while increasing the
thickness by as little as possible.
\begin{itemize}
  \item 50\% increase of number of layers for front panel to gain
   more resistance against plucking stresses
  \item 100\% increase of rib structure height to improve overall
  bending stiffness
  \item addition of five perpendicular ribs to the isogrid as
  shown in Fig.~\ref{fig:isogrid} to enhance spatial isotropy
  \item reduction of adhesive used for bonding to decrease
  plucking
\end{itemize}
The second prototype was delivered by COI in September 2000.
Figure~\ref{fig:mirror2000-peak} shows the focal image of the full
illuminated mirror using the setup shown in
Fig.~\ref{fig:setup-mirror}.
The most striking feature of Fig.~\ref{fig:mirror2000-peak} is the
absence of substructures like the ones in
Fig.~\ref{fig:coi1999-focal}. The overall slope error was
determined to $0.16$\,mrad by the width of the image projections
shown in Fig.~\ref{fig:mirror2000-peak-proj}. It is worth
stressing that design improvements resulted in tenfold increase of
the mirror quality.

\begin{figure}[htb]
  \centering
  \epsfig{file=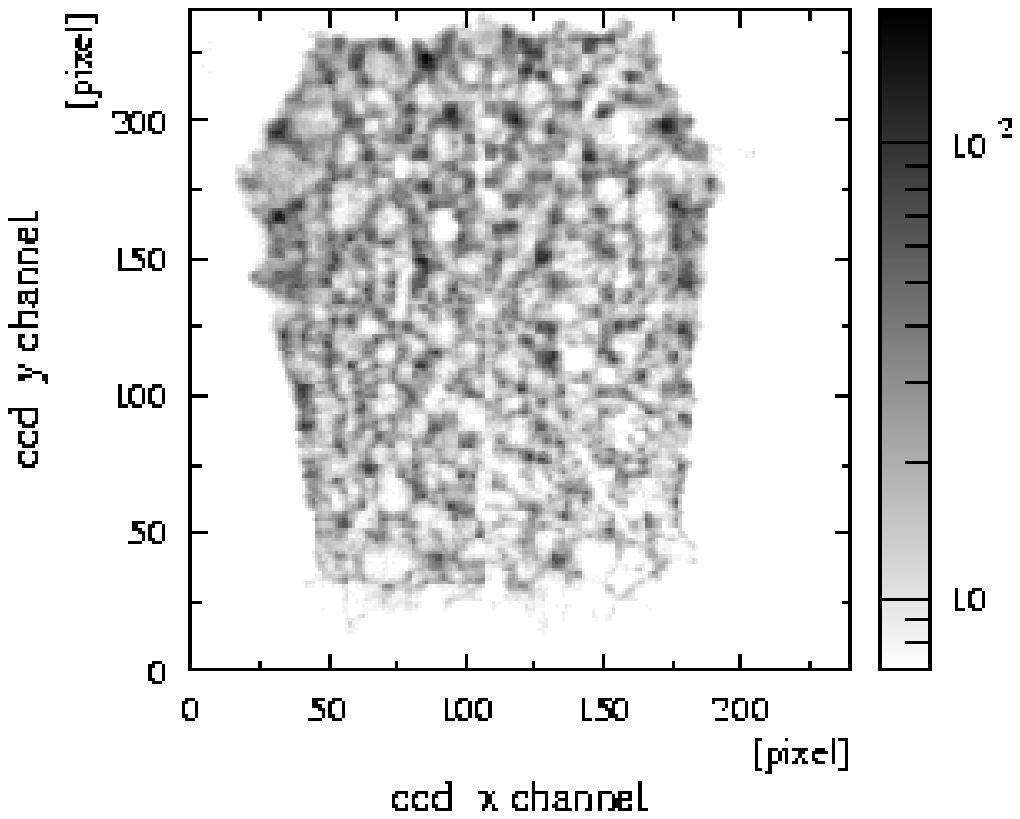}
  \caption[Quasi-Foucault image of the second COI prototype mirror]
  {Surface image of the second COI prototype mirror captured with a HeNe laser
  illumination~\cite{COI:1}. White spots indicate convex and dark spot
  concave distortions.}
  \label{fig:mirror2000-surface}
\end{figure}

The spot size is already close to the limit of the setup defined
by the divergence of the laser beam and the resolution of the CCD
camera. The remaining surface distortions can only be seen in the
Foucault image of the mirror surface shown in
Fig.~\ref{fig:mirror2000-surface}. The image was captured on a
white screen placed at a distance of about 5\,m from the mirror.
The brighter diagonal lines indicate a remaining print-through of
the long continuous isogrid rib possessing the same magnitude as
the local plucking distortions.

\begin{figure}
   \centering
  \mbox{
   \epsfig{file=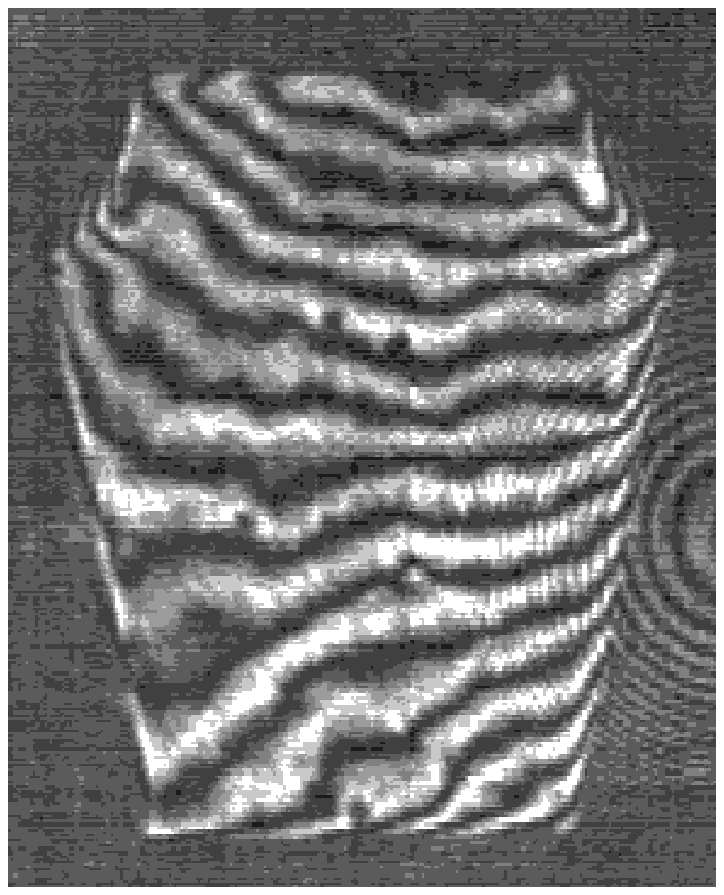,width=.49\textwidth}
   \hfill
   \epsfig{file=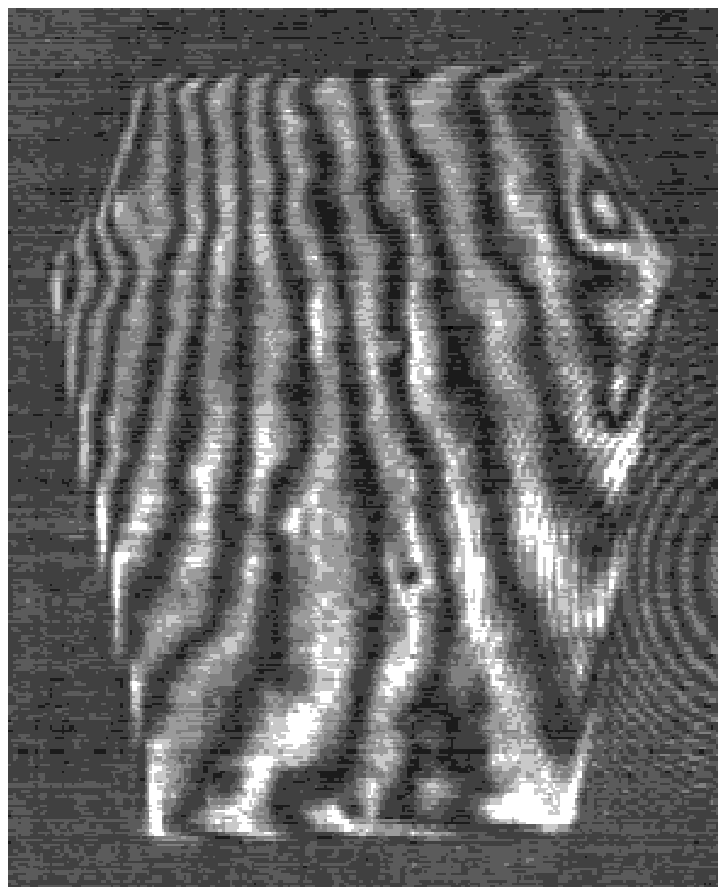,width=.49\textwidth}}
   \caption[Static interferogram of the second COI prototype]
   {Static interferogram for 15 horizontal (left panel) and 15 vertical (right panel)
    fringes across the aperture using a CO$_2$ interferometer~\cite{COI:1}.}
    \label{fig:COI-fringes}

    \vspace*{.8cm}
    \centering
    \epsfig{file=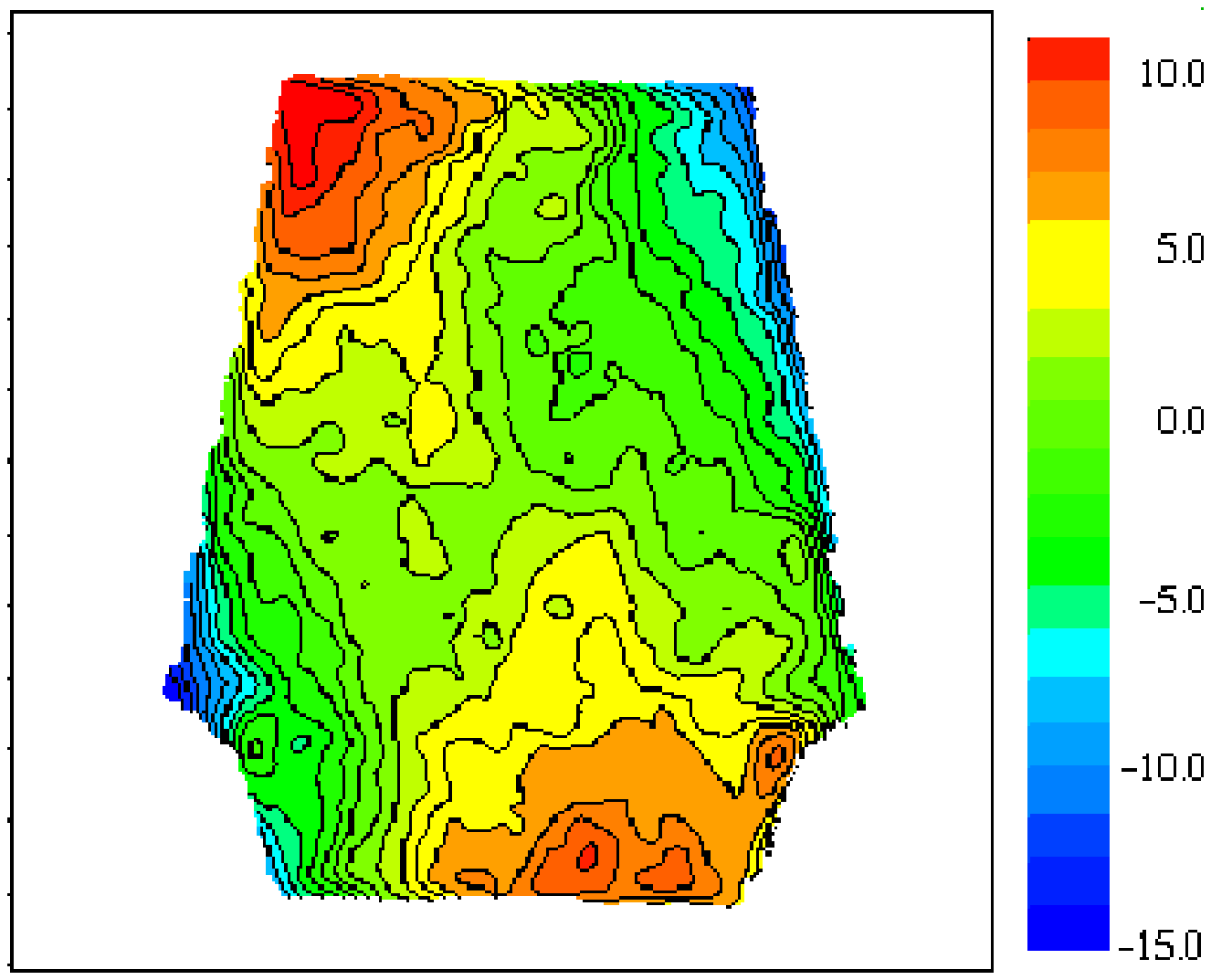,width=10cm}
      \caption[Surface error profile of the second COI proto\-type]
      {Contour plot of the surface error profile in microns~\cite{COI:1}. Positive/negative
      values represent concave/convex aberration feature, respectively.
      }
      \label{fig:COI error-profile}
\end{figure}

 To evaluate the local surface errors quantitatively, the
mirror was measured with phase shifting interferometry by COI
using a CO$_2$ laser with a wavelength of 10.6\,$\mu$m. The phase
shift was observed with respect to the position of a reference
mirror. Figure~\ref{fig:COI-fringes} shows the interference
pattern for 15 horizontal and for 15 vertical fringes.
The mirror surface structure shown in Fig.~\ref{fig:COI
error-profile} was reconstructed by deconvolution of both
interference signals.

A surface irregularity of 4.2\,$\mu$m\,(rms) was measured which is
equivalent to a peak-to-valley deviation of 25.8\,$\mu$m. Besides,
a small edge cusping at the corners was evident resulting in
first-order astigmatism of 4\,$\mu$m peak deviation for an average
angular profile. It is worth stressing that the central
160-mm-radius zone exhibits a very flat average radial profile
with less than 1\,$\mu$m peak deviation.

\section{Summary and outlook}

By use of the Monte Carlo detector simulation based on a realistic
hadronic cocktail it was demonstrated that the meson spectroscopy
capabilities of the CERES detector system would greatly benefit
from a new ultrathin RICH-2 mirror replacing the original thick
glass mirror. The properties and technical requirement of a such a
mirror were specified.

A survey of available mirror substrates led to the conclusion that
only mirrors made of carbon fibers could fulfill these
expectations. The evaluation of CFK mirror samples from different
vendors proved that high UV reflectivity and excellent imaging
quality, as such the most critical requirements, can be achieved
by the appropriate choice of carbon-fiber substrate and mirror
fabrication technology.

A novel approach to the stabilization of ultrathin CFK mirror
shells by an isogrid support structure was developed. It is
especially suitable for meson spectroscopy because the relative
fraction of dielectron pairs remaining in the resonance peak
compared to those shifted in mass due to bremsstrahlung is
enhanced and, thereby, the meson signal-to-background ratio is
improved.

The first prototype incorporating the new isogrid design was
manufactured by Composite Optics Inc.~in October 1999. Optical
measurements revealed large local deformations and a significant
astigmatism indicating insufficient substrate thickness and
support structure stiffness. Based on these findings and bound to
a tight schedule, the CERES collaboration decided that the
considerable risks involved in the implementation of the mirror
for run time in 2000 could not be justified by the expected
benefits. Furthermore, the ongoing study of mirror technologies
was to be finalized.

Several measures were taken in the fabrication of the second
prototype to optimize the mirror design. As a result, the overall
slope error of the mirror was reduced to $0.15$\,mrad, i.e.~ten
times. The residual astigmatism of $4\,\mu$m is negligibly small.
Such an excellent optical quality has never been achieved for an
ultrathin carbon-fiber mirror. The novel replication technique and
the simplicity of the isogrid design result in low manufacturing
cost. Thus, significant cost savings of more than 50\% could be
gained in comparison to the HADES RICH mirror which is fabricated
on the basis of carbon ceramic substrate
technology~\cite{Friese:2000}.

The second prototype arrived only shortly before the main
production run in Fall 2000 ruling out the implementation of a new
full-size mirror in the CERES RICH-2 detector. Nevertheless, the
development of an ultrathin carbon-fiber mirror in only three
years is an accomplishment considering that comparable projects
took more than 5 years until completion~\cite{Zeitelhack:1999ry}.
Future applications are to be expected driven by the recent
renaissance of RICH detectors in heavy ion physics and particle
physics~\cite{Zeitelhack:1999ry,Akiba:1999,DiBari:1999ua,Baum:1999rz}.
A summary of the mirror development is in preparation for
publication~\cite{Hering:2002}.

\chapter{Analysis of Pb-Au collision data at 158 GeV/c per nucleon}

\section{Introduction}

Any experiment attempting to measure low-mass dielectrons in
ultrarelativistic heavy ion collisions has to deal with major
experimental challenges. First, the large number of produced
particles leads to a high detector occupancy resulting in a
serious load on the detectors. Distinction of electrons from the 2
to 3 orders of magnitude more abundant hadrons requires a detector
system with superior particle identification capabilities. Only a
small fraction of all electrons come from {\em nontrivial} sources
other than $\gamma$-conversion and Dalitz decay $\pi^0 \rightarrow
e^{+}e^{-}\gamma$. Finally, uncorrelated electrons and positrons
originating from a large fraction of partially reconstructed pairs
form a huge combinatorial background when combined to pairs.

Although the CERES experiment has been designed to detect
electrons under such conditions, a sophisticated data analysis is
a necessity in order to extract a statistically significant
dielectron signal.

\noindent The data analysis consists of the following steps:
\begin{itemize}
  \item calibration of detector raw data
  \item reconstruction of hits in each detector
  \item combination of hits of all detectors to particle tracks
  including particle identification and momentum determination
  \item rejection of accidentally matching track segments and reduction of
  combinatorial background
  \item single-track efficiency correction by means of Monte Carlo detector simulation
  \item subtraction of the combinatorial background from the pair
  distribution
\end{itemize}

This Chapter is focused on the results of new studies and
developments. The key points of the chain of analysis will be
explained and details will be given where necessary for general
understanding. A thorough description of the previously used
detector calibration and hit reconstruction algorithms can be
found in~\cite{Ceretto:1998,Messer:1998,Voigt:1998,Lenkeit:1998}.
New detector calibrations developed in this paper are included at
pertinent places.

\section{Reasons for re-analysis of the 1996 data set}
\label{sec:deficiencies}

The 1996 data set has been analyzed twice
before~\cite{Sokol:1999,Lenkeit:1998}. The third analysis was
motivated by the following reasons.

First, in previous analyses the signal of correlated
electron-positron pairs was extracted by subtracting the mass
distribution of like-sign pairs from the unlike-sign spectrum.
This procedure assumes the combinatorial unlike-sign background to
exactly resemble the like-sign spectrum. Improper subtraction of
combinatorial background as a cause of the dielectron enhancement
observed was a serious concern, since the signal-to-background
ratio is very small (i.e.~1\,:\,13~\cite{Lenkeit:1998}).
Furthermore, signal and background distribution are similar in
shape. The enhancement observed could be explained by a 3.6\%
increase of the background as noted in~\cite{Lenkeit:1998}. Such a
situation could occur in case of an unrecognized asymmetry of the
detector with regard to detection and reconstruction of
unlike-sign and like-sign pairs.

Second, a smoothing procedure was applied to the like-sign
combinatorial background to reduce statistical error. In fact, the
statistical errors of the individual mass bins were shifted to a
systematic error of the integral distribution. A certain level of
subjectiveness resulting from the particular choice of the fit
function (see Fig.~5.12 in~\cite{Voigt:1998}) increased the
difficulties in understanding the systematic error of the
background subtraction.

In this paper, these particular questions will be addressed in the
discussion of the application of the mixed-event technique, an
alternative method for the construction and subsequent subtraction
of the combinatorial background. The comparison of the two
background subtraction techniques will allow a test and validation
of the assumptions inherent in both methods.

\begin{figure}[htb]
  \epsfig{file=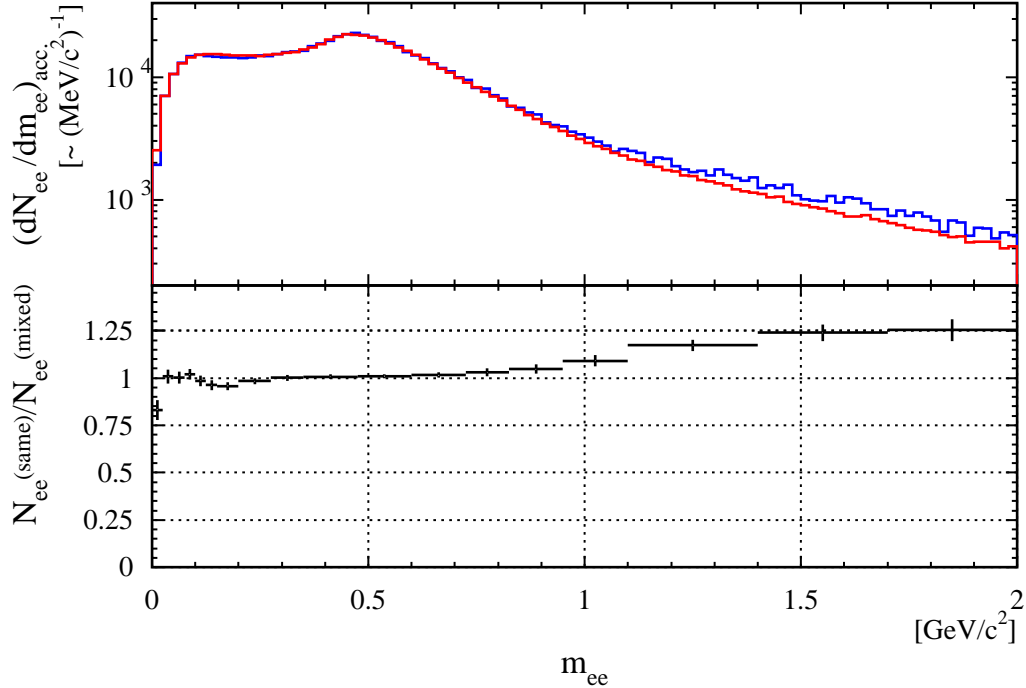}
  \caption[Comparison of mixed-event and same-event
  combinatorial background of the old raw data analysis]
  {Comparison of mixed-event and same-event combinatorial
   background based on the previous raw data analysis~\cite{Ceretto:1998,Messer:1998,Voigt:1998,Lenkeit:1998}.
   Mixed-event and same-event background differ by up to 25\% for mass above
   0.8\,GeV/c$^2$. This deviation is an indication for an artifact of the analysis procedure.}
  \label{fig:old-mixed}
\end{figure}

The first attempt to construct a mixed-event combinatorial
background was based on the results of the previous first-stage
analysis~\cite{Lenkeit:1998}. Figure~\ref{fig:old-mixed} shows the
mixed-event distribution to deviate strongly from the same-event
like-sign background.
In absence of other plausible explanations, a new first-stage
analysis had to be performed to exclude any artifacts of the hit
and track reconstruction or of the event selection applied for
data reduction.

Several minor problems of previous analyses were addressed in the
process of the data re-analysis with the mixed-event technique.
Most notably, a very complex background rejection strategy had
been used in~\cite{Voigt:1998,Lenkeit:1998} which was not easy to
reproduce. The main focus of this paper was placed on the
essentials and clarity with special emphasis on the rejection
strategy.

\section{Hit and track reconstruction}
\subsection{Technical aspects of the raw data analysis}

The first of two steps of analysis, hit and track reconstruction,
requires processing $8000$ GB of raw data, collected in the
recording of $6.0\cdot 10^{7}$ events. It was carried out at the
CERN EFF PC-farm using 40 CPUs for two weeks. The general results
are summarized in Table~\ref{tab:2000 production}.
\begin{table}[htb]
  \centering
   \begin{tabular}{|l|c|}
      \hline
      Analysis stage & Number of events \\
      \hline
      Recorded on tape    & 60.000.000 \\
      Available events    & 41.694.200 \\
      Successfully analyzed & 40.418.548 \\
      \hline
      Events with $\pi$-track & 34.523.208  \\
      Events with e-track &  25.170.135    \\
      Events with dielectron  &  \,\,\,2.380.071  \\
      \hline
  \end{tabular}
  \caption[Results of first-stage raw data analysis]
  {Results of first-stage raw data analysis.}
  \label{tab:2000 production}
\end{table}

Recent progress in data storage technology and the increase of
available computing power allowed to loosen the event selection
and to store all events with at least one electron or
high-momentum pion track. Each event was characterized by the
centrality and by the orientation of the reaction plane used only
in the hadronic flow analysis by~\cite{Slivova:2001}.

\subsection{SDD-hit reconstruction}
 \label{sec:sidc hits}

The major change at the first stage of the analysis was the
implementation of a novel algorithm to reconstruct hits in the SDD
developed by~\cite{Slivova:2001}. Before evaluating the impact of
new software on the dielectron analysis, the general concept of
SDD-hit reconstruction is explained with special emphasis on the
differences between the old and the new software version.

SDDs play a crucial role in the reconstruction of event vertex and
tracks and in the rejection of pairs of close tracks by either
resolving those or using the deposited-energy information for
discrimination. The SDD-hit reconstruction software is used to
determine the hit position taking into account effects such as the
noise of the electronics, pulse shape variations, and the
saturation of the pulse height. Overlapping hits are resolved as
far as possible.

The strategy previously applied was based on the clustering of
pulses of adjacent anodes to hits and determining the hit position
by a center of gravity method~\cite{Chen:1993sh}. Overlapping hits
were split only in anode direction in case of a local minimum
between adjacent pulses in a hit cluster. As noted
in~\cite{Voigt:1998}, this method produces a large fraction of
artificially split single hits deteriorating the single-track
resolution and the close-track rejection power. The new method was
developed to improve this situation~\cite{Slivova:2001}. It
employs a Gaussian pulse fit to find the hit center as well as a
special logic to recognize overlapping hits based on a
double-Gaussian fit of the pulse shape in time-bin direction and a
simple local minimum splitting in anode direction.

A Monte Carlo simulation of the double-hit reconstruction
efficiency by~\cite{Slivova:2001} is plotted in Fig.~\ref{fig:new
SDD hits}.
\begin{figure}
\centering
  \mbox{
   \epsfig{file=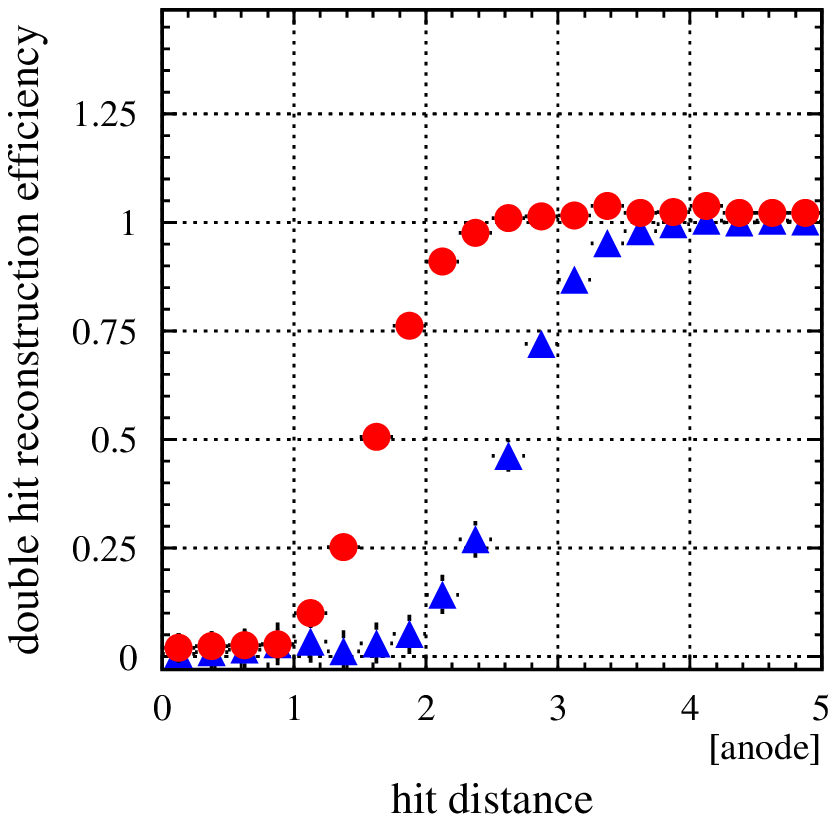,width=.49\textwidth}
   \hfill
   \epsfig{file=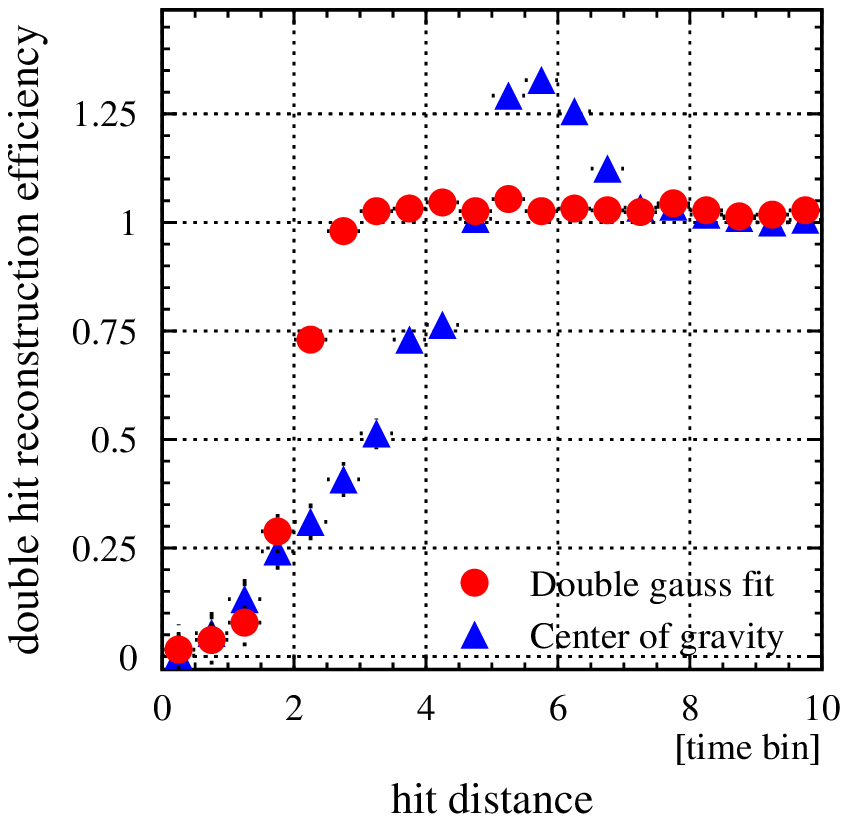,width=.49\textwidth}
  }
  \caption[SDD double-hit resolution]
  {Monte Carlo simulation of the SDD double-hit reconstruction efficiency achieved by means of
  the new\,(red) and the old\,(blue) SDD-hit software~\cite{Slivova:2001}.}
  \label{fig:new SDD hits}

  \vspace*{1cm}
  \epsfig{file=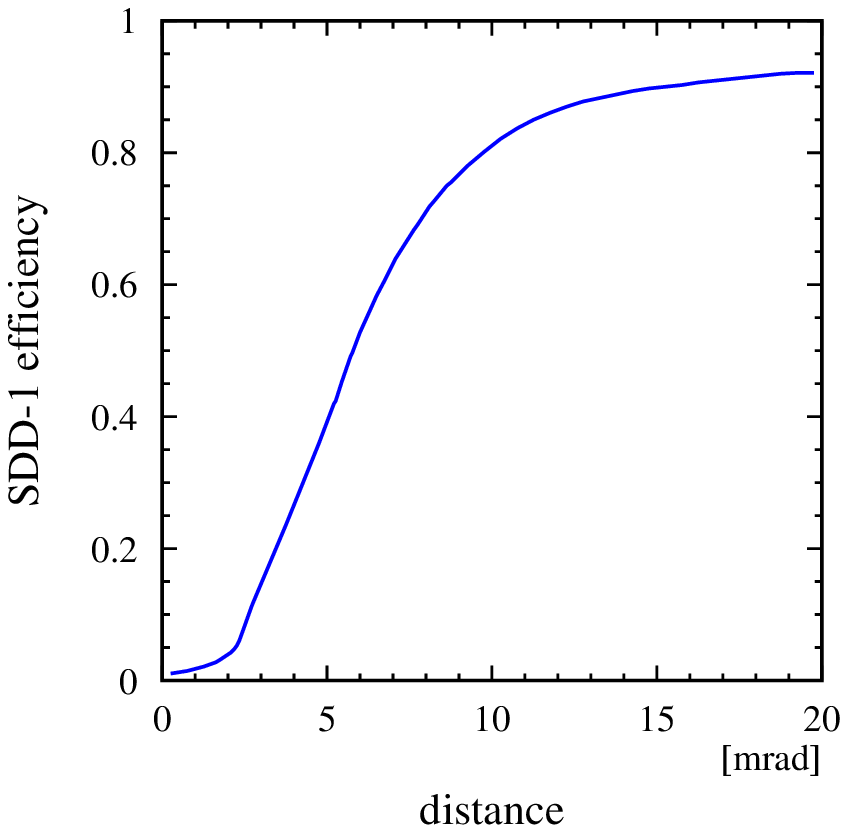,width=220pt}
  \caption[Double-hit reconstruction efficiency of SDD-1]
      {
  Double-hit reconstruction efficiency of SDD-1. Local constancy
  of the hit density provided, the distance between a hit belonging
  to a track and the next closest hit can be described by a
  Gaussian probability distribution. The finite double-hit resolution
  results in a drop in the observed distribution at the point of overlap
  of close hits. Reconstruction efficiency was determined from
  the relative difference between the distance distribution observed and
  a Gaussian distribution fitted to a distance range of
  non-overlapping hits. The effects of artificially split hits were removed
  by a linear approximation.
      }
  \label{fig:si resolu}
\end{figure}
With the old software a large excess of reconstructed pairs
appears in time-bin direction, an indication of artificial
splitting of single hits. This peak has disappeared after
introduction of the new hit reconstruction software.

The double-hit resolution, defined as the distance of
reconstruction of both hits with 50\% probability, is improved by
60\% for anode and time-bin direction. The latter is verified in
this paper by an evaluation of the measured distribution of hits
in the vicinity of a reconstructed track which gives a double-hit
resolution of 5.6\,mrad (see Fig.~\ref{fig:si resolu}).

The Monte Carlo simulation of the double-hit reconstruction
efficiency does not address the problem of artificially split
single hits because a split single hit could be mistaken for a
properly reconstructed double hit. A new approach based on
experimental data was taken in this paper to resolve this issue.

\begin{figure}
  \centering
  \mbox{
   \epsfig{file=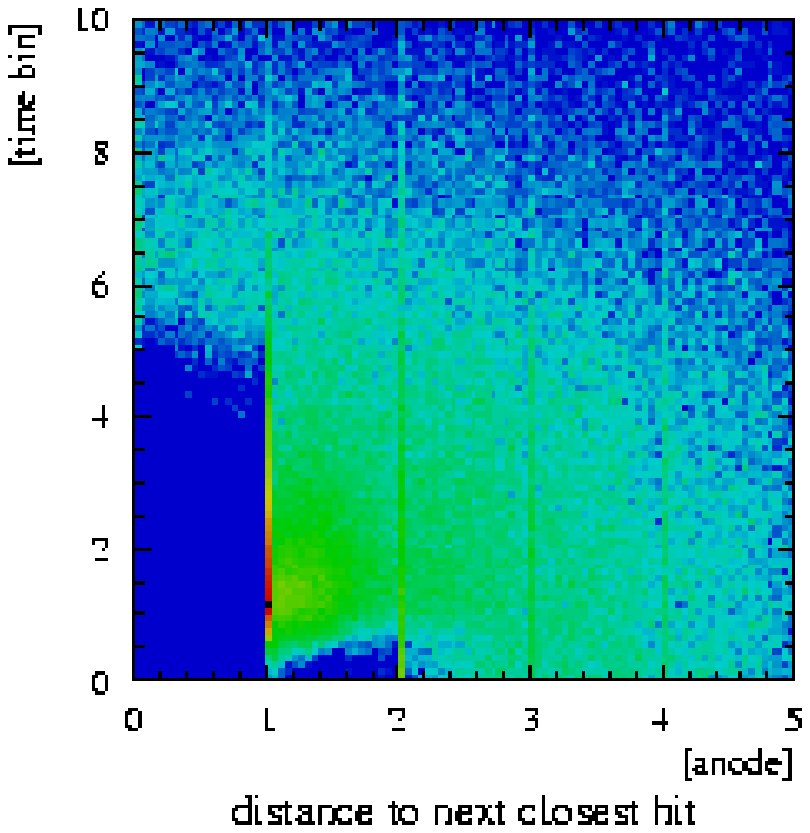,width=.49\textwidth}
   \hfill
   \epsfig{file=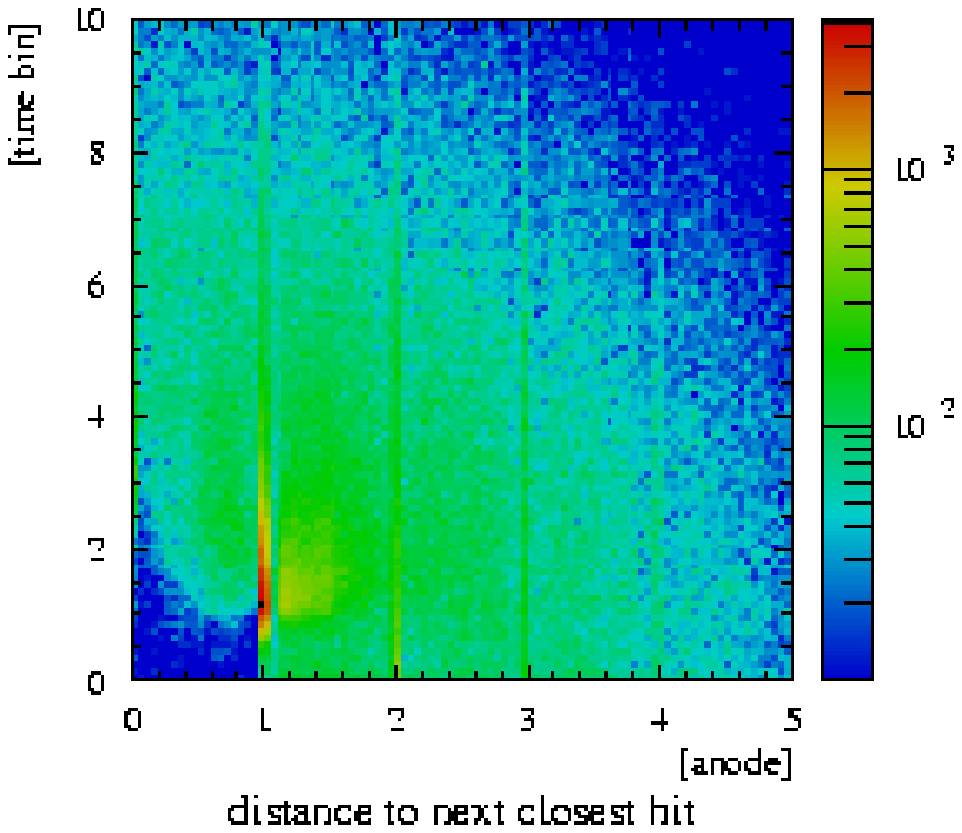,width=.49\textwidth}
  }
  \caption[Density distribution of the next closest hits in SDD-1]
  {Density distribution of the next closest hits in SDD-1. The old version
  of the SDD software (left panel) splits overlapping hits only in anode
  direction. Hence overlapping rings are not completely resolved and the
  double-hit resolution in time-bin direction is very poor. The new software
  version (right panel) splits overlapping hits in both anode and time-bin
  direction invoking a double-Gaussian fit. The peak shown in yellow and
  red color levels indicates artificial hit splitting.
 }
 \label{fig:splitting}
\end{figure}

Without artificially split hits, the SDD-hit density must be
uniform close to a hit belonging to an electron track. Only
information about the next closest hit was stored in the first
stage of analysis. The density distribution of the next closest
hit rapidly decreases with increasing distance to a given track.
It can be approximated by a Gaussian probability distribution. One
part of a hit artificially split belongs to a track while the
other results in an enhanced density of next closest hits. The
finite double-hit resolution causes a depletion in the vicinity of
a hit. The observed density distribution of next closest hits is
plotted in Fig.~\ref{fig:splitting}.
The old software (left panel) shows a huge excess due to
artificial splitting at a hit distance range of 1 to 2 anodes and
1 to 3 time bins. Overlapping hits in time-bin direction were not
split which leads to a depleted hit density up to a distance of 5
time bins. In contrast, the new software shows this area filled
resulting from the additional splitting in time-bin direction.
However, a significant peak indicating artificially split single
hits is still present.

\begin{figure}
 \centering
 \mbox{
   \epsfig{file=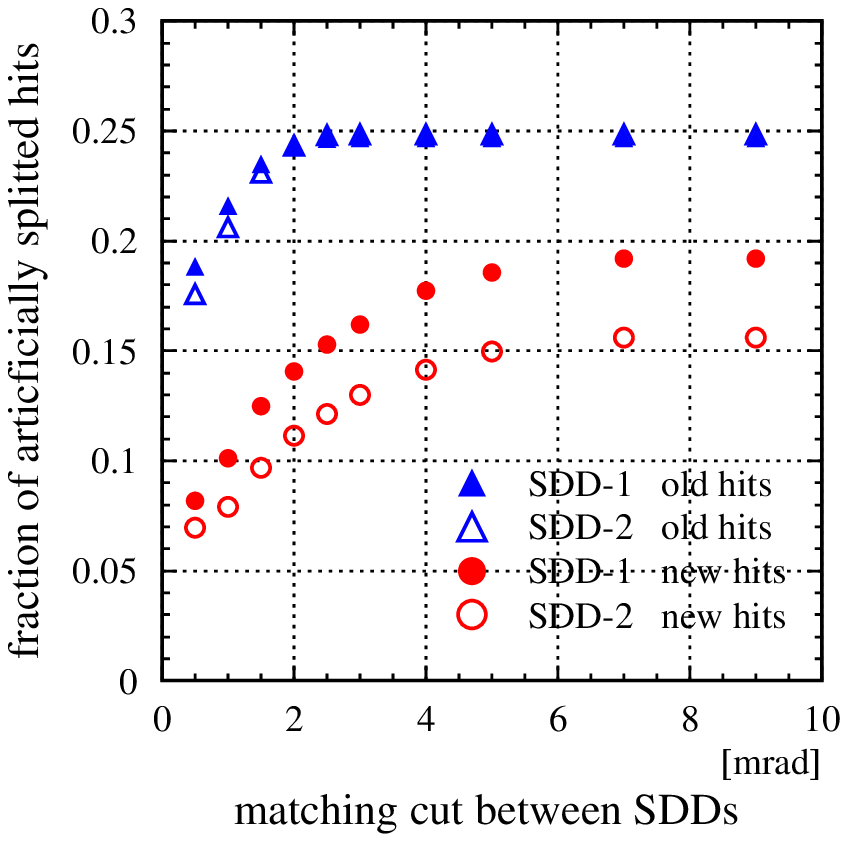,width=.49\textwidth}
   \hfill
   \epsfig{file=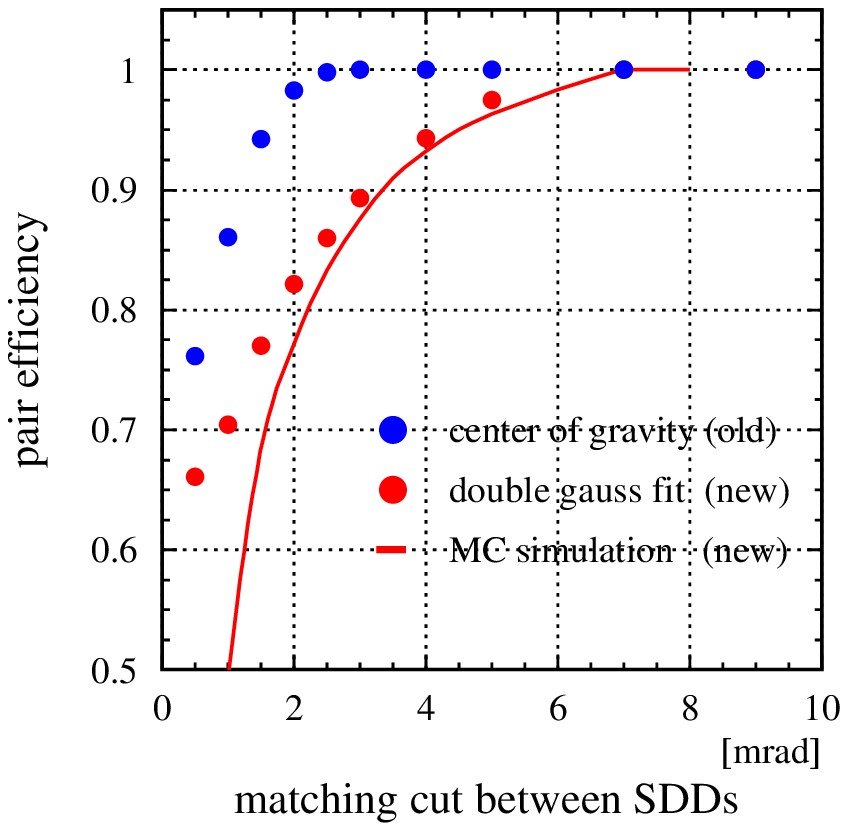,width=.49\textwidth}
  }
  \caption[Impact of artificial hit splitting in the SDD.]
  {Impact of artificial hit splitting in the SDD. Fraction of artificially split hits as a function of the
      matching quality cut of SDD-1 and SDD-2 (left panel).
       Contrast of the pair efficiency loss caused by artificial hit splitting
      in the SDD for new and old software version (right panel). The
      efficiency loss estimated in data was verified by an overlay Monte Carlo simulation.
      Considering the difference of the SDD-matching cuts applied in the previous
      analysis (0.9\,mrad)~\cite{Lenkeit:1998} and this paper (1.3\,mrad), an additional 8\%/16\%
      track/pair reconstruction efficiency loss was estimated for the new
      version.}
 \label{fig:split-rel}
\end{figure}

The fraction of artificially split hits was estimated for a
quantitative study by the integrated yield of the observed excess
relative to the number of regular background hits. Contrasting
both software versions in Fig.~\ref{fig:split-rel} (left panel)
shows the total fraction of artificially split hits to be reduced
from 25\% (old version) to 19\% (SDD-1) and 16\% (SDD-2) (new
version).

The artificial hit splitting in both SDDs occurs independently. In
particular, it is independent of the quality cut applied to the
matching of SDD-1 and SDD-2 hits unless the position of
artificially split hits is shifted. Figure~\ref{fig:split-rel}
(left panel) reveals the relative fraction of split hits in the
new version to decreases rapidly with increasing matching quality
exactly as one would expect if an increasing fraction of
artificially split hits were rejected due to poor matching. This
feature results in an additional pair efficiency loss as
illustrated in Fig.~\ref{fig:split-rel} (right panel). Although
the new software reduces the fraction of split hits considerably,
those that are still artificially split exhibit a larger position
shift (i.e.~comprise smaller hit fractions and/or more
single-anode hits) which in turn results in a 8\%/16\% track/pair
reconstruction efficiency loss compared to the old software.

The study of this effect is not yet completed to date but is
considering the following possible explanations: the Gaussian
double-hit fit is more sensitive to distorted pulse shapes
compared to the center of gravity method and the parameters
controlling the double-hit splitting in anode direction need to be
re-adjusted to accommodate small variation in the drift-time
calibration or the pulse shape between adjacent anodes.

\subsection{Outline of the tracking strategy} \label{sec:tracking}

After completion of the reconstruction of hits from the detector
signals, the spatial hit positions are utilized to construct
particle tracks. The tracking strategy applied in this analysis
closely follows the approach taken in~\cite{Lenkeit:1998}. It
encompasses the following successive steps:
\begin{itemize}
  \item matching of SDD-1 and SDD-2 hits
  \item vertex determination as interception of all SDD track
  segments
  \item reconstruction of RICH-1 Cherenkov rings utilizing SDD
  track segments  as pointer to select proper ring candidates
  \item combination of fitted RICH-1 rings and RICH-2 rings/PD hits
  to RICH-PD track segments taking into account the inhomogeneous B-field
  \item final matching of SDD and RICH-PD track segments to
  complete particle tracks
\end{itemize}
The particular purpose of this strategy is to aid the ring
recognition algorithm in both RICH detectors by external tracking
information of SDDs and PD. All tracking steps were put under
strict scrutiny in search for artifacts that could lead to a
biased track reconstruction. The resulting improvements and other
important modifications are documented below. The matching of the
detectors measuring angles (RICH-1/2) and those measuring spatial
coordinates (SDD, PD) has been changed everywhere. It is now based
on solid angles rather than polar and azimuthal angles not
reflecting spherical symmetry.

\subsection{Reconstruction of SDD track segments}
\label{sec:sidc-tracking}

\begin{figure}
  \centering
  \epsfig{file=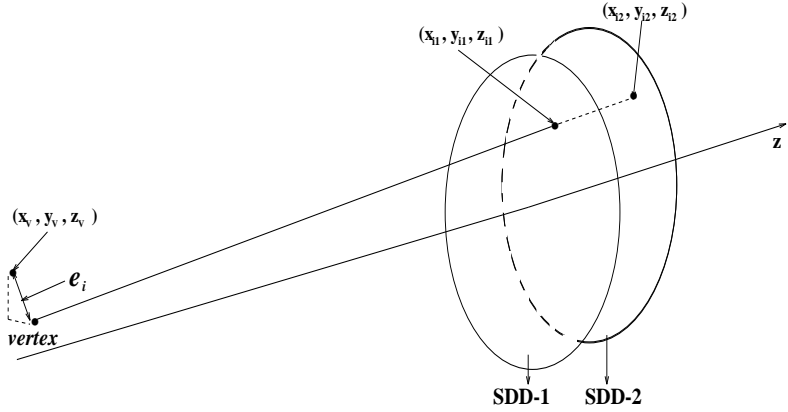,width=12cm}
  \caption[Illustration of the SDD-vertex tracking]
  {Illustration of the SDD-vertex tracking. The vertex is defined as the
   spatial point where the weighted sum of its squared distance $e_i$ to
   all trajectories of matching SDD-1 and SDD-2 hits is minimal.}
  \label{fig:sidc tracking}
\end{figure}

The most likely position of the event vertex is determined by an
optimization procedure (Robust Vertex
Fit~\cite{Agakishiev:1997sq}). The vertex is defined as the
spatial point where the weighted sum of its squared distance to
all trajectories of matching SDD-1 and SDD-2 hits is minimal.
Next, SDD track segments are created. A simple predictor pointing
to SDD-1 is calculated for each hit in SDD-2 (see
Fig.~\ref{fig:sidc tracking}).

Previously, a $4^{\circ}$ tilt of both SDDs~\cite{Messer:1998}
with respect to the beam axis was neglected in the predictor
determination. To include this effect, an improved predictor
method was developed using analytic geometry. Next, a
binary-search algorithm is applied to find the SDD-1 hit closest
to the SDD-2-vertex predictor. This algorithm was modified because
it was failing in certain rare cases. The size of the search
window of the best-matching hit was adjusted to always cover the
maximum matching window of $7$\,mrad.

To avoid ambiguous rejection of track segments within the limit of
the detector resolution, tracks sharing a common hit in SDD-2
(so-called reversed V$_{\rm SDD}$-tracks) were allowed in addition
to track segments sharing a common hit in SDD-1 (V$_{\rm
SDD}$-tracks).

These changes also apply to the second stage of SDD-vertex
tracking where the z-position of the reconstructed vertex is
refined to the exact position of the target disk. An optimized x-y
vertex position is obtained by minimizing the sum residual
distance of the initially found SDD-vertex tracks pointing to the
target. After refinement of the vertex position, an optimized set
of SDD-vertex track segments is obtained following the steps
described above. This procedure improves the SDD-vertex pointing
resolution by $20\%$. Further optimization of the code
implementation results in a $20\%$ reduction of the overall
execution time.

\subsection{Reconstruction of RICH-PD track segments}
 \label{sec:rich tracking}

Particle tracking is closely interlocked with electron
identification and momentum determination as described in
Sec.~\ref{sec:B field} because of the particular setup of the
CERES detector. Electron track reconstruction in the RICH-PD
detector system therefore has to meet all of the following
principal objectives: highest possible tracking efficiency,
sufficiently good pointing accuracy for minimal probability of
accidental matches with fake rings, and precise momentum
determination. Independent fulfillment of these requirements is
restricted by: the momentum dependent multiple scattering (see
Eq.~\ref{equ:scatter}), the $\theta$-deflection caused by the
second-order-field effect (see Fig.~\ref{fig:second order B} in
Sec.~\ref{sec:B field}), and small deviations of the magnetic
field from the nominal B(r)\,$\sim$\,1/r dependence.

Best reconstruction efficiency for low-momentum tracks is vital to
reduce the combinatorial background.  Higher-order effects are
important and are to be treated carefully especially for these
tracks. The previous tracking strategy of
~\cite{Ceretto:1998,Messer:1998,Voigt:1998,Lenkeit:1998} was
replaced by a new tracking strategy based on a detailed simulation
of the CERES detector including all higher-order effects. It
grounds in several basic ideas as follow. First, all previously
fitted RICH-1 rings and all RICH-2-ring candidates are combined
provided they fall within a {\em butterfly}-shaped matching window
described by:
\begin{equation}
 \label{equ:butterfly}
 \Delta\theta_{(\rm RICH-1\,-\,RICH-2)}=\sqrt{\sigma_{\rm resolution}^2+(\sigma_{\rm
 scattering}/p)^2}\;,
\end{equation}
accounting for multiple scattering and detector resolution.
Moreover, $\phi$-deflection in the magnetic field has to be less
than $0.9$\,rad (corresponding to $p$\,$>$\,$150$\,MeV/c).

The expected $\theta$-deflection in RICH-2, caused by the
second-order-field effect, is calculated for each combination of
rings as a function of the $\phi$-deflection and the radial
position of the track. The latter dependencies reflect the
residual non-linear contribution of the magnetic field. The radial
shift of the vertex with respect to the radial symmetry of the
magnetic field is also included.

\begin{figure}[!t]
    \begin{minipage}[t]{.60\textwidth}
        \vspace{0pt}
        \epsfig{file=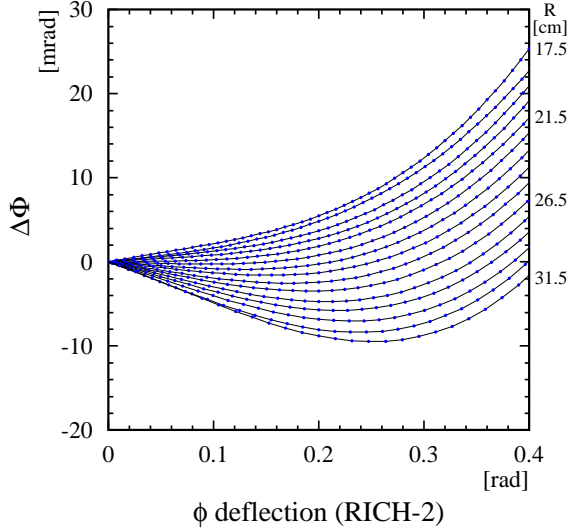,width=\textwidth}
    \end{minipage}%
    \begin{minipage}[t]{.40\textwidth}
      \vspace{0.5cm}
      \caption[New PD $\phi$-predictor function]
      {\newline
       GEANT simulation of the difference between the $\phi$-measurement in PD and RICH-2
       depending on the radial track positions $R$.
       The average deflection measured in the PD was subtracted:
       $\Delta\Phi \equiv \Delta\phi_{\rm SDD-PD}-0.64\Delta\phi_{\rm
       SDD-RICH-2}$. The functional dependence was fitted by a fourth-order polynomial
       to obtain the new PD-$\phi$-predictor function.
      }
      \label{fig:pred func}
    \end{minipage}
\end{figure}

After this correction, the remaining difference in
$\theta$-direction between RICH-1 and RICH-2 is attributed to
multiple scattering and detector resolution. A follow-your-nose
approach was applied to achieve maximum tracking efficiency. With
multiple scattering occurring mainly in the RICH-1 mirror, a
correlated shift in all downstream detectors, namely PD and
RICH-2, must follow. Multiple scattering in azimuthal direction
however cannot be distinguished from the $\phi$-deflection by the
magnetic field and is directly contributing to the momentum
resolution. Next, a predictor is computed into the PD. It includes
multiple scattering in $\theta$-direction and the
second-order-field effect. The predictor function relating RICH-2
and PD coordinates is depicted in Fig.~\ref{fig:pred func}.
It was obtained by tracking $10^6$ particles through the CERES
detector setup using the GEANT simulation
package~\cite{GEANT:1993a}. The expected ring center position in
the RICH-2 detector was determined from the average orientation of
the particles momentum vector while traversing the RICH-2 radiator
volume rather than from the center of the fitted RICH-2 ring. The
latter method becomes inaccurate for low momentum tracks because
the ring shape observed is distorted by the second-order-field
effect.

\begin{figure}[!t]
    \begin{minipage}[t]{.60\textwidth}
        \vspace{0pt}
        \epsfig{file=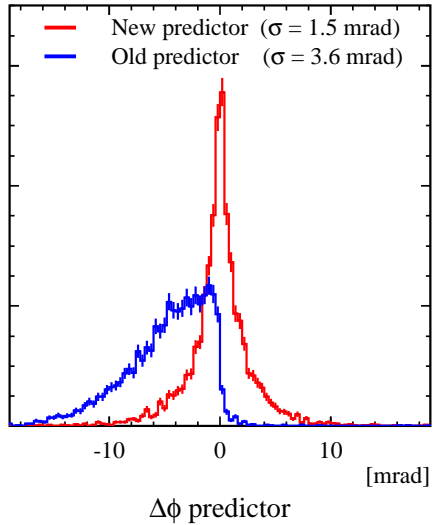,width=\textwidth}
    \end{minipage}%
    \begin{minipage}[t]{.40\textwidth}
      \vspace{0.5cm}
      \caption[Accuracy of the new PD $\phi$-predictor function]
      {\newline
      Deviation of new and old predicted $\phi$-position from the true hit
      position simulated in GEANT including multiple scattering. The
      $\Delta \phi$ distribution of the old version is not only much
      broader but also exhibits a general offset of about $-4$\,mrad
      with respect to the true hit position.
      }
      \label{fig:pd predictor-dphi}
    \end{minipage}
\end{figure}

Figure~\ref{fig:pd predictor-dphi} illustrates the accuracy of the
new PD phi-predictor $\Delta\Phi_{\rm PD}$ compared to the
previously employed version.
The $\Delta \phi$ distribution of the old version is not only much
broader but also exhibits a general offset of about $-4$\,mrad
with respect to the true hit position. The matching of PD hits and
RICH-1/2 segments was little affected because the $\phi$-offset
was partly counterbalanced by a 3\,cm misalignment of the PD
z-position.

In the next tracking step, all RICH-2 ring candidates are fitted,
provided a PD hit was found close to its predictor. All complete
RICH-PD track segment are stored for subsequent matching to the
previously constructed SDD-vertex track segments.

The maximum search window for the closest PD hit was modified to a
fixed solid angle (as opposed to the formerly fixed rectangular
matching window in x-y pad coordinates which does not reflect the
appropriate symmetry of the detector).

Sometimes more than one track segment shares either the same
RICH-1 or RICH-2 ring. The first case represents unresolved
dielectrons (called V-tracks) with a small opening angle for that
the segments have opposite charge. They are kept for further
studies. In previous analyses, all other multiple matches were
rejected by choosing the track with the best $\theta$-match
between RICH-1 and RICH-2 detector. This procedure favors
reconstruction of high-momentum tracks which are less affected by
multiple scattering.  The charge determination is refined at a
later stage of the analysis, based on the more precise measurement
of the deflection between SDD and PD instead of the initial
combination of RICH-1 and RICH-2 detector. The pair-charge
dependence of this rejection introduces a subtle bias for
high-momentum tracks.

With the new tracking, all multiple matches were kept for later
evaluation based on an improved deflection determination and a
further rejection of fake rings by additional quality cuts. The
surviving V-tracks of the same charge were used to evaluate the
fraction of accidental matches of unlike-sign V-tracks. The few
multiple matches that remained unresolved after the quality cut
were finally rejected to avoid a tedious selection procedure.

\begin{figure}[bt]
    \begin{minipage}[t]{.65\textwidth}
        \vspace{0pt}
        \epsfig{file=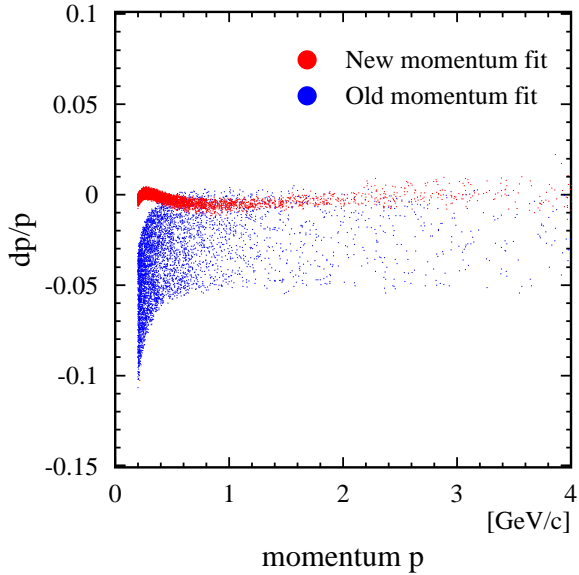,width=\textwidth}
    \end{minipage}%
    \begin{minipage}[t]{.35\textwidth}
      \vspace{0.5cm}
      \caption[Relative error of the reconstructed momentum]
      {\newline
       GEANT simulation of the relative error of the momentum determination
       excluding multiple scattering. The momentum reconstructed with the old software
       exhibits an overall 2.5\% offset which increases for small
       momenta.
      }
      \label{fig:momentum fit}
    \end{minipage}
\end{figure}

By means of GEANT detector simulation, the relation between the
momentum of the electron and its azimuthal deflection in the
magnetic field was determined including higher-order corrections.
Figure~\ref{fig:momentum fit} shows the momentum resolution
contrasting new and old version of the momentum fit. The relative
error of the new method of $1\%$ is small compared to the observed
resolution (see Sec.~\ref{sec:presol}). The $2.5\%$ momentum
offset for the old version is caused by a previously unnoticed
misalignment of the PD z-position. This shift was independently
confirmed by a measurement of reconstructed D-meson mass
~\cite{Petracek:2001}.

Which of the software modifications finally led to the convergence
of same-event and mixed-event background (see Sec.~\ref{sec:mix vs
comb}) could not be verified because the necessary repetitions of
first-stage analysis were prohibited by their excessive need of
computing power, storage space, and time.

\clearpage

\section{Global calibration of the spectrometer}

\subsection{Intercalibration of detectors}

High-precision alignment of all detectors is of crucial importance
for efficient tracking and subsequent rejection of accidental
matches. The external distance measurements of individual detector
with respect to a fixed laboratory frame, made during assembly, do
not have sufficient precision. But they provide initial values and
constraints for an intercalibration with a data sample of
reconstructed high-momentum pions. Compared to electrons,
high-momentum pions ($p$\,$>$\,$4.5$\,GeV/c) are no much affected
by multiple scattering and, therefore, more suitable for precise
detector alignment. Starting with the autocalibration of the SDDs
and the reconstruction of the event vertex, described in
Sec.~\ref{sec:sidc-tracking}, all identified pion tracks are
matched to the independent reference of PD hits and RICH rings.
Each detector was aligned in x-y coordinates with respect to the
center of SDD-1 which was used as reference point. Any residual
offsets in $\theta$-direction can be removed by applying a radial
correction to the local drift velocity of the SDDs.

\begin{figure}[hbt]
  \centering
  \mbox{
   \epsfig{file=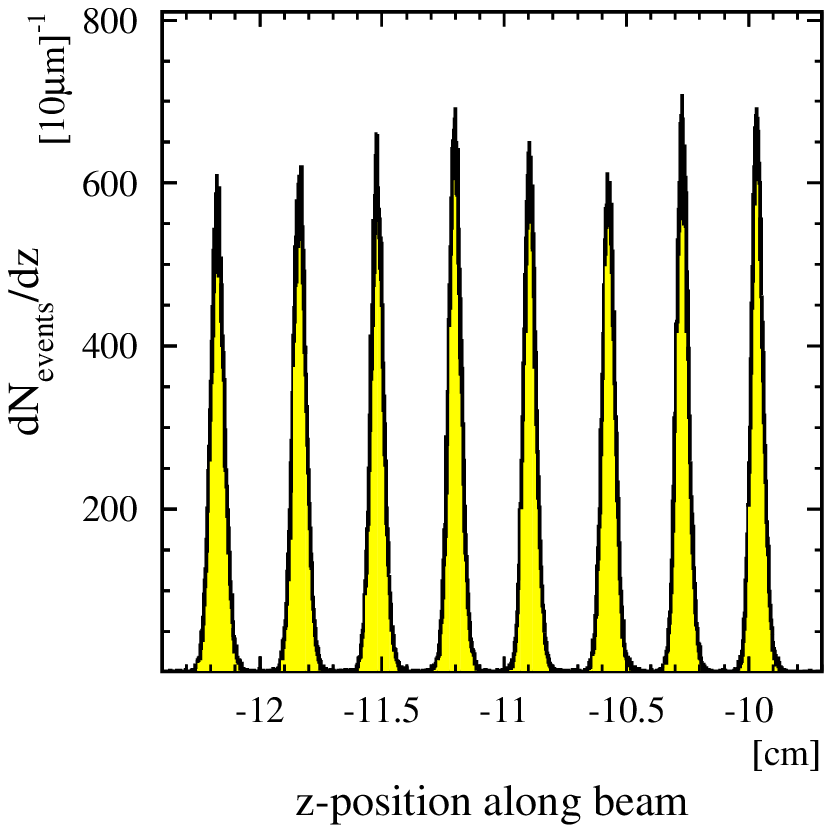,width=.49\textwidth}
   \hfill
   \epsfig{file=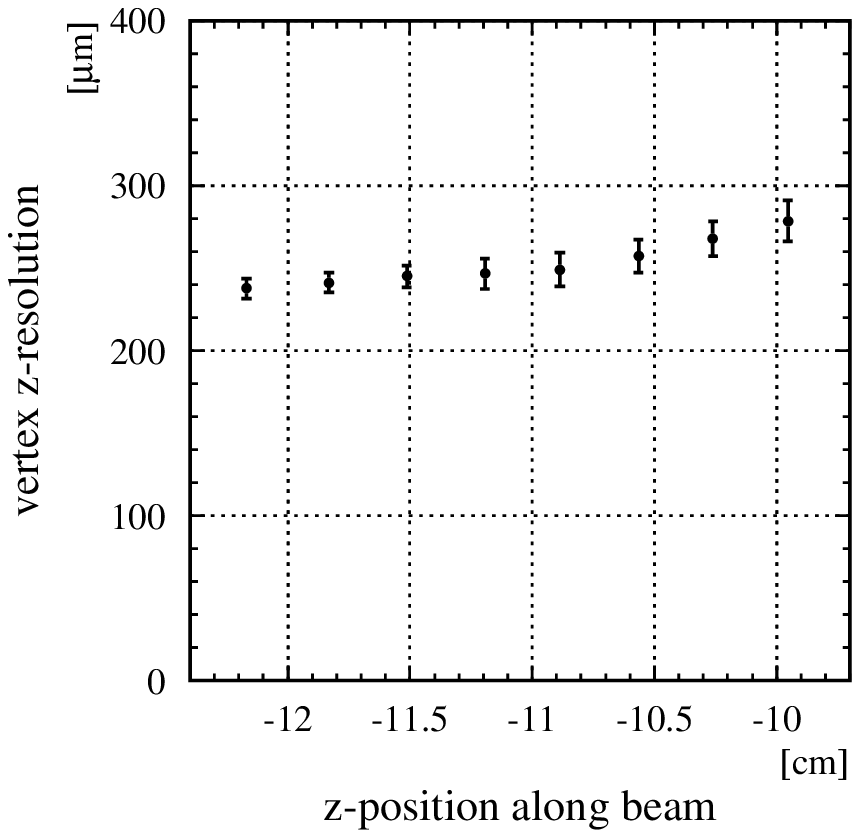,width=.49\textwidth}
  }
  \caption[Quality of the vertex reconstruction]
  {Reconstructed vertex z-positions for run 230 (left panel). The vertex
  distributions of the individual target disks have a width
  of $\sigma=0.24$--$0.28$\,~mm (right panel). The width is small compared to
  the separation of adjacent targets of about 2.4\,mm. The vertex z-resolution decreases
  with increasing distance between target and SDDs for geometry reasons.
  }
\label{fig:vertex}
\end{figure}

Using this calibration strategy, a high-precision alignment of the
'96 detector setup was carried out prior to this paper. Details of
the calibration are documented in~\cite{Ceretto:1998,Messer:1998}.
The attempt to reuse the calibration parameters for this analysis
failed, as they were unable to reconstruct the vertex at the
nominal target positions. An investigation revealed the positions
of the SDD hits, obtained with the new fitting algorithm, to have
been subject to a systematic shift due to the non-Gaussian pulse
shape. The intercalibration of SDD-detector system had to be
redone. A simple study of the calibration parameters was carried
out to avoid an elaborate recalibration of the Silicon-vertex
telescope. The correct spacing of the target disks was regained by
decreasing the outer-radius parameter of SDD-1 by 24\,$\mu$m
resulting in a systematic expansion of the radial scale. The
residual time-dependent offset between the nominal target and the
reconstructed vertex position was used to compute a small
correction factor for the local drift velocity in both SDD
detectors for each run. After this correction, the reconstructed
vertex distributions were centered at the nominal position of the
target disks as shown in Fig.~\ref{fig:vertex} (left panel).
The individual vertex distributions are clearly separated. This
allows for unequivocal identification of the target disk in which
the interaction took place. The SDD-track reconstruction can be
refined with the knowledge of the exact vertex z-position.

The higher double-track resolution of the new SDD-hit
reconstruction software (see Fig.~\ref{fig:new SDD hits}) improved
the resolution of the vertex z-position by 20\,$\mu$m to
260\,$\mu$m (compare~\cite{Messer:1998}). For geometry reasons
explained in~\cite{CERES:02-05} the vertex z-resolution is about
15\% better at the far side of the SDD as seen in
Fig.~\ref{fig:vertex} (right panel).

\begin{figure}
  \centering
  \epsfig{file=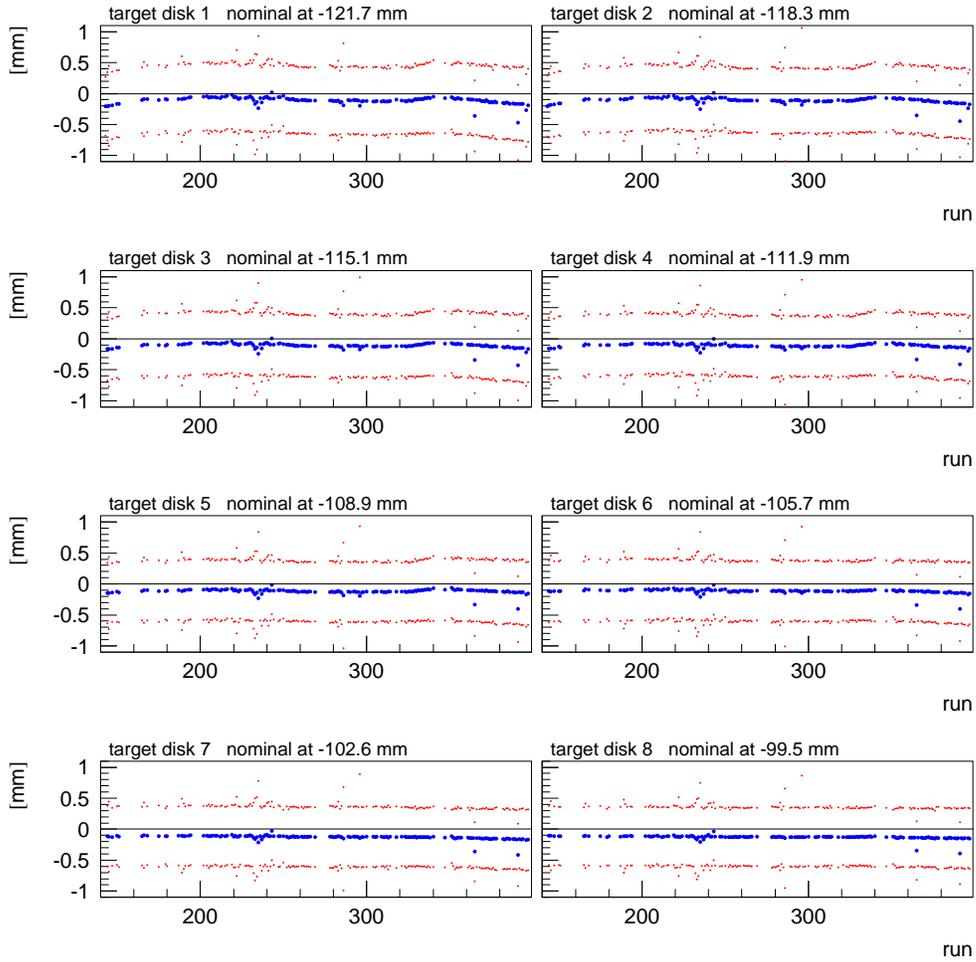,width=14.5cm}
  \caption[Run-to-run stability of the vertex reconstruction]
  {Run-to-run stability of the vertex reconstruction.
  The $\sigma$-width of the vertex fit is indicated by the red dots.
  The few runs exhibiting a larger $\sigma$-width contain less
  than the average number of events
  per run. The residual difference between reconstructed and
  nominal vertex z-position is small compared to the width of the
  distribution and is therefore neglected.}
  \label{fig:r2r vertex disk}
\end{figure}

The vertex z-resolution could be further improved by $15\%$ in
case of best calibration of the SDD-vertex
telescope~\cite{CERES:02-05}. Figure~\ref{fig:r2r vertex disk}
shows the time variation of the mean and the width of the
reconstructed-vertex distribution for each target disk. The
residual offset of about $100$~$\mu$m leads to a very small
systematic shift of the radial($\theta$-) matching of $0.15$\,mrad
which is negligible compared to the resolution of the SDDs.

\subsection{Matching distributions}

Internal consistency and quality of the readjusted calibration can
be evaluated by looking at the residual offset of the centroid,
the shape, and the width of the matching distributions of all
detector combinations.

\begin{figure}[tb]
\centering
   \mbox{
    \epsfig{file=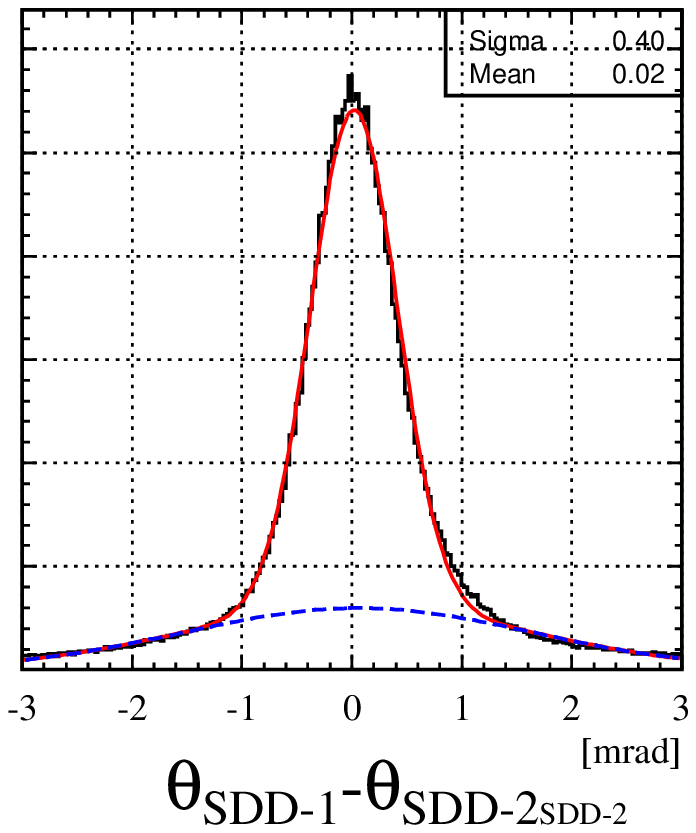,width=.33\textwidth}
    \hfill
    \epsfig{file=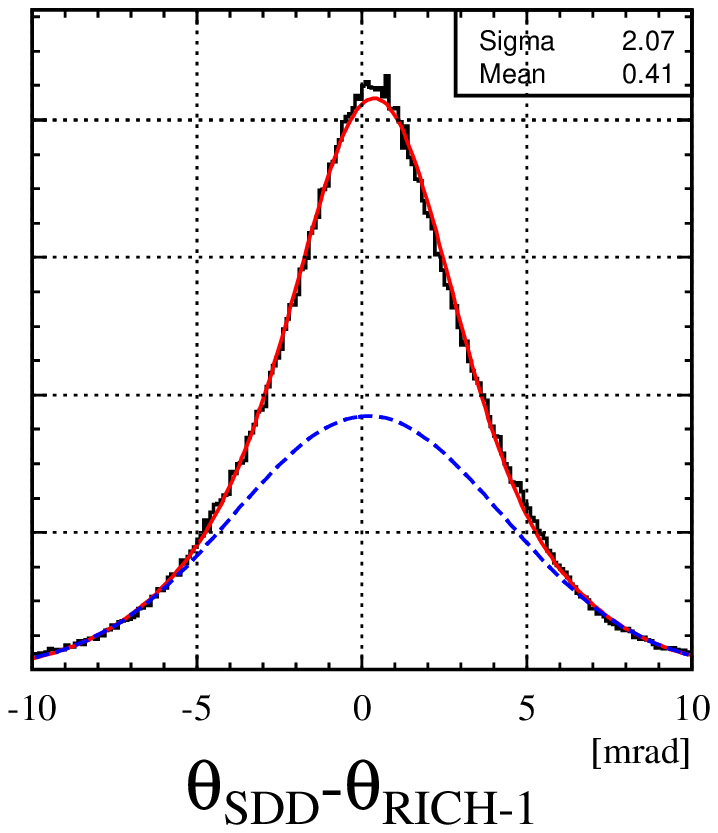,width=.33\textwidth}
    \hfill
    \epsfig{file=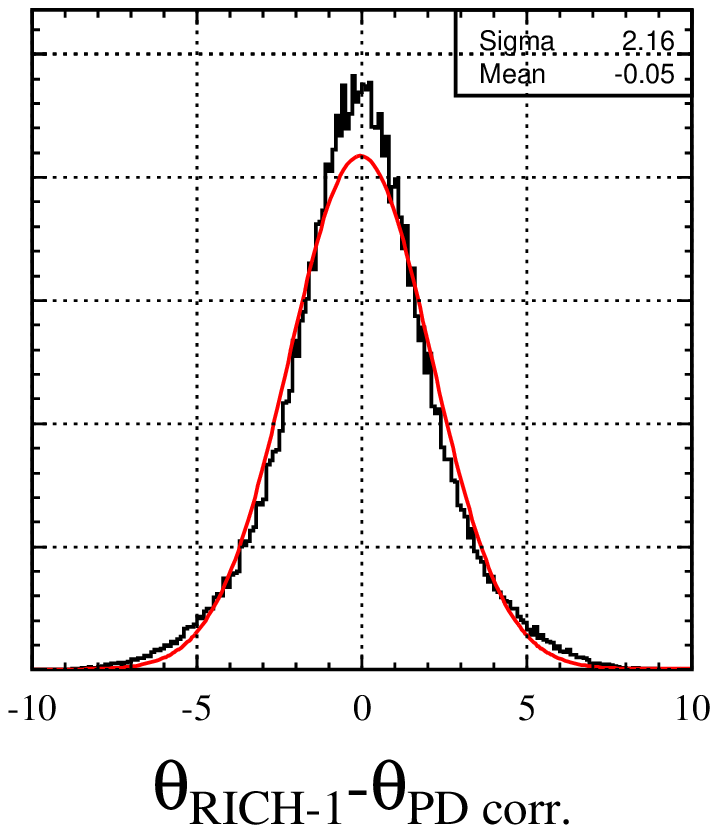,width=.33\textwidth}
   }
   \mbox{
    \epsfig{file=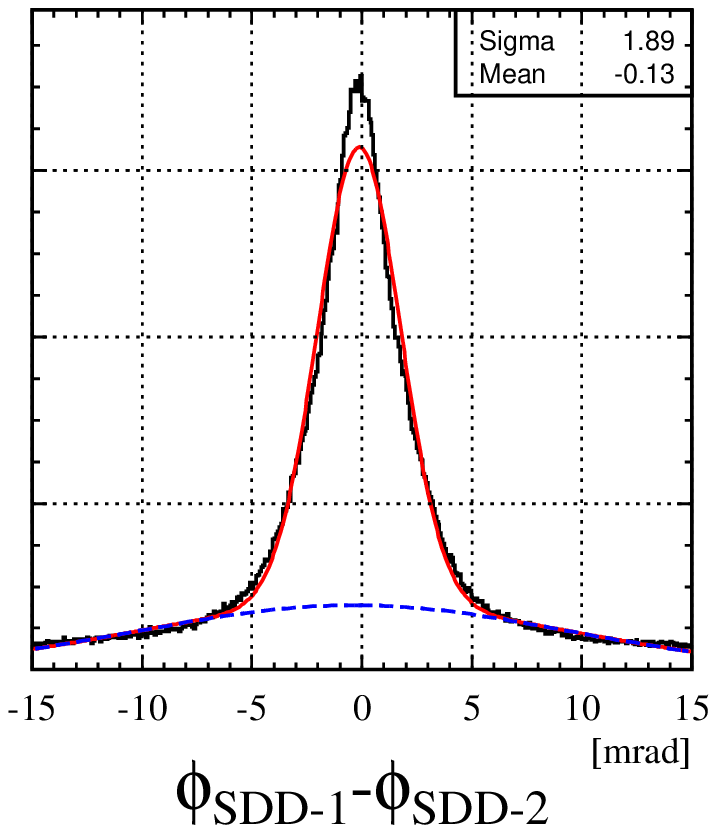,width=.33\textwidth}
    \hfill
    \epsfig{file=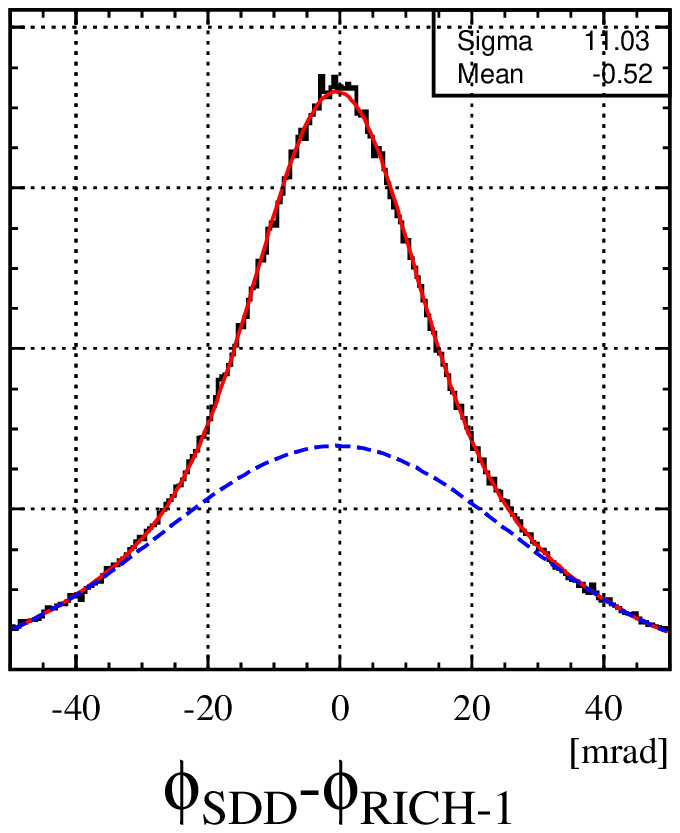,width=.33\textwidth}
    \hfill
    \epsfig{file=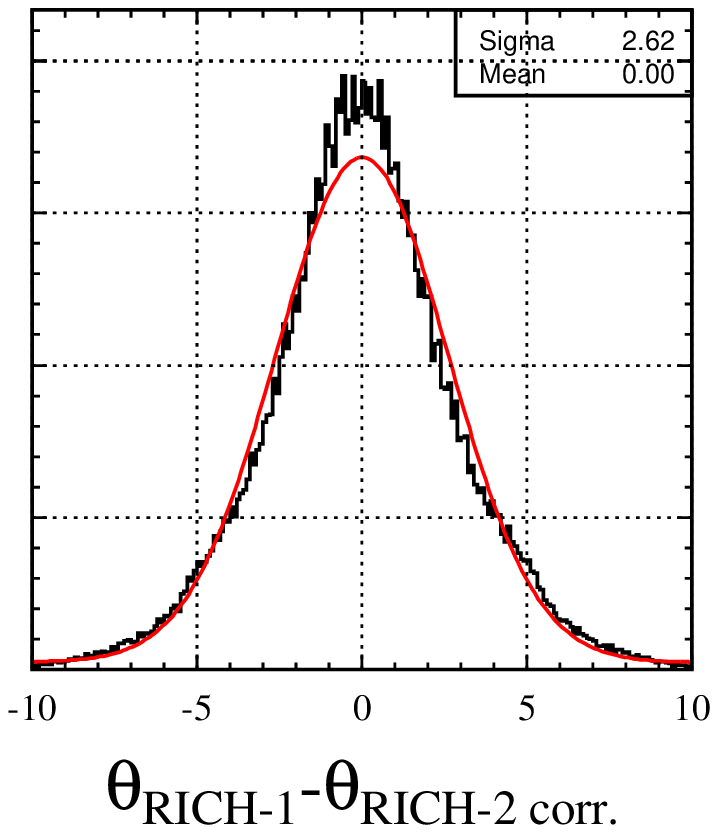,width=.33\textwidth}
   }
   \caption[Detector matching distributions]
   {Matching of SDD-1 and SDD-2 (first column), combined SDDs and PD (second column), RICH-1 and PD/RICH-2 (third column)
   of high-momentum electrons ($p$\,$>$\,$2$\,GeV/c). The
   $\theta$-positions of hits/rings in PD/RICH-2 were corrected for the second-order-field effect.
   The matching distributions were fitted with the sum of two Gaussians
   representing the distributions of true- and {\em fake}-tracks.
   The {\em fake} tracks were rejected in a later stage of analysis.}
 \label{fig:matching}
\end{figure}

Figure~\ref{fig:matching} shows the matching distributions of
high-momentum electrons ($p$\,$>$\,$2$\,GeV/c).
The large background fraction of the SDD matching distribution
(first column in Fig.~\ref{fig:matching}) can be attributed mainly
to single anode hits with very poor resolution in anode direction.
In case of a SDD-vertex track segment with a single anode hit in
either drift chamber, the $\phi$-coordinate of the combined track
was determined by the $\phi$-value of the other detector rather
than the average of both drift chambers in order to improve the
matching with the downstream detectors. The limited reconstruction
capability for partially overlapping rings in RICH-1 originating
from SDD-1 conversions and close $\pi^0$ Dalitz pairs (here not
rejected by other means) results in large fractions of
unrecognized double rings which exhibit a poor matching quality of
SDD and RICH-1 (second column in Fig.~\ref{fig:matching}).

The centroid offset is a measure of the quality of the geometrical
inter-calibration of individual detectors. The observed offset of
less than $0.5$\,mrad for all detector combinations is small
compared to the width of the matching distributions and, thus,
confirms the excellent quality of the calibration.

\begin{figure}[tb]
  \epsfig{file=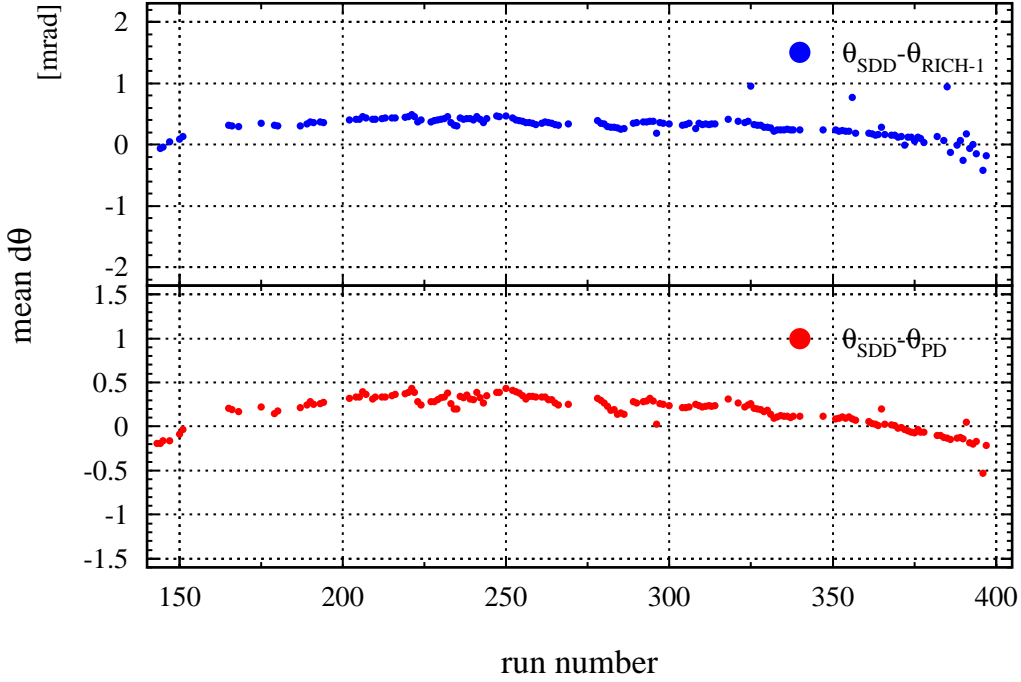}
  \caption[Run-to-run variation of the matching distribution]
  {Run-to-run variation of the centroid of SDD--PD and SDD--RICH $\theta$-matching distribution.
   The similarity of both distributions is prove that the variation is caused by
   a small miscalibration of the SDD-vertex telescope. The residual SDD--RICH-1 offset is negligible
   compared to the width of the matching distribution of
   2.1\,mrad. The SDD--PD $\theta$-match was not used for rejection (see App.~\ref{app:cut
summary}).
  }
  \label{fig:r2r matching}
\end{figure}

The remaining small miscalibration of the SDD-vertex telescope
previously mentioned resulted in a run-to-run variation of the
centroid of SDD--RICH-1 and SDD--PD matching distribution shown in
Fig.~\ref{fig:r2r matching}. The similarity of both distributions
is prove that the variation is indeed caused by the SDD. The
offset of the matching was not corrected because the SDD--RICH-1
offset is small compared to the width of the distribution (see
Fig.~\ref{fig:matching}) and the SDD--PD $\theta$-match was not
used in the analysis (see App.~\ref{app:cut summary}).

In the limit of high momentum, multiple scattering is negligible
and, hence, the width of the matching distribution $\sigma$ is
determined by the contributions of the single detector
resolutions:
\begin{equation}\label{equ:sigma match}
  \sigma_{\rm Match\,Detector\,1-2}=\sqrt{\sigma_{\rm resolution\,Det\,1}^2+\sigma_{\rm resolution\,Det\,2}^2}\;.
\end{equation}
Given the matching distributions of any combination of three
independent detectors, Eq.~\ref{equ:sigma match} can be resolved
to extract the single detector resolution:
\begin{equation}
\label{equ:detect resol}
 \sigma_{\rm resolution\,Det\,1}=\sqrt{\frac{\sigma_{\rm Match\,Detector\,1-3}^2+\sigma_{\rm Match\,Detector\,2-3}^2-\sigma_{\rm Match\,Detector\,
 1-2}^2}{2}}\;.
\end{equation}
Table~\ref{tab:detect resol} summarizes the extracted detector
resolutions.
\begin{table}[htb]
  \centering
   \begin{tabular}{|c|c|}
      \hline
      Detector & resolution [mrad] \\
      \hline
      SDD-1 &  0.28    \\
      \hline
      SDD-2 &  0.28     \\
      \hline
      RICH-1 &    2.03       \\
      \hline
      RICH-2 &    1.66       \\
      \hline
      PD &  0.54     \\
      \hline
    \end{tabular}
  \caption[Detector resolutions]
  {Single-track detector resolutions extracted from
  the width of the $\theta$-matching distributions for various detector combinations.
  The values given for the RICH-1 and the RICH-2 detector represent
  the ring center resolution.}
  \label{tab:detect resol}
\end{table}
The observed resolutions are close to the expected values
calculated from detector
properties~\cite{Lenkeit:1998,Agakishiev:1998nz}.

\subsection{Momentum resolution}
\label{sec:presol}

The experimental momentum resolution is determined by the accuracy
of the measurement of the azimuthal deflection between detectors
before (SDD and RICH-1) and after the B-field (RICH-2, PD):
\begin{equation}
 \label{equ:dp/p}
  \frac{dp}{p}=\frac{d(\phi_{\rm SDD,RICH-1}-\phi_{\rm RICH-2,PD})}{\phi_{\rm SDD,RICH-1}-\phi_{\rm
  RICH-2,PD}}\;.
\end{equation}
It is composed of the single detector resolution and the momentum
dependent multiple scattering. The latter can be inferred from the
experimentally accessible $\theta$-matching distribution:
\begin{equation}
  \Delta\phi(p) \approx
  \Delta\theta(p)=\sqrt{\sigma_{\rm resolution}^2+\sigma^2_{\rm scattering}/p^2}\;.
  \label{equ:dth}
\end{equation}
It should be noted that Eq.~\ref{equ:dth} slightly underestimates
the multiple scattering contribution because it is partly absorbed
in the follow-your-nose tracking approach applied to the RICH-PD
track segments in $\theta$-direction (see Sec.~\ref{sec:rich
tracking}).

\begin{figure}[bt]
  \centering
  \mbox{
   \epsfig{file=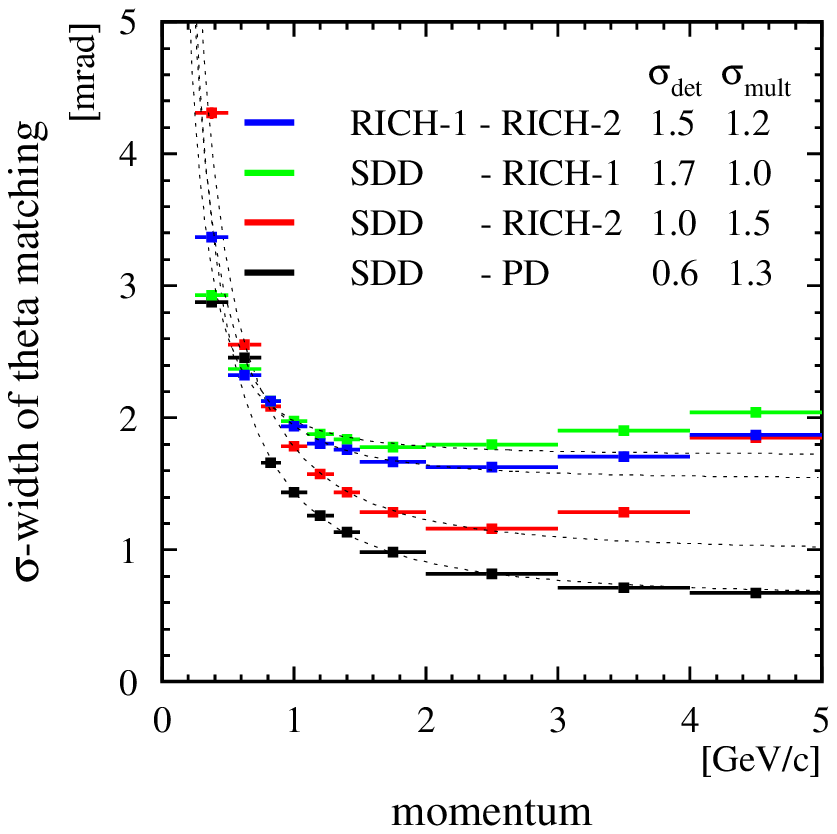,width=.49\textwidth}
   \hfill
   \epsfig{file=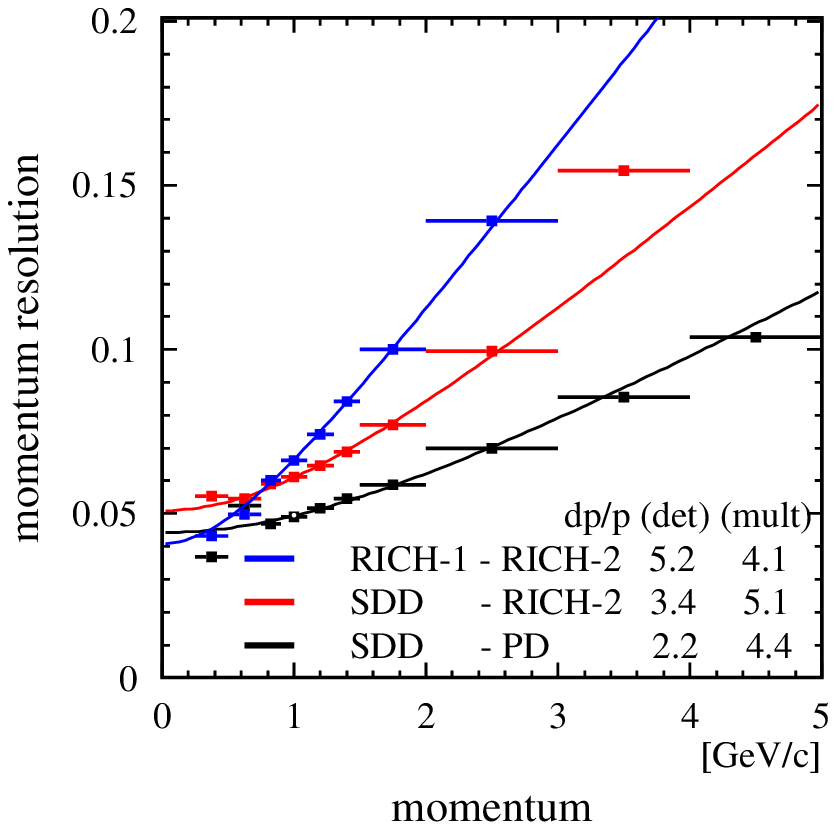,width=.49\textwidth}
  }
  \caption[Momentum resolution]
     {Momentum dependence of the $\theta$-matching of various detector combinations (left panel).
      A contamination of high momentum pions and $\gamma$ conversions leads to
      a seeming deterioration of the resolution for high momenta. Momentum
      resolution of various combinations of detectors
      before and behind the magnetic field (right panel).}
\label{fig:dthe vs p}
\end{figure}

The left panel of Fig.~\ref{fig:dthe vs p} shows the width of the
$\theta-$matching distribution of various detector combinations as
a function of momentum.
To obtain the momentum resolution, Eq.~\ref{equ:dp/p} can be
expressed in terms of Eq.~\ref{equ:dth}:
\begin{equation}
 \frac{dp}{p}(p)=\frac{\sqrt{\sigma^2_{\rm resolution}\,p^2+\sigma_{\rm scattering}^2} }
 {\Delta\phi_{\rm Det1-Det2}}\;.
 \label{equ:mom resol}
\end{equation}
The accuracy of the momentum determination according to
Eq.~\ref{equ:mom resol} is depicted in the right panel of
Fig.~\ref{fig:dthe vs p} for various detector combinations.
Examination reveals the momentum measurement based on the SDD--PD
combination to be best for high momenta while the RICH-1--RICH-2
deflection measurement is best for low momenta. This ground in the
fact that RICH-2 measures the local $\phi$-angle of the particle
trajectory in the RICH radiator after the azimuthal deflection in
the magnetic field. The $\phi$-deflection observed in the RICH-2
detector is about 56\% larger than in the PD detector. For small
deflection, i.e.~high momentum, the resolution is dominated by the
intrinsic detector resolution and favors the SDD--PD combination.
To optimize the mass resolution, the latter effect was taken into
account by simply switching from the RICH-1--RICH-2 momentum
measurement to the SDD--PD combination for momenta smaller than
0.8\,GeV/c. The combined momentum resolution given by:
\begin{equation}
  \frac{dp}{p}(p)= \sqrt{(0.022\pm0.001)^2\cdot p^2\cdot({\rm GeV^/c})^{-2} + (0.041\pm0.002)^2}
 \label{equ:presol}
\end{equation}
agrees with the result of previous studies~\cite{Lenkeit:1998}.


\section{$dN_{\rm ch}/d\eta$ measurement and centrality determination}
\label{sec:nch}

Global observables of a relativistic heavy ion collision such as
the multiplicity, i.e.~total number of emitted particles $N_{\rm
ch}$, or the transverse energy $E_{\rm t}$ carry important
information about the reaction
dynamics~\cite{Stachel:1991pb,Stachel:1993uh}. The centrality of a
collision in particular can be inferred from the particle yield.
The characterization of collisions in terms of centrality and
$N_{\rm ch}$ forms the basis for comparison among various
collision systems and different experiments.

\begin{figure}[tb]
    \begin{minipage}[t]{.65\textwidth}
        \vspace{0pt}
        \epsfig{file=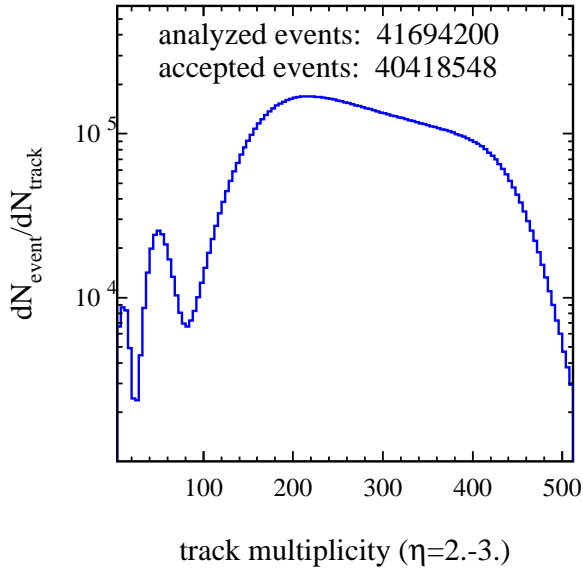,width=\textwidth}
    \end{minipage}%
    \begin{minipage}[t]{.35\textwidth}
      \vspace{0.5cm}
      \caption[Multiplicity distribution]
      {\newline
        Multiplicity distribution for SDD tracks.
      }
      \label{fig:multiplicty}
    \end{minipage}
\end{figure}

In the CERES experiment, the number of charged-particle tracks
$N_{\rm ch}$ is measured with SDDs in the pseudorapidity range
from 2 to 3. The distribution of $N_{\rm ch}$ per event obtained
is shown in Fig.~\ref{fig:multiplicty}.
The cutoff towards lower multiplicity corresponds to the trigger
threshold of 100 hits in the multiplicity detector MD. The peak at
$N_{\rm ch}$\,$=$\,$60$ is caused by non-target interaction. These
event are later removed by an offline centrality cut.

\begin{figure}[htb]
  \epsfig{file=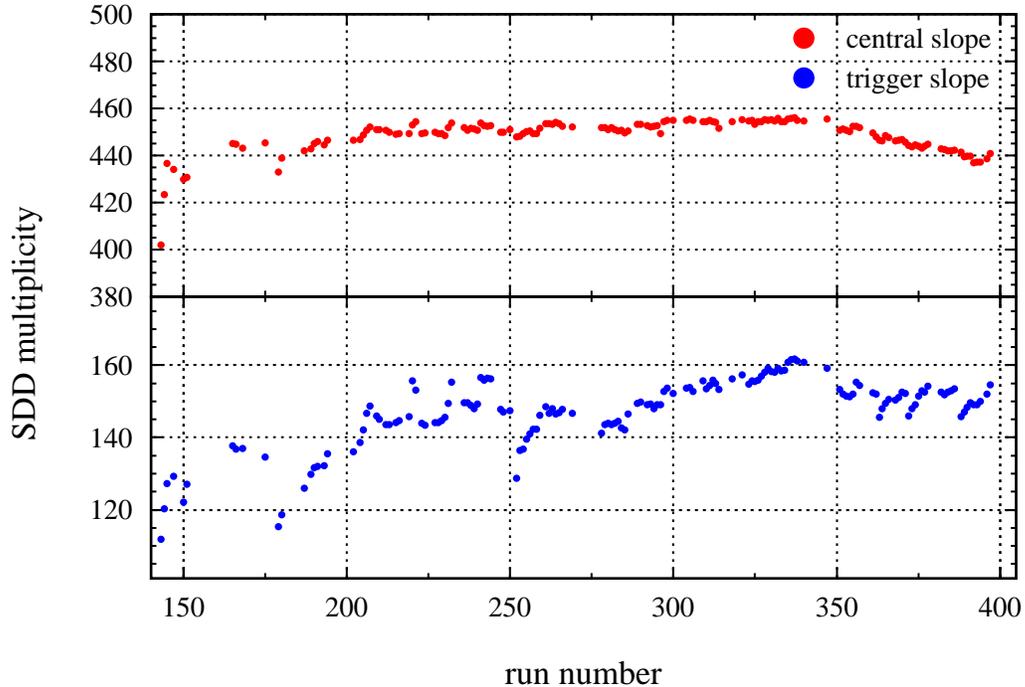}
  \caption[Run-to-run variation of the measured SDD track multiplicity]
  {Run-to-run variation of the position of the trigger slope (lower panel)
  and the central slope (upper panel).
  The position of the slope was defined as $N_{\rm meas}$ at half of the maximum
 (see Fig.~\ref{fig:multiplicty}).}
  \label{fig:r2r-nch}
\end{figure}

Figure~\ref{fig:r2r-nch} shows the position of the trigger slope
(left edge) as well as of the central slope (right edge) of the
multiplicity distribution to change with time. The apparent
decrease of the central-slope position was caused by the gradual
deterioration of one high-voltage sector in the SDD (see
Fig.~\ref{fig:anode dedx} in Sec.~\ref{sec:dedx}). The remaining
variation was attributed to a slight temperature dependence of the
SDD reconstruction efficiency. The data was corrected for both
effects.
The large fluctuation of the trigger slope is most likely caused
by an unstable voltage supply for the multiplicity detector. It
results in a variation of the initial centrality selection and,
therefore, does not directly impact on the measured multiplicity
except for the weighted multiplicity average (less than 1\%). The
limited statistics of the data sample did not allow to enable full
exclusion of the low-multiplicity range affected by the trigger
fluctuation.

The measured multiplicity distribution needs to be corrected for
the reconstruction efficiency of the SDD\@ to obtain the true
number of emitted charged particles. The efficiency correction was
derived from a Monte Carlo simulation of 10000 realistic UrQMD
events~\cite{Bass:1996ud}. It describes passage of all particles
through target and detectors taking into account energy loss,
$\gamma$-conversions, $\delta$-radiation, particle decays, and all
detector properties including electronic noise~\cite{Lenkeit:1998,
Slivova:2001}. All simulated events were reconstructed as genuine
raw data events.

Delta electron tracks do not point to the vertex and can be
recognized and removed to the level of less than one track per
event~\cite{Damjanovic:2001}. The beam pile-up was considered to
be small.

\begin{figure}
  \centering
  \mbox{
   \epsfig{file=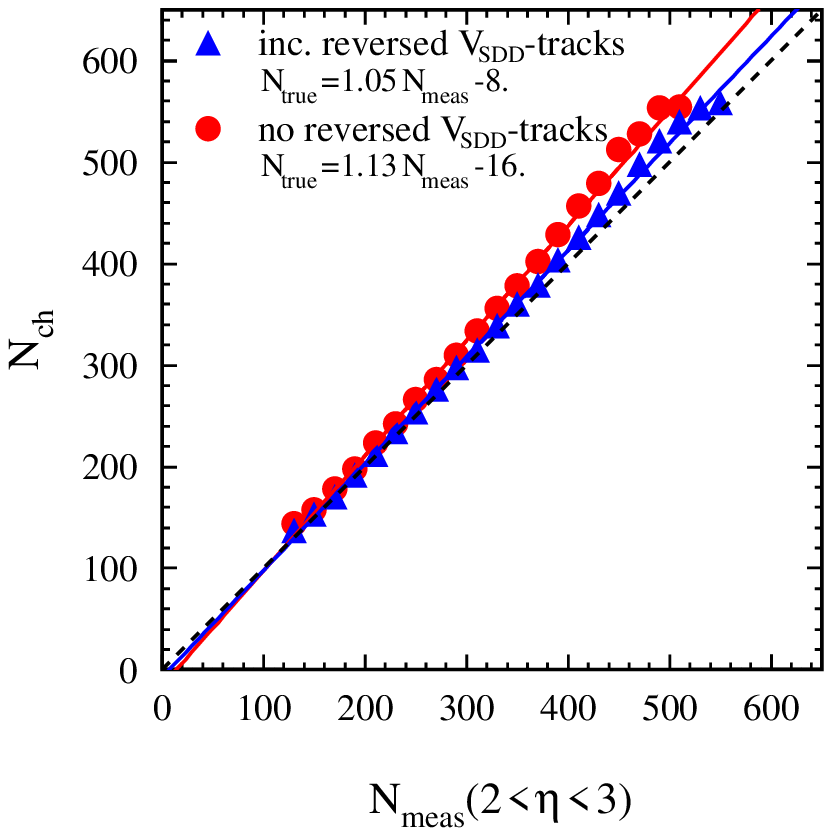,width=.49\textwidth}
   \hfill
   \epsfig{file=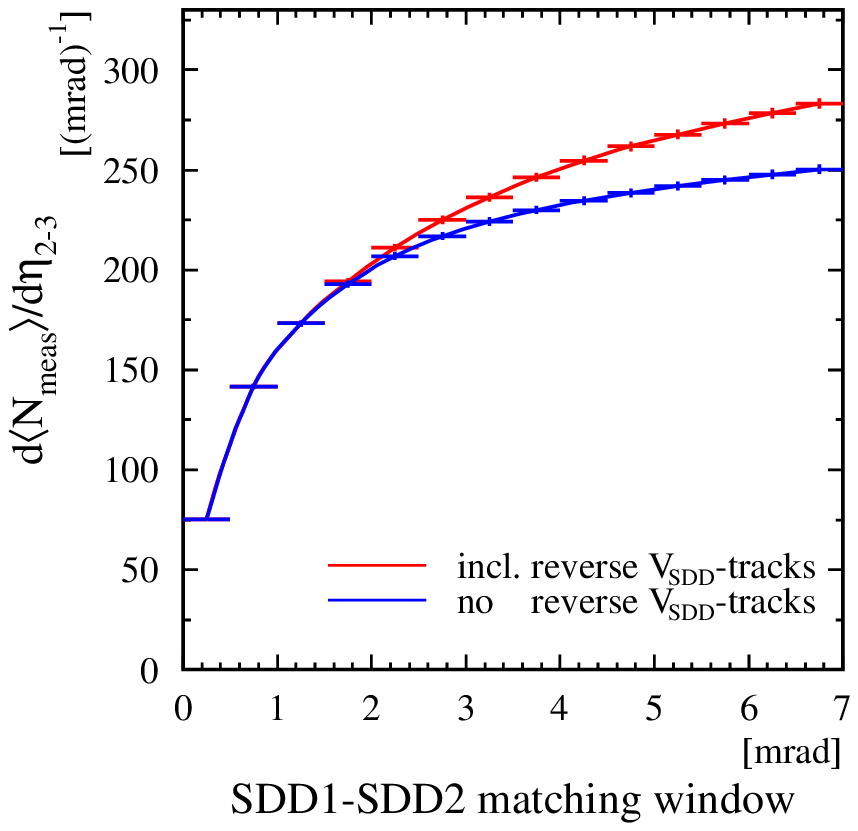,width=.49\textwidth}
  }
  \caption[$N_{\rm ch}$-efficiency correction]
  {Charged-particle reconstruction efficiency simulation (left panel). The true number of charged
  particles $N_{\rm ch}$ is a linear function of the measured
  track multiplicity $N_{\rm meas}$: $N_{\rm ch}$\,=\,(1.05$\pm$0.01)\,$\cdot$\,$N_{\rm
  meas}$\,+\,(8.$\pm$2.). The efficiency correction of $N_{\rm meas}$ is larger
  for the old version without reversed V$_{\rm SDD}$-tracks ($N_{\rm ch}$\,=\,1.13\,$\cdot$\,$N_{\rm
  meas}$\,+\,15.). Impact of reverse V$_{\rm SDD}$-tracks on the measured average track
  multiplicity as a function of the SDD-1--SDD-2 matching cut (right panel).
  Reversed V$_{\rm SDD}$-tracks are a source of {\em fake} tracks
  for a matching window larger than 2\,mrad.}
   \label{fig:multi corr}
  \epsfig{file=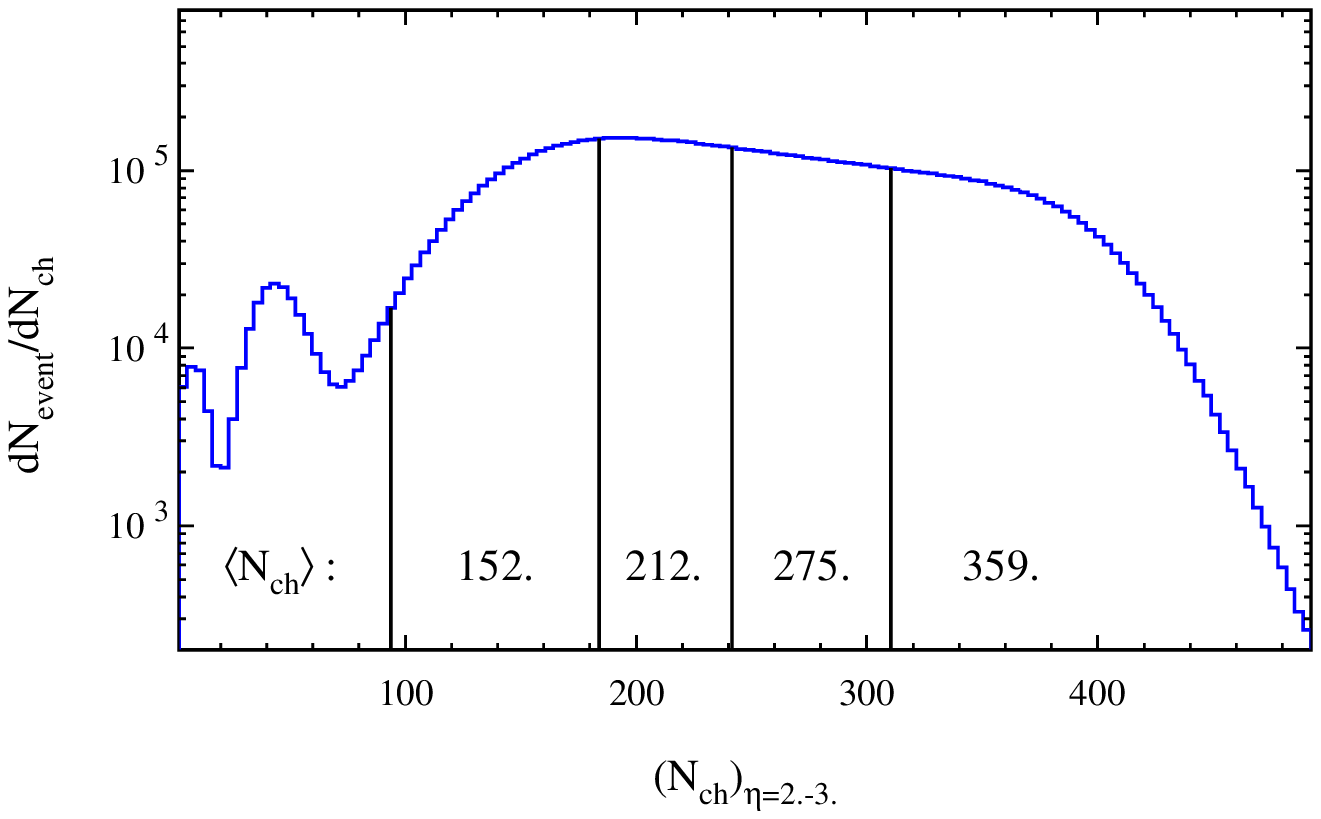,width=12.7cm}
  \caption[$N_{\rm ch}$ distribution]
  {Efficiency corrected $N_{\rm ch}$ distribution. The multiplicity
  distribution was divided into four bins of equal statistics for the study of
  the centrality dependence of the dilepton production.}
  \label{fig:nch}
\end{figure}

The comparison of the number of reconstructed tracks with the
number of initial tracks, illustrated in Fig.~\ref{fig:multi corr}
(left panel), shows the obtained correction factor of 5\% to be
smaller then the 20\% upward scaling found in the previous
analysis~\cite{Lenkeit:1998}.
The difference is attributed to a use of a relatively large
matching window between SDD-1 and SDD-2\,(7\,mrad) and the
inclusion of reversed V$_{\rm SDD}$-tracks (i.e.~two tracks
sharing the same hit in SDD-2) in the SDD tracking which results
in a 15\% increase of random background matches as shown in
Fig.~\ref{fig:multi corr} (right panel).

The average charge-particle multiplicity obtained with the Monte
Carlo correction for the two cases with and without reversed
V$_{\rm SDD}$-tracks differs by 10\% which suggests that Monte
Carlo simulation does not fully describe these subtle differences
in the tracking. Therefore, it seemed best to account for the
reversed V$_{\rm SDD}$-track contribution by applying an
additional $-15$\% correction to the measured multiplicity. The
fully corrected multiplicity distribution is shown in
Fig.~\ref{fig:nch}.

The systematic error of this method is very difficult to evaluate,
as the correction relies solely upon the Monte Carlo simulation. A
rough estimate can be obtained by attributing all variations in
the $N_{\rm ch}$ measurement of various track reconstruction
methods to the systematic error. A comparison of the new and old
SDD software gives a difference of 5\%. Additionally, the Monte
Carlo simulation lacks a description of the observed efficiency
variations (4\%), the V$_{\rm SDD}$-track contributions (8\%), and
realistic SDD pulse shapes including artificial hit splitting. The
sum of all contributions gives an estimate of the upper limit of
about 11\% for the relative systematic error.

\begin{figure}[!t]
    \begin{minipage}[t]{.65\textwidth}
        \vspace{0pt}
        \epsfig{file=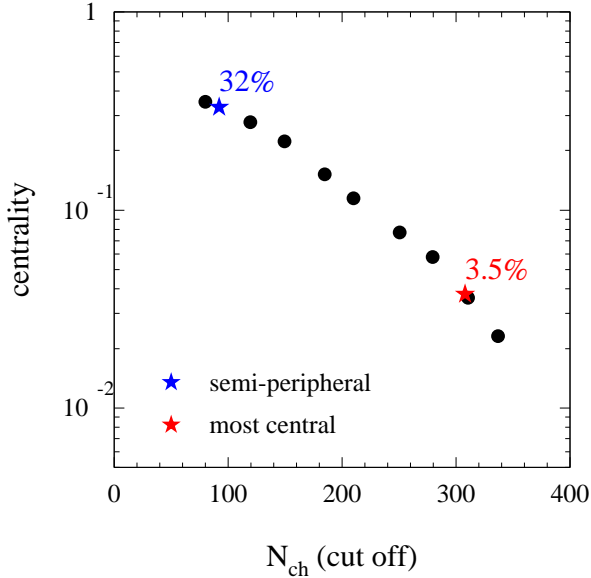,width=\textwidth}
    \end{minipage}%
    \begin{minipage}[t]{.35\textwidth}
      \vspace{0.5cm}
      \caption[Centrality as a function of the cut on
      charged-particle mul\-ti\-plic\-i\-ty]
      {\newline
      Centrality as a function of the cut on
      charged-particle multiplicity ($2 $\,$<$\,$ \eta $\,$<$\,$ 3$)~\cite{Ceretto:1998,Weber:1997}.
      }
      \label{fig:cross sec}
    \end{minipage}
\end{figure}

The centrality of a collision refers to the fraction {\em X} of
the total geometric cross section $\sigma_{\rm geom}$:
\begin{equation}
 \sigma_{\rm geom} = \pi\,\left( R_{\rm \, projectile} + R_{\rm \, target}
 \right) \,;\, R \approx 1.2~{\rm fm}\,A^{\frac{1}{3}}\;,
 \label{equ:sigma geom}
\end{equation}
where {\em R} and {\em A} denote the radius and the mass number of
the colliding ions, respectively. The total cross section observed
can be related to the detected number of charged particles for a
minimum bias event selection:
\begin{equation}
 \sigma = \frac{N_{\rm min\,bias}}{N_{\rm target}\,N_{\rm
 beam\,counter}}\;.
 \label{equ:min bias}
\end{equation}
Here $N_{\rm target}$, $N_{\rm beam\,counter}$, and $N_{\rm min\,
bias}$ are the number of target nuclei per unit area, the number
of beam particles, and the number of observed minimum bias events,
respectively. Integration of Eq.~\ref{equ:min bias} gives the
centrality {\em X} for a certain $N_{\rm ch}$ cutoff:
\begin{equation}
 X = \frac{1}{\sigma_{\rm geom}} \int^{\infty}_{N_{\rm
 ch}(\rm cutoff)}\left(\frac{d\sigma}{dN_{\rm ch}}\right)\,dN_{\rm ch}.
 \label{equ:geo fraction}
\end{equation}
The resulting relation is plotted in Fig.~\ref{fig:cross sec}.
The $N_{\rm ch}$ cutoff of 104 applied in the off-line analysis
corresponds to a relative cross section of (32$\pm$6)\%. The
selection of most central events according to $N_{\rm
ch}$\,$>$\,310 equals a relative cross section of (3.5$\pm$0.6)\%.
The relative cross section exhibits a large uncertainty because
the radius of the colliding nuclei in Eq.~\ref{equ:sigma geom} is
not precisely known.


\section{Simulation of the combinatorial background}
 \label{sec:mix-intro}

\subsection{Sources of combinatorial background}

There are two dominant sources of dilepton in the mass region of
$m_{\rm ee}$\,$<$\,$2~{\rm GeV/c}^2$: first, leptonic and
semi-leptonic decays of scalar and pseudo-scalar
mesons~\cite{PDBook} and, second, photon conversions in the target
and the first silicon drift chamber with
$\pi^0\rightarrow\gamma\gamma$ being the predominant source of
photons.

These decays lead to the production of electron-positron pairs,
henceforth referred to as unlike-sign pairs. Production of
electron-electron or positron-position pairs, so-called like-sign
pairs, requires higher-order processes. The strongest of them is
the $\pi\rightarrow e^+ e^+ e^- e^-$ decay, which is not only
suppressed by a factor of $\approx$\,380~\cite{PDBook} relative to
the $\pi_0$ Dalitz decay but also charge symmetric. It is
negligible for this analysis.

The experiment measures a certain number of positron tracks {\em
n$_+$} and of electron tracks {\em n$_-$} in each event. Exclusive
measurement is not possible because most dileptons are produced
within the collision zone. Considering all combinations of
observed electron and positron tracks of an event, it is
impossible to decide which unlike-sign pair originates from a
single decay and which is an accidental combination of individual
tracks of separate decay processes.

Therefore, two classes of unlike-sign pairs can be distinguished:
the actual signal of correlated dielectrons $S_{+-}$ and the
so-called {\em combinatorial} background pairs $B_{+-}$\@. The
total observed unlike-sign pair distribution $N_{+-}$ can be
expressed as:
\begin{equation}
 N_{+-}^{\rm total}=S^{\rm corr.}_{+-}+B_{+-}^{\rm comb.}\,.
 \label{equ:pair signal}
\end{equation}
Both signal and background exhibit a continuous spectrum. The
combinatorial unlike-sign background can be estimated by the
same-event like-sign method or by the mixed-event technique.

The first method is based on the fact that the same-event
combinatorial like-sign background is identical to the
combinatorial unlike-sign background in the absence of correlated
like-sign pairs from physics origin and under the assumption of
acceptance and efficiency for electrons and positrons being the
same. In the mixed-event technique, tracks from different events
are combined to yield the combinatorial unlike-sign background.

All published dielectron invariant mass spectra of the CERES
collaboration~\cite{Agakishiev:1998vt,Drees:1995us,Ullrich:1996wt,Agakishiev:1998au,Lenkeit:1999xu}
were obtained with the same-event like-sign background method.
Previous attempts to employ a mixed-event background were
unsuccessful. The first detailed comparative study of both methods
will be presented in Sec.~\ref{sec:comb} and~\ref{sec:mixed}. The
comparison of both simulated backgrounds can yield valuable
insight, test inherent model assumption, and ultimately increase
confidence in the background subtraction procedure.


\subsection{Same-event combinatorial background}
 \label{sec:comb}

The multiplicity of electrons $N_-$ or positrons $N_+$ produced in
a collision can be described by a Poisson probability distribution
$P$ (see Fig.~\ref{fig:ntracks}):
\begin{figure}[tb]
    \begin{minipage}[t]{.65\textwidth}
        \vspace{0pt}
        \epsfig{file=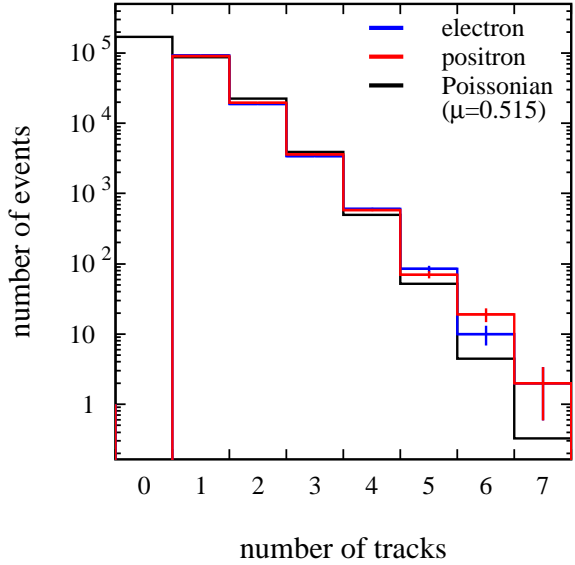,width=\textwidth}
    \end{minipage}%
    \begin{minipage}[t]{.35\textwidth}
      \vspace{0.5cm}
      \caption[Track multiplicity]
      {\newline
      Track multiplicity of run 144. The distributions of
      electrons and positrons are almost identical. Both are well
      described by a Poissonian distribution with a mean value of
      $\mu$\,=\,0.515. The variation of the track multiplicity distribution
      within a charged-particle multiplicity bin is neglected~\cite{Hegyi:2000sp}.}
      \label{fig:ntracks}
    \end{minipage}
\end{figure}
\begin{equation}
 P(N_{+/-}=k) =
 \frac{\overline{N}_{+/-}^k}{k!}\,\exp{\left(-\overline{N}_{+/-}\right)}\;.
\label{equ:npoisson}
\end{equation}
The probability {\em B} to observe $n_{+/-}$ tracks out of
$N_{+/-}$ initial particles is distributed binomially:
\begin{equation}
 B(n_{+/-}=k) = \frac{N_{+/-}}{k!\,(N_{+/-}-k)!}
 ({\varepsilon}_{+/-})^{k}\,(1-{\varepsilon}_{+/-})^{N_{+/-} -
 k}\;.
 \label{equ:binomal}
\end{equation}
The probability to observe a lepton track $\varepsilon$ is a
product of the probability of a particle falling into the
acceptance and the reconstruction efficiency depending on various
single-track selection cuts. $\varepsilon_+$ and $\varepsilon_-$
are treated separately to account for possible charge asymmetries
of the acceptance or the reconstruction. Making use of
Eq.~\ref{equ:binomal} one obtains the average number of
reconstructed tracks:
\begin{eqnarray}
 \overline{n_+} & = & \varepsilon_+N_+ \;,\;
 \overline{(n_+)^2}=\varepsilon_+(1-\varepsilon_+)N_+ + \varepsilon_+^2
 N_+^2\,,\\
 \overline{n_-} & = & \varepsilon_-N_- \;,\;
 \overline{(n_-)^2}=\varepsilon_-(1-\varepsilon_-)N_- + \varepsilon_-^2
 N_-^2\nonumber\,.
 \label{equ:binom-mean}
\end{eqnarray}
In first-order approximation, the mean number of pairs per event
with $N_+$ positrons and $N_-$ electrons is given by:
\begin{eqnarray}
 \overline{n_{++}} & = & \kappa_{++} \sum_{k=0}^{N_+}
 \frac{k(k-1)}{2}\,B(n_+=k) = \frac{1}{2}\kappa_{++}\varepsilon_+^2
 N_+(N_+-1)\;, \label{equ:p mean}\\
 \overline{n_{--}} & = & \kappa_{--}
 \sum_{k=0}^{N_-} \frac{k(k-1)}{2}\,B(n_-=k) = \frac{1}{2}\kappa_{--}\varepsilon_-^2 N_-(N_--1)
 \;,\nonumber\\
 \overline{n_{+-}} & = & \kappa_{+-} \sum_{k=0}^{N_+}\sum_{l=0}^{N_-} k\,B(n_+=k)\,l\,B(n_-=l)
 = \kappa_{+-}\varepsilon_+\varepsilon_- N_+N_-\,.\nonumber
\end{eqnarray}
The factor $\kappa$ denotes two-track efficiency, introduced by
physics correlation, detector, or analysis for each sort of
charged pairs. Making use of Eq.~\ref{equ:p mean}, the number of
pairs averaged over all events becomes:
\begin{eqnarray}
 \langle n_{++}\rangle & = & \sum_{N_+=0}^{\infty} \overline{n_{++}}\,P(N_+)\\
 & = & \frac{1}{2}\kappa_{++} \varepsilon_+^2 \sum_{N_+=0}^{\infty} N_+(N_+-1)\,P(N_+)\nonumber \\
 & = & \frac{1}{2}\,\kappa_{++}\,\varepsilon_{+}^2
 {\left(\overline{N_+}\right)}^2\;,\nonumber\\
  \label{equ:n-like-av}
 \langle n_{--}\rangle & = & \sum_{N_-=0}^{\infty}
\overline{n_{--}}\,P(N_-)
 = \frac{1}{2}\,\kappa_{--}\,\varepsilon_{-}^2
 {\left(\overline{N_-}\right)}^2\;,\\
 \langle n_{+-}\rangle & = & \sum_{N_+=0}^{\infty}\sum_{N_-=0}^{\infty}
 \overline{n_{+-}}\, P(N_+)\,P(N_-)
 = \kappa_{+-}\,\varepsilon_{+}\,\varepsilon_{-}
 \overline{N_+}\,\overline{N_-}\;.
  \label{equ:n-unlike-av}
\end{eqnarray}
Equation~\ref{equ:n-unlike-av} represents the unknown unlike-sign
combinatorial background. By comparison of $\langle n_{++}\rangle$
and $\langle n_{--}\rangle$ with $\langle n_{+-}\rangle$, it
becomes obvious that the geometric mean of the like-sign
background is an excellent approximation of the unlike-sign
combinatorial background:
\begin{eqnarray}
 \underbrace{\langle n_{+-}\rangle}_{\rm unlike-sign\,bg}
 & \equiv & \underbrace{2\,\sqrt{\langle n_{++}\rangle \,\langle n_{--}\rangle}}_{\rm
 like-sign\,bg}\;,
\nonumber \\
 \kappa_{+-}\,\varepsilon_{+}\,\varepsilon_{-}
 \overline{N_+}\,\overline{N_-} & \equiv & \sqrt{\kappa_{++}\,\kappa_{--}}\,
 \varepsilon_+\,\varepsilon_-\,\overline{N_+}\,\overline{N_-}\;. \label{equ:combbk}
\end{eqnarray}
This proposition is fulfilled if the two-track efficiency $\kappa$
is pair charge independent. A charge asymmetry of the single-track
reconstruction probability $\varepsilon$, on the other hand, will
not alter this result. A more detailed derivation including an
extension to the case of the simultaneous occurrence of correlated
and independent lepton sources can be found
in~\cite{Gazdzicki:2000fy}.

\begin{figure}[tb]
  \centering
  \epsfig{file=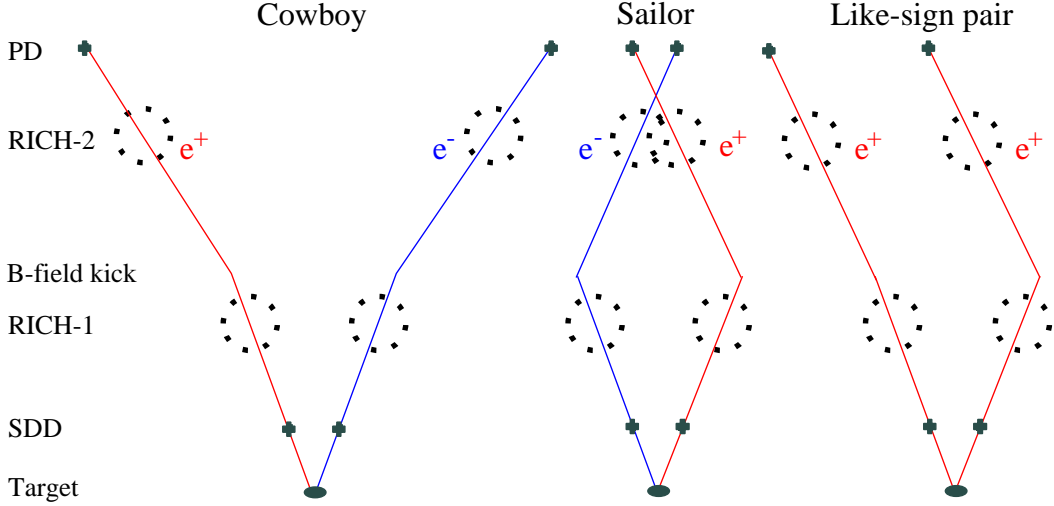,width=14cm}
  \caption[Pair-charge symmetry breaking in the magnetic field]
  {The magnetic field breaks the symmetry of like-sign
   and unlike-sign pairs, for unlike-sign pairs comprise a ``cowboy'' and a ``sailor'' configuration.}
  \label{fig:cowboy}
\end{figure}

For the detector the assumption about the pair charge independence
of the efficiency is certainly true for large pair opening angles.
In this case, the tracks are well separated in all detectors. It
can, however, be contested if the pair-charge symmetry of the
detector is broken by the magnetic field. While the like-sign
pairs with a small opening angle remain always close in space, the
unlike-sign pairs form either a so-called ``cowboy'' or ``sailor''
configuration as illustrated in Fig.~\ref{fig:cowboy}. Thus, a
finite two-track resolution would affect like-sign and unlike-sign
pairs differently.

\begin{figure}[tb]
  \centering
  \mbox{
   \epsfig{file=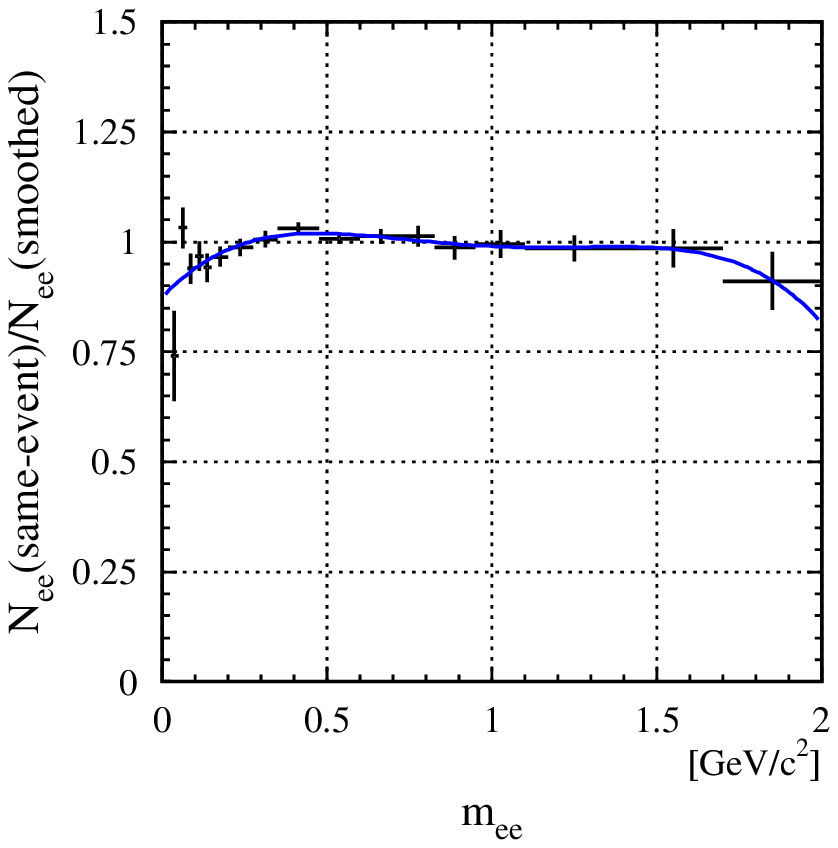,width=.49\textwidth}
   \hfill
   \epsfig{file=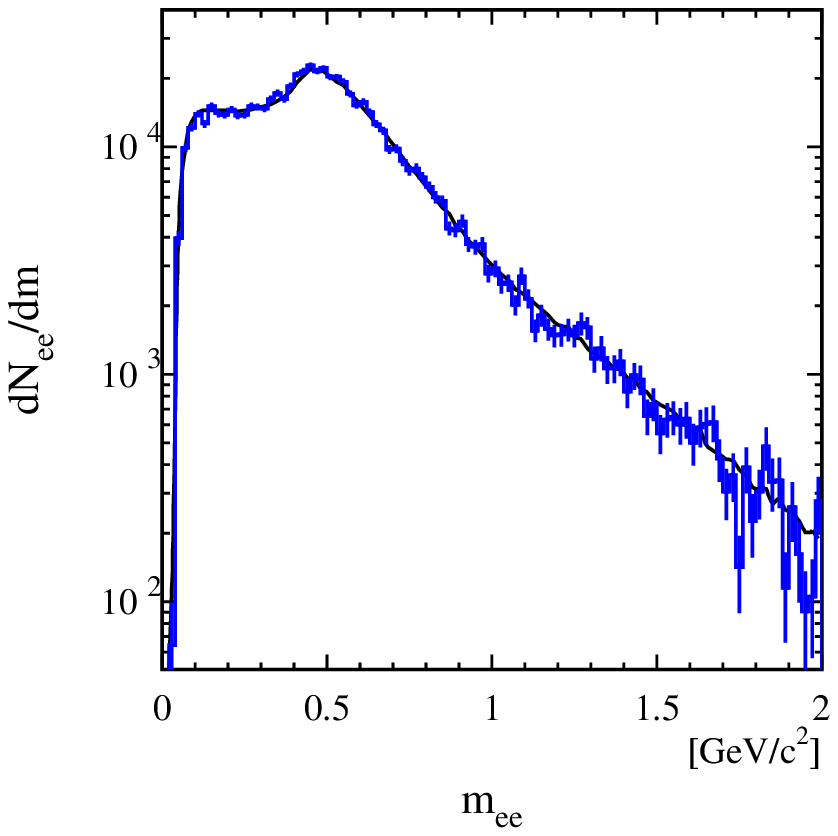,width=.49\textwidth}
  }
  \caption[Smoothed like-sign background distribution]
  {Ratio of the like-sign background to the smoothed background (left panel).
   The 4$^{\rm th}$-order-polynomial correction function for
   the smoothed like-sign background distribution was included.
   Comparison of the observed like-sign distribution and
   the corrected smoothed background (right panel). All rejection cuts were
   applied.}
\label{fig:smooth fit}
\end{figure}

The second disadvantage of the like-sign background estimation
method is the large statistical error $ \sigma(\sqrt{4\langle
n_{++} \rangle\langle n_{--} \rangle})$ of the simulated
combinatorial background:
\begin{eqnarray}
 \sigma\left({\sqrt{4\langle n_{++} \rangle\langle n_{--} \rangle}}\right)& =
 &
 \frac{2\sigma_{\langle n_{++}\rangle}\sigma_{\langle n_{--}
 \rangle}}{\sqrt{\sigma^2_{\langle n_{++} \rangle}+\sigma^2_{\langle
 n_{--}\rangle}}}\\
 & \approx & \sqrt{\langle n_{\rm like} \rangle} \qquad (\langle n_{like} \rangle \approx 2\langle n_{++} \rangle
 \approx 2\langle n_{--} \rangle) \nonumber \;.
 \label{equ:comb error}
\end{eqnarray}
A Poissonian distribution for the statistical error and
statistical independence of the $(++)$ and $(--)$ pair samples
were assumed. By use of Eqs.~\ref{equ:pair signal}
and~\ref{equ:comb error}, the relative statistical error of the
correlated dilepton signal $\Delta S_{+-}/S_{+-}$ can be expressed
as:
\begin{eqnarray}
\frac{\Delta S_{+-}}{S_{+-}} & = &
\frac{\sqrt{\sigma^2_{N_{+-}}+\sigma^2_{B_{+-}}}}{N_{+-}-B_{+-}}
\approx \frac{\sqrt{2}\,\sigma_{B_{+-}}}{N_{+-}-B_{+-}}
\\
& \approx & \frac{\sqrt{2\langle n_{\rm like }\rangle}}{\langle
n_{+-}\rangle-2\sqrt{\langle n_{++}\rangle \langle
n_{--}\rangle}}\nonumber\;.
 \label{equ:sig-error}
\end{eqnarray}
To circumvent the statistical limitations of the same-event
like-sign background, a smoothing procedure was applied in
previous analyses~\cite{Voigt:1998,Lenkeit:1998}. An approximation
of the functional shape of the same-event combinatorial background
was obtained by sampling the final $\theta$-, $\phi$-, and
$p_{\bot}$-distribution of single tracks and calculating the
invariant mass of each pair of simulated tracks. Using this
procedure, a background distribution can be obtained that is
basically free of statistical errors. To account for impact
parameter dependence, the tracks were divided into four
multiplicity and four theta bins. Finally, a mass dependent
correction factor was applied to the smoothed invariant-mass
background distribution. It is extracted by fitting the ratio of
the observed combinatorial like-sign background to the smoothed
background while keeping the integral of the measured distribution
as shown in Fig.~\ref{fig:smooth fit} (left panel). The resulting
invariant-mass distribution of the smoothed background is plotted
in the right panel of Fig.~\ref{fig:smooth fit}.

\begin{figure}[tb]
    \begin{minipage}[t]{.65\textwidth}
        \vspace{0pt}
        \epsfig{file=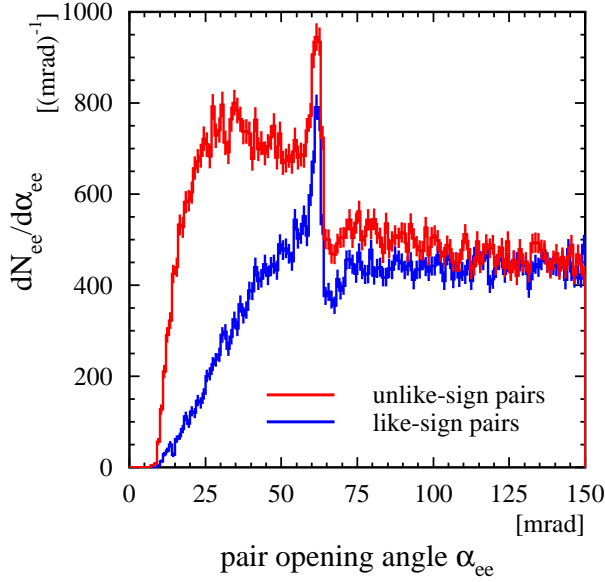,width=\textwidth}
    \end{minipage}%
    \begin{minipage}[t]{.35\textwidth}
      \vspace{0.5cm}
      \caption[Pair-opening-angle distribution]
      {\newline
       The pair-opening-angle distribution shows nearly touching or slightly overlapping RICH rings
       to be ``pulled'' or ``pushed'' to a touching-ring configuration with
       an opening angle of 60\,mrad. Additionally, the reconstruction efficiency
       increases as the photon hits of one ring promote the reconstruction of an other
       touching ring.
      }
      \label{fig:touching rings}
    \end{minipage}
\end{figure}

Effectively, this procedure hides the statistical error by
shifting it to the systematic error of the background subtraction.
An additional drawback of this method results from the fact that
any scattering observed between adjacent mass bins is solely
attributed to statistical fluctuations. This is a rather weak
assumption because artifacts of the RICH-ring reconstruction
algorithm are known to alter the touching-ring configuration (see
Fig.~\ref{fig:touching rings}). This leads to structures in the
pair-opening-angle distribution which are directly translated into
localized variation in the invariant-mass spectrum. Any
correlation of this type would be masked by the smoothing.

To exclude these potentially dangerous effects and to assess the
systematic errors involved, it would be of utmost importance to
independently verify this method.

\subsection{Mixed-event technique}
 \label{sec:mixed}

Alternatively to the use of the independent like-sign pairs of
each event, unlike-sign pairs, obtained by combination of opposite
charged tracks of different events, are inherently independent.
This procedure is commonly known as the mixed-event technique. The
straightforward modification of Eq.~\ref{equ:p mean} gives the
mean number of observed mixed unlike-sign pairs of two randomly
selected events A and B with initial multiplicity $N^A_{+/-}$ and
$N^B_{+/-}$:
\begin{equation}
 \overline{n_{+-}^{\rm mixed}} = \sum_{k=0}^{N^A_+}
 \sum_{l=0}^{N^B_-} k\,l\,B(n_+^A=k)\,B(n_-^B=l)+ \sum_{k=0}^{N^A_-} \sum_{l=0}^{N^B_+}
 k\,l\,B(n_-^A=k)\,B(n_+^B=l)\;,
 \label{equ:mix-p+-mean}
\end{equation}
where each track of event A is combined with all tracks of event B
with opposite charge. Again, the averaging of many pairs of events
with different initial particle multiplicity yields:
\begin{eqnarray}
 \langle n_{+-}^{mixed}\rangle^{AB}& = & \sum_{i=1}^{\infty}\sum_{k=1}^{\infty}
 \overline{n_{+-}^{\rm mixed}}\, P^A(i)\,P^B(k) \nonumber \\
& = &\varepsilon_{+}^A\,\varepsilon_{-}^B
 \overline{N_+^A}\,\overline{N_-^B}+\varepsilon_{+}^B\,\varepsilon_{-}^A
 \overline{N_+^B}\,\overline{N_-^A}\;.
 \label{equ:mixed n+-av}
\end{eqnarray}
It is important to note the single-track events to be contributing
to this average. In contrast, the same-event technique requires at
least two initial tracks per event. To obtain the mixed-event
background, the mixed unlike-sign distribution (see
Eq.~\ref{equ:mixed n+-av}) has to be normalized with the number of
mixed-event pairs $N_{\rm mixed}$:
\begin{equation}
 \langle n_{+-}^{\rm mixed}\rangle =
 \frac{\,N_{\rm exp}}{N_{\rm mixed}}\langle n_{+-}^{\rm mixed}\rangle^{AB}
 = \overline{\varepsilon_{+}}\,\overline{\varepsilon_{-}}
 \overline{N_+}\,\overline{N_-}\;,
\label{equ:mixed norm n+-}
\end{equation}
where N$_{\rm exp}$ denotes the total number of fully analyzed
events including those where no electron track was found.
Equation~\ref{equ:mixed norm n+-} defines the mixed unlike-sign
background.

All attempts to construct the mixed-event unlike-sign background
on the basis of the previous raw data analysis failed because the
single-track events have been left out. Furthermore, only events
containing at least one dielectron recognized either as a
so-called {\em Dalitz} or open pair were stored for further
analysis (see~\cite{Voigt:1998} for detailed description).
Inspection of Eq.~\ref{equ:mixed n+-av} and~\ref{equ:mixed norm
n+-} reveals that the complete rejection of single-track events
together with a signal pair dependent background discrimination
must lead to a biased event sample and, hence, to a potentially
distorted mixed background (see Fig.~\ref{fig:old-mixed}).

A comparison of Eq.~\ref{equ:n-unlike-av} and~\ref{equ:mixed norm
n+-} shows $\langle n_{+-}^{\rm mixed}\rangle$ to be equivalent to
the much sought-after independent unlike-sign background:
\begin{eqnarray}
 \langle n_{+-} \rangle & \equiv & \langle n_{+-}^{\rm mixed}\rangle\;,\nonumber\\
 \kappa_{+-}\,\varepsilon_{+}\,\varepsilon_{-}
 \overline{N_+}\,\overline{N_-} & \equiv & \overline{\varepsilon_{+}}\,\overline{\varepsilon_{-}}
 \overline{N_+}\,\overline{N_-}\;,
 \label{equ:mixed norm n+-2}
\end{eqnarray}
\noindent provided:
\begin{eqnarray}
 \varepsilon_{+}\,\varepsilon_{-} & = & \overline{\varepsilon_{+}}\,\overline{\varepsilon_{-}}\;,
 \label{equ:mixed cond1}\\
  \kappa_{+-} & = & 1\;.\label{equ:mixed cond2}
\end{eqnarray}
Experimentally, the condition~\ref{equ:mixed cond1} can be
approximated by restricting the event mixing to classes of events
with similar properties and subsequent averaging of all classes.
Technically, all events were divided into sub-samples of 4
multiplicity bins and 158 time bins. The temporal sub-samples were
found to be important because pressure, temperature, and detector
parameters changed considerably during the 6 weeks of data
recording.

The condition~\ref{equ:mixed cond2} means an infinite two-track
resolution. While this assumption holds for large pair opening
angle (i.e.~$\alpha_{\rm pair}$\,$>$\,$80$\,mrad), it is obviously
not fulfilled for close pairs. For those the correlation factor
$\kappa_{+-}$ drops below one and, hence, the mean number of
background pairs is overestimated by the mixed-event background.
Note, that the correlation factor $\kappa_{+-}$ should not depend
much on charged-particle multiplicity because the two-track
resolution is dominated by the double-ring resolution of RICH-1
and RICH-2 detector, and the observed background in both RICH
detectors is related to electronics noise and scattered beam
particles but not to the event multiplicity.

In general, there are two main advantages of the mixed-event
technique. First, it is also applicable to the deconvolution of
like-sign correlated signals (e.g. open charm
detection~\cite{Crochet:2001qd}) which, however, is of no
importance for this analysis. Second, the statistical error of the
mixed-event background distribution can be reduced simply by
increasing the number of mixed events {\em $N_{\rm mixed}$} in
Eq.~\ref{equ:mixed norm n+-}. It is important that {\em $N_{\rm
mixed}$} is chosen in such a way that the probability to select
the same-event more than once remains small. Otherwise, the result
will be hampered by the auto-correlation of the mixed events.
Consequently, the ultimate increase of statistics is limited by
the size of the sub-samples used for mixing.

For technical reasons the ratio of the number of mixed pairs to
the number of same-event like-sign pairs was fixed rather than the
event ratio. A simulation based on a toy model confirmed that a
pair-mixing ratio of up to $n_{\rm mixed}/n_{\rm like}=20$ is safe
with respect to any
auto-correlation~\cite{Voloshin:1994:mw,Miskowiec:2000}.

In case of a small pair-mixing ratio, the statistical error of the
mixed background $\sigma(\langle n_{+-}^{\rm mixed} \rangle)$ can
be expressed in terms of the statistical error of the same-event
like-sign background:
\begin{equation}
  \sigma_{\langle n_{+-}^{\rm mixed} \rangle}=\sqrt{\frac{\,n_{\rm like}}{n_{\rm
  mixed}}}\sqrt{\langle n_{+-}^{\rm mixed} \rangle}\;.
\end{equation}
Compared to the same-event like-sign background, the statistical
error of the mixed background is reduced by about a factor of 5
for a standard mixing ratio of 20. For this example the
statistical error of the signal is reduced by 30\% compared to the
same-event method. If the background had no statistical
uncertainty at all, a maximum reduction of 42\% could be achieved
according to Eq.~\ref{equ:sig-error}.

\begin{figure}[tb]
    \begin{minipage}[t]{.65\textwidth}
        \vspace{0pt}
        \epsfig{file=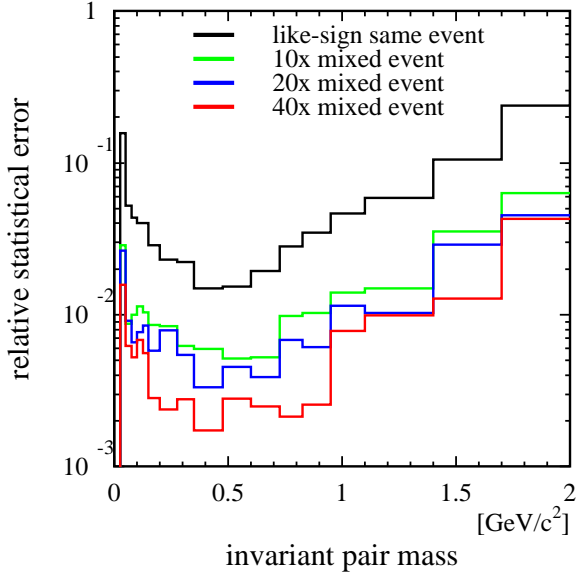,width=\textwidth}
    \end{minipage}%
    \begin{minipage}[t]{.35\textwidth}
      \vspace{0.5cm}
      \caption[Estimate of the statistical error of the mixed-background dis\-tri\-bu\-tion]
      {\newline
      Estimate of the statistical error of the mixed-event background
      for mixing ratios of 10, 20, and 40.
      }
      \label{fig:mix-mass-error}
    \end{minipage}
\end{figure}

The mass dependence of the relative statistical error was
estimated by the statistical fluctuation of the mixed-event
background for randomly selected event sub-samples.
Figure~\ref{fig:mix-mass-error} shows the average deviation of the
mixed-background distribution from the mean value.
Taking the example of a mixing ratio of 20 the relative
statistical error of the mixed background increases continuously
from 0.5\% in the mass range of 0.2\,$<$\,$m_{\rm
ee}$\,$<$1\,GeV/c$^2$ to about 5\% at 2\,GeV/c$^2$.

\begin{figure}
  \centering
  \epsfig{file=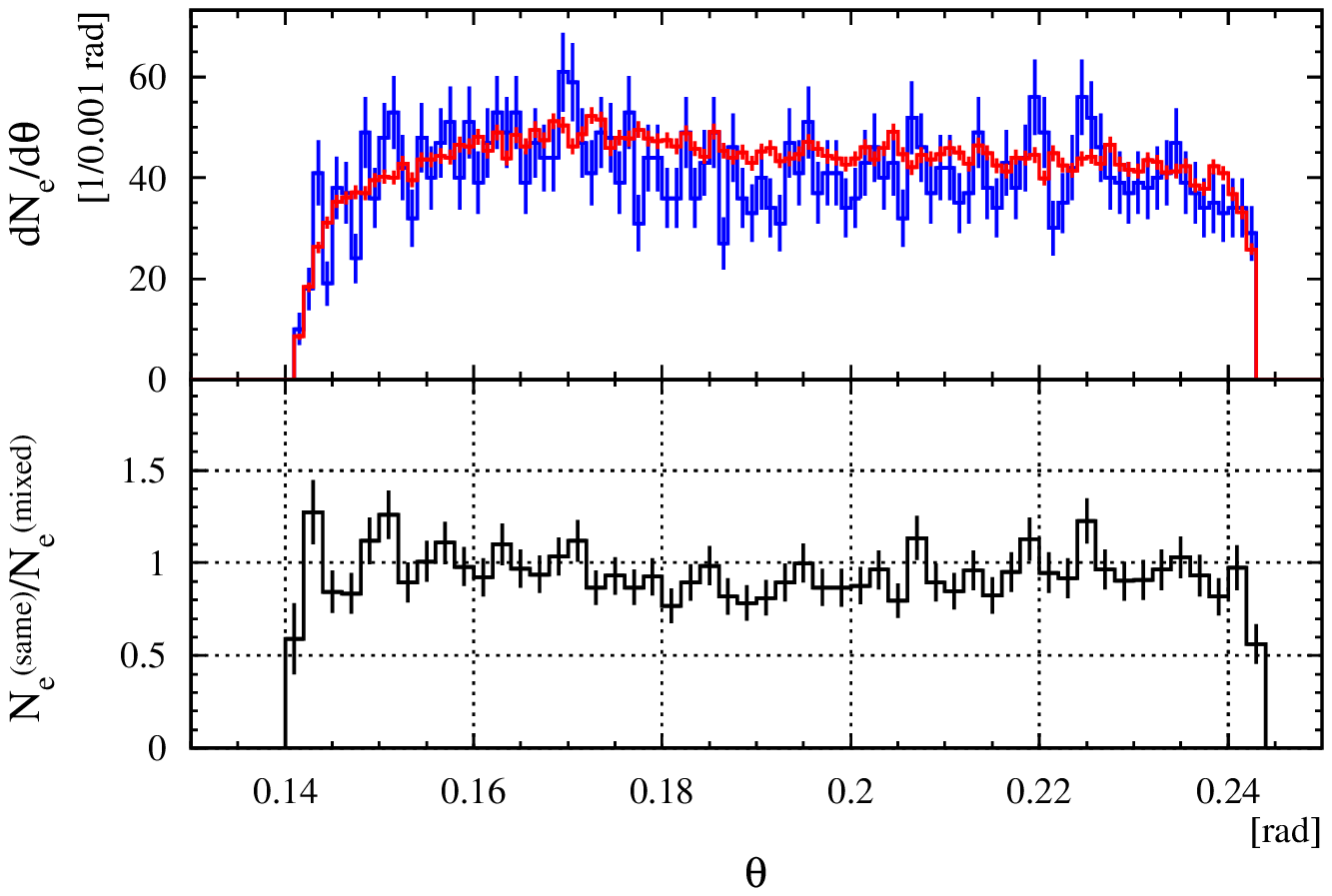}
  \epsfig{file=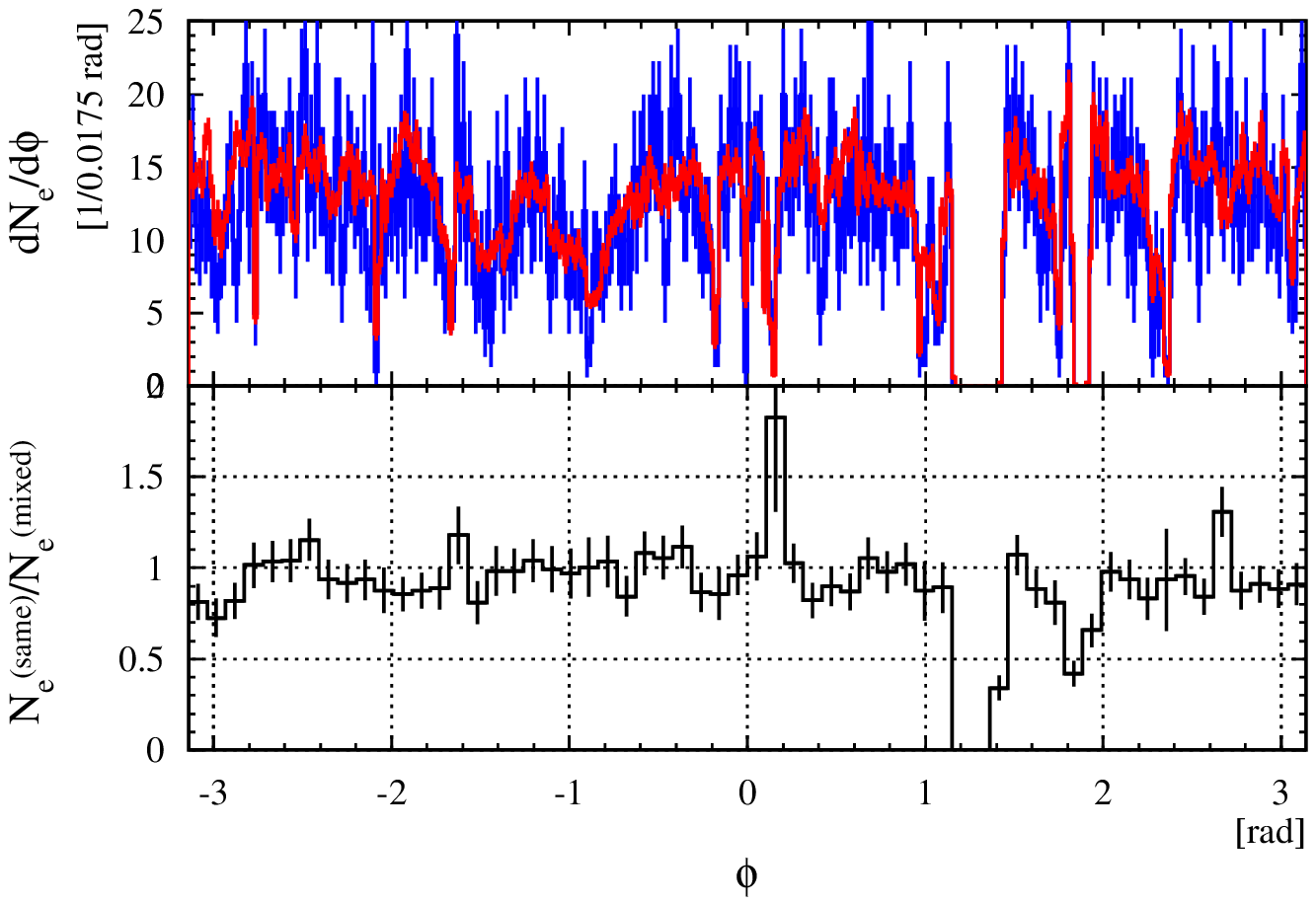}
  \caption[Comparison of same-event and mixed-event single-track $\theta$- and $\phi$-distribution]
  {Comparison of same-event (blue line) and mixed-event (red line)
  $\theta$- and $\phi$-distribution of single tracks shown in the upper and the lower panel, respectively.}
  \label{fig:mix-theta}
\end{figure}

\begin{figure}[htb]
  \epsfig{file=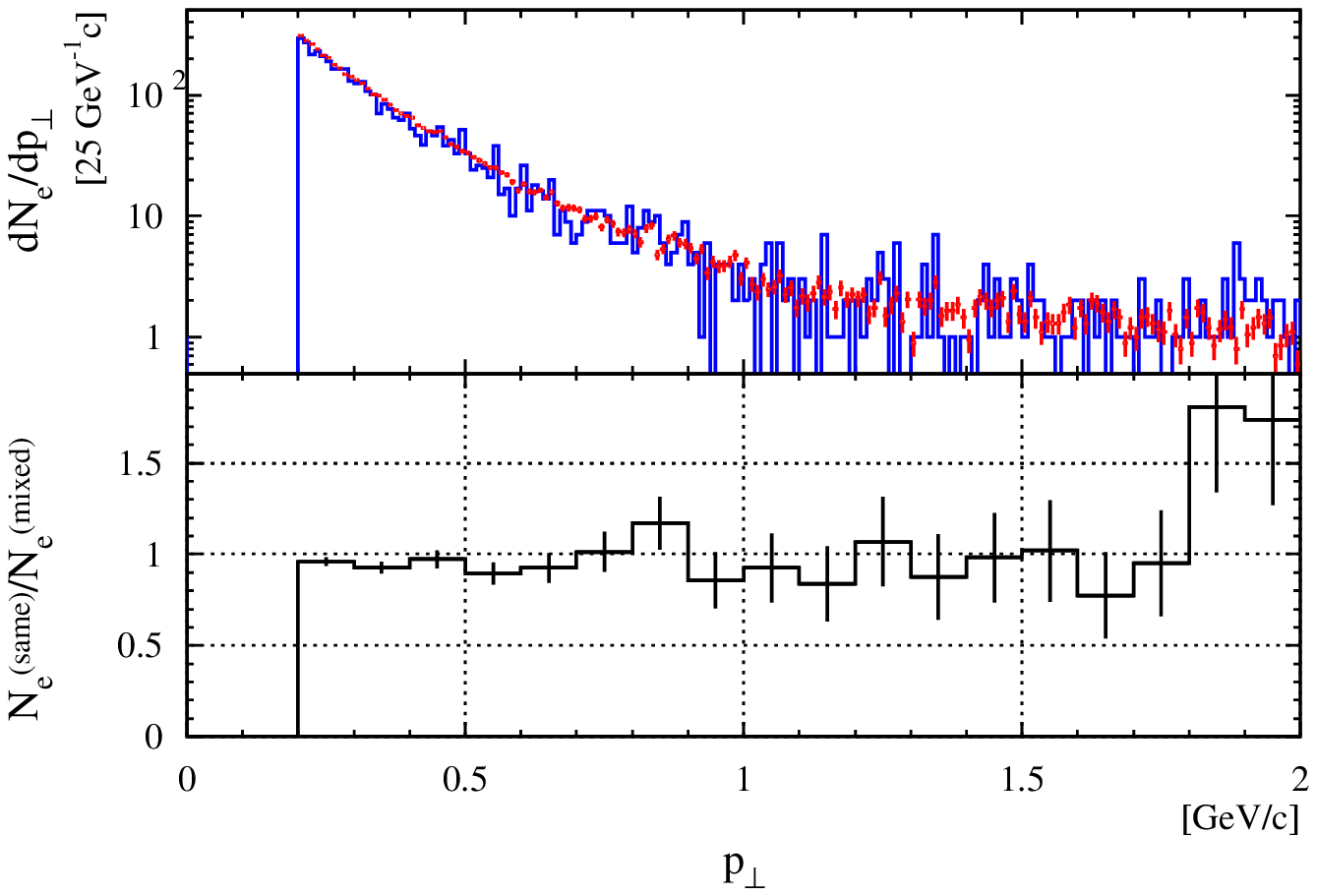}
  \caption[Comparison of same-event and mixed-event single-track $p_{\bot}$-dis\-tri\-bu\-tion]
  {Comparison of same-event (blue line) and mixed-event (red line)
  $p_{\bot}$-distribution of single tracks.}
  \label{fig:mix-pt}
\end{figure}

According to condition~\ref{equ:mixed cond1}, the shape of the
mass distribution of the mixed unlike-sign background is very
sensitive to variations of single-track reconstruction efficiency
or acceptance with time. It was carefully checked that the
properties of the single tracks contained in the mixed-event and
the same-event selection match well. The single-track
distributions of same-event and mixed-event background agree well
with each other, as apparent in Fig.~\ref{fig:mix-theta}.
The small differences in the $\phi$-distribution are restricted to
low-efficiency regions in the RICH detectors and acceptance holes
of the SDDs. These can be attributed to the inherent averaging of
discontinuities by the mixed-event technique. The impact of this
effect can be neglected because the occupancy is low in these
regions. The close resemblance of both $p_\bot$-distributions of
mixed and same-event selections, as seen in Fig.~\ref{fig:mix-pt},
is proof for local efficiency changes in RICH-2 or PD detector not
to alter the momentum determination.

\subsection{Comparison of mixed-event and same-event background}
\label{sec:mix vs comb}

\begin{figure}
  \centering
  \epsfig{file=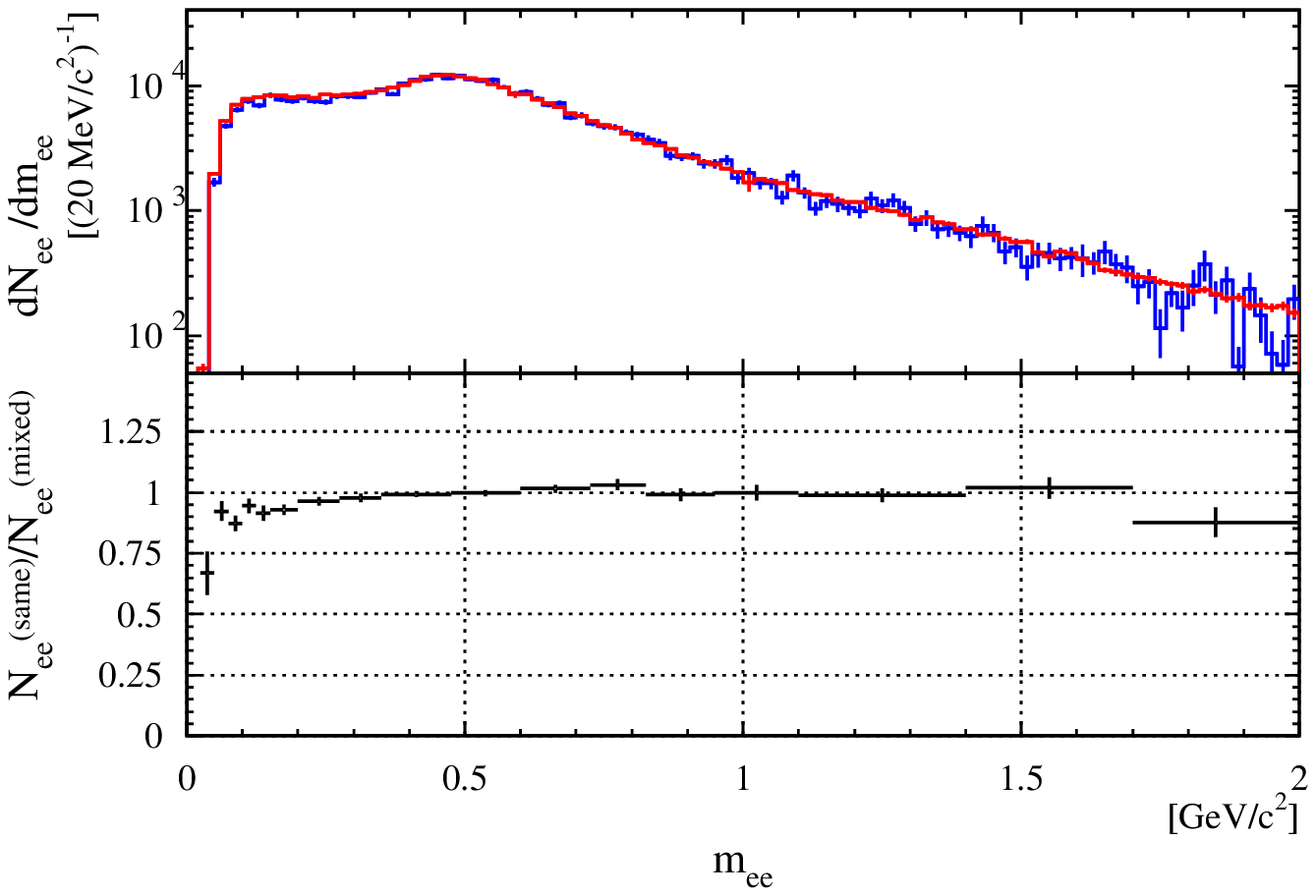,width=13cm}
  \caption[Comparison of same-event like-sign and mixed-event unlike-sign background]
  {Comparison of same-event like-sign (blue line) and mixed-event unlike-sign (red line) background
  (all rejection cuts applied).}
  \label{fig:mix-comb-ratio}
  \epsfig{file=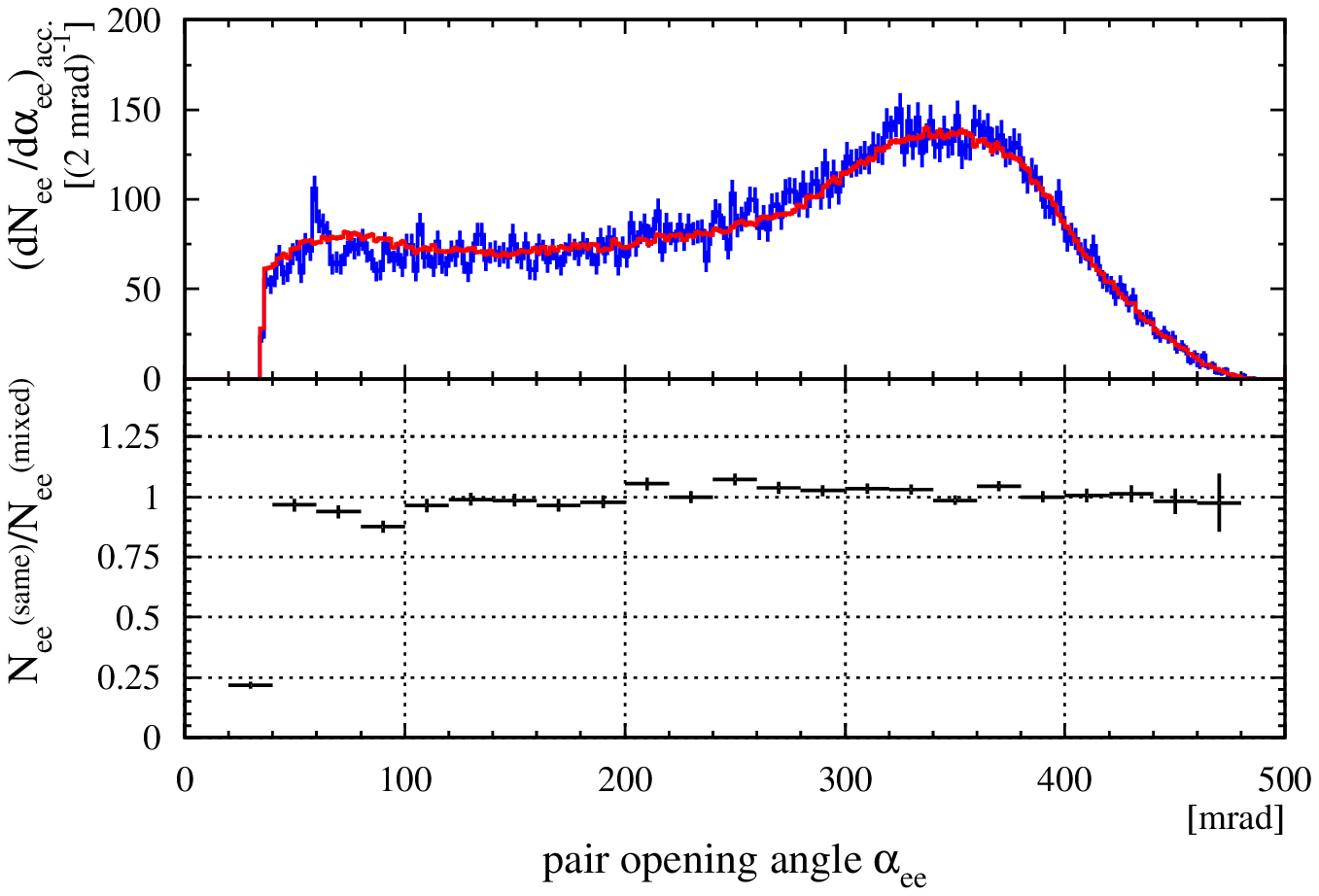,width=13cm}
  \caption[Comparison of same-event and mixed-event pair-opening-angle distribution]
  {Comparison of same-event (blue line) and mixed-event (red line) pair-opening-angle distribution.}
  \label{fig:mix open}
\end{figure}

After the construction of the same-event and the mixed-event
background in Sec.~\ref{sec:comb} and~\ref{sec:mixed}, the
inherent assumptions of both methods can be verified by comparison
of the invariant-mass distributions plotted in
Fig.~\ref{fig:mix-comb-ratio}.
The agreement of the shape of both distributions is striking
considering that the number of background pairs varies by more
than 2 order of magnitude in the observed mass region. Combining
Eqs.~\ref{equ:combbk} and~\ref{equ:mixed norm n+-} gives:
\begin{eqnarray}
 \underbrace{2\,\sqrt{\langle n_{++} \rangle \langle n_{--}} \rangle}_{\rm same-event\,bg}  & \stackrel{?}{=} &
 \underbrace{\langle n_{+-}^{mixed}\rangle}_{\rm mixed\,bg} \nonumber \\
 \varepsilon_{+}\,\varepsilon_{-}\,\sqrt{\,\kappa_{++}\kappa_{--}}\,
 \overline{N_+}\,\overline{N_-} & \stackrel{?}{=} &
 \overline{\varepsilon_{+}}\,\overline{\varepsilon_{-}}
 \overline{N_+}\,\overline{N_-}\;.
 \label{equ:bk equiv}
\end{eqnarray}
Thus, the observed equivalence requires that:
\begin{equation}
 \varepsilon_{+}\,\varepsilon_{-} =
 \overline{\varepsilon_{+}}\,\overline{\varepsilon_{-}}\qquad\mbox{and}\qquad
 \sqrt{\,\frac{{\rm d}\kappa_{++}}{{\rm d}m_{\rm ee}}\,\frac{{\rm d}\kappa_{--}}{{\rm d}m_{\rm ee}}} = 1\;.\label{equ:bk cond}
\end{equation}
According to Eq.~\ref{equ:bk cond}, the mixed-event average of the
single-particle detection probabilities resembles the same-event
average, and the two-track-correlation factor $\kappa$ must be
very close to one. This is a remarkable result. It proves the
particular choice of mixing-sub-sample size to be appropriate to
ensure sufficient temporal stability of the track efficiency and
of the pair acceptance with respect to pair mass. Note, that the
relative error of the mixed- to same-event ratio is dominated by
the statistical error of the same-event background as expected for
a relative mixing ratio of 20.

The 10\% drop of the ratio in Fig~\ref{fig:mix-comb-ratio} for
pairs with mass below 350\,MeV/c$^2$ is caused by the finite
two-track resolution of the same-event background. The finite
spatial resolution of both RICH
detectors~\cite{Voigt:1998,Lenkeit:1998} and the artifacts of the
ring reconstruction for touching rings introduce a correlation for
pairs with small opening angles, as seen Fig.~\ref{fig:mix open},
that depends on the RICH-ring distance.

Although the different behaviour of like-sign and unlike-sign
pairs in the magnetic field (see Fig.~\ref{fig:cowboy}) does not
allow to directly conclude that:
\begin{equation}
 \kappa_{+-}(m_{\rm ee})\stackrel{?}{=}\,\sqrt{\,\kappa_{++}(m_{\rm ee})\kappa_{--}(m_{\rm ee})}
 = 1\;,
 \label{equ:alpha cond}
\end{equation}
it is still a very good approximation because the pair efficiency
$\kappa$ is dominated by the RICH-1 detector due to its comparably
low spatial resolution (Tab.~\ref{tab:detect resol}). Most
important, this effect is limited to the low-mass region. In
principle, the double-track reconstruction efficiency can be
implemented into the mixed-event background but this procedure is
obstructed by the complex experimental conditions (e.g.
multiple-anode hits in SDDs and overlapping RICH-rings), eluding a
precise description by the Monte Carlo simulation. Not correcting
for the double-track efficiency results in a systematic error of
the dilepton signal induced by the mixed-background subtraction.
However, it is limited to 10\% because of the large
signal-to-background ratio (S\,:\,B\,$\approx$\,1) in the low-mass
region.

It is worth mentioning that the mixed like-sign invariant-mass
distributions, namely $\langle n_{++}\rangle$ and $\langle
n_{--}\rangle$, agree very well with each other, as apparent from
the ratio plotted in Fig.~\ref{fig:mix ++,--}.
\begin{figure}[tb]
    \begin{minipage}[t]{.65\textwidth}
        \vspace{0pt}
        \epsfig{file=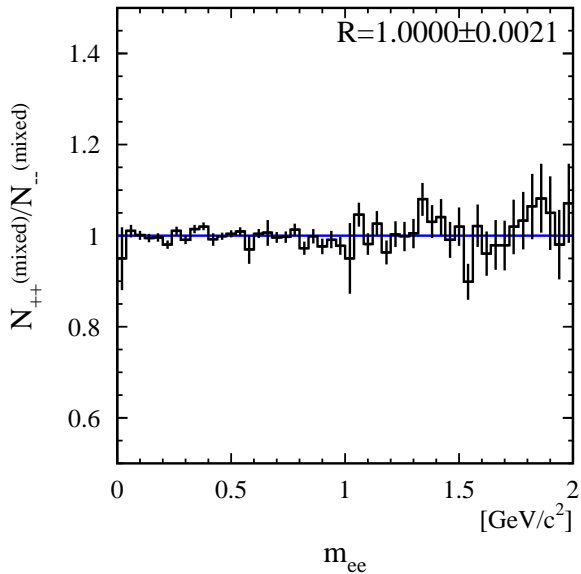,width=\textwidth}
    \end{minipage}%
    \begin{minipage}[t]{.35\textwidth}
      \vspace{0.5cm}
      \caption[Ratio of $(++)$- to $(--)$-mixed invariant mass dis\-tri\-bution]
      {\newline
       Ratio of $(++)$- to $(--)$-mixed invariant mass distribution.
      }
      \label{fig:mix ++,--}
    \end{minipage}
\end{figure}
Rewriting Eq.~\ref{equ:mixed norm n+-} in terms of $(++)$- and
$(--)$-pairs gives the condition for equivalence:
\begin{eqnarray}
 \langle n_{++}^{\rm mixed} \rangle & \equiv & \langle n_{--}^{\rm mixed} \rangle \nonumber\\
 \overline{\varepsilon_{+}}\,\overline{\varepsilon_{+}}\,\overline{N_+}\,\overline{N_+} & \equiv &
 \overline{\varepsilon_{-}}\,\overline{\varepsilon_{-}}\,\overline{N_-}\,\overline{N_-}\;.
\label{equ:mix ++,--}
\end{eqnarray}
Provided that charge symmetry enforces identical initial
multiplicity of $\overline{N_+}$ and $\overline{N_-}$ (see
Sec.~\ref{sec:mix-intro}), the average single-track detection
probability $\overline{\varepsilon}$ must be identical for both
charges, exactly as one would expect for the $\phi$-symmetry of
the CERES detector.

\begin{figure}[tb]
    \begin{minipage}[t]{.65\textwidth}
        \vspace{0pt}
        \epsfig{file=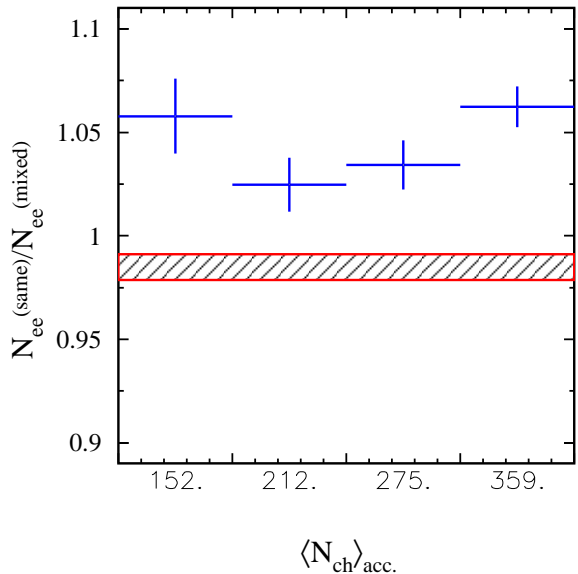,width=\textwidth}
    \end{minipage}%
    \begin{minipage}[t]{.35\textwidth}
      \vspace{0.5cm}
      \caption[Total normalization of mixed-event background]
      {\newline
      Ratio of the mixed-event unlike-sign background, normalized by the number
      of mixed-event pairs, to the same-event like-sign background for 4 multiplicity bins.
      Both spectra were corrected for single-track efficiency.
      }
      \label{fig:mix-norm}
    \end{minipage}
\end{figure}

While the arguments above were solely based on the shape of the
mixed-event invariant-mass distribution, its total normalization
is indispensable for a correct background subtraction. As noted
in~\cite{Lenkeit:1998}, an underestimation of the independent
background of about 4\% would exhaust the strength of the observed
dilepton excess. As already explained in Sec.~\ref{sec:mixed}, the
integrated mixed-event background is expected to slightly
overestimate the same-event background due to the two-track
efficiency losses. In contrast, the ratio plotted in
Fig.~\ref{fig:mix-norm} shows that the same-event background
overshoots the mixed-event background.
Correcting both distributions for the single track efficiency
determined in Sec.~\ref{sec:track eff} is not sufficient to
recover the correct total normalization. There is an indication
that this effect is related to large localized event-to-event
reconstruction efficiency changes in both RICH detectors. If a
continuous pad-wise gain calibration could resolve this problem,
remains to be seen in future analysis. Unfortunately, a slight
multiplicity dependence of the effect (see
Fig.~\ref{fig:mix-norm}) does not allow for a sufficiently
accurate re-normalization. It should be mentioned that, if this
effect were an artifact of the same-event background, it would
result in a twofold increase of pair signal which can be clearly
excluded.

To avoid these difficulties, the normalization of mixed-event
background was fixed with respect to the total number of observed
same-event background pairs with mass above 0.35\,GeV/c$^2$. This
selection explicitly excludes the low-mass range, where the
mixed-event background is not identical to the same-event
background (see Fig.~\ref{fig:mix open}). As a consequence, 20\%
of the gain in statistics of the mixed-event background had to be
sacrificed. The error of the total normalization is now limited by
the statistical error of the same-event like-sign background
integrated above $m_{\rm ee}$\,$>$\,$0.35$GeV/c$^2$.

To summarize, it is demonstrated that the mixed-event background
resembles the same-event like-sign background. Theoretical and
experimental consideration lead to the conclusion that both
accurately simulate the much sought-after independent unlike-sign
background. It is worth stressing that the observed likeness
excludes the existence of any significant pair correlations and,
therefore, strongly supports the validity of the background
subtraction procedure.

\clearpage

\section{Reduction of combinatorial background}
\label{sec:rejection}

\subsection{Rejection strategy}
\label{sec:rej strategy}

As set out in Sec.~\ref{sec:mix-intro}, the correlated-dilepton
signal has to be extracted from the observed number of unlike-sign
pairs by the subtraction of the independent combinatorial
background pairs (see Eq.~\ref{equ:pair signal}). In fact, the
contributions of the $\pi_0$~Dalitz decay, the $\gamma$ conversion
decay, and the single tracks of partially reconstructed pairs
would overwhelm the number of signal pairs by three orders of
magnitude for no further rejection. Moreover, the huge relative
statistical error of the signal, given by Eq.~\ref{equ:sig-error},
would invalidate any measurement owing to the very small number of
expected signal pairs. For this reason the recognition and
subsequent rejection of tracks stemming from the above mentioned
sources is of utmost importance for the reduction of the
combinatorial background.

Any rejection of tracks based on a certain cut criteria must
balance the obtained background rejection power $\varrho$ and the
unavoidable loss of efficiency $\varepsilon$. Both are related to
the relative statistical error of the pair signal by:
\begin{eqnarray}
 \frac{dS}{S} = \frac{\sqrt{S+2B^{\rm comb}}}{S} & \approx &
 \frac{\sqrt{2B^{\rm comb}}}{S} \\
 & \approx & \frac{\sqrt{1-\varrho}}{\varepsilon} \frac{ \sqrt{2
 B_{\rm initial}^{\rm comb}}}{S_{\rm initial}} \nonumber\;,
 \label{equ:ds/s opt}
\end{eqnarray}
where $\varrho$ denotes the probability to reject a combinatorial
background pair and $\varepsilon$ the pair reconstruction
efficiency for a particular set of cuts. According to
Eq.~\ref{equ:ds/s opt}, it is possible to optimize the statistical
significance of the observed signal by minimization of the ratio
$(1-\varrho)/\varepsilon^2$ with respect to all applied cuts.
Carried out on data, this procedure has the inherent danger of
selecting a statistical upward fluctuation~\cite{Sokol:1999}.
Therefore, the reconstruction efficiency was determined
independently in a Monte Carlo simulation of the detector
system~\cite{Lenkeit:1998}.

As already noted in Sec.~\ref{sec:deficiencies}, the rejection
strategy chosen in the previous
analyses~\cite{Sokol:1999,Lenkeit:1998} involved several
correlated multi-parameter cuts. The resulting optimization
procedure is very complex and difficult to reproduce. One of the
goals of this work was to simplify the rejection strategy by
focusing on a few powerful cuts and to gain better understanding
of those.

It is useful to contemplate the characteristics of $\pi_0$-Dalitz
and $\gamma$-conversion decays for the following discussion. The
yield of electron tracks originating from the $\pi_0$ decays is
given by:
\begin{equation}
 \frac{dN_{\rm e}}{dN_{\rm ch}}={\rm BR}_{\pi^0\rightarrow e^+e^-\gamma}\cdot\frac{dN_{\pi^0}}{dN_{\rm ch}}\cdot\frac{dN_{\rm ee}}{dN_{\pi^0}}
 =0.43\cdot0.01198\cdot2.=1.1\cdot10^{-2}\;.
 \label{equ:ndalitz}
\end{equation}
Here ${\rm BR}_{\pi^0\rightarrow e^+e^-\gamma}$ and
$\frac{dN_{\pi^0}}{dN_{\rm ch}}$ are the branching ratio and the
ratio of the expected number of $\pi_0$~mesons to the number of
observed charged particles within the acceptance of the CERES
spectrometer, respectively. Equation~\ref{equ:ndalitz} results in
an average of 1.4 electrons tracks per event for a mean value of
N$_{\rm ch}$\,=\,138 ($\eta$\,=\,2.1--2.65). The additional
contributions of the $\eta_0$ and $\eta'$ Dalitz decay are
negligible (see Sec.~\ref{app:genesis}).

All photons created either by meson Dalitz decay or in the initial
collision can convert into dilepton pairs while traversing, first,
the target and, then, the downstream detectors. The induced yield
is determined by the dominating meson decays:
\begin{eqnarray}
 \frac{dN_{\rm e}}{dN_{\rm ch}} = 2 \cdot \frac{dN_{\pi^0}}{dN_{\rm ch}} \cdot
 \frac{dN_{\gamma}}{dN_{\pi_0}}\cdot\frac{7}{9}\cdot\frac{X}{X_0}& = &
 2\cdot0.44\cdot2.25\cdot\frac{7}{9}\cdot\frac{X}{X_0} = 1.5\cdot\frac{X}{X_0}
 \\
 & = & 1.5\cdot \frac{\rm 0.0025\,cm}{\rm 0.35\,cm} =
 1.1\cdot10^{-2}\,\,\,({\rm target})\nonumber\\
 & = & 1.5\cdot \frac{\rm 0.028\,cm}{\rm 9.36\,cm} =
 4.5\cdot10^{-3}\,\,\,\mbox{(SDD-1)} \nonumber\;,
 \label{equ:nconv}
\end{eqnarray}
where $X/X_0$ is the thickness of the material in terms of its
radiation length. The conversions in the target and in the SDD-1
lead to a mean track-multiplicity of 1.5 and 0.6, respectively,
for a mean value of N$_{\rm ch}$\,=\,138. Any conversions
occurring downstream of SDD-2 are rejected by requiring a particle
hit in both SDDs.

Both conversion and Dalitz pairs are distinguished from all other
sources by their very low mass of $m_{\rm
ee}$\,$<$\,0.2\,GeV/c$^2$ (see Fig.~\ref{fig:cocktail} in
App.~\ref{app:genesis}). As the low-momentum tracks with
$p_{\bot}$\,$<$\,60\,MeV/c cannot be reconstructed in the magnetic
field (see Sec.~\ref{sec:rich tracking}), many of those pair are
only partially reconstructed and, hence, the mass remains unknown.
Owing to this, the rejection is primarily based on the
pair-opening-angle characteristic. Figure~\ref{fig:dielectron
opang} illustrates that most Dalitz and conversion pairs have an
opening angle of less than $35$\,mrad which makes it an equally
distinguishing feature as the pair mass.
\begin{figure}
  \centering
  \mbox{
   \epsfig{file=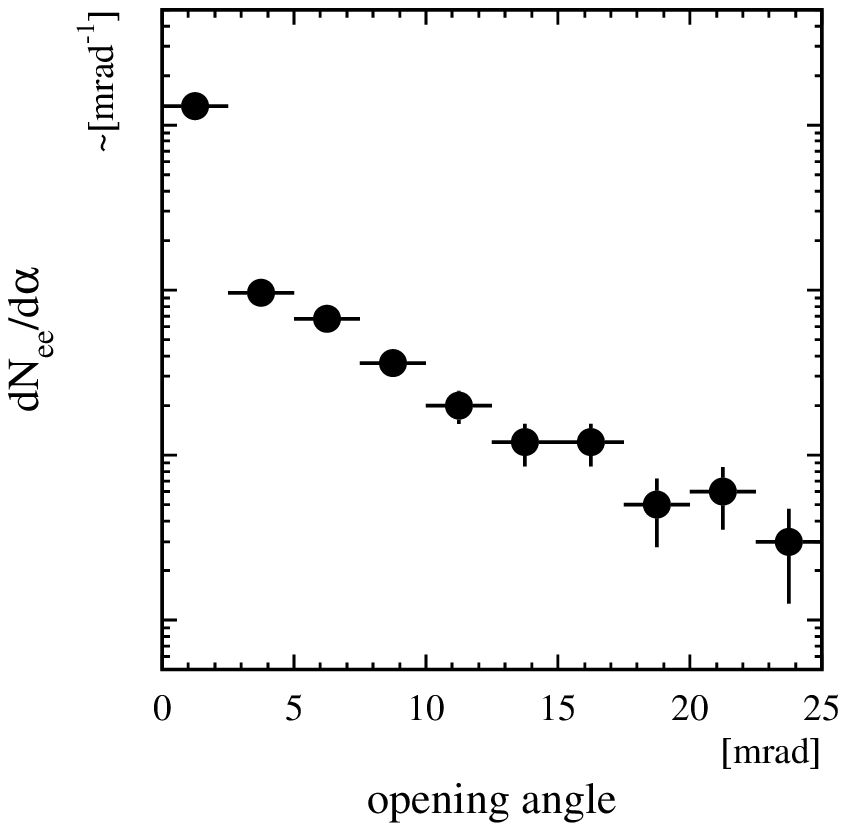,width=.49\textwidth}
   \hfill
   \epsfig{file=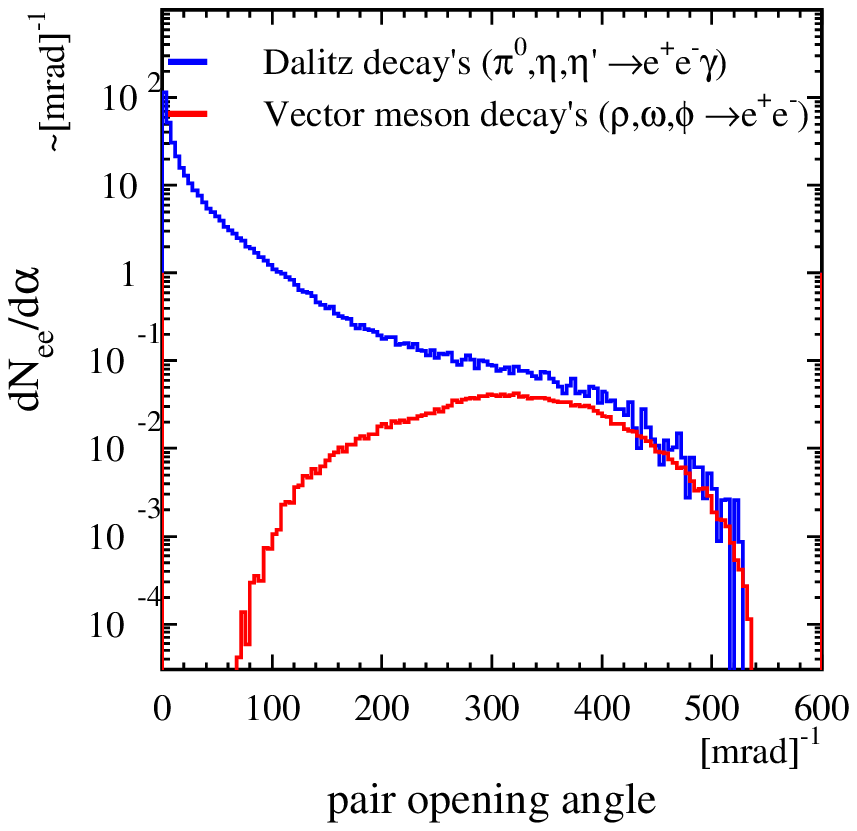,width=.49\textwidth}
  }
  \caption[Simulated opening-angle distribution of dielectron sources]
  {Simulated opening-angle distribution of target and SDD-1 conversions (left panel) and of Dalitz and
  vector meson decays (right panel). The spectrum of conversion pairs
  was obtained by Monte Carlo simulation of Pb-Au collisions created
  by the URQMD model~\cite{Bass:1996ud}. The opening-angle distribution of pairs originating
  from Dalitz and vector meson decays was modeled with the GENESIS event generator~\cite{Sako:2000}.}
  \label{fig:dielectron opang}

  \vspace*{.8cm}
  \epsfig{file=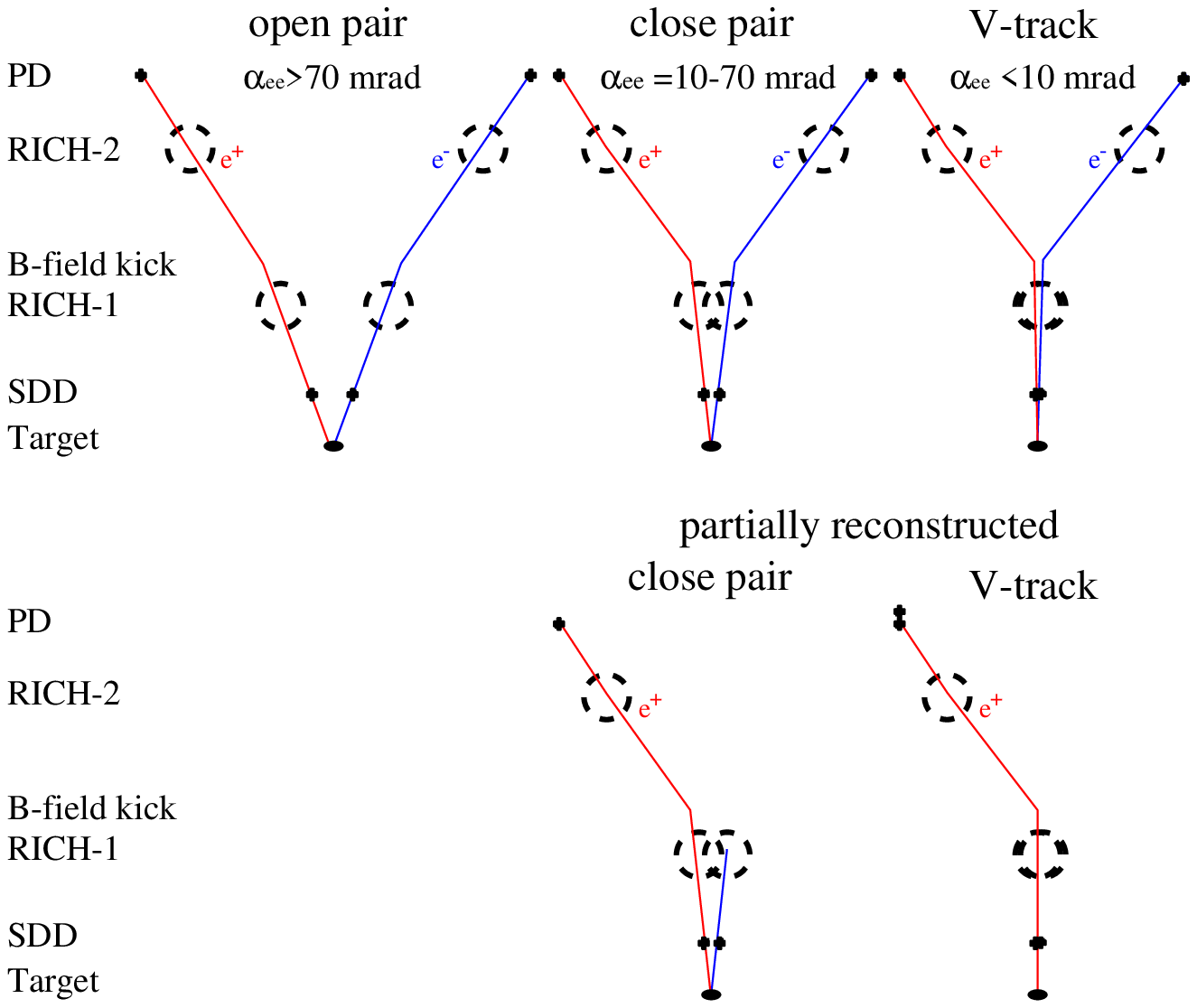,width=11cm}
  \caption[Schematic view of the distinguished dielectron configurations]
  {Schematic view of the distinguished dielectron configurations.}
  \label{fig:pair config}
\end{figure}
Thus, even partially reconstructed Dalitz and conversion pairs can
still be recognized as such, if an additional hit in the SDDs or a
ring in the RICH-1 detector was found in the close vicinity of the
reconstructed track. Figure~\ref{fig:pair config} illustrates the
notations for the common pair configurations.

\begin{figure}[htb]
  \epsfig{file=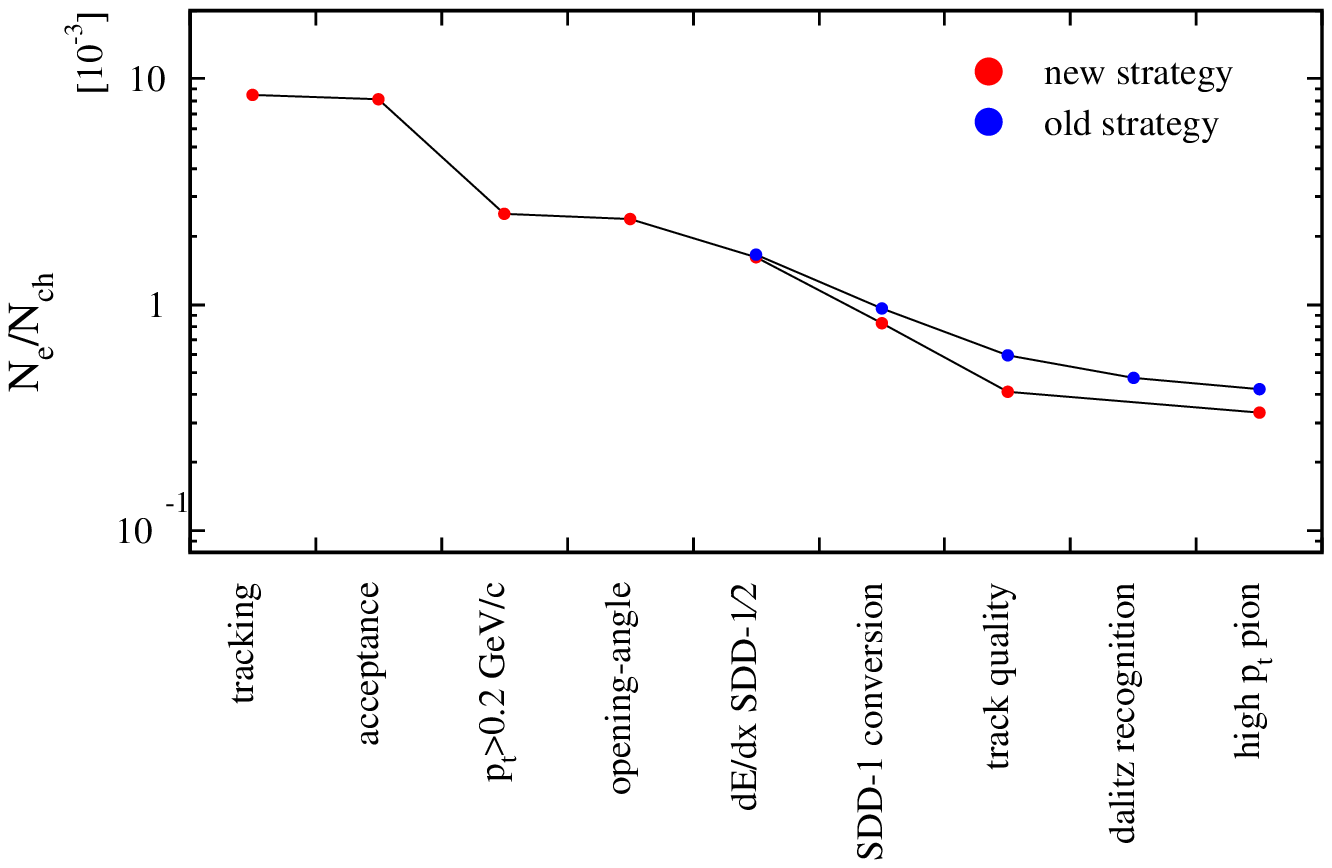}
  \caption[Combinatorial background reduction by rejection cuts]
  {Combinatorial background reduction by rejection cuts.
  The values of the previous analysis were included for comparison (see Fig.~6.5 in~\cite{Lenkeit:1998}).}
  \label{fig:bk rejection cuts}
\end{figure}

In anticipation of the following detailed study of each rejection
cut, a summary of the complete background rejection strategy is
already given in Fig.~\ref{fig:bk rejection cuts}.
In contrast to the previous analysis (see Fig.~6.5
in~\cite{Lenkeit:1998}), the new strategy encompasses only five
rejection cuts discarding two, but at the same time improves the
rejection power by a factor of 1.4\@.

The so-called Dalitz cut was discarded for two reasons. First, it
rejected little background and, second, single tracks were
rejected relying on the properties of low-mass unlike-sign
"signal" pairs which cannot be simulated by the mixed-event
technique because all mixed pairs are uncorrelated by definition.
The cut also induces a subtle background correlation explained in
the following. All pairs of a given event that share a track with
an unlike-sign Dalitz pair, identified by a pair mass below
150\,MeV/c$^2$ and an opening angle of less than 50\,mrad, are
rejected. Only 50\% of the identified pairs are truly Dalitz
pairs, as the signal-to-background ratio is about 1:1. A
misidentified Dalitz pair rejects one unlike-sign and one
like-sign pairs in the example of a three-track event. In most
cases, the unlike-sign pair rejected must have a larger mass than
the misidentified Dalitz pair. However, no such restriction
applies to the rejected like-sign pair. As a result, the
unlike-sign background is slightly overestimated/underestimated by
the like-sign background for large/small mass, respectively. For
the second discarded cut, namely the close-partial-track cut, it
was found that the gain in rejection power did not offset the
decrease in efficiency, thereby, lowering the significance of the
signal.

A comprehensive summary of all rejection cuts is presented in
Table~\ref{tab:cuts} in App.~\ref{app:cut summary}. Note, that the
order of the rejection-cuts is important as most of those are
partially correlated. \clearpage

\subsection{Double-d{\em E}/d{\em x} rejection in SDD-1 and SDD-2}
 \label{sec:dedx}

\begin{figure}
  \centering
  \epsfig{file=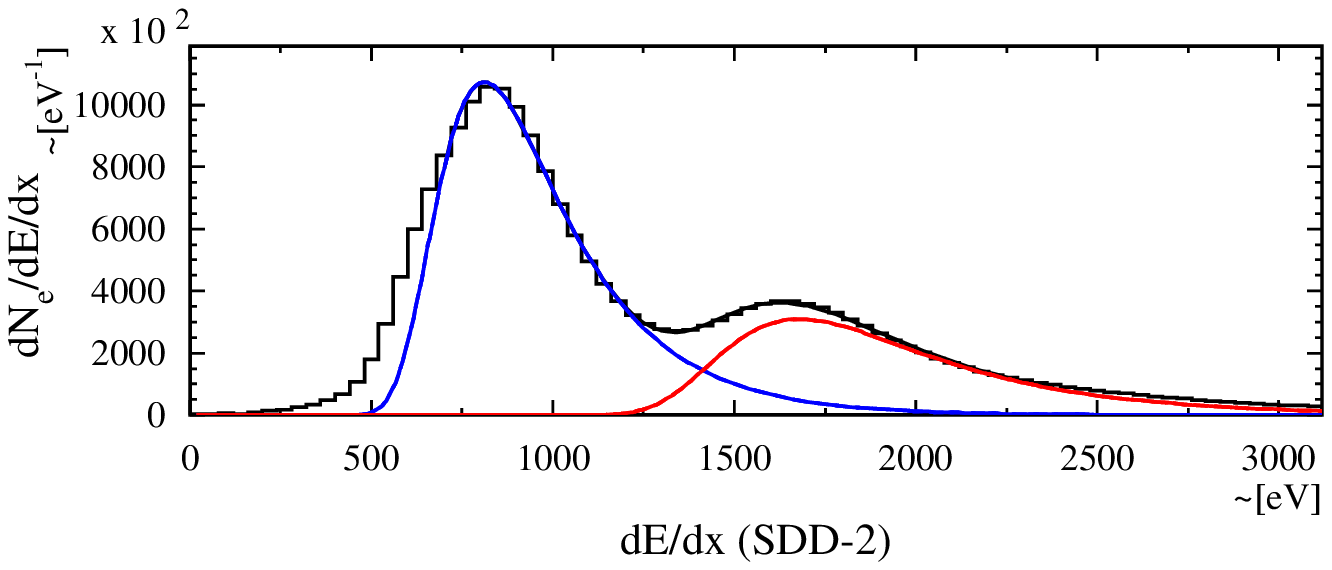}
  \caption[Double-Landau fit of the d{\em E}/d{\em x} distribution in SDD-2]
  {Double Landau fit of the d{\em E}/d{\em x} distribution in SDD-2. The d{\em E}/d{\em x}
  distribution of single and double hits is indicated by a blue and red line, respectively.
  The apparent difference for tracks with d{\em E}/d{\em x} below
  700 FADC counts is caused by artificially split hits which are shifted to lower d{\em E}/d{\em x} (see
  Sec.~\ref{sec:sidc hits}).
  }
  \label{fig:dedx fit}

  \vspace*{1cm}
  \epsfig{file=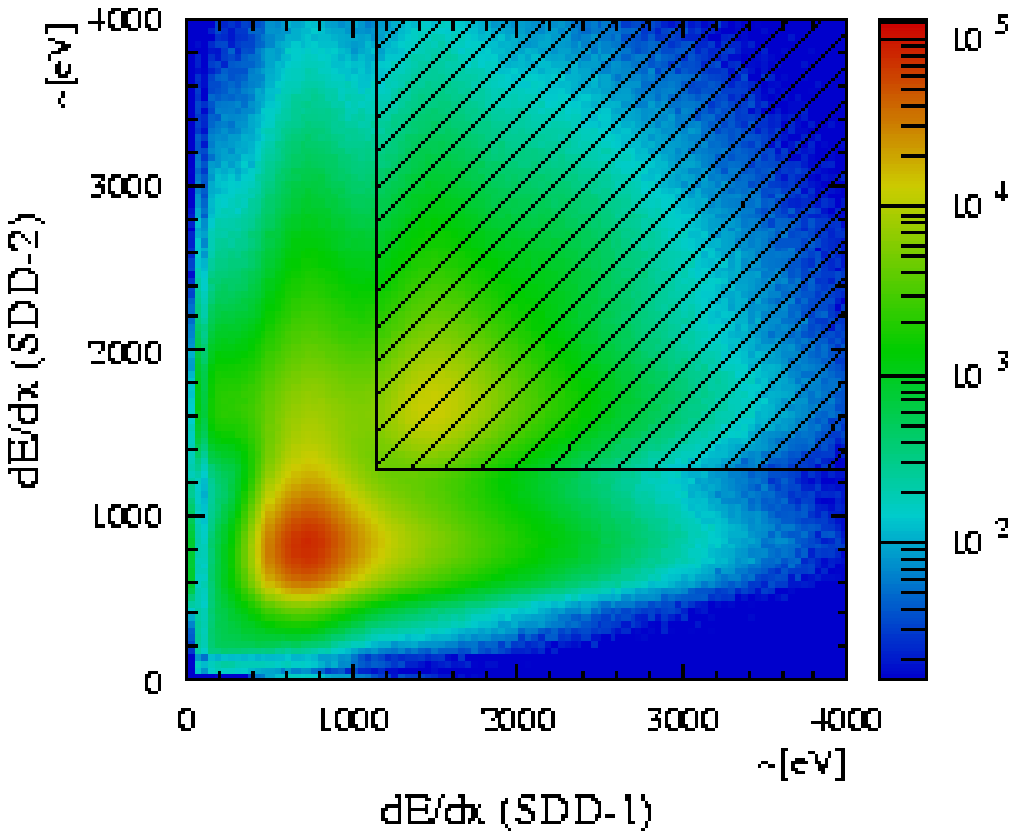}
  \caption[SDD-1 vs. SDD-2 d{\em E}/d{\em x} distribution]
  {SDD-1 vs. SDD-2 d{\em E}/d{\em x} distribution. The hatched area is rejected by the double-d{\em E}/d{\em x} cut.}
  \label{fig:2d dedx cut}
\end{figure}

The energy loss of electrons {\em E}$_{\rm\,loss}$ in the SDDs can
be approximated by a Landau distribution~\cite{PDBook}:
\begin{equation}
 \frac{d^2N}{dEdx} \sim \exp{(-\frac{x}{2}+{\rm e}^{-x})}\qquad\mbox{with}\qquad x = \frac{E_{\rm loss}-E_{\rm
 max}}{\sigma}\;,
 \label{equ:landau}
\end{equation}
where {\em E}$_{\rm\,max}$ and $\sigma$ denote the most probable
energy loss and the width of the distribution, respectively. For a
thin detector the distribution is skewed towards high energies
(the so-called Landau tail) due to the large fluctuations of the
number of collisions involving large energy transfer. As apparent
in Fig.~\ref{fig:dedx fit}, the Landau distribution describes the
measured data very well, even though the observed shape is a
convolution with various electronics and detector characteristics
as described in~\cite{Weber:1997}.

Dilepton pairs with an opening angle of less than 2.5\,mrad cannot
be resolved in the SDDs and therefore deposite twice the energy of
a single track. The corresponding double-amplitude peak is clearly
seen Fig.~\ref{fig:dedx fit}. The target conversions and close
Dalitz pairs distinguished by their small opening angle are most
efficiently rejected by a correlated cut requiring high hit
amplitude in both SDDs as outlined in Fig.~\ref{fig:2d dedx cut}.
In SDD-1, the pair-opening-angle range to be rejected was
artificially increased by addition of the amplitude of the next
closest hit in a 5\,mrad range to further enhance the rejection
power.

In view of the fact that in the previous analysis the double-d{\em
E}/d{\em x} peak (in Fig.~\ref{fig:2d dedx cut}) was already
rejected at the first stage of the data analysis, it might be
conceivable that important calibrations were casuistical or simply
overlooked. Most important, the so-called ballistic-deficit
correction accounts for the fact that the measured number of FADC
counts decreases as a function of the radial position of the hit.
The radial width of the electron cloud increases with the drift
time due to diffusion according to~\cite{Woerner:1990}:
\begin{equation}
 \sigma^2_r(r)=\sigma^2_{r,r_0}+2\,D\,t_{\rm drift}(r)\;,
 \label{equ:sidc drift}
\end{equation}
where $\sigma_{r,r_0}$ and {\em D} are the initial radial width of
the hit and the diffusion constant, respectively. The response
function of the SDD pre-amplifier depends on the width of the
input signal and, therefore, the measured output signal decreases
for hits at the lower radius. This effect is further amplified by
fact that the amplitude threshold, applied in the hit
reconstruction for the noise suppression, introduces a relative
amplitude loss that is largest for hits with a large width. This
implies that the SDD-hit amplitude has to be re-calibrated
inasmuch as the amplitude threshold is changed.

\begin{figure}[tb]
  \centering
  \mbox{
   \epsfig{file=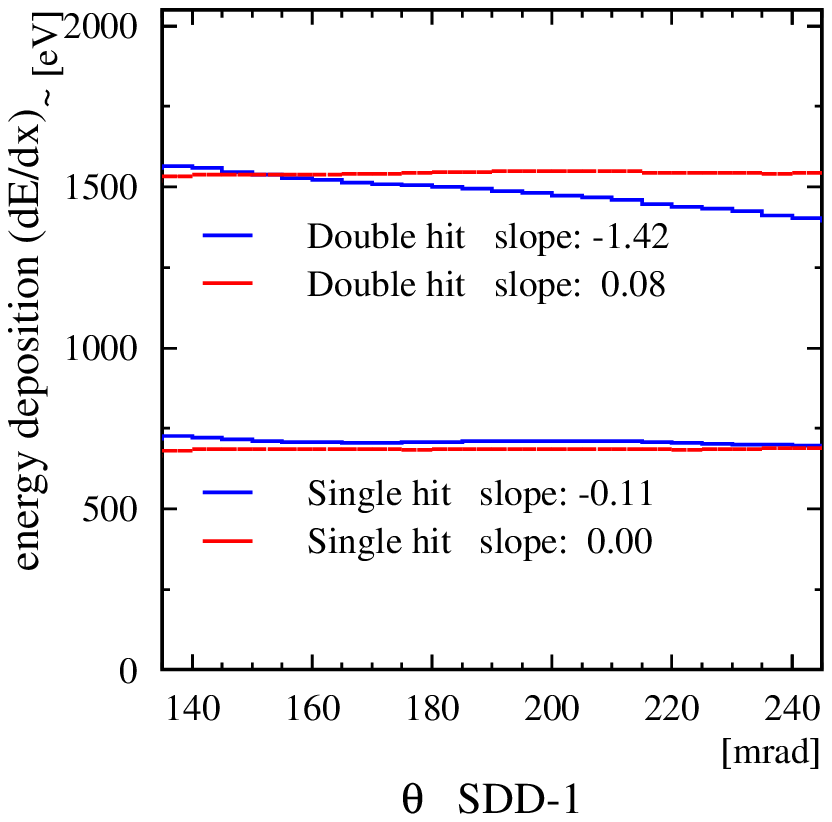,width=.49\textwidth}
   \hfill
   \epsfig{file=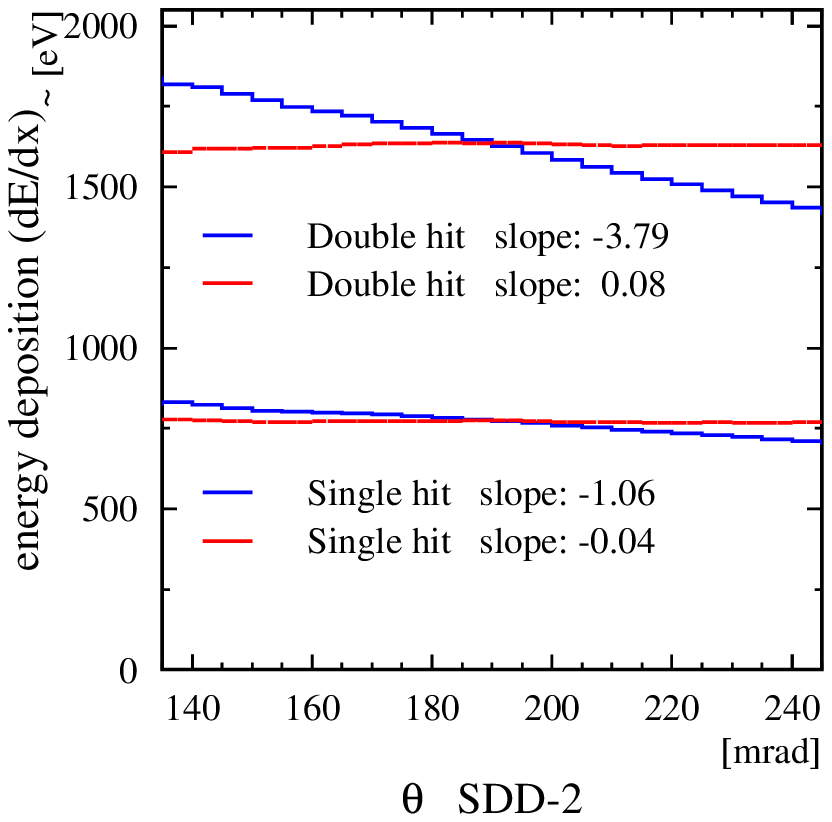,width=.49\textwidth}
  }
  \caption[SDD-ballistic-deficit correction]
  {Most probable number of FADC counts of SDD-hits measured as function of the
   radial($\theta$-) hit position for single and double hits. To measure d{\em E}/d{\em x} correctly, the
   number of FADC counts needs to be corrected for the ballistic
   deficit and the amplitude threshold effect. However, a significant decrease of the d{\em E}/d{\em
   x} measured is observed towards large radii after applying the correction
   function of the previous analysis~\cite{Lenkeit:1998} (blue
   line). The constants of d{\em E}/d{\em x} desired was regained
   after the recalibration (red line).}
  \label{fig:dedx-theta}
\end{figure}

\begin{figure}[htb]
  \epsfig{file=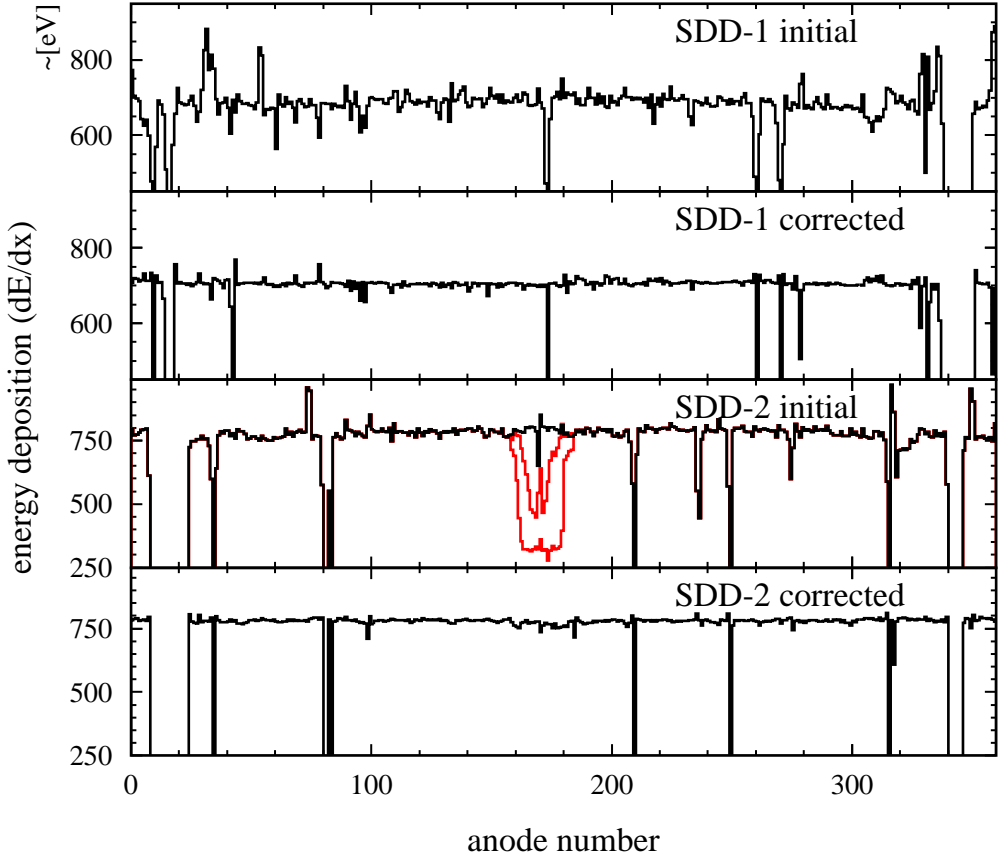}
  \caption[Anode-wise recalibration of d{\em E}/d{\em x} distribution]
  {Anode-wise recalibration of the d{\em E}/d{\em x} distribution.
   A hardware problem in SDD-2 appearing in run 308 resulted in an increasing
   deterioration of the d{\em E}/d{\em x} measurement of 20 anodes (red
   line).}
  \label{fig:anode dedx}
\end{figure}

For the previous calibration shown in Fig.~\ref{fig:dedx-theta}
the mean hit amplitude decreases by as much as 30\% towards the
outer radius of the SDDs, indicating an over-correction of the
ballistic deficit and the amplitude threshold effect.
As the relative change in $\theta$-direction varies also with the
absolute amplitude value, the new $\theta$-dependent calibration
function was interpolated for hits with amplitudes between those
of single- and double-amplitude hits. An additional correction
factor was applied for each anode to account for small amplitude
variations illustrated in Fig.~\ref{fig:anode dedx}.
Small temperature variations of the SDD system during the 6 weeks
of data acquisition altered the observed mean amplitude by up to
2\%~\cite{Hering:2001a} and were corrected for each run.

As a result of the recalibration, the width of the Landau
distribution decreased by about 30\%, thereby improving the
separation of single- and double-d{\em E}/d{\em x} distribution.

\begin{figure}[tb]
  \centering
  \epsfig{file=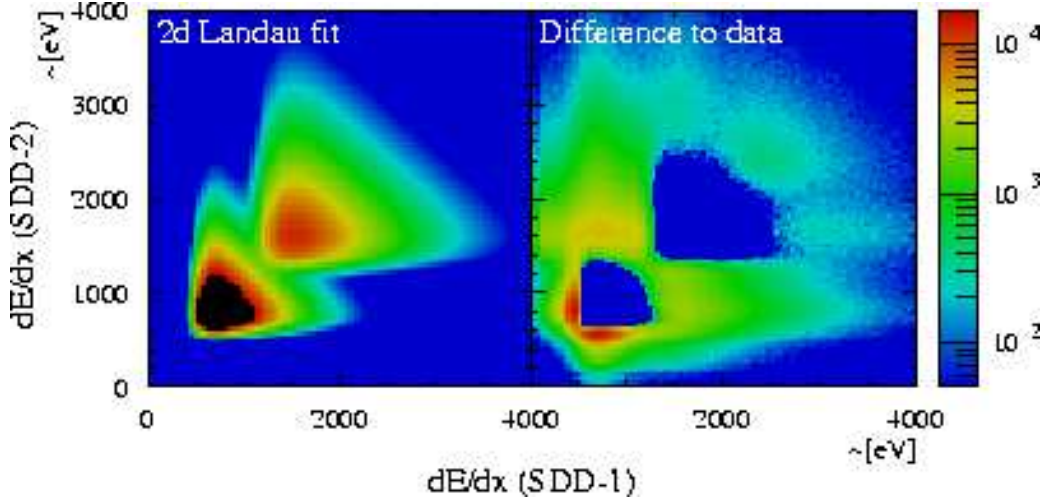}
  \caption[2d-double-Landau fit of d{\em E}/d{\em x} distribution]
  {2-dimensional-double-Landau fit of d{\em E}/d{\em x} distribution.}
  \label{fig:2d dedx fit}
\end{figure}

To optimize the cut values of the double-d{\em E}/d{\em x}
rejection, a simple method solely based on the track properties
was developed. In contrast to the Monte Carlo simulation used
in~\cite{Lenkeit:1998}, this approach sidesteps the difficulties
involved in the simulation of the very complex SDD
characteristics. Assuming that the amplitude distributions of
single- and double-amplitude hits were exactly known,
Eq.~\ref{equ:ds/s opt} could be employed to find the optimum
values of the d{\em E}/d{\em x}-contour cut as shown in
Fig.~\ref{fig:2d dedx cut}. First, the probability to observe a
track with single or double amplitude is defined as:
\begin{equation}
 P_{\rm single}= \frac{N^{\rm track}_{\rm single}}{N^{\rm
 track}}\qquad\mbox{and}\qquad
 P_{\rm double}= \frac{N^{\rm track}_{\rm double}}{N^{\rm track}}\; .
 \label{equ:dedx track}
\end{equation}
Applying the cut, the probability of a single track surviving that
cut defines the efficiency $\varepsilon$ and the rejection power
$r$ is given by the probability to reject a double-d{\em E}/d{\em
x} track:
\begin{equation}
 \varepsilon  = \frac{N^{\rm track}_{\rm single}({\rm surviving})}{N^{\rm track}_{\rm
 single}}\qquad\mbox{and}\qquad
  r  = \frac{N^{\rm track}_{\rm double}({\rm rejected})}{N^{\rm track}_{\rm
 double}}\;.
 \label{equ:dedx prob}
\end{equation}
The probability that a signal pair survives the cut can be
expressed in terms of the single-track efficiency as:
\begin{equation}
 P^{\rm ee}_{\rm signal}= {\varepsilon}^2\,P_{\rm single}^2\;.
 \label{equ:2dedx si}
\end{equation}
All combinatorial pairs containing at least one double-amplitude
track are regarded as background. The probability to find such a
pair is given by:
\begin{equation}
 P^{\rm ee}_{\rm background}=(1-r)^2\,P_{\rm double}^2 + 2\,{\varepsilon}\,(1-r)\,P_{\rm single}\,P_{\rm
 double}\;.
 \label{equ:2dedx bk}
\end{equation}
Expressing Eq.~\ref{equ:ds/s opt} in terms of Eqs.~\ref{equ:dedx
track}, \ref{equ:2dedx si}, and~\ref{equ:2dedx bk} yields the
optimization function:
\begin{equation}
 \max \left(\frac{B+S}{S^2}\right)\sim
 \max \left(\frac{(1-r)^2}{{\varepsilon}^4}+\frac{1}{\varepsilon^2}\,\left(\frac{N^{\rm track}_{\rm single}}{N^{\rm track}_{\rm double}}\right)^2
 +\frac{2(1-r)}{\varepsilon^3}\,\frac{N^{\rm track}_{\rm single}}{N^{\rm track}_{\rm
 double}}\right)\;.
 \label{equ:2dedx opt}
\end{equation}
Next, the observed d{\em E}/d{\em x} distribution of
Fig.~\ref{fig:2d dedx cut} was fitted with a two-di\-men\-sional
double-Landau distribution. The fit result is shown in
Fig.~\ref{fig:2d dedx fit}.
The wide peak at a double amplitude in SDD-2 and about single
amplitude in SDD-1, seen in the distribution of the residual
difference between the fit and the data, is obviously related to
$\gamma$~conversion decays in SDD-1. These were excluded from the
fit to be treated in a separate rejection cut. Although the fit
function underestimates the tails of the data distribution, it was
made sure, that the most important region between the peaks of
both distributions is described accurately enough and the residual
does not affect the optimization.

\begin{figure}[!t]
  \centering
  \epsfig{file=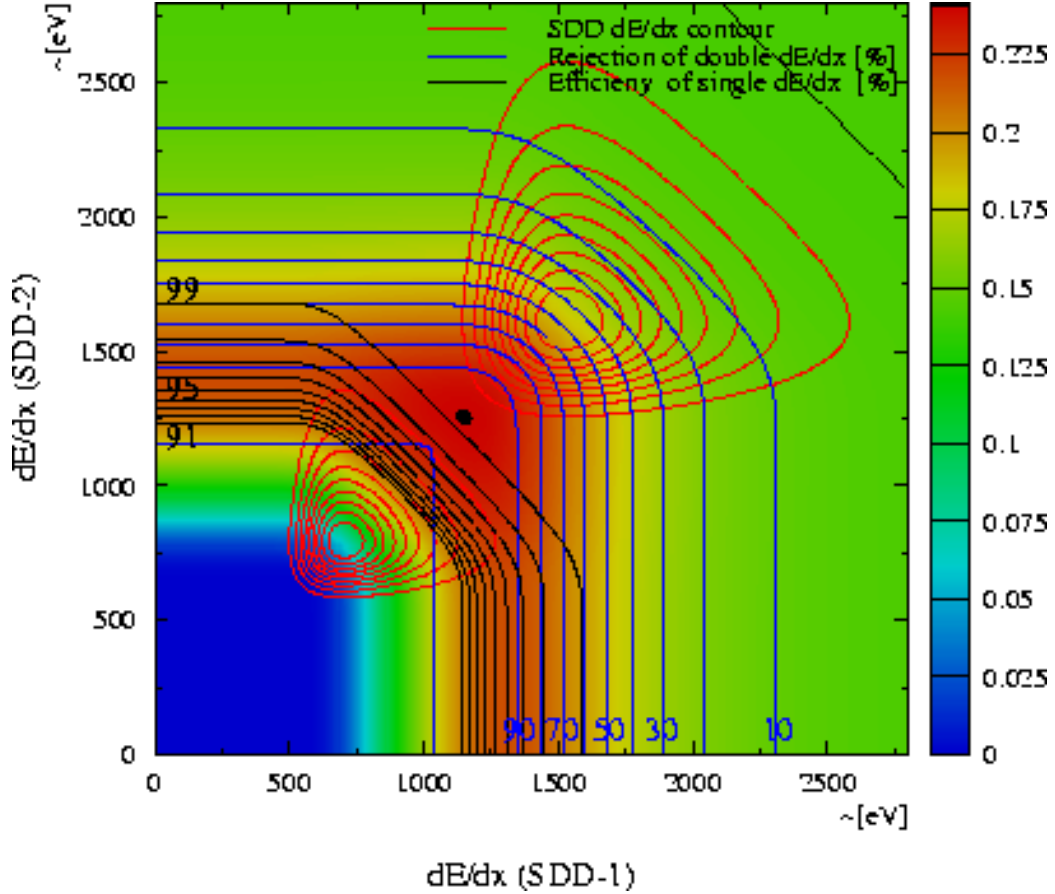}
  \caption[Efficiency
  and rejection contours of the double-d{\em E}/d{\em x} cut]
  {Efficiency and rejection contours of the double-d{\em E}/d{\em x}
  cut overlaid on the optimization function. The value of the
  optimization function (see Eq.~\ref{equ:2dedx opt}) illustrated by
  color levels peaks at the cut values of d{\em E}/d{\em x}$_{\rm
\,SDD-1}$\,$>$\,$1130\pm20$ and d{\em E}/d{\em x}$_{\rm \,
SDD-2}$\,$>$\,$1240\pm20$ corresponding to a rejection power of
about 95\% and an efficiency of 99\%, respectively.}
  \label{fig:dedx eff}
\end{figure}

The contours of cut values with equal efficiency and equal
rejection are plotted in Fig.~\ref{fig:dedx eff}.
The value of the optimization function (see Eq.~\ref{equ:2dedx
opt}) also shown in Fig.~\ref{fig:dedx eff} peaks at the cut
values of d{\em E}/d{\em x}$_{\rm \,SDD-1}$\,$>$\,$1130\pm20$ and
d{\em E}/d{\em x}$_{\rm \, SDD-2}$\,$>$\,$1240\pm20$ corresponding
to a rejection power of about 95\% and an efficiency of 99\%,
respectively. The maximum is relatively broad and, therefore, the
significance of the signal must be insensitive to slight
variations of the cut values. Applying the d{\em E}/d{\em x}
rejection cut to the data improves the signal-to-background ratio
by a factor of two.

\subsection{SDD-1 conversion rejection}

\begin{figure}[htb]
  \centering
  \epsfig{file=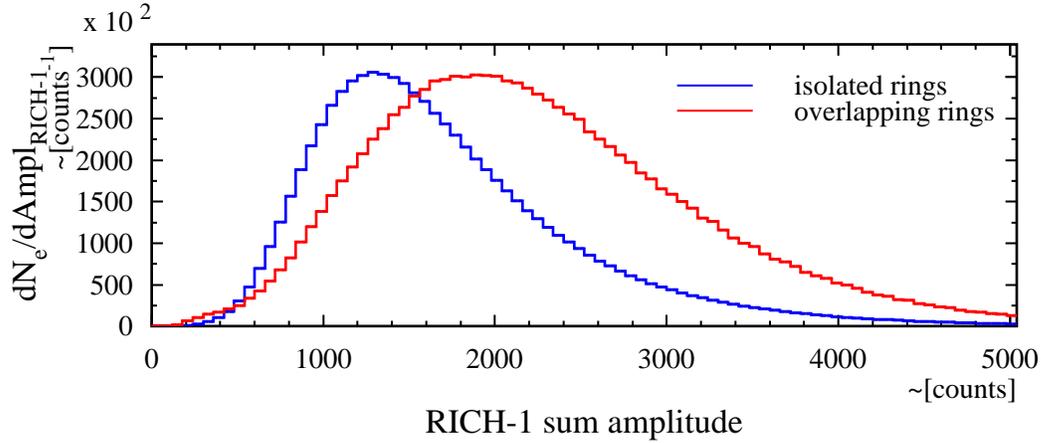}
  \caption[RICH-1 sum-amplitude of isolated- and overlapping-rings]
  {RICH-1 sum-amplitude of isolated- and overlapping-rings. Isolated rings were selected
  by requiring a single-track d{\em E}/d{\em x} value in both SDDs and no V-track signature
  in the RICH detectors (see Fig.~\ref{fig:pair config}). V-tracks exhibiting a
  double-track d{\em E}/d{\em x} value in both SDDs were regarded as overlapping rings.}
  \label{fig:r1sum}
\end{figure}

The main characteristics of SDD-1 $\gamma$-conversions are the
following: a single-amplitude hit in SDD-1, a double-amplitude hit
in SDD-2 or a second hit in the close vicinity, and an
overlapping- or double-ring in RICH-1 detector. Therefore, the
SDD-1 conversion cut rejects tracks with a double amplitude in
SDD-2 (including a summation of the amplitude of the next closest
hit within 7.5\,mrad) and a sum amplitude in RICH-1 larger than
1550 counts. An investigation revealed that the rejection power of
this cut is limited mainly by the poor separation of isolated- and
overlapping-rings in RICH-1 detector (only 50\% see
Fig.~\ref{fig:r1sum}).
\begin{figure}
  \centering
  \epsfig{file=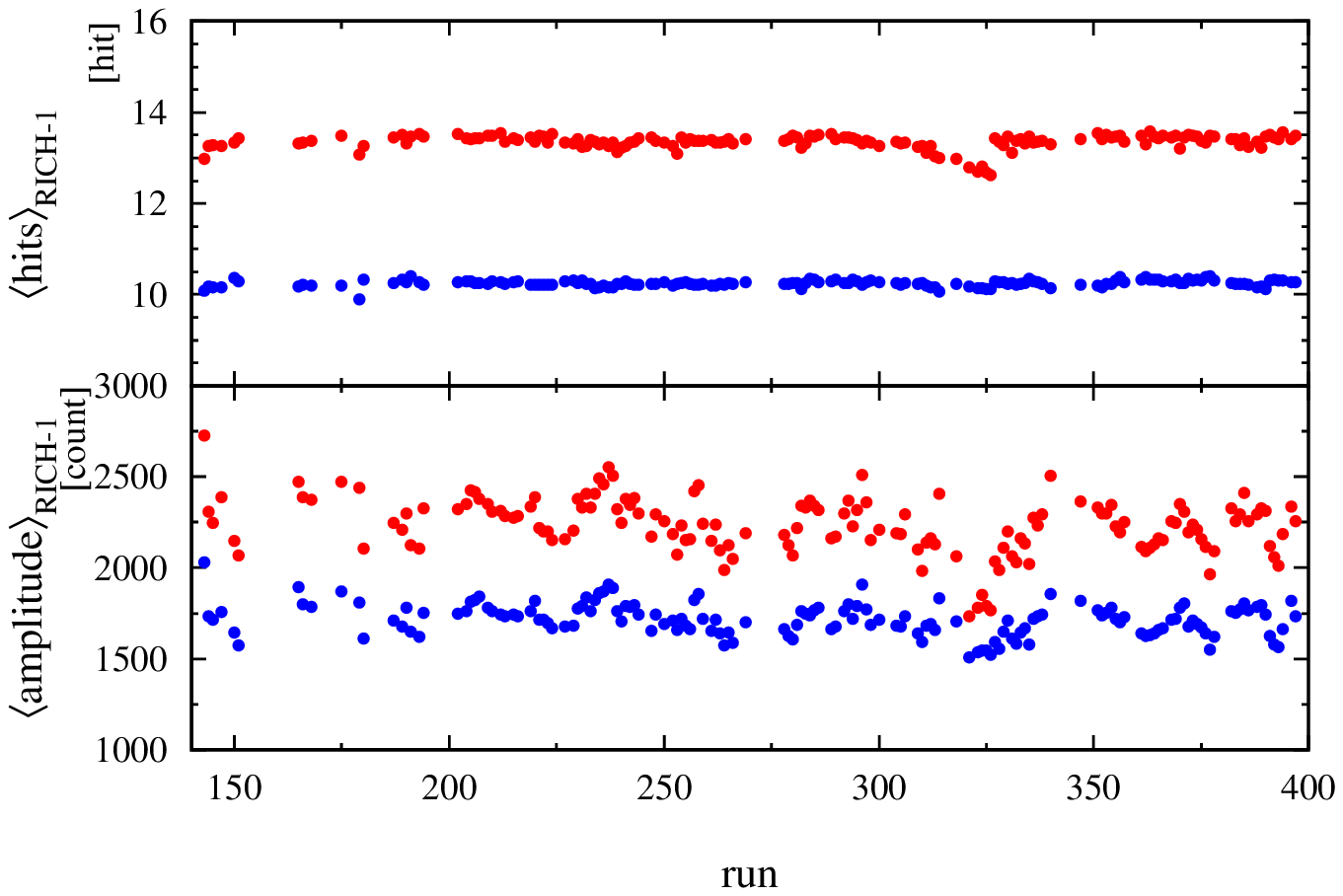}
  \caption[Run-to-run variation of the number of hits per ring
   and the ring sum-amplitude of isolated- and overlapping-rings in RICH-1]
  {Run-to-run variation of the most probable number of hits per ring
   and the most probable ring sum-amplitude of isolated- and overlapping-rings
   in the RICH-1 detector indicated by blue and red points, respectively. Isolated rings were selected
   by requiring a single-track d{\em E}/d{\em x} value in both SDDs and no V-track signature
   in the RICH detectors (see Fig.~\ref{fig:pair config}). V-tracks exhibiting a
   double-track d{\em E}/d{\em x} value in both SDDs were regarded as overlapping rings.
   }
  \label{fig:r2r-r1}

  \vspace*{.2cm}
  \begin{minipage}[t]{.63\textwidth}
       \vspace{0pt}
       \epsfig{file=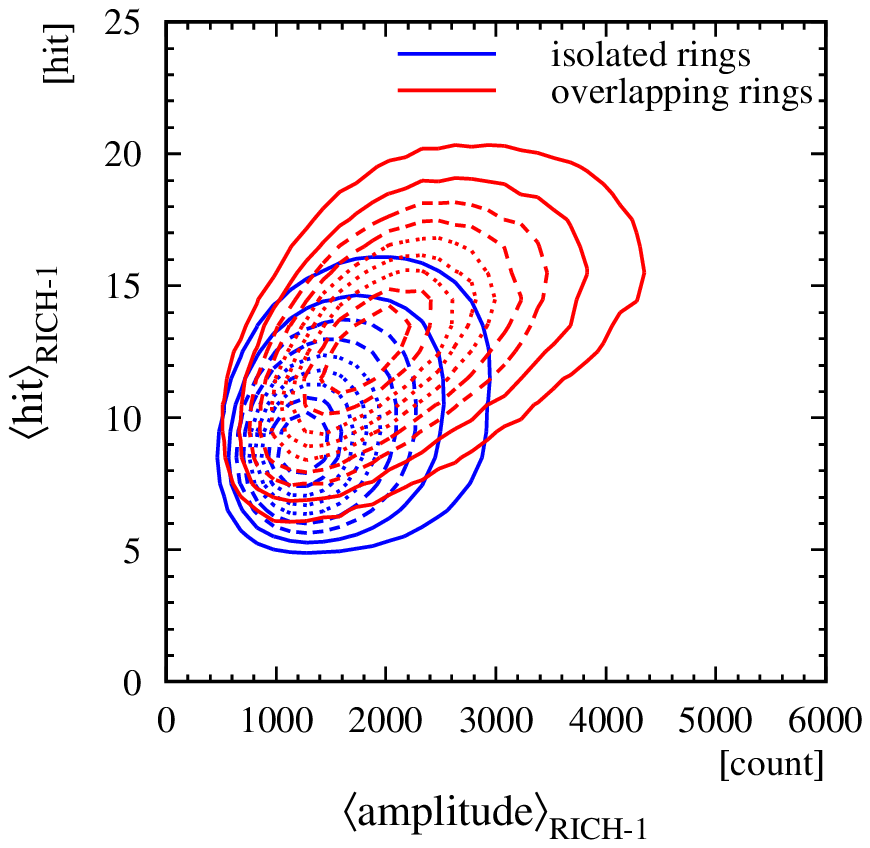,width=\textwidth}
  \end{minipage}%
  \begin{minipage}[t]{.37\textwidth}
      \caption[RICH-1 sum-amplitude vs. mean number of hits per ring of isolated- and overlapping-rings]
      {\newline
      RICH-1 sum-amplitude vs. mean number of hits per ring of isolated- and
      overlapping-rings. Isolated rings were selected by requiring a single-track d{\em E}/d{\em x}
      value in both SDDs and no V-track signature in the RICH detectors (see Fig.~\ref{fig:pair config}).
      V-tracks exhibiting a  double-track d{\em E}/d{\em x} value in both SDDs
      were regarded as overlapping rings.
      }
      \label{fig:2d r1sum nph}
    \end{minipage}
\end{figure}
In parts, this can be attributed to the fact that the selected
double rings are indeed only partly overlapping ring with an
opening angle of less than about 10\,mrad. Depending on the actual
center of the reconstructed ring, a certain fraction of the
overlapping rings will not be covered by the summation mask, which
is a ring with about 6\,mrad width. A simple calculation shows the
maximum sum-amplitude separation between isolated- and
overlapping-rings to be about a factor of 1.7\@.

The mean amplitude per ring varies by 25\% with time as apparent
in Fig.~\ref{fig:r2r-r1}, indicating that the gain of RICH-1
detector was not always properly readjusted to account for
variations of atmospheric pressure. The mean number of hits per
ring depicted in Fig.~\ref{fig:r2r-r1} is an alternative
observable to distinguish isolated from overlapping rings. It is
much less sensitive to gain variations but the separation of
isolated- and overlapping-rings is equally poor. If a correlated
cut including the number of hits per ring and the amplitude sum
shown in Fig.~\ref{fig:2d r1sum nph} would improve the situation
remains to be seen. The above described version of the SDD-1
conversion cut rejects about 60\% of the like-sign background.

\subsection{Track quality}

Several track quality criteria help to reject so-called fake
tracks reconstructed from accidentally matching hits and rings. A
detailed description can be found
in~\cite{Ceretto:1998,Messer:1998,Lenkeit:1998}. The cut values
chosen in this analysis are largely identical to the previous
analyses except for those affected by the improved SDD resolution.

\begin{figure}[!tb]
  \centering
  \mbox{
   \epsfig{file=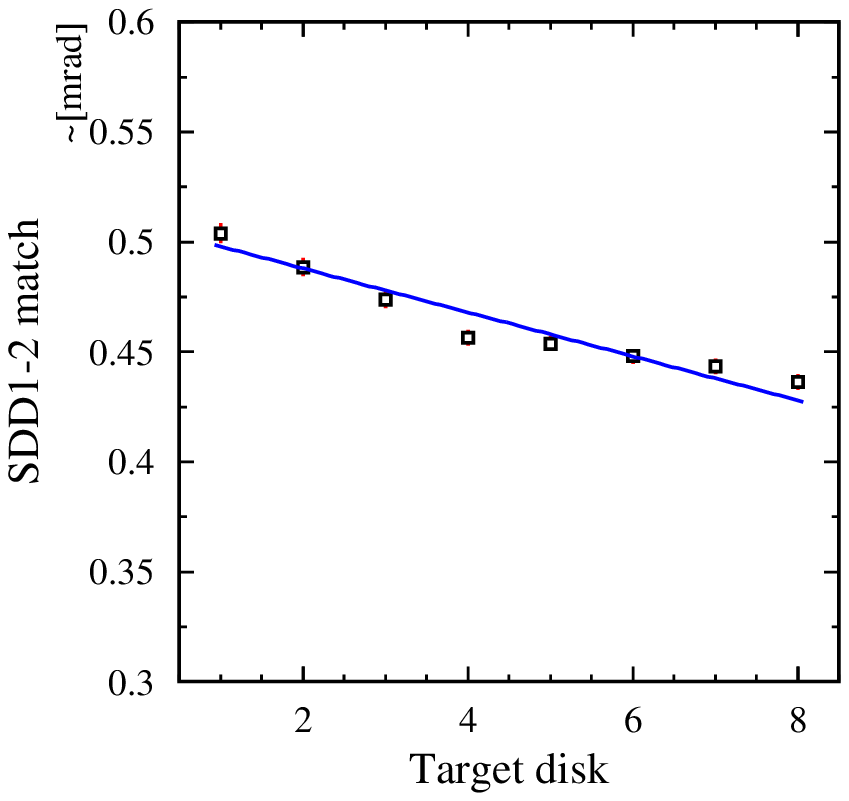,width=.49\textwidth}
   \hfill
   \epsfig{file=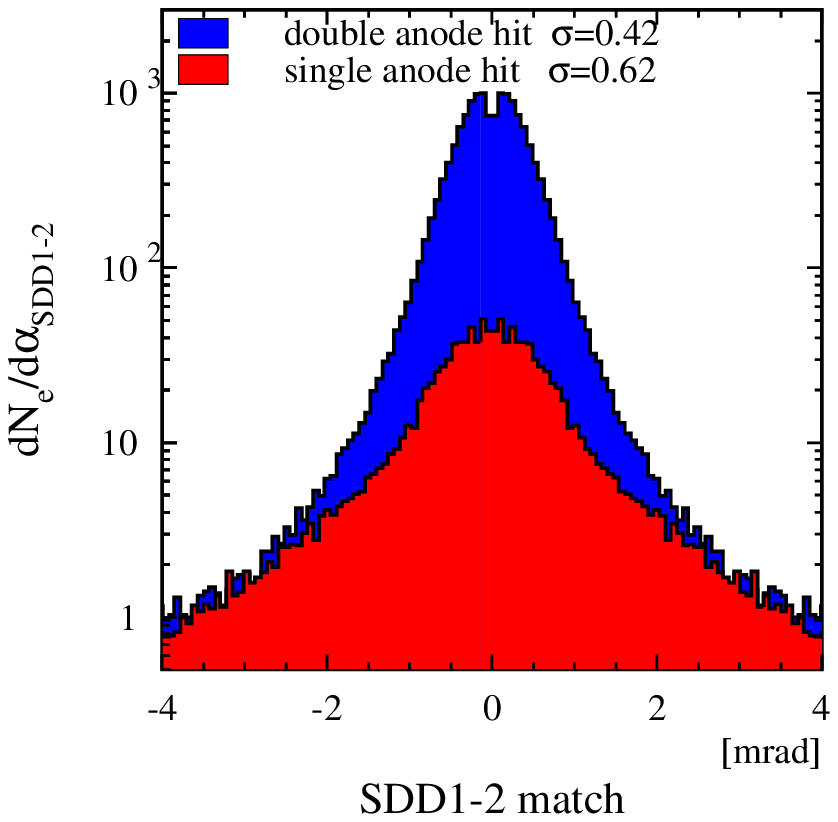,width=.49\textwidth}
  }
  \caption[Matching quality of SDD-1 and SDD-2]
  {Width of the matching distribution of SDD-1 and SDD-2 as a function of the target
  disk (left panel). Matching distribution of SDD-1 and SDD-2 of single- and
  double-anode hits (right panel).
  All tracks with a SDD-matching quality of less than 1.3\,mrad were rejected.}
  \label{fig:s12-target}
\end{figure}

Most important, the SDD-1--SDD-2 matching cut was refined to
1.3\,mrad including the dependence on z-position of the event
vertex~\cite{CERES:02-05}. The data plotted in the left panel of
Fig.~\ref{fig:s12-target} shows the 15\% resolution increase
expected between the first and the last target disk.
It is worth stressing that the SDD-matching cut plays an important
role in the rejection of target conversions and close Dalitz
decays because it acts like an opening-angle cut for partially
reconstructed pairs (i.e.~one SDD hit missed). The great
disadvantage of this cut is to reject a substantial fraction of
tracks comprising single-anode hits which exhibit a very poor
matching resolution (right panel of Fig.~\ref{fig:s12-target}).

The rejection of displaced artificially split hits results in a
pair-efficiency loss of about 16\% which cannot be recovered
without sacrificing rejection power. As a result, the SDD-matching
cut had to be loosened by 50\% compared to the previous
analysis~(0.9\,mrad) in order to maximize the statistical
significance of the open-pair signal.

\begin{figure}
  \epsfig{file=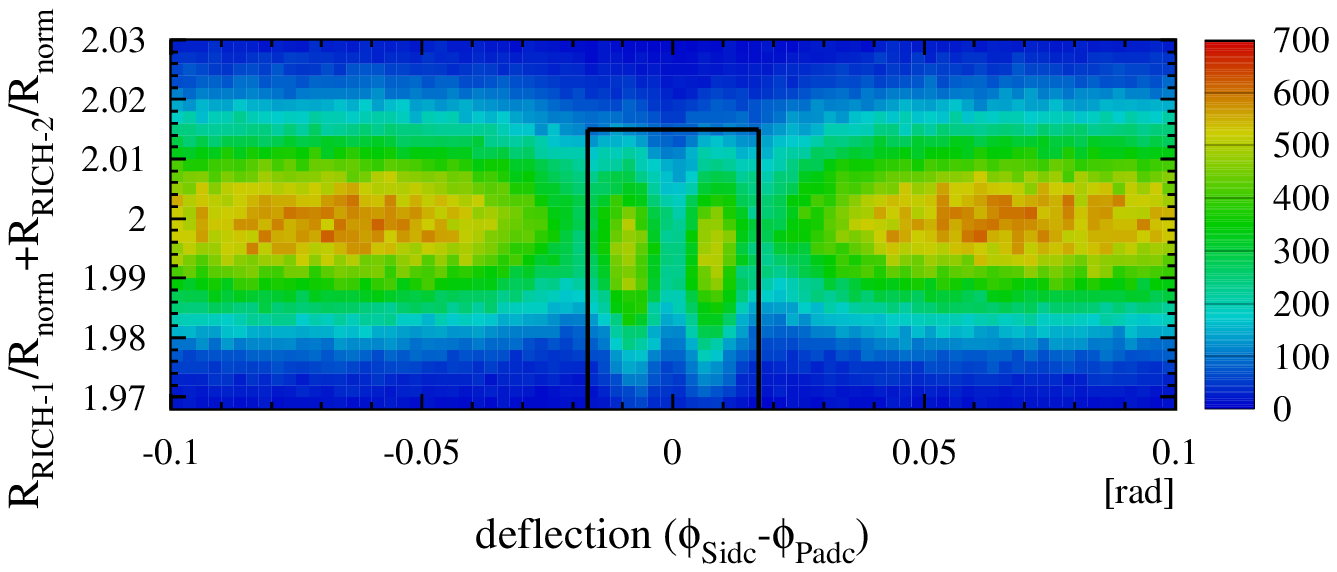}
  \caption[Rejection of misidentified high-momentum pions]
  {Rejection of misidentified high-momentum pions (see App.~\ref{app:cut summary}).}
  \label{fig:pi-cut}

  \vspace*{1cm}
  \epsfig{file=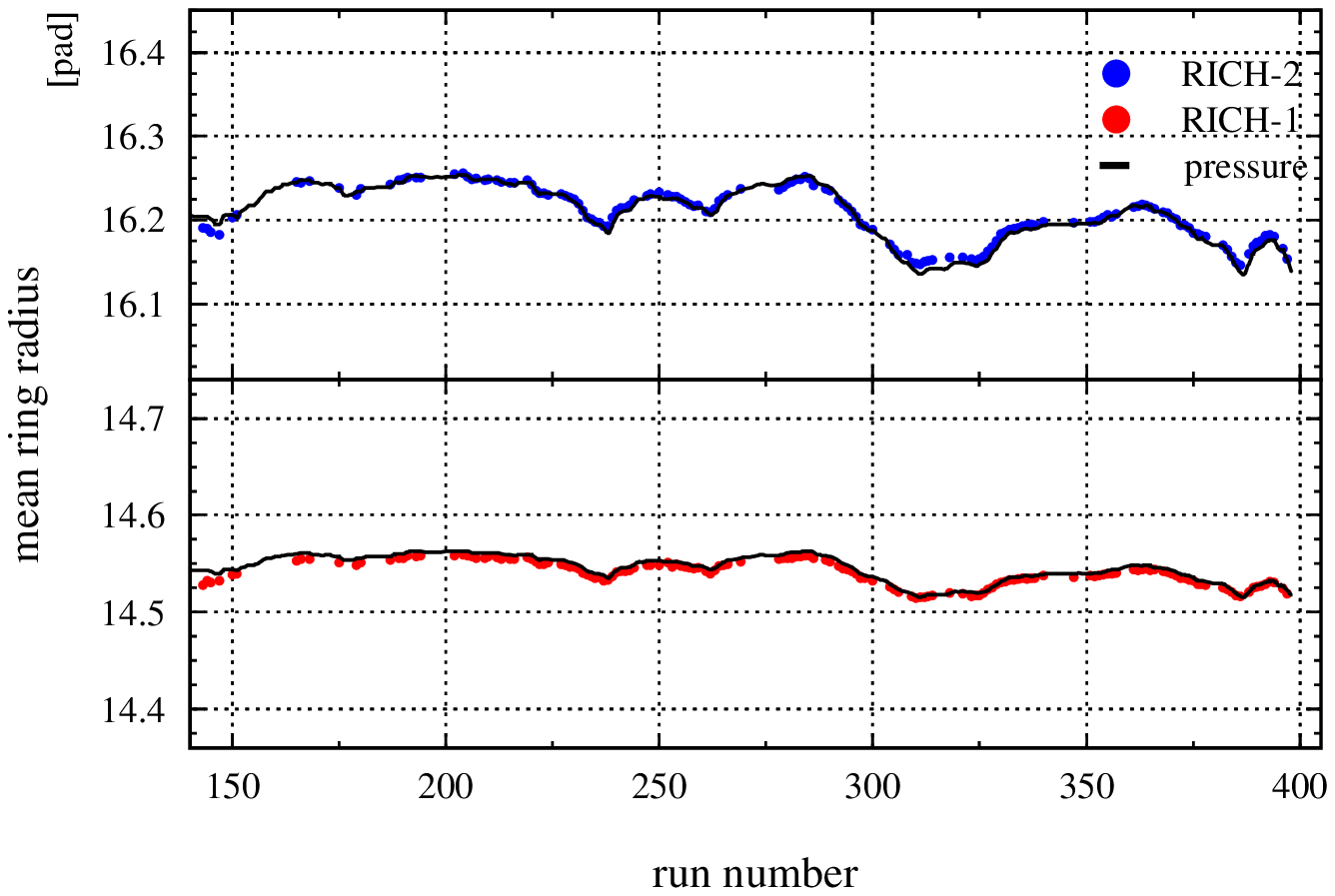}
  \caption[Run-to-run variation of RICH-ring radius]
  {Run-to-run variation of RICH-ring radius overlayed with the relative change of the atmospheric pressure.}
  \label{fig:r12rad}
\end{figure}

Additionally, misidentified charged pions contaminating the sample
should be rejected. Considering that only pions with a momentum of
more than 4.5\,GeV/c produce rings in the RICH detectors,
high-momentum pions are rejected by the characteristics of a small
deflection in the magnetic field in combination with a smaller
ring radius compared to electrons. Figure~\ref{fig:pi-cut} shows
that the misidentified pions can be clearly distinguish from
high-momentum electron tracks.
The outline of the applied cut is overlayed in
Fig.~\ref{fig:pi-cut}.

It was noticed for the first time that the nominal ring radius in
both RICH detectors changes by about 0.8\% due the variation of
the atmospheric pressure as illustrated in Fig.~\ref{fig:r12rad}.
The ratio of the measured ring radius to the nominal ring radius
of both RICH detectors is added for the rejection cut, thereby
amplifying the pressure dependence. If not corrected, this effect
reduces the rejections power of the cut by about 12\% and the
reconstruction efficiency by 1.3\%.

\subsection{Detector acceptance}

Table~\ref{tab:accep} summarizes the geometrical acceptance of the
individual detectors. Even though the final pair acceptance is
limited by the RICH-2 detector to the pseudo-rapidity range of 2.1
to 2.65 (corresponding to a $\theta$-range of 141 to 240\,mrad),
the larger acceptance of the SDDs and of the RICH-1 detector is
useful for the rejection of conversion and Dalitz pairs that fall
only partially into the final acceptance. It is worth mentioning,
that the second-order-field effect deflects tracks traversing the
magnetic field towards the beam line, i.e.~lower theta (see
Fig.~\ref{fig:second order B} in Sec.~\ref{sec:B field}). The
resulting momentum-dependent restriction of the pair acceptance
becomes significant for tracks with a momentum of less than
150\,MeV/c.
\begin{table}[tb]
  \centering
  \begin{tabular}{|l|c|c|c|c|}
  \hline
  Detector & $\theta$-min [mrad] & $\theta$-max [mrad] & $\eta$-min & $\eta$-min
  \\ \hline
  SDD-1 & 90. & 300. & 1.86 & 3.1 \\ \hline
  SDD-2 & 90. & 300. & 1.86 & 3.1 \\ \hline
  RICH-1 & 141. & 290. & 1.90 & 2.65 \\ \hline
  RICH-2 & 141. & 240. & 2.1 & 2.65 \\ \hline
  PD     & 135. & 252. & 2.05 & 2.69 \\ \hline \hline
  Combined & 141. & 240. & 2.1 & 2.65 \\ \hline
  \end{tabular}
  \caption[Acceptance of CERES detectors]
  {Summary of the acceptance of all CERES detectors.
  The value cited for RICH-1/2 detector corresponds to the $2/3$ ring acceptance.
  The second-order-field effect restricts the inner edge of
  the RICH-2 acceptance for $p_{\bot} $\,$<$\,$ 150$\,MeV/c.}
  \label{tab:accep}
\end{table}

The track reconstruction limits the acceptance to a momentum range
of $0.17 $\,$<$\,$ p $\,$<$\,$ 9$\,GeV/c. The lower limit is
imposed by the second-order-field effect, which distorts the
RICH-2 rings, and the rapidly rising probability to pick up
accidental matches. The high-momentum limit results from two
effects: first, the charge determination becomes ambiguous for
very small deflection in the magnetic field because of the finite
detector resolution and, second, the particle identification is
lost due the contamination of high-momentum pions.

\begin{figure}[htb]
    \begin{minipage}[t]{.65\textwidth}
        \vspace{0pt}
        \epsfig{file=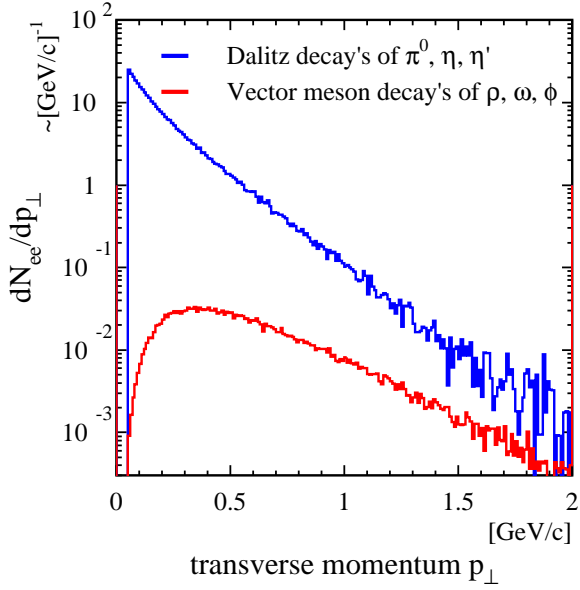,width=\textwidth}
    \end{minipage}%
    \begin{minipage}[t]{.35\textwidth}
      \vspace{0.5cm}
      \caption[Transverse-momentum distribution of Dalitz decays and vector meson
  decays]
      {\newline
      Transverse-momentum distribution of Dalitz decays and vector-meson
      decays included in the hadronic cocktail simulated by GENESIS~\cite{Sako:2000}.}
      \label{fig:source pt}
    \end{minipage}
\end{figure}

A further restriction of the low-transverse-momentum acceptance is
the most powerful tool to reduce the combinatorial background. The
transverse-momentum distribution of the most interesting
vector-meson decays peaks at about 350\,MeV/c while the
distribution of the {\em trivial} Dalitz decays rises
exponentially for small momenta as depicted in
Fig.~\ref{fig:source pt}.

The generally applied transverse-momentum cut of
$p_{\bot}$\,$>$\,$200$\,MeV/c rejects more than 60\% of the
background at a cost of less than 10\% efficiency of the vector
mesons. To study the low-momentum aspects of the dielectron
spectrum, the transverse-momentum cut was lowered to
$p_{\bot}$\,$>$\,$100$\,MeV/c.

Finally, pairs with an opening angle of $\alpha_{\rm
ee}$\,$>$\,$35$\,mrad were selected to further suppress
dielectrons stemming from $\gamma$-conversion and Dalitz decays.
According to Fig.~\ref{fig:dielectron opang}, the opening angle of
these background sources is predominately below 35\,mrad.

\section{Reconstruction-efficiency determination}

\label{sec:track eff}

The probability to observe a collision-produced dilepton is
limited by the efficiency of the detection in each detectors, the
track reconstruction, and the background reduction. To determine
the absolute yield of dileptons, the observed number of correlated
unlike-sign pairs given by Eq.~\ref{equ:pair signal} must be
corrected with the actual reconstruction efficiency
$\varepsilon_{\rm ee}$ of each pair:
\begin{equation}
 N_{\rm ee}=\sum_{i=1}^{\rm N^{\,}_{\rm event}}\sum_{k=1}^{\rm N_{\rm
 pair}^{i}}
 \frac{1}{\varepsilon_{\rm ee}^{i,k}}\;.
 \label{equ:e pair}
\end{equation}
It is worth stressing that the efficiency correction of each pair
cannot be generically replaced by the average reconstruction
efficiency because:
\begin{equation}
 \sum_{i=1}^{\rm N} \frac{1}{\varepsilon_{\rm ee}^i}\neq {\rm N} \left\langle
 \frac{1}{\varepsilon_{\rm ee}}\right\rangle\;.
 \label{equ:ave e}
\end{equation}
In the following it is assumed, that the pair efficiency can be
factorized by the single-track detection probability and a small
correction factor accounting for pair correlations as is
indirectly proven in the discussion of the mixed-event background
in Sec.~\ref{sec:mixed}. In view of the careful calibration of the
detector properties with respect to the pressure and temperature
variations during the run time, the residual time-dependence of
the reconstruction efficiency is neglected. Therefore, the pair
reconstruction efficiency can be expressed in terms of the
single-track phase space parameters (here $\theta$, $\phi$, and
$p$) and the number of observed charged particles $N_{\rm ch}$
which is related to the centrality of the collision:
\begin{eqnarray}
 \label{equ:ee fact}
 \varepsilon_{\rm ee} & = & \varepsilon_{\rm e}^{\rm track\,1}\,
 \varepsilon_{\rm e}^{\rm track\,2}\,\kappa_{\rm ee}\\
 \varepsilon_{\rm e} & = &\varepsilon_{\rm e}(N_{\rm ch},\theta,\phi,p)\nonumber\\
 \kappa_{\rm ee} & = &\kappa_{\rm ee}(\alpha_{\rm ee},N_{\rm ch},
 \theta_1,\phi_1,p_1,\theta_2,\phi_2,p_2)\;.\nonumber
\end{eqnarray}
There are two ways to determine the pair reconstruction
efficiency. First, it can be estimated by a comparison of
expected- and observed-yield of pairs with mass below
200\,MeV/c$^2$. The dilepton production in this mass range is
dominated by the Dalitz decay of $\pi_0$-, $\eta$-, and
$\eta$'-mesons. The yield of these contributions was measured with
an accuracy of better than
10\%~\cite{Agakishiev:1998mv,Agakishiev:1998mw} and is well
described by the simulation of the cocktail of the hadronic
sources including the acceptance and the momentum resolution of
the CERES detector (see App.~\ref{app:genesis}). Although this
method does not allow extraction of differential pair efficiency
distributions, it is reckoned as a valuable reference for the
average pair efficiency.

The second option to determine the pair reconstruction efficiency
is a Monte Carlo simulation of the complete detector system,
including the track reconstruction and the background rejection.
It involves the following steps:
\begin{itemize}
  \item simulation of a huge number of dielectrons with realistic kinetic
  properties according to the decays of the known hadronic sources (App.~\ref{app:genesis})
  \item use of the GEANT software to simulated the passage of
  the generated pairs through the CERES-detector system. GEANT
  simulates all particle interactions with detector materials
  such as multiple scattering and bremsstrahlung. The hit
  positions of all particles as well as their energy
  deposition was determined by the response function of the
  individual detectors taking internal physics and electronics effects into
  account. For the first time, the emission and subsequent tracking of Cerenkov photons included
  the optical properties of the RICH detectors.
  \item embedding of detector responses obtained into the
  raw data of genuine events
  \item performance of the first-stage data analysis on so-called overlay events
  \item application of all background rejection cuts
\end{itemize}
After the last step, the differential reconstruction efficiency is
determined by the probability to identify the simulated track in
the vicinity of its original position in each detector.

Extensive study of this method had proven that all relevant
detector characteristics are reasonably well described by the
simulation~\cite{Lenkeit:1998}. Since then, the simulation was
refined in many details. Small inconsistencies between the GEANT
detector emulation and the Monte Carlo simulation could be
resolved.

As a consequence of the modified tracking strategy and the new
SDD-hit reconstruction, the Monte Carlo simulation had to be
readjusted to reproduction of the distributions observed in data
(see~\cite{Slivova:2001} for details).

A number of $2\cdot 10^6$ simulated tracks were embedded into raw
data events on a one-to-one basis for the study of the
single-track reconstruction efficiency. The raw events were chosen
from different parts of the run time to average detector
aberrations. A simulated particle was counted as successfully
detected, if all hits of a reconstructed track were within a range
of three times the detector resolution to the particle's original
direction. Deciding if a particle was truly lost or merely
scattered is ambiguous to some extend because genuine and embedded
hits cannot be distinguished on the detector level. A comparison
of events, analyzed with and without an additional embedded track,
showed this systematic error to be in the order of 2\%. The
statistical error of the expected yield of dielectons being large,
a relative error of less than 10\% is acceptable for the
differential shape of the efficiency distribution.

\begin{figure}[bt]
  \centering
  \mbox{
   \epsfig{file=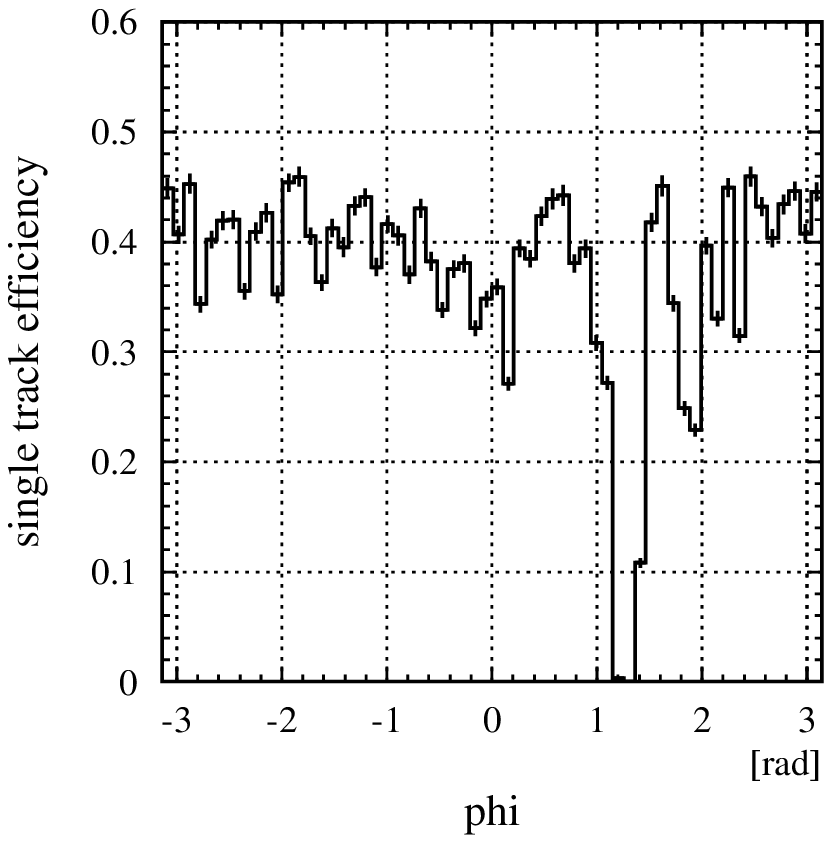,width=.49\textwidth}
   \hfill
   \epsfig{file=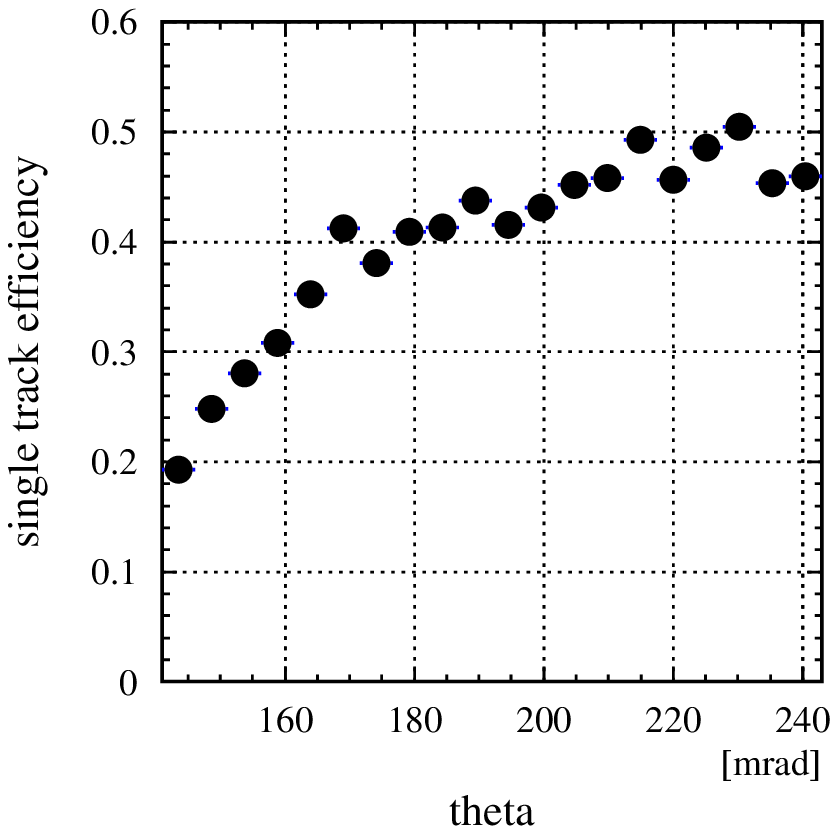,width=.49\textwidth}
  }
  \caption[$\phi$- and $\theta$-dependence of single-track reconstruction efficiency]
  {Single-track reconstruction efficiency as a function of the azimuthal angle $\phi$ (left panel) and the polar angle
  $\theta$ (right panel). The variations in the $\phi$
  distribution result from dead or inefficient anodes in the SDDs
  and efficiency variations in the RICH detectors caused by
  discharges and hot spots. The hole at 1.3\,rad is related to a region of
  dead anodes in the SDD-2. The $\phi$-dependence of the efficiency was not
  corrected apart from the influence of the dead anodes, as the complicated interplay of
  several effects is not well understood. The efficiency drop towards small $\theta$-values as apparent in
  the right panel reflects the increasing hit/ring density close to mid-rapidity at 110\,mrad and
  the decreasing number of hits per ring in RICH-2 detector at the inner edge of the acceptance.
  }
  \label{fig:eff phi}
\end{figure}

The projections of the multi-dimensional efficiency function $
\varepsilon_{\rm e}$ of Eq.~\ref{equ:ee fact} were plotted for
following discussion. The $\phi$-dependence depicted in the left
panel of Fig.~\ref{fig:eff phi} is almost flat, except for a large
hole at 1.3\,rad. This is caused by a region of dead anodes in the
SDD-2. The $\phi$-dependence of the efficiency was not corrected
apart from the influence of the dead anodes, as the complicated
interplay of several effects is not well understood.

Figure~\ref{fig:eff phi} (right panel) shows the
$\theta$-dependence to be roughly constant above 170\,mrad but to
decreases by more than a factor of two towards the inner edge of
the acceptance. This efficiency drop reflects the increasing
hit/ring density close to mid-rapidity at 110\,mrad and the
decreasing number of hits per ring in RICH-2 detector. At the
inner edge about one third of the area of a RICH-2 ring is already
outside of the detector acceptance.

\begin{figure}[bt]
  \centering
  \mbox{
   \epsfig{file=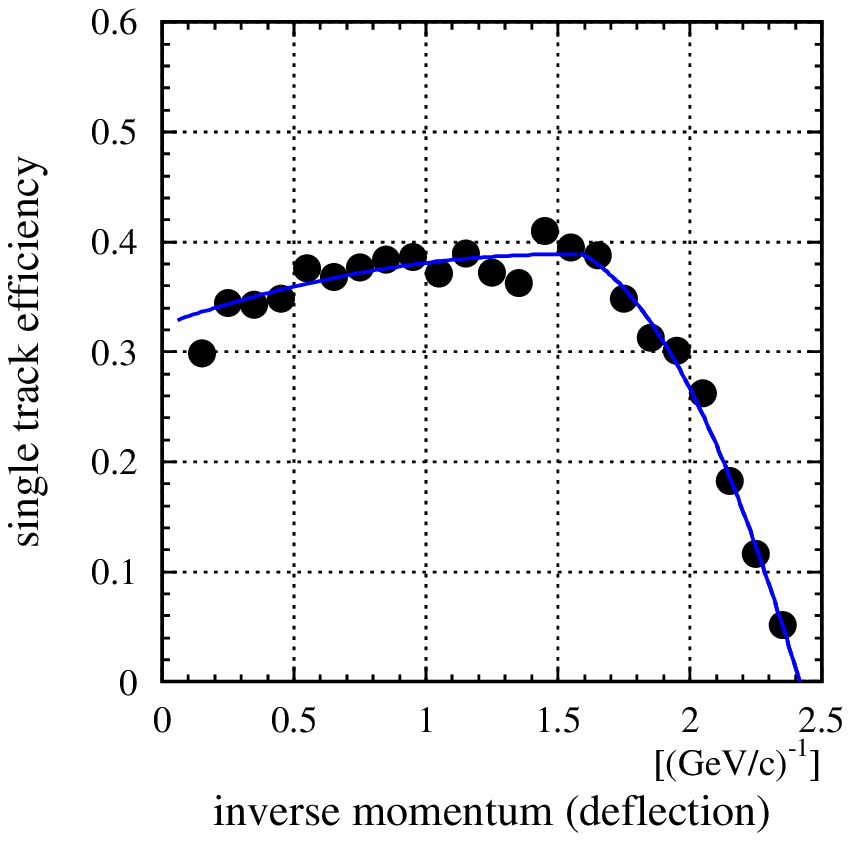,width=.49\textwidth}
   \hfill
   \epsfig{file=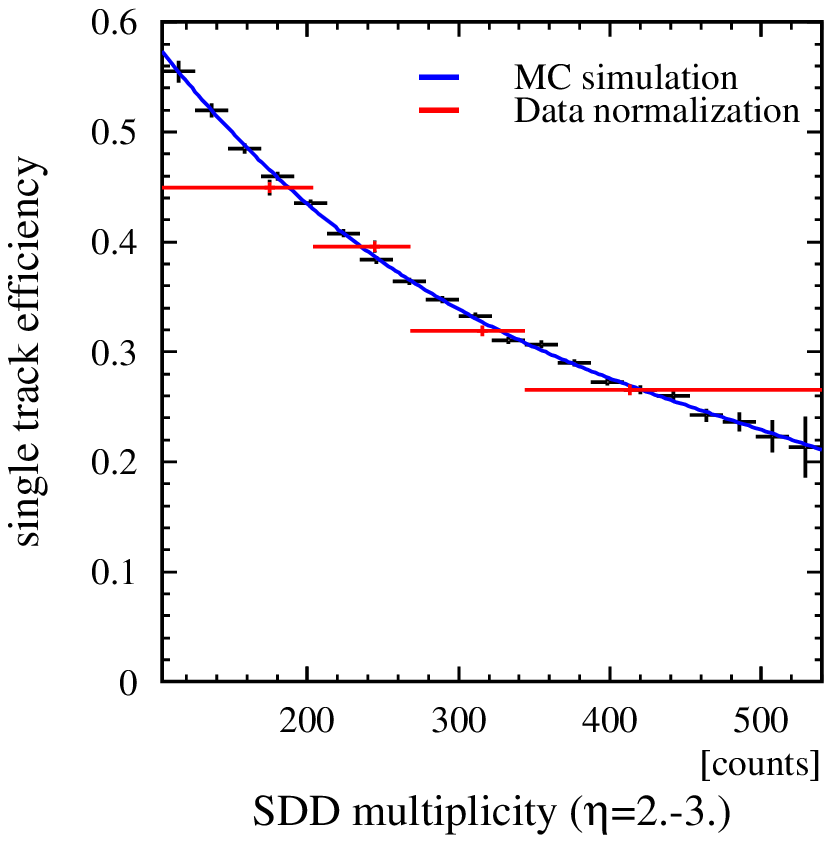,width=.49\textwidth}
  }
  \caption[Momentum- and multiplicity-dependence of single-track re\-con\-struc\-tion efficiency]
  {Single-track reconstruction efficiency as a function of the inverse momentum $1/p$ (left panel)
  and the charged-particle multiplicity (right panel). The inverse momentum is proportional
  to the azimuthal deflection $\Delta\phi$ in the magnetic field between RICH-1 and RICH-2 detector
  (see Eq.~\ref{equ:deflect}). The blue line in the left panel shows a polynomial fit to the simulated
  data to be used later for correcting the momentum dependence of the reconstruction efficiency.
  The MC-simulated multiplicity efficiency distribution (blue line)
  depicted in the right panel agrees well
  with the average single-track efficiency (red points) determined by the ratio of the measured low-mass yield
  to the expected yield of the hadronic sources.}
  \label{fig:eff p}
\end{figure}

Figure~\ref{fig:eff p} (left panel) demonstrates the
reconstruction efficiency to be approximately independent of
momentum for a deflection smaller than 230\,mrad
(i.e.~$1/p<1.6$\,GeV$^{-1}$c).
It drops rapidly towards larger deflection corresponding to
$p_{\bot}$\,$<$\,200\,MeV/c, which however is only important for
the $p_{\bot}$\,$>$\,100\,MeV/c selection. The slight decrease
towards smaller deflection, i.e.~larger momentum, can be
attributed to the momentum dependence of the butterfly-shaped
matching window of the tracking between RICH-1, RICH-2, and PD
detector.

The efficiency of the track reconstruction additionally depends
strongly on the charged-particle multiplicity as shown in the
right panel of Fig.~\ref{fig:eff p}. The efficiency obtained by
the Monte Carlo simulation agrees well with the values extracted
with the first method, i.e.~the ratio of the observed number of
low-mass pairs to the expected hadronic yield of each multiplicity
bin.

In general, the multi-dimensional efficiency function
Eq.~\ref{equ:ee fact} cannot be described by a simple product of
its single-parameter projections
(i.e.~$\varepsilon(\phi)$,\,$\varepsilon(\theta)$,\,$\varepsilon(p)$,
and\,$\varepsilon(N_{\rm ch})$) discussed above because some of
those may not be independent. In the following, the relation
between the single-parameter efficiency projections is studied to
find a simple representation for the reconstruction efficiency.

\begin{figure}
  \centering
  \mbox{
   \epsfig{file=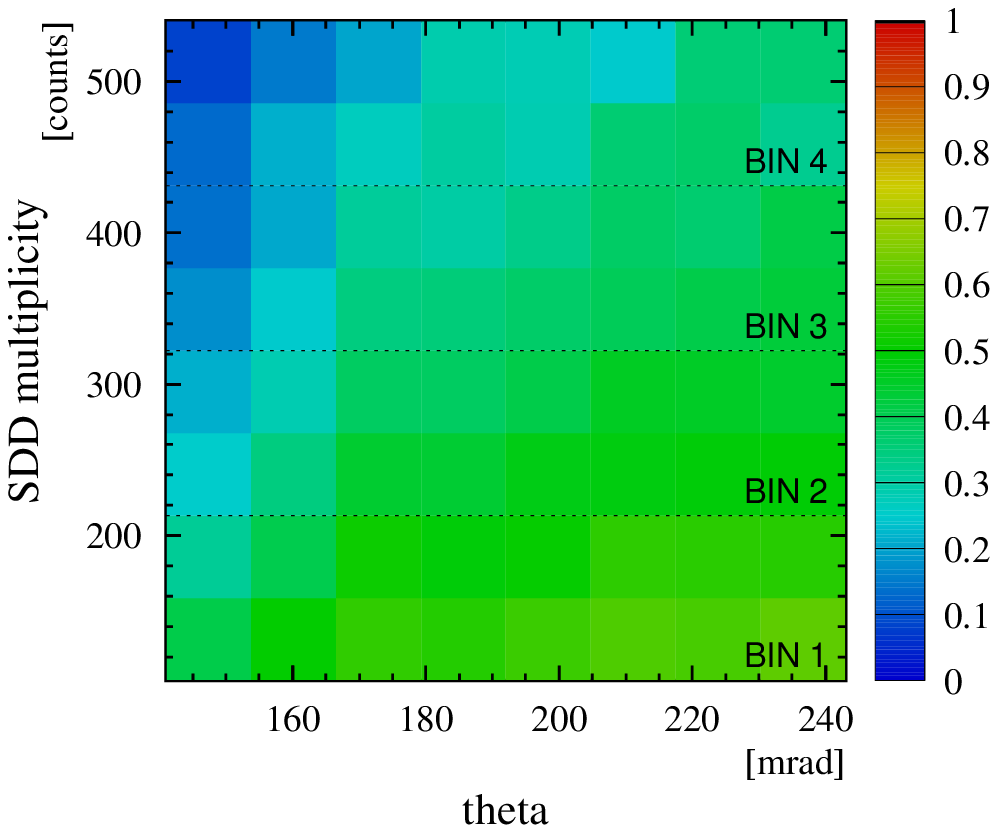,width=.49\textwidth}
   \hfill
   \epsfig{file=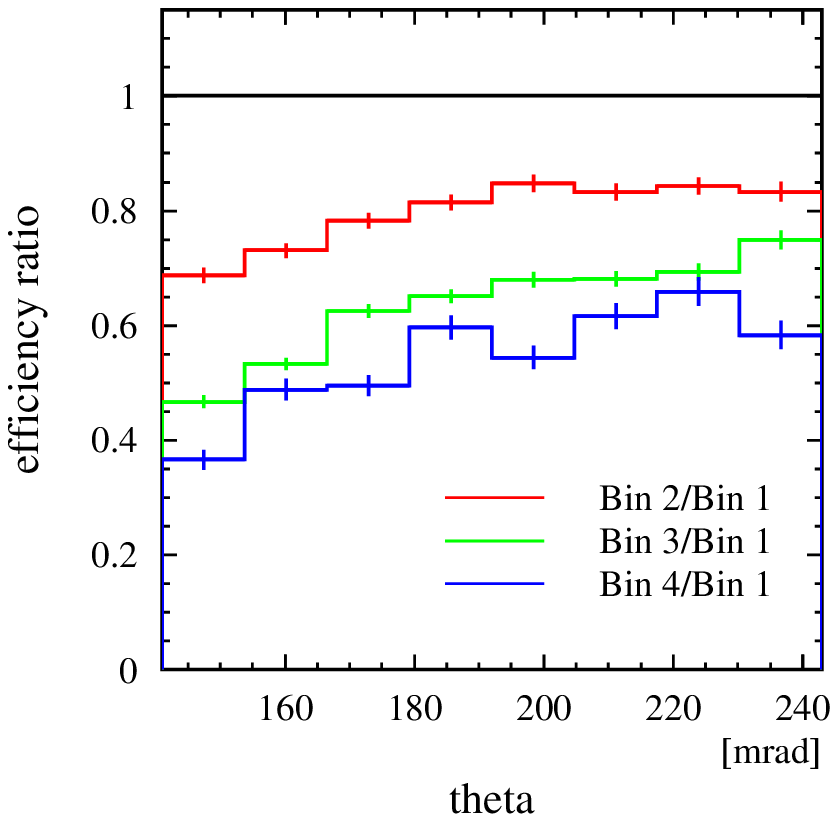,width=.49\textwidth}
  }
  \caption[Single-track reconstruction efficiency as a function of $\theta$ and SDD multiplicity]
  {Single-track reconstruction efficiency as a function of $\theta$ and
   SDD multiplicity (left panel). Relative change of the $\theta$-dependence of the efficiency with SDD
   multiplicity (right panel). The $\theta$- and the multiplicity-dependence of the efficiency are correlated
   as the slope of $\theta$-dependence increases with rising multiplicity.}
  \label{fig:eff th-nch}

  \vspace*{.8cm}
  \mbox{
   \epsfig{file=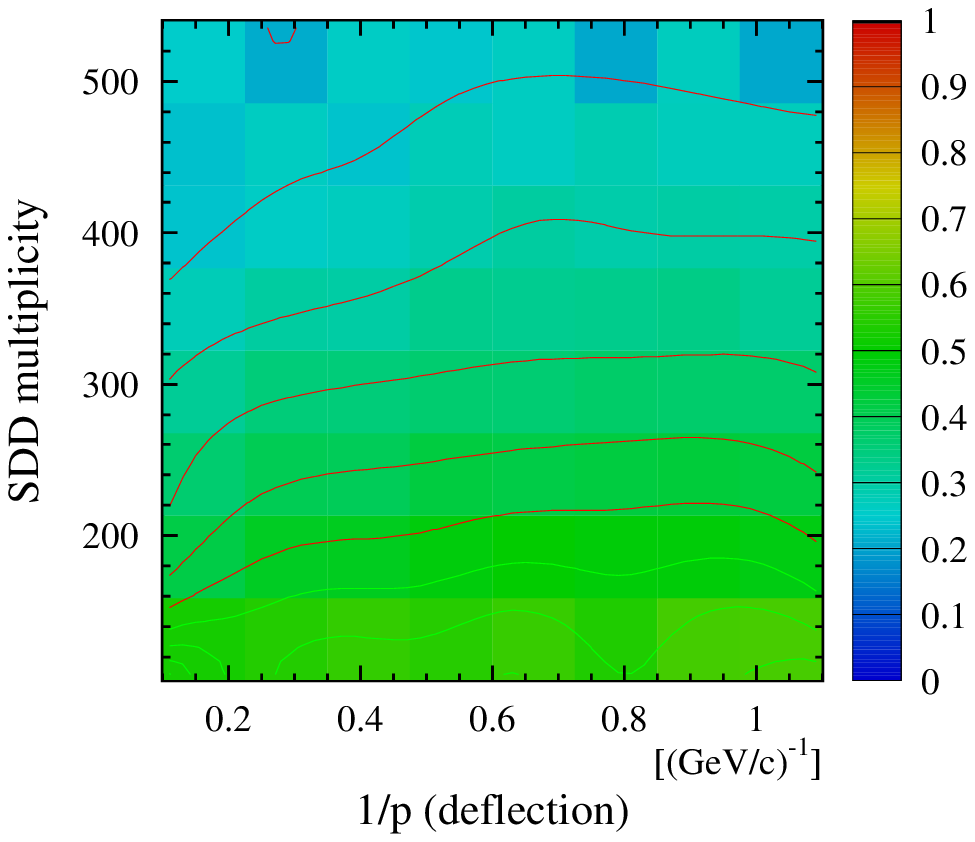,width=.49\textwidth}
   \hfill
   \epsfig{file=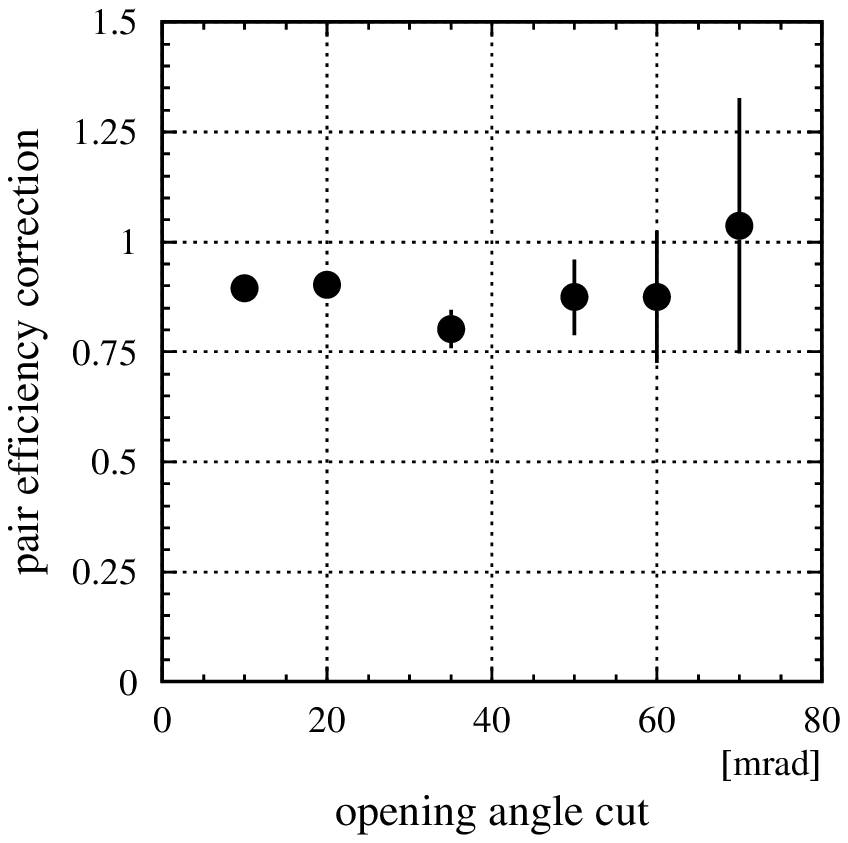,width=.49\textwidth}
  }
  \caption[Single-track reconstruction efficiency as a function of inverse momentum and SDD multiplicity
          and pair reconstruction efficiency as a function of the opening-angle cut]
  {Single-track reconstruction efficiency as a function of inverse momentum and SDD
  multiplicity (left panel). The inverse momentum is proportional
  to the azimuthal deflection $\Delta\phi$ in the magnetic field between RICH-1 and RICH-2 detector
  (see Eq.~\ref{equ:deflect}. Overlay Monte Carlo simulation of pair reconstruction
  efficiency as a function of the opening-angle cut (right panel). The simulated pairs
  consisted of $\pi^0$ and $\eta'$ mesons. Only pairs with an opening angle larger than
  $\alpha_{\rm ee}$ were accepted.}
  \label{fig:eff p-nch}
\end{figure}

Given the finite detector resolution, the reconstruction
efficiency of a track is subject to the density of close hits or
rings in the respective detectors. The hit density is determined
by multiplicity- and $\theta$-distribution of the charged
particles produced in the collision, comprising mostly pions. The
rapidity density d$N_{\rm ch}$/d$y$ was observed to be constant in
the CERES acceptance~\cite{Cooper:1999ij}. The energy of the pions
produced in Pb-Au collisions at 158\,GeV/c per nucleon is much
larger than their rest mass. Therefore, the rapidity $y$ of a
particle can be approximated by the pseudo-rapidity $\eta$ which
in turn depends on only on the polar angle
$\theta$~\cite{pbm:2000}. As a result, the local hit density per
unit area is given by a simple product of two functions of the
charged-particle multiplicity and the rapidity, respectively. In
general, it is not possible to disentangle multiplicity- and
rapidity-dependence of the reconstruction efficiency accordingly
because the efficiency is assumed to be a non-linear function of
the local hit density.
 The 2-dimensional contour depicted in Fig.~\ref{fig:eff th-nch} (left
panel) shows the efficiency to drop more rapidly at the inner edge
of the acceptance. This is more clearly seen in Fig.~\ref{fig:eff
th-nch} (right panel) - the slope of the $\theta$-dependence
increases with rising multiplicity. This effect has the important
consequence of the pair efficiency to be increasing with large
opening angles because these pairs fall only into the acceptance
if both tracks are close to the upper $\theta$-limit of
acceptance.

Figure~\ref{fig:eff p-nch} (left panel) shows the momentum
dependence of the efficiency to be correlated weakly with the
charged-particle multiplicity. As a result, the applied
single-track efficiency correction can be decomposed as:
\begin{equation}
 \varepsilon_{\rm e}=\epsilon(p)\,\epsilon(\theta,N_{\rm ch})\;.
 \label{equ:eff e}
\end{equation}
To investigate the residual pair efficiency (see Eq.~\ref{equ:ee
fact}), a number of $1\cdot 10^6$ simulated dielectron pairs,
produced in $\pi_0$ and $\eta$ Dalitz decays, were embedded into
genuine events. If two tracks were found according to the criteria
mentioned above, the pair was counted as successfully
reconstructed. The residual pair efficiency correction was
determined as:
\begin{equation}
 \kappa_{\rm ee}=\frac{\varepsilon_{\rm ee}^{\rm MC}}{ \varepsilon_{\rm e}^{\rm
 track\,1}\,\varepsilon_{\rm e}^{\rm track\,2}}\;,
\end{equation}
where $\varepsilon_{\rm e}$ was computed by Eq.~\ref{equ:eff e}.

A residual efficiency correction depending on pair properties is
expected to depend at most on the pair opening angle for the case
of touching or overlapping RICH rings. However, this effect shown
in the right panel of Fig.~\ref{fig:eff p-nch} turns out to be
small. This was already expected from the discussion of pair
correlations in the context of the background subtraction.

The observed yield corrected for Monte Carlo pair efficiency is
about $30\%$ lower than the expected yield of all hadronic sources
of pairs with mass below 200\,MeV/c$^2$. This discrepancy is not
surprising. Although the Monte Carlo method allows to study all
aspects of the analysis, its benefit to the absolute efficiency
determination is limited by the multitude and high complexity of
the involved detector and analysis dependencies. Particularly, the
RICH-ring reconstruction algorithm and the hit reconstruction in
the SDD are sensitiv to marginal changes of the parameters of the
read-out electronics and of the environment conditions and are
obstructing an adequate description by the Monte Carlo simulation.

A solution was to correct all data first with the Monte Carlo pair
efficiency to account for the differential efficiency dependencies
and then to normalize the mass spectrum to the expected dielectron
yield of pairs with mass below 200\,MeV/c$^2$ simulated by
GENESIS~\cite{Sako:2000}.

\clearpage
\section{Physics results and discussion}

\subsection{Results of the new analysis of the 1996 data set}
\label{sec:results}

Background rejection and subtraction was applied to $4.1\cdot
10^7$ recorded events, as described in Sec.~\ref{sec:mix-intro}
and~\ref{sec:rejection}. The results were a final data sample
comprising 3537$\pm$103 dielectrons for $m_{\rm
ee}$\,$<$\,$0.2$\,GeV/c$^2$ and 1305$\pm$210 dielectrons for
$m_{\rm ee}$\,$>$\,$0.2$\,GeV/c$^2$, reconstructed with a
signal-to-background ratio of  1:0.96 and  1:13.9, respectively.
The signal refers to pairs with an opening angle larger than
$35$\,mrad and a minimum transverse momentum of
$p_\bot$\,$>$\,$200$\,MeV/c of both tracks. The spectrometer
acceptance covers the pseudo-rapidity range of
$2.1$\,$<$\,$\eta$\,$<$\,$2.65$.

\begin{figure}[!htb]
  \centering
  \epsfig{file=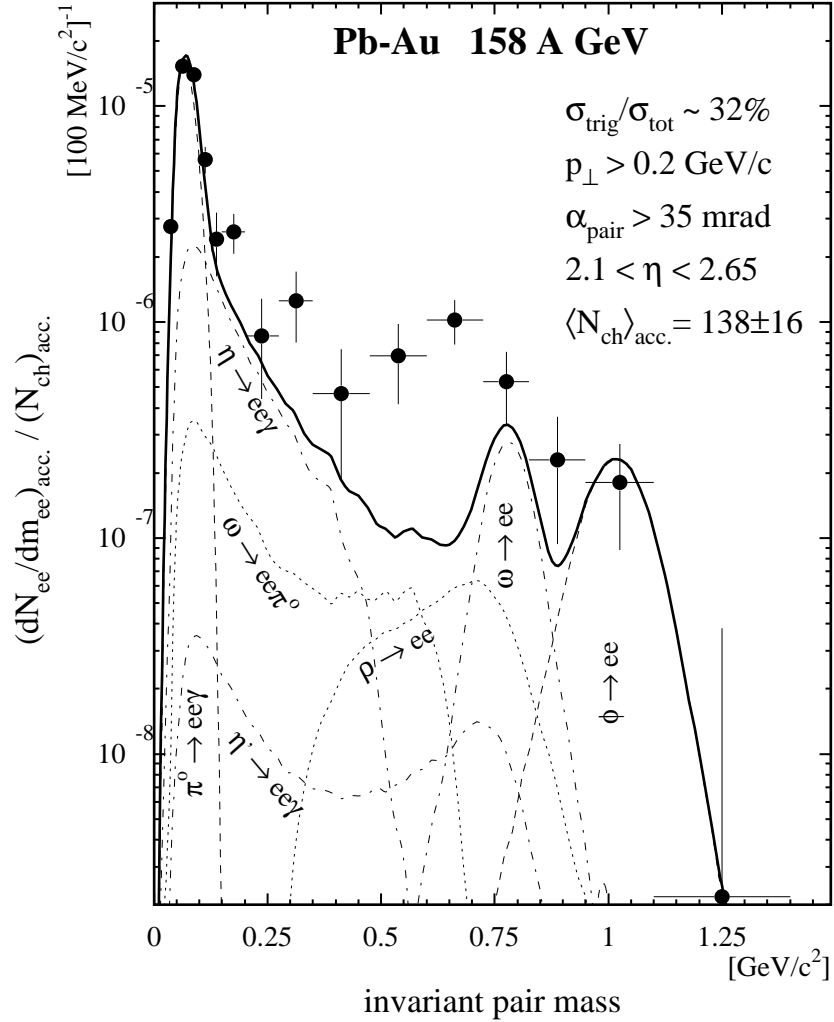}
  \caption[Dielectron-invariant-mass spectrum for $p_{\bot}$\,$>$\,0.2\,GeV/c]
  {Inclusive dielectron-invariant-mass spectrum of Pb-Au
  collisions at 158\,GeV/c per nucleon divided by the number of charged particles.
  The solid line represents the expected yield of all hadronic
  sources~\cite{Sako:2000}.}
  \label{fig:mass-pt-0.2}
\end{figure}

The resulting invariant-mass spectrum of Pb-Au collisions at
158\,GeV/c per nucleon is presented in Fig.~\ref{fig:mass-pt-0.2}.
The pair yield plotted has been divided by the charged-particle
multiplicity in the CERES acceptance. It was measured to be
$N_{\rm ch}=250\pm30$ (see Sec.~\ref{sec:nch}) for the larger
pseudo-rapidity range of $2$\,$<$\,$\eta $\,$<$\,$3$ which
translates into an average multiplicity of $\langle N_{\rm
ch}\rangle_{\rm acc.}=138\pm16$ for the nominal acceptance. The
spectrum was corrected for single-track efficiency as described in
Sec.~\ref{sec:track eff} and, subsequently, normalized to the
expected yield from hadronic decays for $m_{\rm
ee}$\,$<$\,$0.2$\,GeV/c$^2$ using the GENESIS event generator
(see~App.~\ref{app:genesis}). Comparison of observed and expected
pair yield results in an average 6\% reconstruction efficiency of
low-mass pairs ($m_{\rm ee} $\,$<$\,$0.2$\,GeV/c$^2$), assuming
that the total yield is saturated by the contribution of known
hadronic decays. This pair efficiency is 30\% lower than in the
previous analysis~\cite{Lenkeit:1998}. The newly implemented
splitting of overlapping hits in the SDD results in an unexpected
efficiency loss of pairs estimated to be 16\% (see
Sec.~\ref{sec:sidc-tracking}). The insufficient size of the
matching window between RICH-1, RICH-2, and PD detector applied in
first-stage analysis caused an additional pair efficiency loss of
about 10\% (see Fig.~\ref{fig:eff p}).

The statistical error of the signal includes the combinatorial
background contribution which corresponds to the like-sign
same-event background for pairs with mass below 0.35\,GeV/c$^2$
and to the mixed-event background for pairs with mass above
0.35\,GeV/c$^2$ (see Sec.~\ref{sec:mix vs comb}).

Three sources contribute to the systematic uncertainties of the
data: the extrapolated low-mass yield of the hadronic cocktail,
the average charged-particle density $\langle N_{\rm ch}\rangle$,
and the pair reconstruction efficiency.

Uncertainty in the low-mass yield of the hadronic cocktail arises
from error of the branching ratios and relative production cross
sections of $\pi_0$, $\eta$, and $\eta'$, as well as the error
related to the parameterizations of the $p_{\bot}$\,input
distributions estimated to be about 25\% in
App.~\ref{app:genesis}.

The error of the average charged-particle density measurement
results from the following factors: the accuracy of the
reconstruction efficiency determined by Monte Carlo simulation,
the linearity of the rapidity distribution in the range of
$2$\,$<$\,$\eta$\,$<$\,$3$, the beam pile-up, the production of
$\delta$ electrons, and the run-to-run variation of the trigger
threshold and of the reconstruction efficiency. All other
contributions are absorbed into the error of the relative
normalization to the cocktail, except the time variations. The
latter is estimated to be about 5\%.

The systematic error of the pair reconstruction efficiency is
given by the uncertainty of the Monte Carlo description, namely:
the detector response functions, the alignment of the overlay
tracks to the event vertex, run-to-run variations of the gain in
the UV detectors, and run-to-run temperature dependent changes of
the SDD and RICH detector properties. As the data is normalized to
the low-mass yield of the hadronic cocktail, the uncertainty in
the pair reconstruction efficiency affects the results via its
variation with invariant mass. This contribution is estimated to
be about 10\%.

The combined systematic uncertainty of the absolute yield is about
40\%. It was verified that this value is in accordance with the
systematic error estimated by the change of the dielectron yield
with respect to small variations of the rejection cuts. The yields
of pairs with mass above $m_{\rm ee}$\,$>$\,$250$\,MeV/c$^2$
obtained from independent analyses of the two data sets of
opposite field direction agree with each other within the limit of
the statistical error: $N_{\rm ee}=(5.3\pm3.2)\cdot 10^{-6}$ and
$N_{\rm ee}=(4.8\pm2.9)\cdot 10^{-6}$ for positive and negative
B-field, respectively.

Following previous analyses, the data is compared to the expected
yield of hadronic decays as simulated with an improved version of
the GENESIS event generator (see App.~\ref{app:genesis}). The
hadronic cocktail was folded with the experimentally measured
spatial and momentum resolution. Note, that for all previous CERES
publications, the predicted yield is 30\% too low for $m_{\rm
ee}$\,$<$\,$0.2$\,GeV/c$^2$ due to a heretofore unnoticed
computing problem within GENESIS (see App.~\ref{app:genesis}).

The most striking feature of Fig.~\ref{fig:mass-pt-0.2} is the
large excess of observed dielectrons with respect to the
contributions of the hadronic decays. Starting at an invariant
mass of about twice the pion mass, the data begins to deviate from
the cocktail encompassing all the range up to the $\omega$
resonance. For even higher mass the statistical error of the data
becomes large, but the observed spectrum seems to concur with the
expected decay contribution of the $\phi$-meson. Integration of
the measured yield of pairs with mass above $m_{\rm
ee}$\,$>$\,$250$\,MeV/c$^2$ renders a relative enhancement by a
factor of $3.0\pm1.3$(stat.)$\pm1.2$(syst.) with respect to the
expected hadronic cocktail.

In~\cite{Voigt:1998,Lenkeit:1998}, a strong increase of the
dielecton enhancement was reported for small transverse pair
momentum ($q_{\bot}$\,$<$\,$0.5$\,GeV/c). However, the statistical
significance was limited due to the transverse-momentum cut of
$p_\bot$\,$>$\,$200$\,MeV/c imposed on single tracks. To verify
and substantiate this remarkable observation, the data analysis
was extended to very low momentum tracks with
$p_\bot$\,$>$\,$100$\,MeV/c for the first time. This extension
also made possible the study of a previously inaccessible region
of phase space.

\begin{figure}[!htb]
  \centering
  \epsfig{file=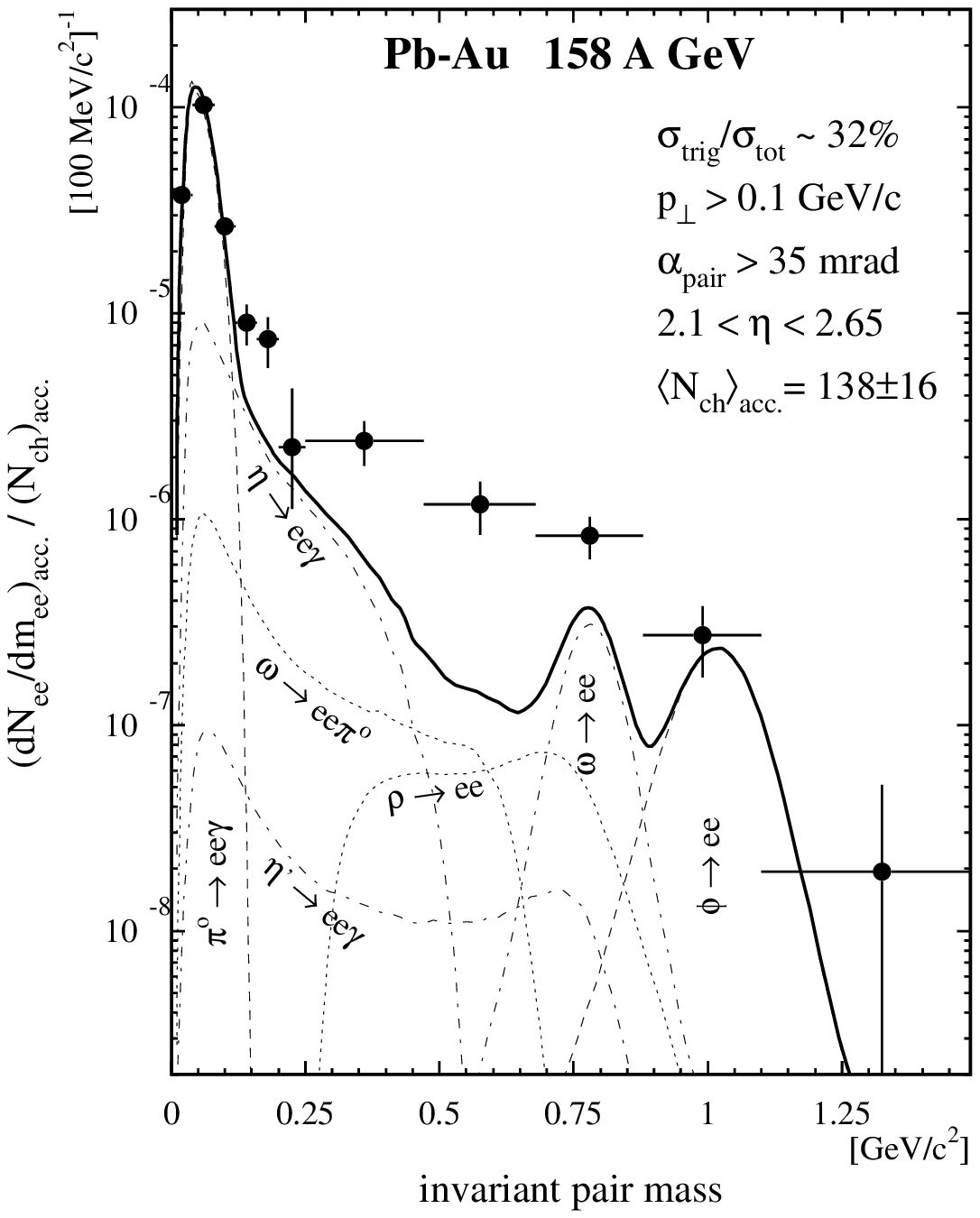}
  \caption[Dielectron-invariant-mass spectrum for $p_{\bot}$\,$>$\,0.1\,GeV/c]
  {Inclusive dielectron invariant mass spectrum for $p_{\bot}$\,$>$\,0.1\,GeV/c
  divided by the number of charged particles.
  The solid line represents the expected yield of all hadronic
  sources~\cite{Sako:2000}.}
  \label{fig:mass-pt-0.1}
\end{figure}

The net signal increases dramatically for a lower
transverse-momentum cut of $p_\bot$\,$>$\,$100$\,MeV/c:
$19212\pm291$(stat.) and $2018\pm382$(stat.) for $m_{\rm
ee}$\,$<$\,$0.2$\,GeV/c$^2$ and $m_{\rm
ee}$\,$>$\,$0.2$\,GeV/c$^2$, respectively. The
signal-to-background ratio, however, deteriorates by almost a
factor of two to 1\,:\,1.74 and 1\,:\,36 for low-mass and
high-mass pairs, respectively. Figure~\ref{fig:mass-pt-0.1} shows
the efficiency corrected dielectron invariant mass spectrum for
$p_\bot$\,$>$\,$100$\,MeV/c.
Compared to the $p_\bot$\,$>$\,$200$\,MeV/c spectrum, the
abundance of low-mass dielectrons, mostly stemming from $\pi^0$
and $\eta$ Dalitz decays, increases by a tenfold, while the
high-mass region of the spectrum remains little affected. The
observed excess relative to the hadronic cocktail is similar to
the $p_\bot$\,$>$\,$200$\,MeV/c selection, as expected for an
inclusive spectrum. Integration of the measured yield above
$m_{\rm ee}$\,$<$\,$0.250$\,GeV/c$^2$ gives a relative enhancement
factor of $3.7\pm1.0$(stat.)$\pm1.5$(syst.).

Study of the multiplicity dependence provides insight into the
dielectron production mechanism. For pairs originating from the
decay of hadrons in the final state the yield must scale linearly
with the number of particles produced upon freeze out of the
fireball. Dielectron production, resulting from $\pi\pi$
annihilations in the hot and dense hadron gas formed in the early
stages of nuclear collisions, is expected to increases
quadratically with particle density~\cite{Cerny:1986gt}:
\begin{equation}
 \frac{dN_{\rm
ee}}{d\eta} \sim \left(\frac{dN_{\rm ch}}{d\eta}\right)^{\alpha}
\qquad\mbox{with}\qquad \alpha=2\;.
 \label{equ:nch scaling}
\end{equation}
Other proposed collision scenarios involve a scaling behaviour
characterized by values of $\alpha=1.1$~\cite{Cleymans:1998es} or
$\alpha=1.3$~\cite{Feinberg:1976ua} for the dependence on
charged-particle multiplicity.

\begin{figure}[!htb]
  \epsfig{file=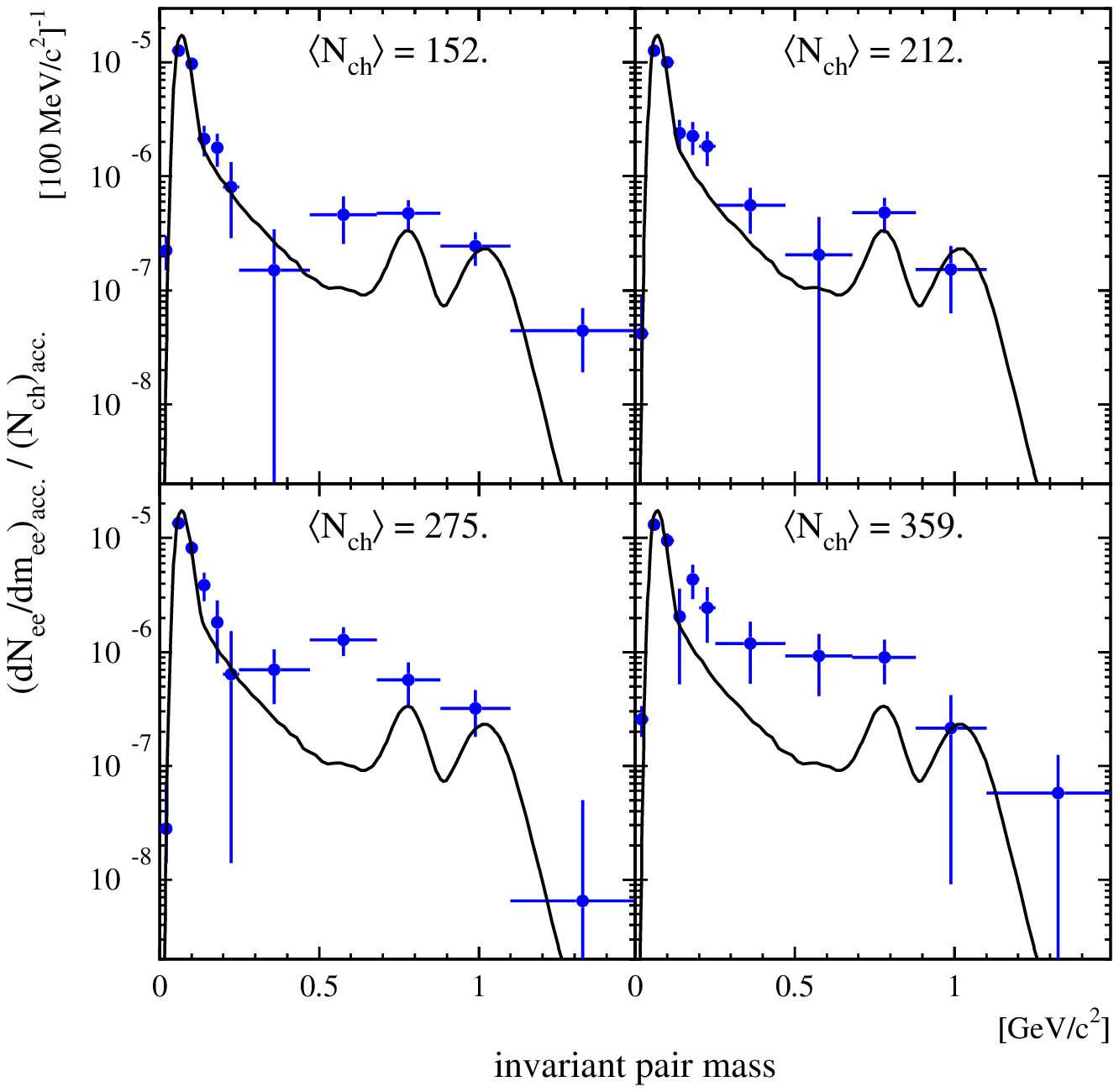}
  \epsfig{file=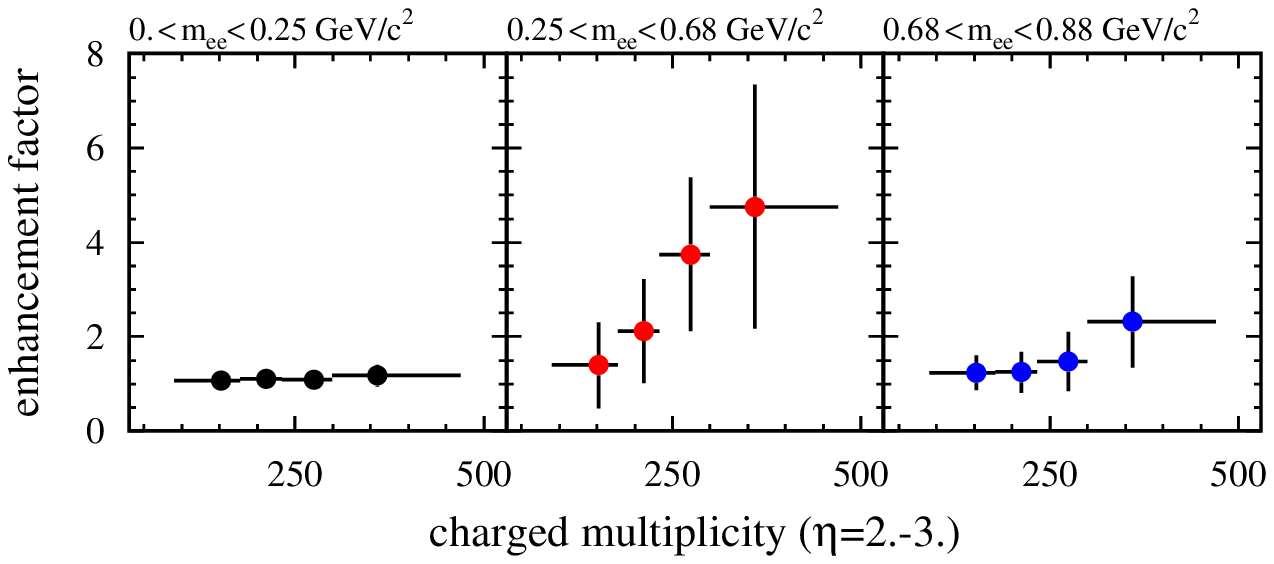}
  \caption[$N_{\rm ch}$-dependence of invariant-mass spectrum for $p_{\bot}$\,$>$\,$0.2$\,GeV/c]
  {$N_{\rm ch}$-dependence of the invariant mass spectrum for $p_{\bot}$\,$>$\,$0.2$\,GeV/c (upper
  panel). The solid line represents the expected yield of all hadronic
  sources~\cite{Sako:2000}. $N_{\rm ch}$-dependence of the enhancement factor for three
  different invariant mass bins (lower panel).}
  \label{fig:mass-mbin-pt-0.2}
\end{figure}

The 1996 Pb-Au data sample, encompassing a centrality range of the
top 32\% of the geometric cross section, was divided into 4
multiplicity bins of equal statistics as shown in
Fig.~\ref{fig:nch} of Sec.~\ref{sec:nch}. The upper panel of
Fig.~\ref{fig:mass-mbin-pt-0.2} shows the invariant mass spectrum
measured for each multiplicity bin.
The dielectron yield exhibits a strong increase with multiplicity
for the mass range of 200--700\,MeV/c$^2$ indicating a non-trivial
origin of the enhancement.

\begin{figure}[!htb]
  \epsfig{file=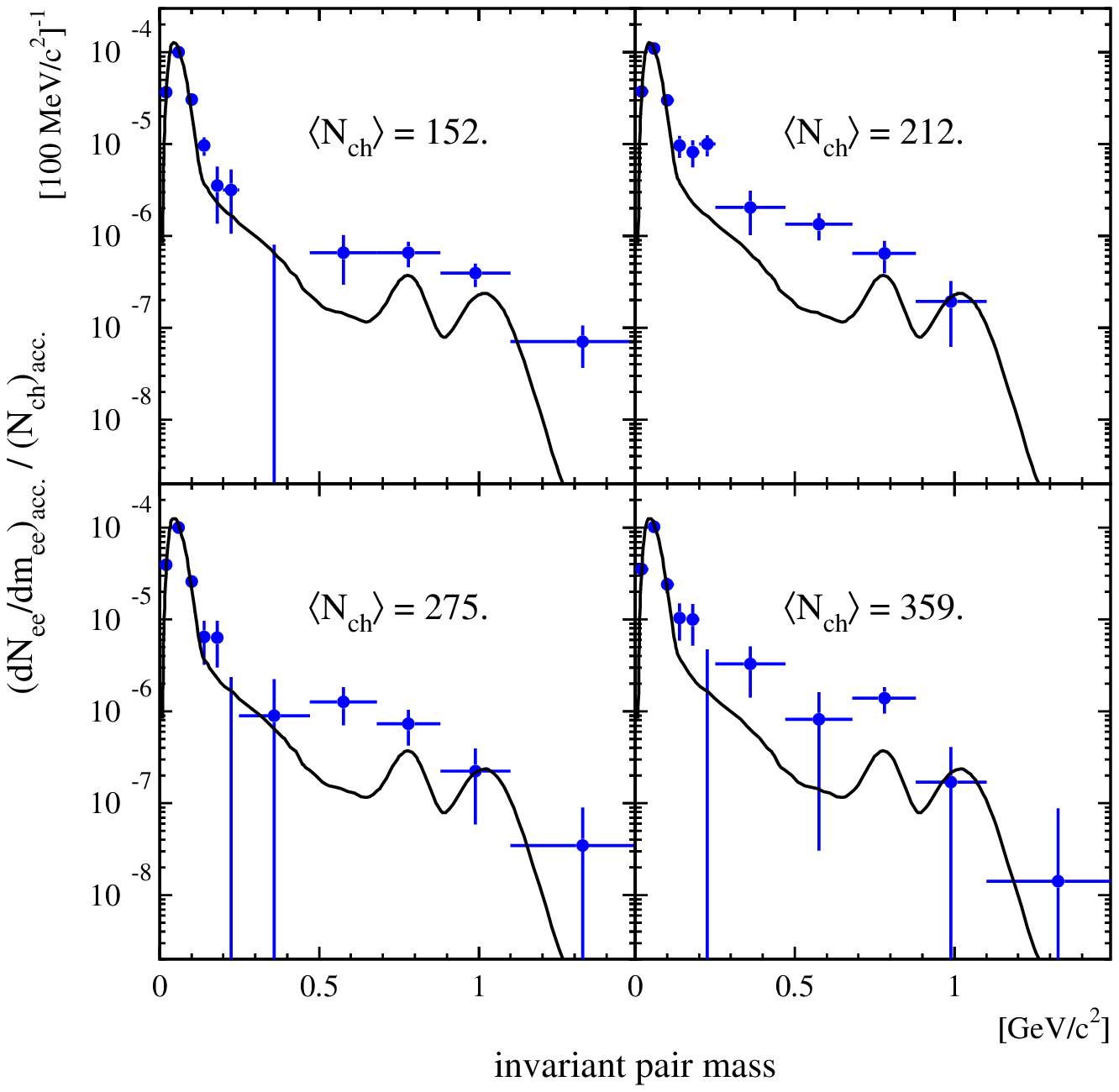}
  \epsfig{file=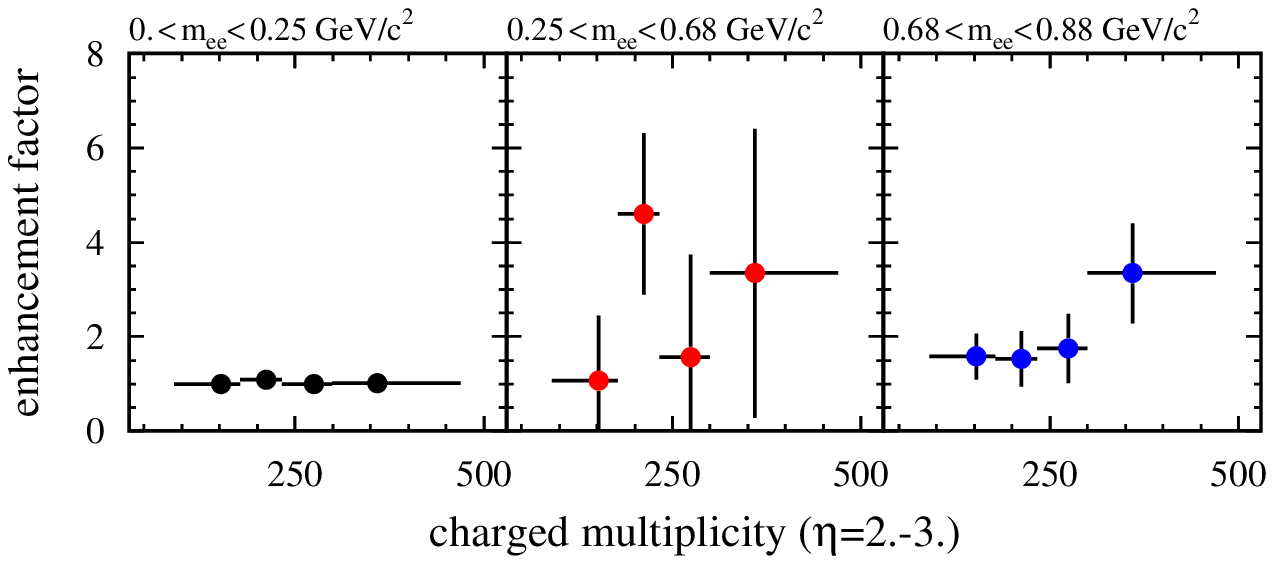}
  \caption[$N_{\rm ch}$-dependence of invariant-mass spectrum for $p_{\bot}$\,$>$\,$0.1$\,GeV/c]
  {$N_{\rm ch}$-dependence of the invariant mass spectrum for $p_{\bot}$\,$>$\,$0.1$\,GeV/c (upper panel).
  The solid line represents the expected yield of all hadronic
  sources~\cite{Sako:2000}.
  $N_{\rm ch}$-dependence of the enhancement factor for three different mass bins (lower panel).
  }
  \label{fig:mass-mbin-pt-0.1}
\end{figure}

To illustrate this dependence more clearly, the relative
enhancement factor defined as the ratio of the dielectron yield to
the hadronic cocktail is plotted in the lower panel of
Fig.~\ref{fig:mass-mbin-pt-0.2} for three invariant mass bins. The
enhancement factor stays constant for pairs with mass below
250\,MeV/c$^2$ as expected for dielectrons originating from
$\pi^0$, $\eta$, and $\eta'$ Dalitz decays. The linear increase
($\alpha\approx 2$), apparent for the mass range of
250--680\,MeV/c$^2$, provides strong evidence for two-body
annihilation processes as the major source of the observed pairs.
Again, an almost constant enhancement factor is observed for pairs
with $\omega$-meson mass and above, typical of final-state hadron
decays. The mass spectra including very low momentum tracks with
$p_{\bot}$\,$>$\,$100$\,MeV/c display the same feature (see
Fig.~\ref{fig:mass-mbin-pt-0.1}), even though the conclusion of a
non-trivial origin of the enhancement is less compelling due to
the large statistical errors.

\begin{figure}[!htb]
  \epsfig{file=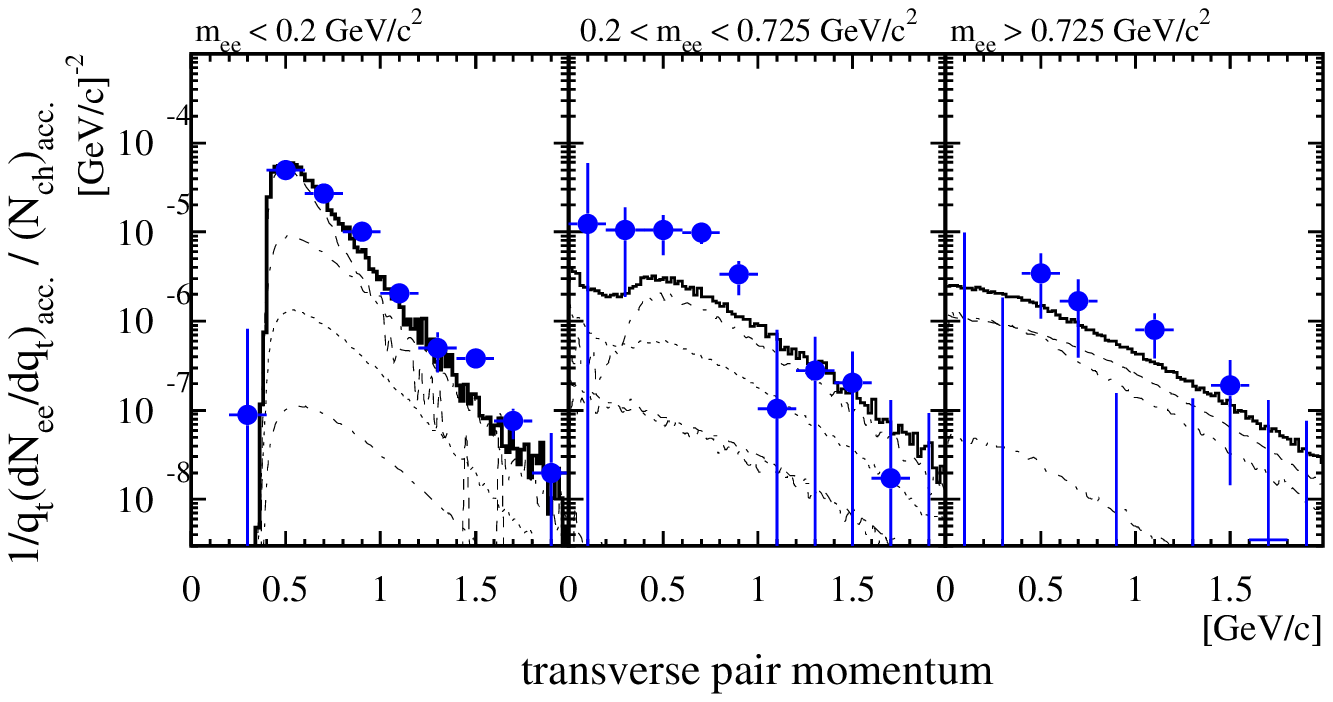}
  \epsfig{file=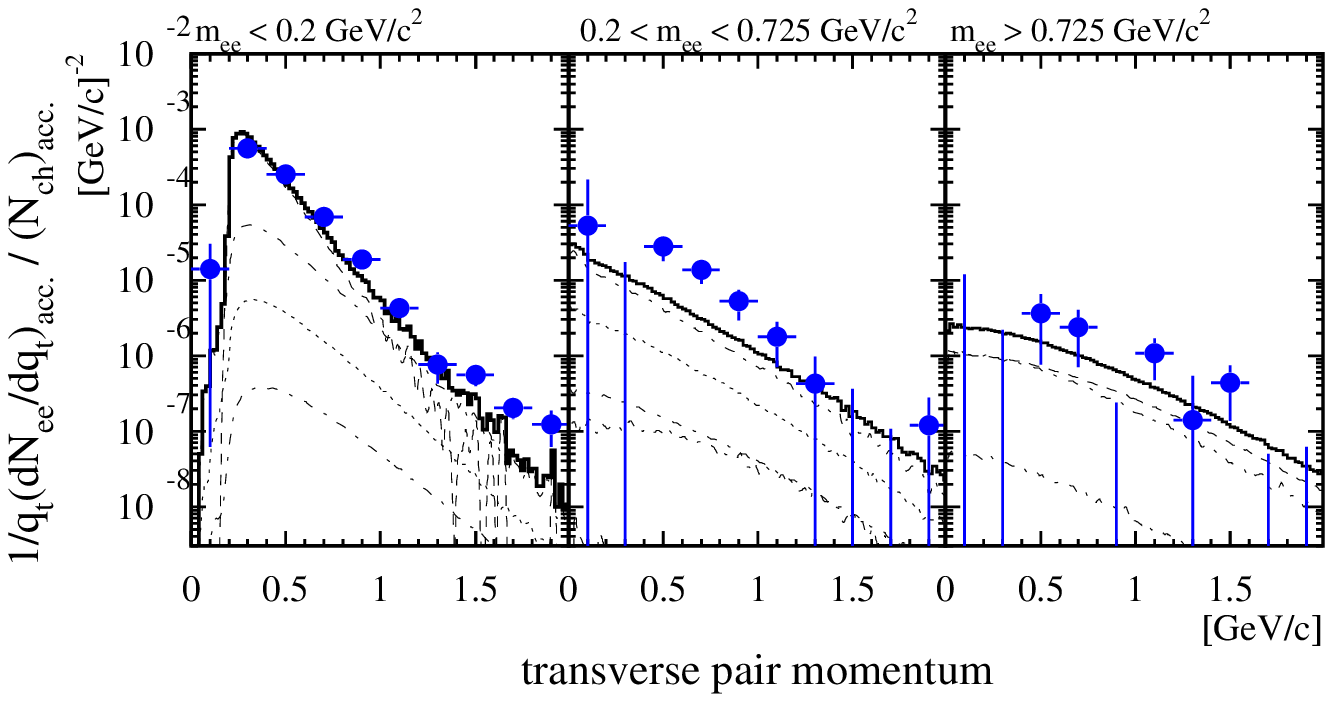}
  \caption[Transverse-pair-momentum spectra]
  {Transverse-pair-momentum spectra for $p_{\bot}$\,$>$\,$200$\,MeV/c
  (upper panel) and $p_{\bot}$\,$>$\,$100$\,MeV/c (lower panel).
  The solid line represents the expected yield of all hadronic
  sources~\cite{Sako:2000}.}
  \label{fig:pairpt}
\end{figure}

The dielectron transverse momentum $q_{\rm t}$, i.e.~the total
momentum of the pair perpendicular to the beam axis of the
colliding nuclei, is an additional observable for discrimination
of different production mechanisms. Figure~\ref{fig:pairpt}
presents a comparison of the measured lorentz-invariant $q_{\rm
t}$ spectra with the hadronic cocktail for three invariant mass
bins.
Inspection of the mass range of 200--725\,MeV/c$^2$ reveals that
the excess is most pronounced for small transverse pair momenta
below 1\,GeV/c. Most interesting, the limitation of the pair
acceptance by the $p_{\bot}$\,$>$\,$200$\,MeV/c cut, as apparent
from the dip of the cocktail spectrum at small $q_{\rm t}$, is
also visible in the data. Given that the excess persists for the
extension of the acceptance towards smaller transverse momentum
($p_{\bot}$\,$>$\,$100$\,MeV/c), it cannot possibly be related to
a deficient understanding of the pair acceptance of the
spectrometer. No significant deviations from the hadronic cocktail
are observed for pairs with lower/higher invariant mass.

\begin{figure}
  \epsfig{file=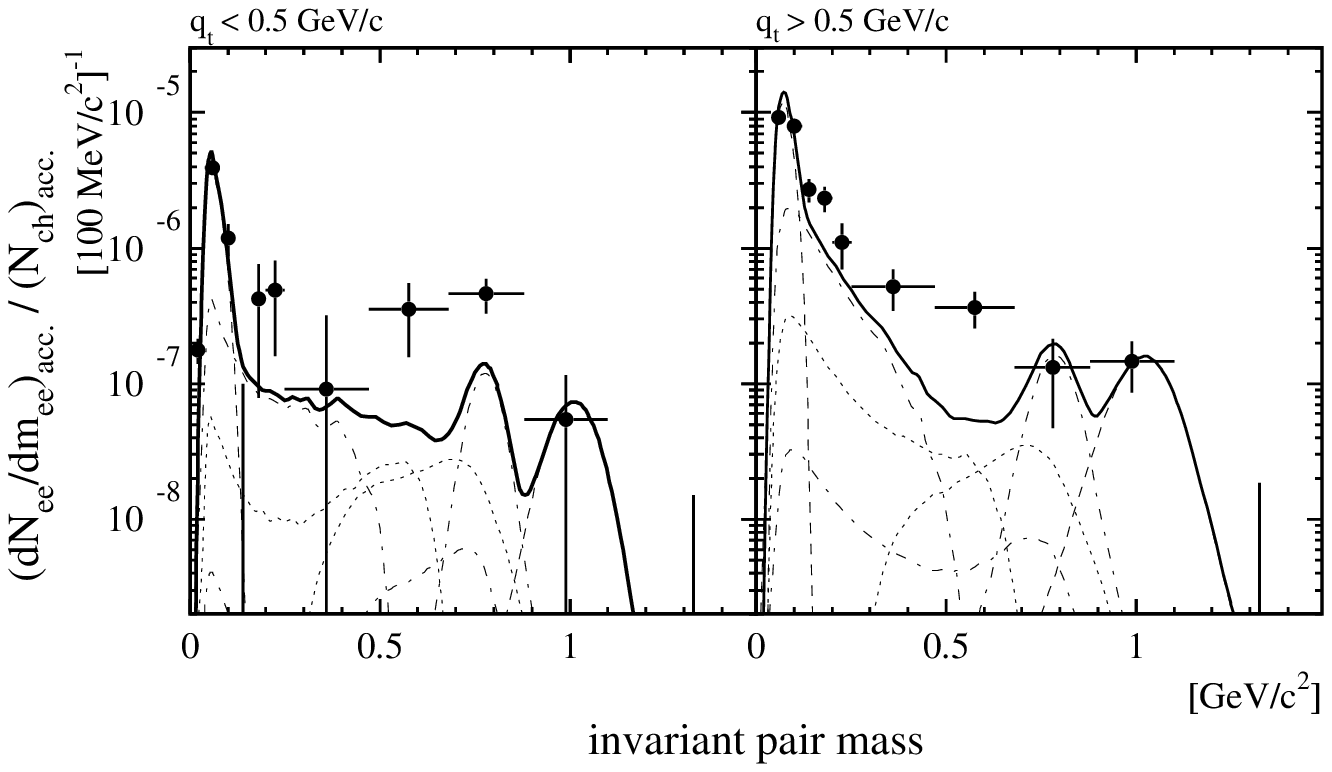}
  \epsfig{file=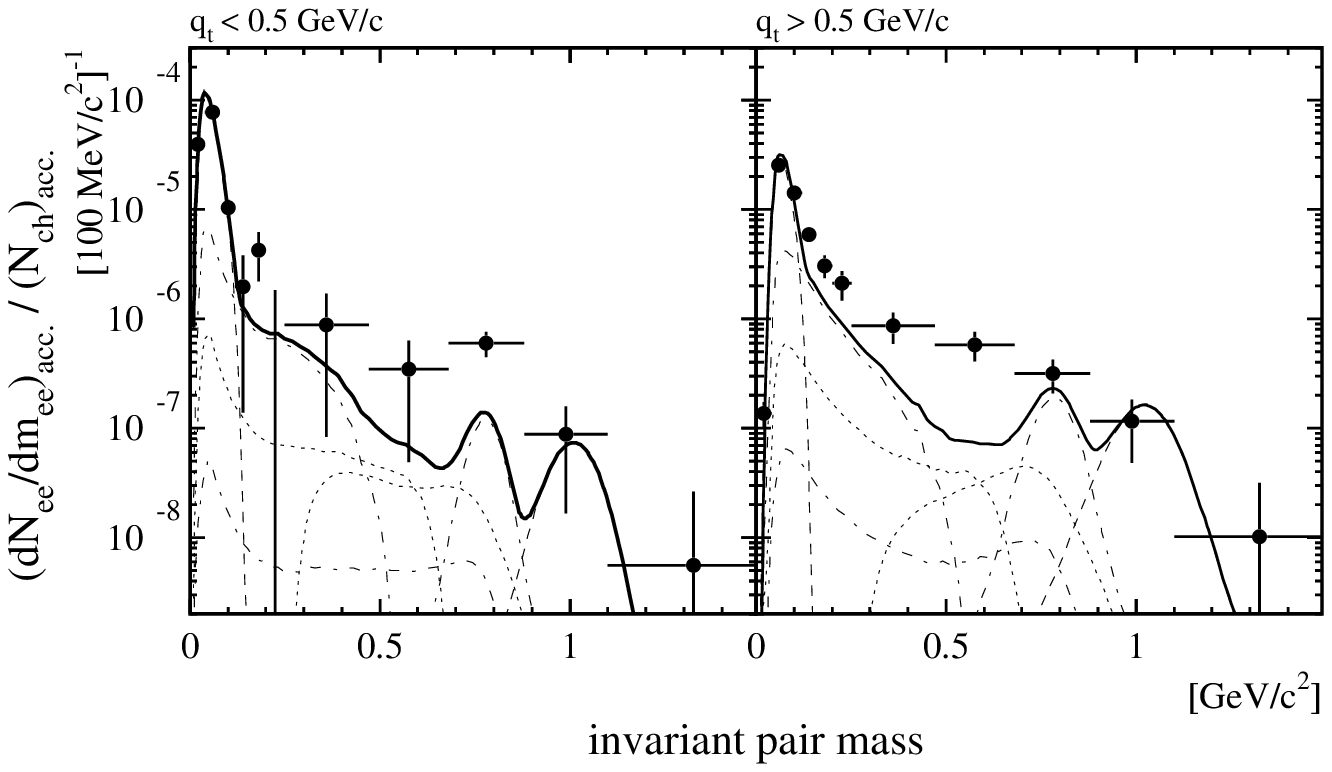}
  \caption[Transverse-pair-momentum dependence of the invariant-mass spectrum]
  {Transverse-pair-momentum dependence of the invariant-mass spectrum for
  $p_{\bot}$\,$>$\,$200$\,MeV/c (upper panel) and $p_{\bot}$\,$>$\,$100$\,MeV/c
  (lower panel). The solid line represents the expected yield of all hadronic
  sources~\cite{Sako:2000}.}
  \label{fig:mass-pairpt}
\end{figure}

For an alternative representation the invariant mass spectrum is
presented separately for transverse pair momentum below and above
500\,MeV/c. Both spectra plotted in Fig.~\ref{fig:mass-pairpt} are
distinctly different.
The difference between the data and the hadronic cocktail becomes
more pronounced for the $q_{\rm t}$\,$<$\,$500$\,MeV/c selection -
now extending up to the $\omega$ resonance, while the excess is
largely reduced for $q_{\rm t}$\,$>$\,$500$\,MeV/c - but still
significant in the mass region of 500--680\,MeV/c$^2$. It is worth
stressing that the spectrum for $q_{\rm t}$\,$<$\,$500$\,MeV/c is
greatly improved by the reduced transverse-momentum cut of
$p_{\bot}$\,$>$\,$100$\,MeV/c. Integration of the
$p_{\bot}$\,$>$\,$200$\,MeV/c data sample above 250\,MeV/c$^2$
invariant mass produces an enhancement factor of
$3.6\pm2.4$(stat.) and $1.6\pm0.8$(stat.) for $q_{\rm
t}$\,$<$\,$500$\,MeV/c and $q_{\rm t}$\,$>$\,$500$\,MeV/c,
respectively.

\clearpage
\subsection{Comparison to other CERES results}

In this section, the results presented in Sec.~\ref{sec:results}
will be discussed and compared to the results of previous CERES
studies of Pb-Au collisions at 158\,GeV/c per
nucleon~\cite{Voigt:1998,Sokol:1999,Lenkeit:1998,Natascha:1998}.

\begin{figure}[htb]
    \begin{minipage}[t]{.65\textwidth}
        \vspace{0pt}
        \epsfig{file=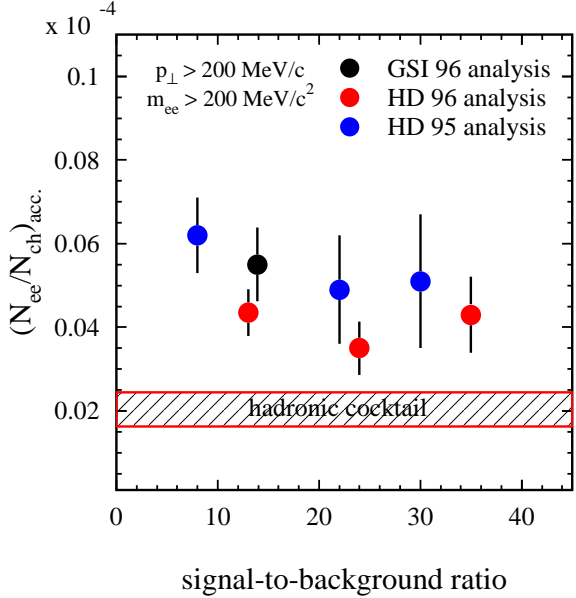,width=\textwidth}
    \end{minipage}%
    \begin{minipage}[t]{.35\textwidth}
      \vspace{0.5cm}
      \caption[CERES measurements of the dielectron yield of Pb-Au collisions at 158\,GeV/c per nucleon]
      {\newline
     The observed yield~\cite{Voigt:1998,Lenkeit:1999xu} exceeds the expected
     contributions of hadronic sources by ca. a
     factor of 2.5 for m$_{\rm ee}$\,$>$\,$200$\,MeV/c$^2$.
     As expected, the dielectron yield
     measured at different stages of the analysis
     does not depend on the signal-to-background ratio.}
      \label{fig:yield}
    \end{minipage}
\end{figure}

The most prominent observation of a strongly enhanced dielectron
production for invariant mass above 200\,MeV/c$^2$ was confirmed
by this re-analysis. A collation of all measurements in
Fig.~\ref{fig:yield} shows the observed dielectron yields to be
consistent within the statistical errors.
The important discovery of a stronger than linear rise of the
dielectron production rate with multiplicity was substantiated for
the mass range of the largest enhancement
(0.25--0.725\,GeV/c$^2$). The level of agreement is remarkable
given the fact that this new analysis was based on a completely
different background subtraction technique, an improved tracking
strategy, and a refined efficiency determination.

\begin{figure}[!t]
  \centering
  \epsfig{file=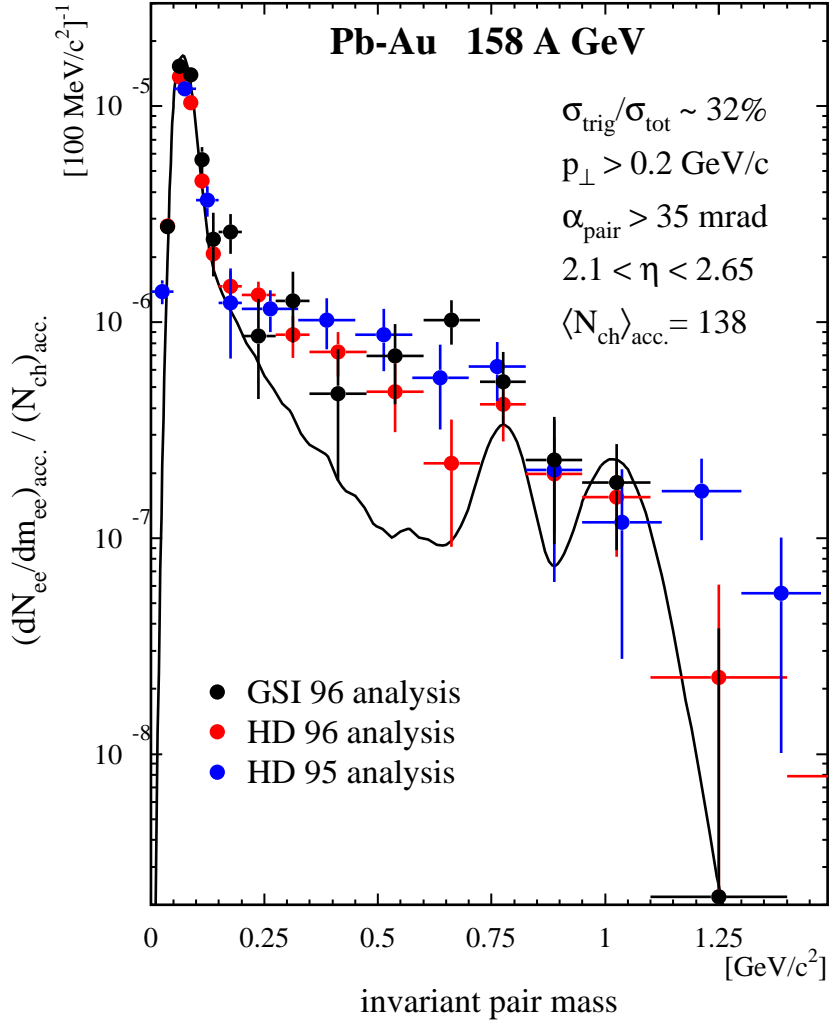}
  \caption[Comparison of the invariant-mass spectrum with previous anal\-y\-ses]
  {Comparison of the invariant-mass spectrum ($p_{\bot}$\,$>$\,$200$\,MeV/c) with
  the results of previous analyses by the Heidelberg group (HD)~\cite{Voigt:1998,Lenkeit:1998}.
  The solid line represents the expected yield of all hadronic
  sources~\cite{Sako:2000}.}
  \label{fig:mass-pt-0.2-comp}
\end{figure}

The following detailed comparison is focused on the latest and
most advanced previous study~\cite{Lenkeit:1998} which also served
as a starting point of this paper. Subjecting the invariant mass
spectra plotted in Fig.~\ref{fig:mass-pt-0.2-comp} to direct
comparison reveals differences that need to be addressed.

First, the spectra differ by about 35\% for low-mass pairs
($m_{\rm ee}$\,$<$\,$200$\,MeV/c$^2$). This is because the
efficiency used to correct the data in the previous analysis was
obtained from a Monte Carlo simulation while in this study the
spectrum was normalized to the expected yield at the $\pi^0$ peak
($m_{\rm ee}$\,$<$\,$200$\,MeV/c$^2$). The resulting discrepancy
is still within the range of the systematic error of both
measurements. The results for pairs with larger mass agree well
with each other except for the data points at $m_{\rm ee} =
400$(650)\,MeV/c$^2$. The apparent differences are statistically
significant. The previous analysis presented
in~\cite{Lenkeit:1998} and this paper are based on the same data
set; therefore, the observed discrepancy is a measure for the
systematic error of the analysis.

Second, the statistical errors plotted are larger for the new
analysis because of the lower reconstruction efficiency, but
similar to those of the old analysis before applying the
background smoothing which involved a certain level of
subjectiveness in the choice of the best background fit function.

The factorization of the $N_{\rm ch}$ and $\theta$ efficiency
dependence and the 2.5\% offset in the momentum determination
result in a systematic error of the old analysis which is small
compared to statistical uncertainties.

\begin{figure}[!t]
  \centering
  \epsfig{file=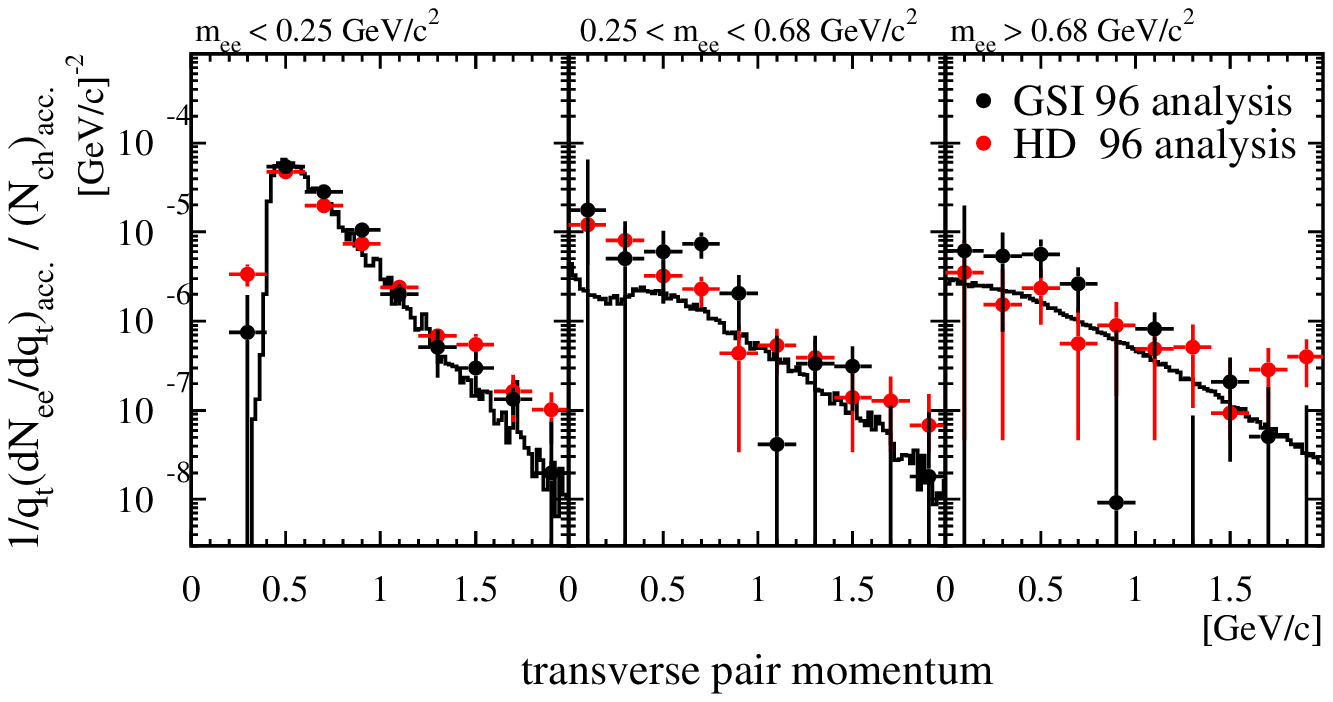}
  \epsfig{file=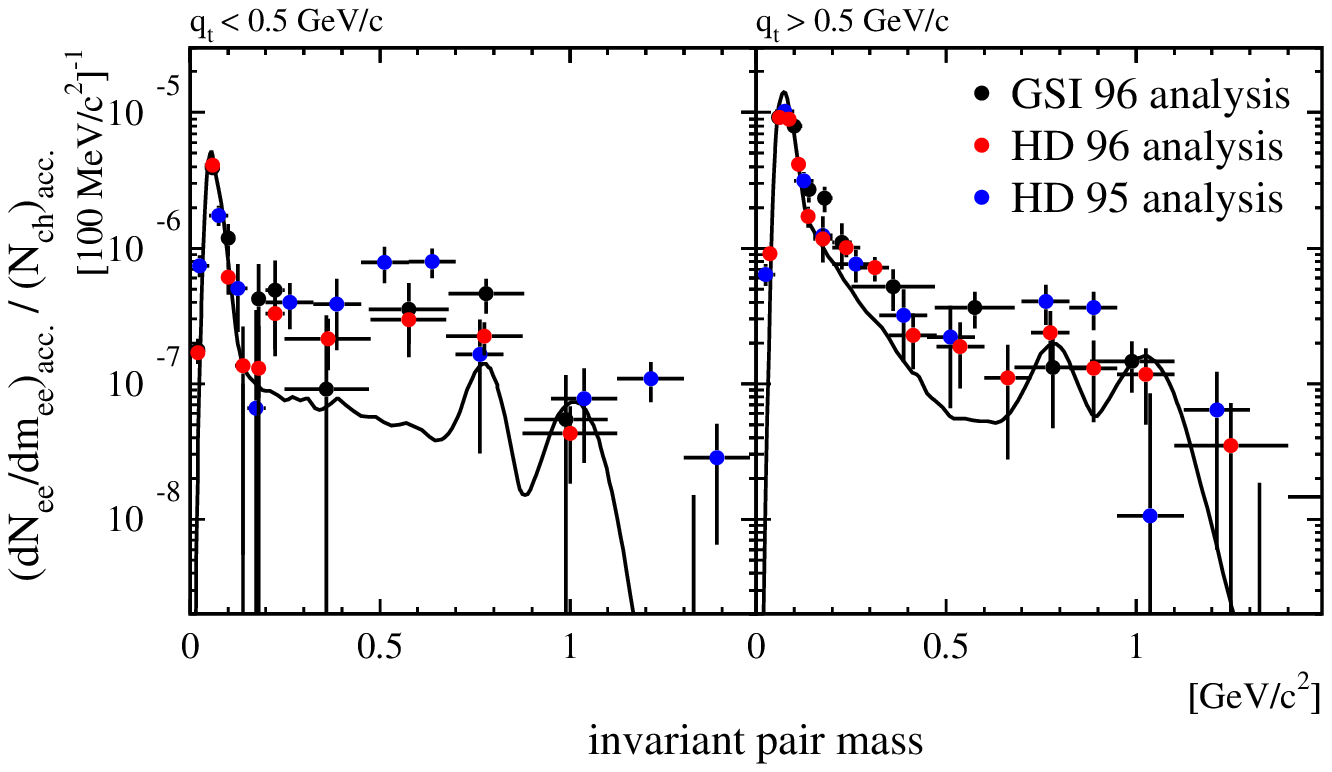}
  \caption[Comparison of $q_{\rm t}$-spectra and $q_{\rm t}$-dependence
           of the invariant mass spectrum with previous analyses]
  {Comparison of the $q_t$-spectra for $p_{\bot}$\,$>$\,$200$\,MeV/c
   with the previous analysis~\cite{Lenkeit:1998} (upper panel).
   Comparison of the invariant mass spectrum for $p_{\bot}$\,$>$\,$200$\,MeV/c
   with previous analyses by the Heidelberg group
   (HD)~\cite{Voigt:1998,Lenkeit:1998} for $q_{t}$\,$<$\,$500$\,MeV/c and
   $q_{t}$\,$>$\,$500$\,MeV/c (lower panel).
  }
  \label{fig:qt-comp}
\end{figure}

Figure~\ref{fig:qt-comp} (upper panel) compares the transverse
pair momentum spectra for $p_{\bot}$\,$>$\,$200$\,MeV/c.
In contrast to the previous analysis, an excess is already
observed for a pair momentum of about 1\,GeV/c in the mass region
of 200--725\,MeV/c$^2$ which is even more pronounced for the
$p_{\bot}$\,$>$\,$100$\,MeV/c sample. The difference is most
likely an artifact of the factorization of the $N_{\rm ch}$- and
$\theta$-efficiency dependence. In central collisions, where the
largest excess is observed, factorizing underestimates the slope
of the $\theta$ efficiency dependence (see Fig.~\ref{fig:eff
th-nch}). Therefore, tracks at small theta are suppressed and
tracks at large theta are enhanced. Consequently, the pairs
contributing to the excess are artificially enhanced for very low
transverse pair momentum and suppressed for larger transverse pair
momentum.

The same feature can be seen in the comparison of the invariant
mass spectra for transverse pair momentum smaller and larger than
500\,MeV/c in Fig.~\ref{fig:qt-comp} (lower panel). The
enhancement observed in~\cite{Lenkeit:1998} is larger(smaller)
than what was found in this study for $q_{\rm
t}$\,$<$\,$0.5$\,GeV/c ($q_{\rm t}$\,$>$\,$0.5$\,GeV/c),
respectively.

\subsection{Theoretical interpretations}
\label{sec:theory}

In Sec.~\ref{sec:results}, it was demonstrated that the hadronic
cocktail does not suffice to explain the observed dielectron
yield. The stronger than linear rise of the enhancement with the
number of charged particles in the final state points to a
two-body annihilation process. The high abundance of pions at SPS
energies makes pion annihilation
$\pi^{+}\pi^{-}\rightarrow\rho\rightarrow e^{+}e^{-}$ the most
likely explanation. In a hot and dense hadronic medium, the
annihilation process is subject to modifications induced by
interactions with surrounding hadrons and/or partial restoration
of chiral symmetry as set out in Sec.~\ref{sec:theo intro}. The
system spends most of the time in its hadronic phase, even though
evidence was found for a phase transition in Pb-Au collisions to a
quark-gluon plasma~\cite{Heinz:2000bk}. The expected contribution
from quark-quark annihilation is very small compared to
conventional sources of dielectrons~\cite{Rapp:1999zw}.

A complete description of dilepton production in heavy ion
collisions requires modeling of the time evolution of the
collision system. In general, theoretical collision models can be
divided into hydrodynamical
approaches~\cite{Srivastava:1999bt,Hung:1997mq,Baier:1997if,Murray:1998wz},
transport
models~\cite{Bratkovskaya:1997qe,Bleicher:2000xh,Li:1996mi,Brown:2001nh,Schneider:2000cd},
and thermal fireball
models~\cite{Rapp:1997fs,Rapp:1999us,Lee:1998um,Gallmeister:1999dj}.
The main advantage of hydrodynamical simulations is the capability
of incorporating phase transitions in a well defined way via the
equation of state. In contrast, transport models are better at the
implementation of rescattering and absorption processes. The
phenomenological fireball models allow for simple comparison of
underlying microscopic models.

To explain the CERES dielectron data of nucleus-nucleus
collisions, various options were proposed including  Brown-Rho
scaling~\cite{Brown:1991kk,Brown:1998ca,Song:1996af,Song:1999tn},
collision broadening of the $\rho$-meson spectral
function~\cite{Friman:1997tc,Rapp:1997fs}, open charm
production~\cite{Braun-Munzinger:1998xy}, chiral meson
mixing~\cite{Teodorescu:2000mg,Florkowski:2000mx,Herrmann:1993za},
quark-quark annihilation~\cite{Schneider:2000cd}, and thermal
plasma radiation~\cite{Kampfer:2001zw}. A recent review can be
found in~\cite{Rapp:1999ej}.

The recent version of the thermal fireball simulation by
Rapp~\cite{Rapp:2001} was chosen as a representative model to
compare to the experimental results. It is generally recognized as
a comprehensive and reliable simulation of the dilepton production
in heavy ion collisions. Several scenarios of in-medium
modifications for the $\pi\pi$ annihilation process are among the
specifics of the model used. The time evolution is treated in a
thermal fireball approach. The experimentally determined initial
conditions ($T_{\rm ini}$\,=\,190\,MeV, $\varrho_{\rm
ini}$\,=\,2.55$\varrho_0$), the hadro-chemical freezeout ($T_{\rm
fo}$\,=\,115\,MeV, $\varrho_{\rm fo}$\,=\,0.33$\varrho_0$), as
well as a finite pion chemical potential $\mu_{\pi}$ are included.

\begin{figure}[!tb]
  \centering
  \epsfig{file=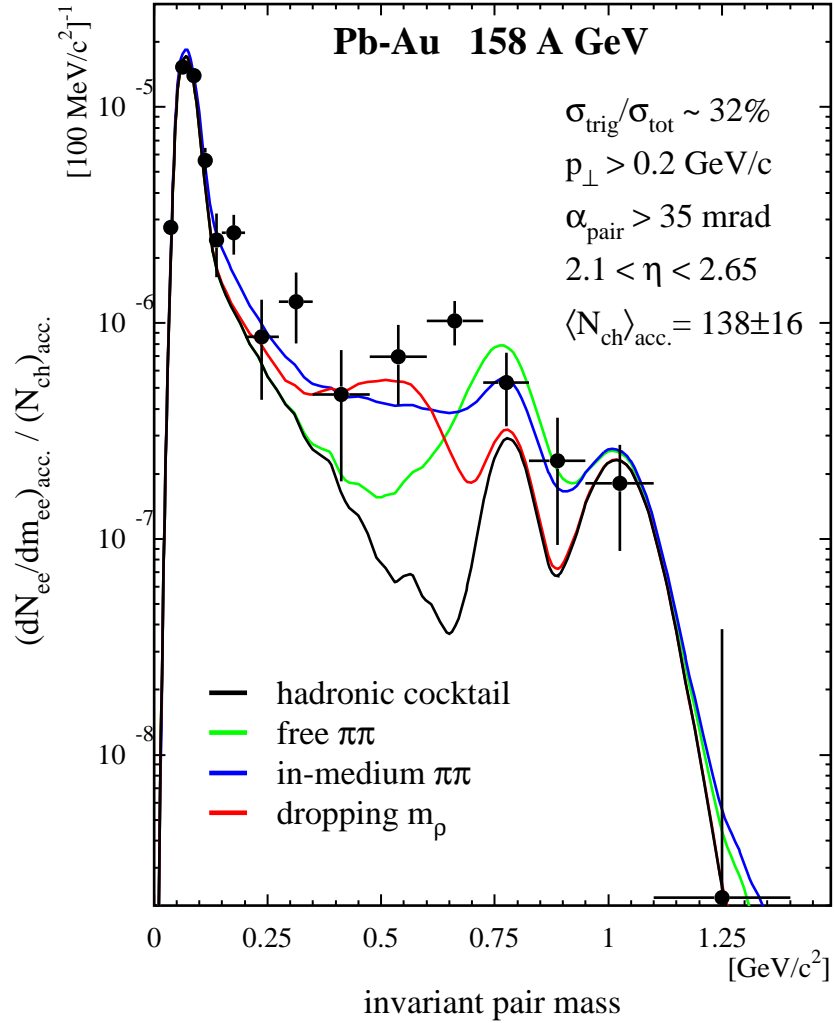}
  \caption[Comparison of the invariant mass spectrum for $p_{\bot}$\,$>$\,$0.2$\,GeV/c with theoretical models]
  {Comparison of the invariant mass spectrum for $p_{\bot}$\,$>$\,$0.2$\,GeV/c with theoretical models~\cite{Rapp:2001}.
  The dielectron yield predicted for $\pi\pi$ annihilation
  was added to the standard cocktail of the hadronic sources
  (without $\rho$-meson contribution). Experimental acceptance as well as momentum
   resolution were applied to the model calculations.}
  \label{fig:mass-pt-0.2-theory}
\end{figure}

Figure~\ref{fig:mass-pt-0.2-theory} compares the inclusive
invariant mass spectrum for $p_{\bot}$\,$>$\,$0.2$\,GeV/c with
three different theoretical scenarios.
First, the $\pi\pi$ annihilation with vacuum spectral function
gives too much yield at the $\rho/\omega$-peak and hardly fills
the hole between $m_{\rm ee}$\,=\,0.2--0.7\,GeV/c$^2$. It can be
concluded that in-medium modification must play an important role.

Second, the dropping $\rho$ mass scenario according to Brown-Rho
scaling (or Hatsuda-Lee sum rules) fits much better to
experimental data but underestimates the observed yield at the
peak of the free vacuum $\rho$-meson.  This scenario entails a
reduction of the $\rho$-meson width as well as a sharp threshold
at twice the pion mass for the onset of the enhancement. Brown-Rho
scaling is based on phenomenological implementation of the
restoration of chiral symmetry in the framework of an effective
field theory. In this case, the dependence of the in-medium $\rho$
mass on temperature $T$ and density $\varrho_{\rm B}$ is given by:
 \begin{equation}
 m_{\rho}^{\ast}=
 m_{\rho}\left(1-C\frac{\varrho_{\rm B}}{\varrho_0}\right)\left(1-\left(\frac{T}{T_{\rm c}^{\chi}}\right)^2\right)^{\alpha}
 \label{equ:BR scaling}
\end{equation}
with $C=0.15$, $T_c^{\chi}=200$\,MeV, and $\alpha=0.3$ (QCD sum
rule estimate).

Third, both $\pi$ and $\rho$ properties are modified in the medium
due to rescattering (collisional broadening of the spectral
function). The resulting spectrum is very similar to the dropping
$\rho$ mass scenario for mass below 600\,MeV/c$^2$. However, more
strength is expected at the vacuum $\rho$-meson peak resulting in
a better agreement with the data. The ingredients of this model
are chiral reduction~\cite{Steele:1997tv}, many-body calculation
of the $\rho$-meson spectral function~\cite{Rapp:1997fs}, rhosobar
excitations on thermally excited baryon resonances, and a complete
assessment of mesonic contributions.

\begin{figure}[!tb]
  \centering
  \epsfig{file=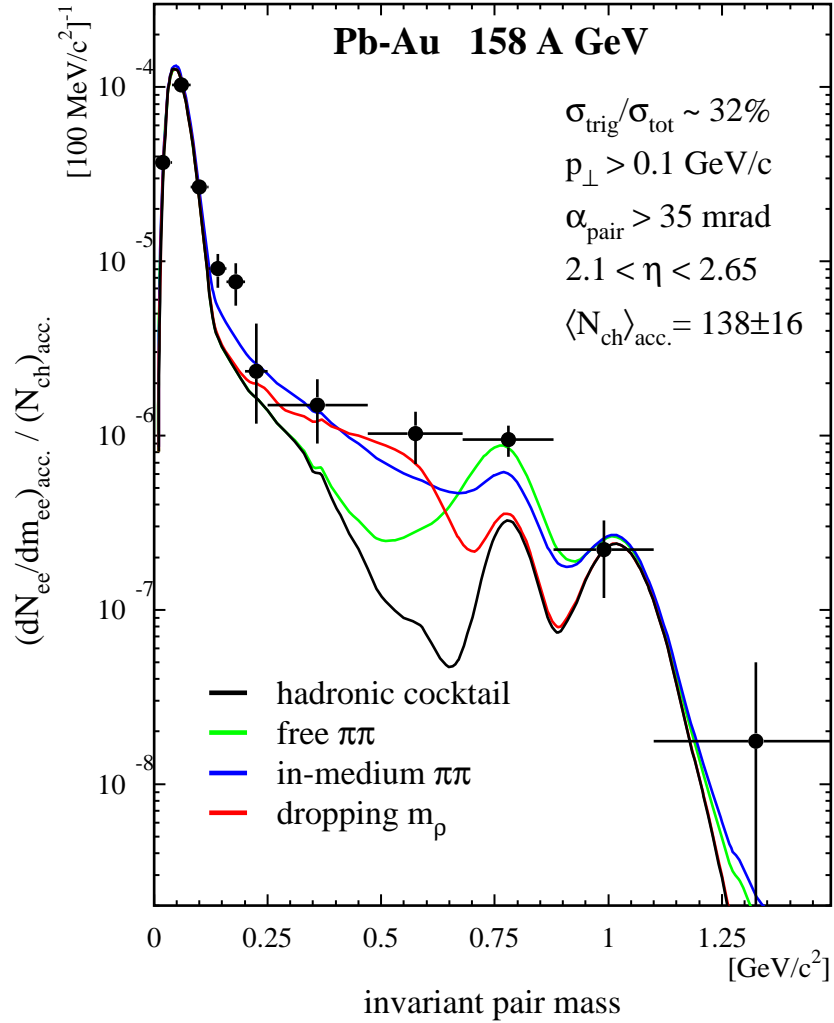}
  \caption[Comparison of the invariant mass spectrum with theoretical models for $p_{\bot}$\,$>$\,$0.1$\,GeV/c]
  {Comparison of the invariant mass spectrum with theoretical models for $p_{\bot}$\,$>$\,$0.1$\,GeV/c~\cite{Rapp:2001}.
  The dielectron yield predicted for $\pi\pi$ annihilation
  was added to the standard cocktail of the hadronic sources
  (without $\rho$-meson contribution). Experimental acceptance as well as momentum
   resolution were imposed on the simulated cocktail data.}
  \label{fig:mass-pt-0.1-theory}
\end{figure}

Both the dropping $\rho$ mass scenario and the in-medium
broadening give reasonable account of the dielectron enhancement
in the 0.3--0.6 GeV/c$^2$ region. This is true also in the case of
extension of the acceptance to single track
$p_{\bot}$\,$>$\,$0.1$\,GeV/c as apparent from
Fig.~\ref{fig:mass-pt-0.1-theory}.
Again, the scenario of in-medium broadening of the $\rho$ spectral
function seems more plausible.

Although the first CERES data for the centrality dependence of the
mass spectrum was presented several years ago, no systematic
theoretical calculations are available yet.

\begin{figure}
  \centering
  \epsfig{file=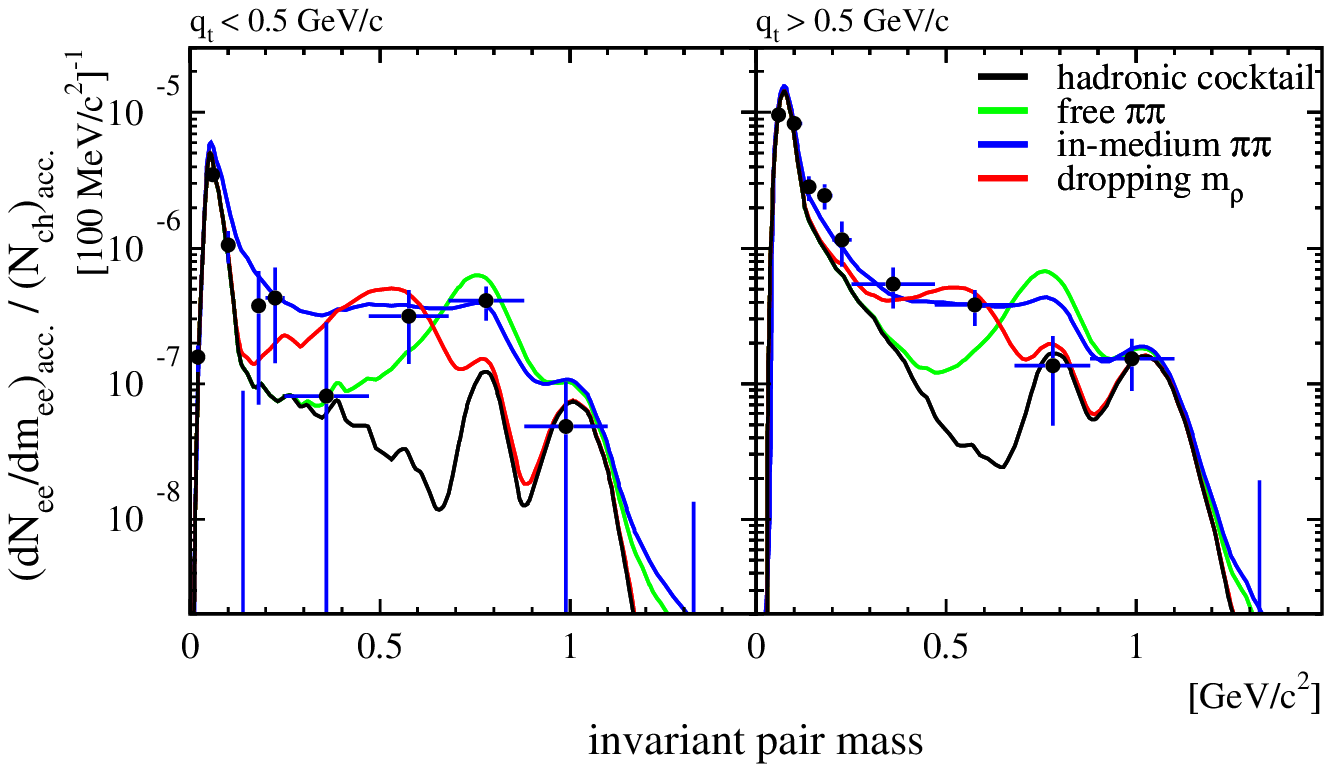}
  \epsfig{file=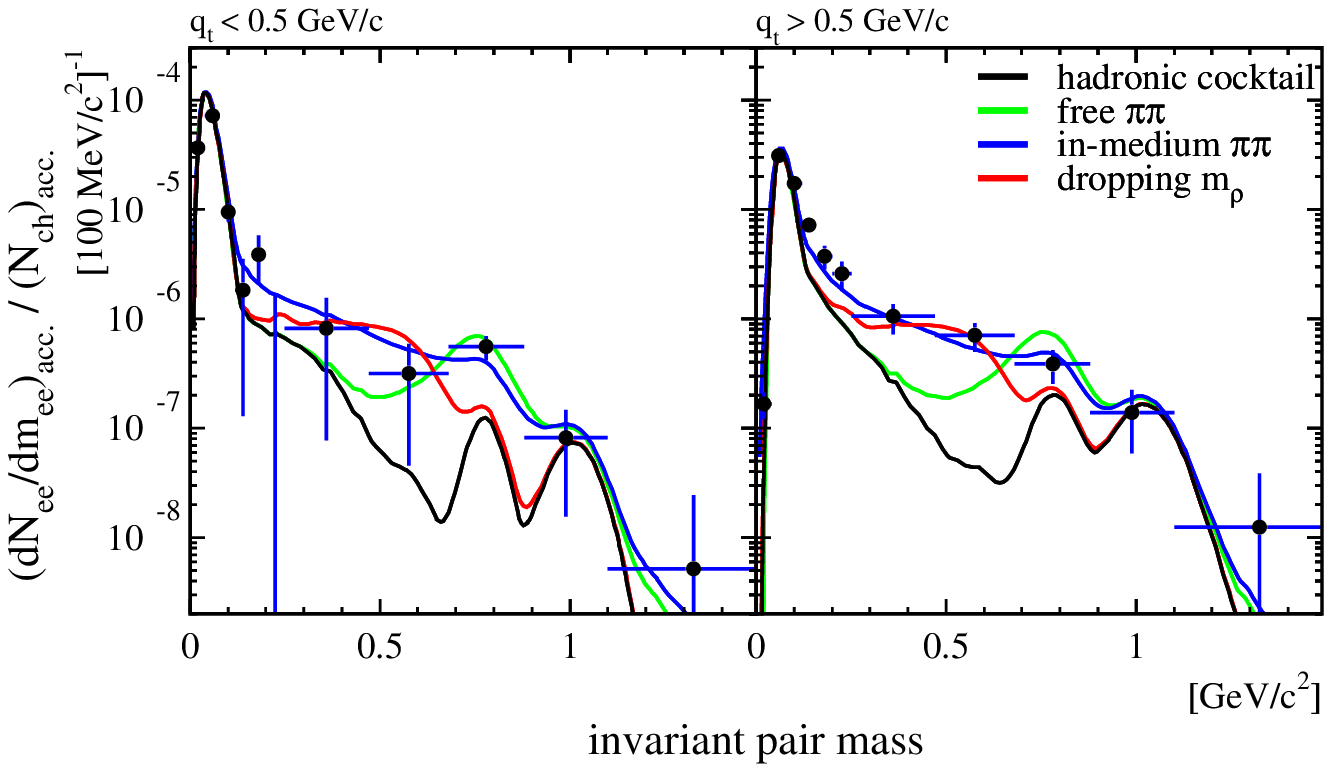}
  \caption[Comparison of the transverse pair momentum dependence of the invariant-mass spectrum with theoretical models]
  {Comparison of the transverse pair momentum dependence of the
  invariant mass spectrum with theoretical models for
  $p_{\bot}$\,$>$\,0.2\,GeV/c (upper panel) and
  $p_{\bot}$\,$>$\,0.1\,GeV/c (lower panel)~\cite{Rapp:2001}.
  The dielectron yield predicted for $\pi\pi$ annihilation
  was added to the standard cocktail of the hadronic sources
  (without $\rho$-meson contribution). Experimental acceptance as well as momentum
   resolution were imposed on the simulated cocktail data.}
  \label{fig:mass-pairpt-theory}
\end{figure}

Figure~\ref{fig:mass-pairpt-theory} compares the invariant mass
spectrum for two distinct transverse pair momentum selections with
model calculations.
The free $\pi\pi$ scenario without in-medium modifications clearly
fails to account for the increase of the dielectron yield for low
transverse momenta.

The observed transverse-momentum dependency can arise from the
fact that Lorentz invariance is broken in the thermal frame.
Therefore, the in-medium propagator, describing the dynamics of
the meson, can depend on energy and momentum separately.
Transverse and longitudinal modes emerge as polarization states
that are no longer isotropic. Moreover, the thermal occupancy is
sensitive to a reduction of the $\rho$-meson mass:
\begin{equation}
 f^{\rho}(q_0)=\sqrt{
(m^{\ast}_{\rho})^2+\overrightarrow{\mathbf{q}}^2} \;.
 \label{equ:qt rho}
\end{equation}
The three-momentum dependence becomes more pronounced for smaller
in-medium $\rho$ mass. It leads to a relative enhancement of the
$\rho$-meson for small three-momenta or equivalent small $q_{\rm
t}$.

Although the model predictions for the dropping $\rho$ mass and
the collision broadening scenario differ significantly in this
particular representation, the large statistical errors of the
data do not permit to distinguish both. The apparent difficulties
of both scenarios to account for the large yield observed at the
free vacuum $\rho$-peak could also point to an underestimation of
the $\omega$-contribution which is not well determined so far.

\begin{figure}[!tb]
  \centering
  \epsfig{file=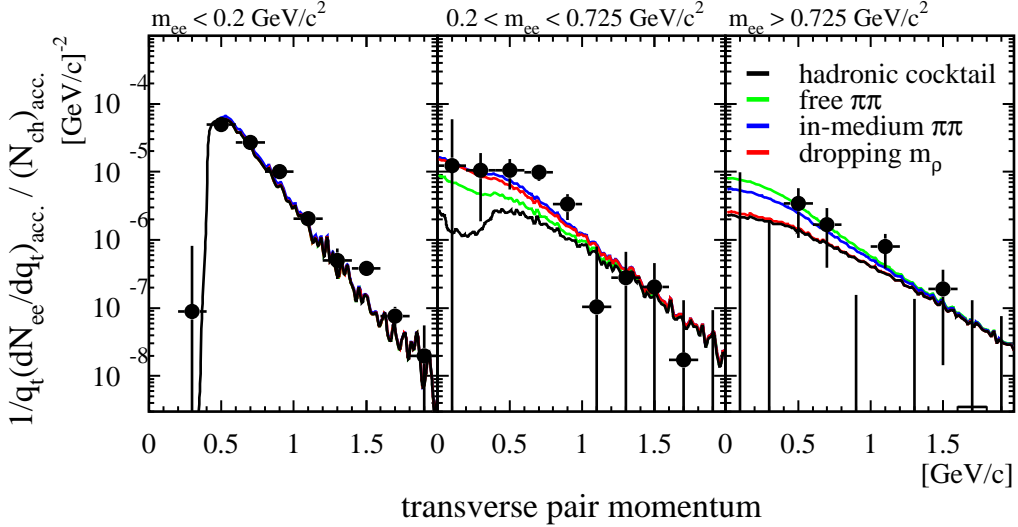}
  \caption[Comparison of the transverse pair momentum spectra with theoretical models]
  {Comparison of the transverse pair momentum spectra with theoretical models for
  $p_{\bot}$\,$>$\,0.2\,GeV/c~\cite{Rapp:2001}.
  The dielectron yield predicted for $\pi\pi$ annihilation
  was added to the standard cocktail of the hadronic sources
  (without $\rho$-meson contribution). Experimental acceptance as well as momentum
   resolution were imposed on the simulated cocktail data.}
  \label{fig:pairpt-theory}
\end{figure}

Finally, a comparison of the transverse momentum spectra for
$p_{\bot}$\,$>$\,0.2\,GeV/c with the model calculations is
presented in Fig.~\ref{fig:pairpt-theory}.

To conclude, the present data excludes the most simple scenario
with vacuum $\rho$-mass. The statistical errors, however, do not
permit to distinguish the two approaches which include in-medium
modifications.

The difference between both models is most evident in the region
of the $\omega$-resonance, namely a factor of two. Excluding one
or the other with a confidence level of 96\% would require a
relative statistical error of less than 10(15)\%. To achieve this
level of accuracy for the $3.3\cdot10^7$ central Pb-Au collisions
recorded in 2000 would require a signal-to-background ratio of
larger than 1:5(1:10) assuming a realistic pair-reconstruction
efficiency of 22\%. This is a very challenging task but the
reduction of the systematic errors, i.e.~multiplicity and
efficiency determination, to same level of accuracy might prove
even more difficult.

\section{Summary and outlook}

The most important result of this new analysis of the dielectron
production in Pb-Au collisions at 158\,GeV/c per nucleon is the
independent confirmation of the previous result: a significant
excess of dielectrons observed in the mass range of $200 $\,$<$\,$
m_{\rm ee} $\,$<$\,$ 700$\,MeV/c$^2$ compared to the expected
yield of hadronic sources. In the same mass region, there is
evidence for a stronger than linear rise of the yield with
charged-particle multiplicity. The transverse-momentum spectra
show an enhancement for the transverse pair momentum below
1\,GeV/c that increases towards small $q_t$. The contribution of
very low momentum pairs, i.e.~single-track momentum of
$p_{\bot}$\,$>$\,$100$\,MeV/c, was studied for the first time. It
was found that the excess increases towards small transverse pair
momentum. The use of the mixed-event technique for background
subtractions has ruled out the possibility of artifacts of the
same-event background subtraction as a source of the dielectron
enhancement. The refined calibration of SDD and RICH allowed a
better rejection of the combinatorial background. However, this
improvement was partially cancelled by a 30\% efficiency loss due
to new software for SDD-hit reconstruction and tracking.

The comparison with theoretical model calculations shows that the
observed dielectron yield cannot be explained by the known
hadronic sources including contributions from free pion
annihilation. Indeed, only scenarios invoking in-medium
modification of the $\rho$- and/or $\pi$-meson can account for the
observed yield as well as the spectral shape.  Both the dropping
$\rho$ mass and the collision broadening scenario are viable for
the present data, even though the second option seems more
plausible.

A precision measurement of the yield and spectral shape of the
heavy vector mesons is indispensable to distinguish the different
scenarios. It requires a high statistics data sample combined with
a much better momentum resolution. The upgrade of the CERES
experiment with a radial TPC is expected to fulfill both
requirements. First studies~\cite{Schmitz:2001} achieved a
momentum resolution of $dp/p=\sqrt{(0.027)^2+(0.024\cdot
p\cdot{\rm GeV}^{-1}{\rm c})^2}$ for the reconstruction of
$\lambda$-mesons. Further improvements are expected after a
refined calibration of the TPC.

In the fall of 2000, the upgraded CERES experiment was operated
with a very good performance. A sample of $3.3\cdot10^7$ central
Pb-Au collisions at 158\,GeV/c per nucleon was recorded. The
progress made so far promises to fulfill the high expectations for
a precision measurement of the low-mass dilepton spectrum.

Other regions of the nuclear matter phase diagram will be explored
by experiments measuring dilepton production in heavy ion
collisions, most notable HADES~\cite{Muntz:1999td} and
PHENIX~\cite{Morrison:1998qu}. Hopefully, the combination of all
results will soon allow to conclude about possible restoration of
chiral symmetry in hot and dense nuclear matter and its relation
to the QGP phase transition.

\begin{appendix}

\chapter{New GENESIS event generator}
 \label{app:genesis}

The GENESIS event
generator~\cite{Voigt:1998,Sako:2000,Lenkeit:1998,Irmscher:1993,Ullrich:1994}
is a tool to simulate the relative abundance of dielectrons
produced by hadron decays in proton-proton (pp), proton-nucleus
(pA), and nucleus-nucleus (AA) collisions. The invariant mass
range covered by the CERES acceptance ($m_{\rm
ee}$\,$<$\,$2$\,GeV/c$^2$) is dominated by the decay of light
scalar and vector mesons comprising $\pi^0,\eta,\eta', \rho^0,
\omega$, and $\phi$~\cite{PDBook}. Open charm production is
negligible~\cite{Braun-Munzinger:1998xy}. To create this so-called
{\em hadronic cocktail}, pA and AA collisions are treated as a
superposition of individual nucleon-nucleon collisions. The
hadronic cocktail provides a reference for the comparison with the
yield observed in pA and AA collisions. Any deviations would
indicate a violation of the scaling behaviour and/or in-medium
effects.

The simulation requires prior knowledge of the differential
production cross section, the widths of all decays including
dielectrons in the final state, and a description of decay
kinematics for all relevant particles. Differential cross sections
are unknown for most light mesons (except $\pi^0$, $\eta$, and
$\eta'$). The absolute meson yield of pA collisions can be
inferred from measurements in pp collisions at comparable impact
energies (for a compilation see~\cite{Agakishiev:1998mw}).

\begin{figure}[htb]
  \centering
  \epsfig{file=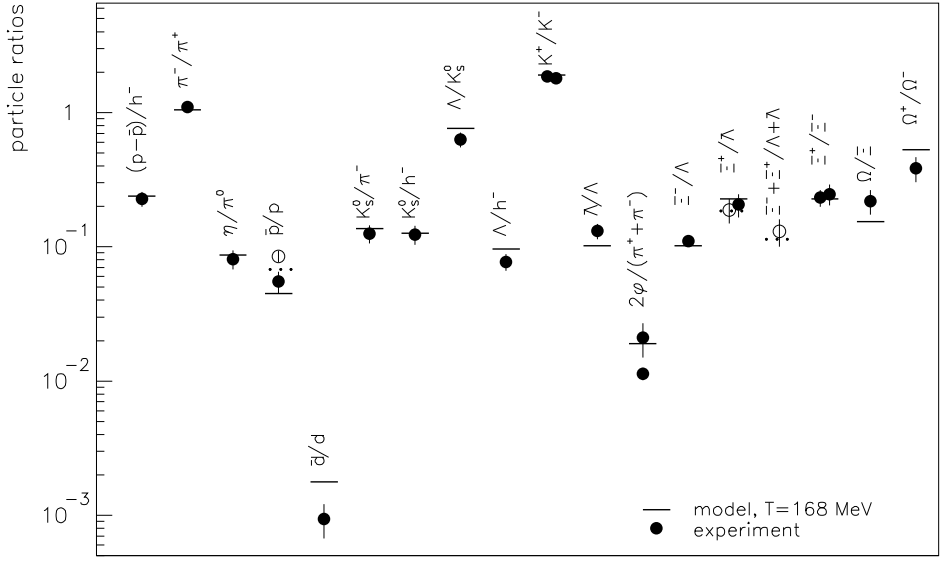,width=14cm}
  \caption[Comparison of the observed particle ratios with the prediction of the thermal model]
  {Comparison of the observed particle ratios with the prediction of the thermal model~\cite{Braun-Munzinger:1999qy}. }
  \label{fig:thermal}
\end{figure}

Proton-nucleus Collisions are modeled by a superposition of
nucleon-nucleon collisions and the yield thereof is assumed to
scale with the mean charged-particle multiplicity of a collision
system. The relative cross sections ($\sigma/\sigma_{\pi^0}$) for
Pb-Au collisions are taken from a thermal
model~\cite{Braun-Munzinger:1999qy}. The model describes particle
production in heavy ion collisions accurately, as demonstrated in
Fig.~\ref{fig:thermal}.
The cross sections and branching ratios contained in GENESIS are
summarized in Table~\ref{tab:cross section}.
\begin{table}
  \centering
  \begin{tabular}{|c|c|c|c|r|} \hline
    Particle & Decay & $\sigma/\sigma_{\pi^0}$ (p+A) & $\sigma/\sigma_{\pi^0}$ (Pb+Au) & $BR$\hspace{0.8cm} \\ \hline
    $\pi^{0}$ & $e^+e^-\gamma$   & 1     & 1     & $1.198 \times 10^{-2}$\\ \hline
    $\eta$    & $e^+e^-\gamma$   & 0.053 & 0.085 & $5.0 \times 10^{-3}$  \\ \hline
    $\rho^0$  & $e^+e^-$         & 0.065 & 0.094 & $4.44 \times 10^{-5}$ \\ \hline
    $\omega$  & $e^+e^-\pi^{0}$  & 0.065 & 0.069 & $5.9 \times 10^{-5}$  \\ \hline
    $\omega$  & $e^+e^-$         & 0.065 & 0.069 & $7.15 \times 10^{-5}$ \\ \hline
    $\phi$    & $e^+e^-$         & 0.0033& 0.018 & $3.09 \times 10^{-4}$ \\ \hline
    $\eta'$   & $e^+e^-\gamma$   & 0.009 &0.0078 & $5.6 \times 10^{-4}$  \\ \hline
  \end{tabular}
  \caption[Relative production cross section and branching ratio of light me\-sons]
  {Relative production cross section $\sigma/\sigma_{\pi^0}$ and branching ratio $BR$ of light mesons used in
  new GENESIS~\cite{Sako:2000}. The relative cross sections for Pb-Au collisions are taken from the thermal
  model~\cite{Braun-Munzinger:1999qy}.}
  \label{tab:cross section}
\end{table}
For comparison with experimental data, the cocktail is divided by
the total number of charged particles within the nominal detector
acceptance. It is directly related to number of produced neutral
pions via the ratio $\langle N_{\pi^0}/ N_{\rm
ch}\rangle$\,=\,0.44~\cite{Sako:2000}.

The properties of the parent particles are determined by
transverse momentum and rapidity distribution. The kinematic
distributions of pions were measured by
WA98~\cite{Aggarwal:1998vh,Aggarwal:2001gn},
NA44~\cite{Kaneta:1999}, and NA49~\cite{Appelshauser:1998vn}. The
transverse-momentum spectrum of charged pions by NA44 was used to
extrapolated the transverse-mass spectrum by WA98 towards small
transverse mass including the additional contribution of the
$\eta\rightarrow 3\pi^0$ decay. The parameterization is documented
in~\cite{Sako:2000}. The inverse slope parameter of the NA44
$\pi^0$ transverse mass spectrum changes by about 5\% with
centrality. This dependence is not yet implemented in GENESIS and,
therefore, contributes 5\% to the inherent systematic error. WA98
quotes a systematic error of 10\% for pairs with transverse mass
above $m_{\bot}$\,$>$\,$400$\,MeV/c$^2$. The systematic error of
the combined spectrum is estimated to be about 10\%.

\begin{figure}[htb]
    \begin{minipage}[t]{.65\textwidth}
        \vspace{0pt}
        \epsfig{file=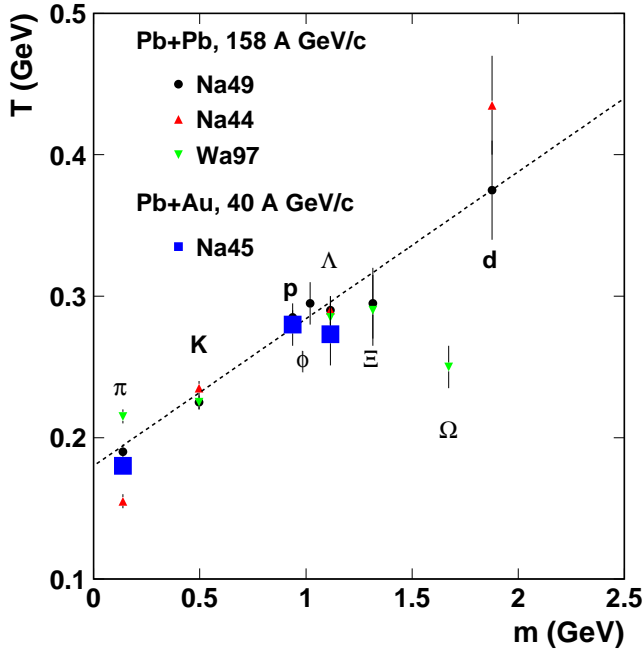,width=\textwidth}
    \end{minipage}%
    \begin{minipage}[t]{.35\textwidth}
      \vspace{0.5cm}
      \caption[$m_\bot$-scaling of hadron spectra]
      {\newline
      Inverse slope parameter $T$ fitted to hadron spectra from central
      Pb-Pb collisions at the
      SPS~\cite{Appelshauser:1998vn,Bearden:1998vp,Bearden:1998wi,Appelshauser:1998va,Andersen:1998vu,Andersen:1998ph,Aggarwal:1998vs}.
      }
      \label{fig:mt spectrum}
    \end{minipage}
\end{figure}

The transverse-mass distributions of all other mesons are
described by exponential distributions. The inverse slope
parameter {\em T} increases systematically with particle mass (see
Fig.~\ref{fig:mt spectrum}) as a result of collective flow. It is
parameterized as:
\begin{equation}
 T = 0.175~{\rm GeV}+0.115\cdot m\qquad(c=1)\,.
\end{equation}

The rapidity distributions are assumed to resemble those of the
negative hadrons measured by NA49~\cite{Appelshauser:1998vn}. The
width is adjusted according to the maximum kinematic rapidity
limit.

The cross section is factorized for calculation of the final decay
kinematics into a contribution of a QED point source and an form
factor {\em F} describing the internal structure of the meson:
\begin{equation}
 \frac{{\rm d}\sigma}{{\rm d}q^2} = \left(\frac{{\rm d}\sigma}{{\rm d}q^2}\right)_{\rm point\,source}\cdot F(q^2)^2\,.
\end{equation}
The shape of the form factors are based on the measurements of the
Lepton-G collaboration and theoratical model calculations.  An
extensive summary of all relevant form factors is presented
in~\cite{Sako:2000}. According to the detailed discussion of the
systematic errors presented in~\cite{Agakishiev:1998mv}, the
uncertainty in the branching ratios and the form factors
contribute about 15\% below 450\,MeV/c$^2$, 30\% in the mass range
of 450--750\,MeV/c$^2$, and 6\% above 750\,MeV/c$^2$.

Finally, the dielectrons generated are subject to detector
acceptance and finite momentum resolution.

\begin{figure}[htb]
    \begin{minipage}[t]{.65\textwidth}
        \vspace{0pt}
        \epsfig{file=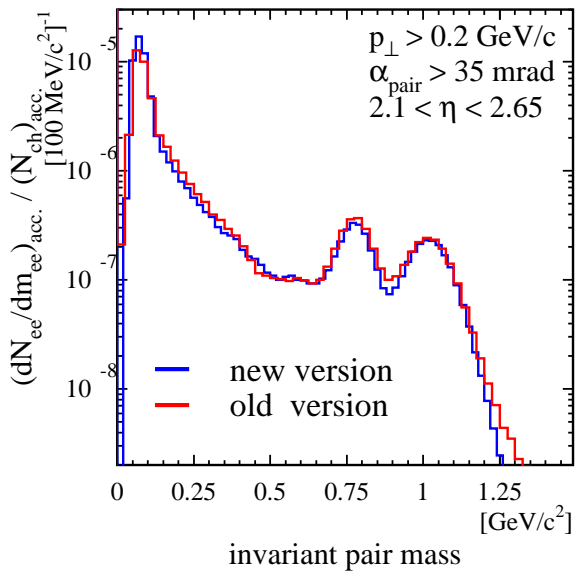,width=\textwidth}
    \end{minipage}%
    \begin{minipage}[t]{.35\textwidth}
      \vspace{0.5cm}
      \caption[Impact of the corrected GENESIS acceptance filter on the hadronic cocktail]
      {\newline
      Impact of the corrected GENESIS acceptance filter on the hadronic
      cocktail. The previously used cocktail~\cite{Lenkeit:1998} underestimates
      the yield in the low-mass region by 35\%.
      }
      \label{fig:p bug}
    \end{minipage}
\end{figure}

A comparison of the invariant mass spectra (see Fig.~\ref{fig:pA
data} in Sec.~\ref{sec:ceres prog}) shows the hadronic cocktail to
describe the observed yield in p-Be and p-Au collisions well
within the systematic errors of about 20\%. As a result of a
recent review and extension of the GENESIS code~\cite{Sako:2000},
the cocktail has decreased by up to 20\% in the mass region of
$0.15$\,$<$\,$m_{\rm ee}$\,$<$\,$1.5$\,GeV/c$^2$ (see
Fig.~\ref{fig:p bug}).

The low-mass yield ($m_{\rm ee}$\,$<$\,$0.2$\,GeV/c$^2$) was
previously underestimated by 35\% as a result of a problem in the
procedure used to apply the momentum resolution.

The cocktail for p-Be and p-Au collisions is much less affected
(5\%) due to the lower transverse momentum cut
($p_\bot$\,$<$\,$50$\,MeV/c). With hindsight to this fact, the
validity of the statement about the total exhaustion of the
observed dilepton yield by the known hadronic sources remains
sustained.

\begin{figure}[htb]
  \centering
  \epsfig{file=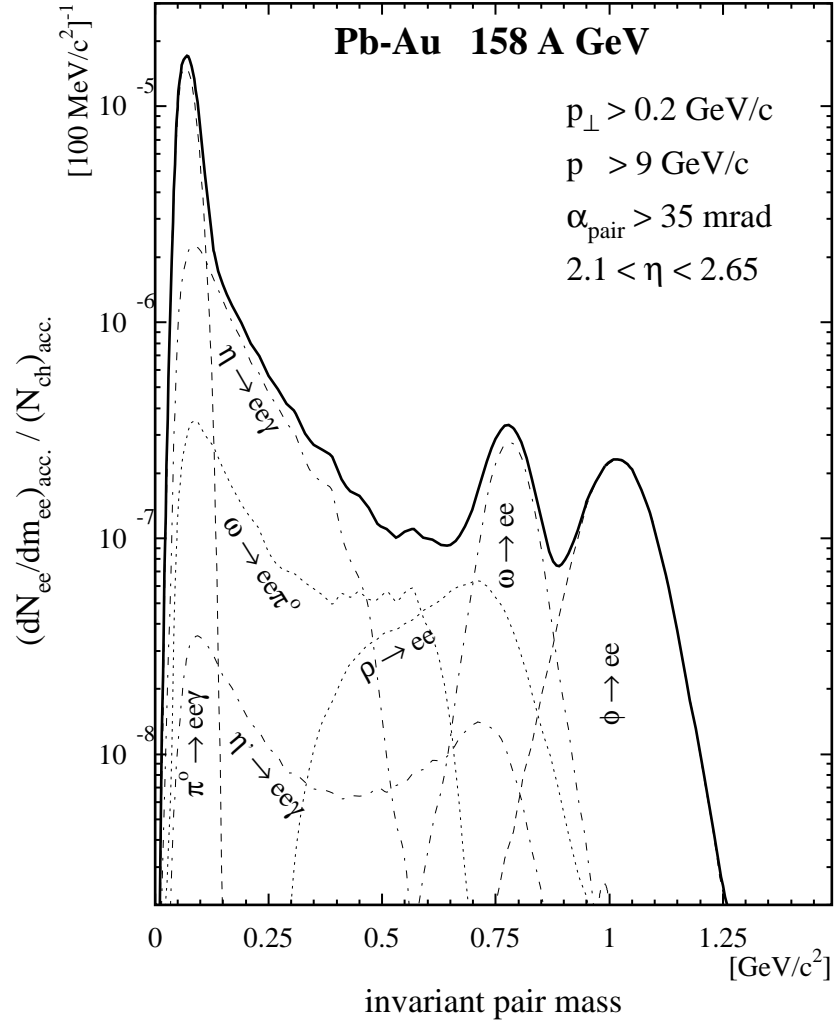}
  \caption[Genesis cocktail of known hadronic sources for Pb-Au collisions at 158\,GeV/c per nucleon]
  {Genesis cocktail of the known hadronic sources for Pb-Au collisions at 158\,GeV/c per nucleon.
  Integration of spectrum gives a total yield of  $(N_{\rm
ee}/N_{\rm ch})_{\rm acc.} = 1.21\cdot 10^{-5}$ which divides into
$(N_{\rm ee}/N_{\rm ch})_{\rm acc.} = 1.01\cdot 10^{-5}$ for
$m_{\rm ee}$\,$<$\,$0.2$\,GeV/c$^2$ and $(N_{\rm ee}/N_{\rm
ch})_{\rm acc.} = 2.04\cdot 10^{-6}$ for $m_{\rm
ee}$\,$>$\,$0.2$\,GeV/c$^2$. The experimental momentum resolution
of $dp/p=\sqrt{(0.041)^2+(0.022\cdot p\cdot{\rm GeV}^{-1}{\rm
c})^2}$ was applied.}
  \label{fig:cocktail}
\end{figure}

The new hadronic cocktail for Pb-Au collisions at 158\,GeV/c per
nucleon is plotted in Fig.~\ref{fig:cocktail}.
Integration of spectrum gives a total yield of  $(N_{\rm
ee}/N_{\rm ch})_{\rm acc.} = 1.21\cdot 10^{-5}$ which divides into
$(N_{\rm ee}/N_{\rm ch})_{\rm acc.} = 1.01\cdot 10^{-5}$ for
$m_{\rm ee}$\,$<$\,$0.2$\,GeV/c$^2$ and $(N_{\rm ee}/N_{\rm
ch})_{\rm acc.} = 2.04\cdot 10^{-6}$ for $m_{\rm
ee}$\,$>$\,$0.2$\,GeV/c$^2$. In contrast to~\cite{Lenkeit:1998},
the yield of pairs with very high mass ($m_{\rm
ee}$\,$>$\,$1.5$\,GeV/c$^2$) is slightly reduced as result of the
upper momentum cut of $p$\,$<$\,$9$\,GeV/c imposed for the
discrimination of high-momentum pions.

\chapter{Summary of rejection strategy}

 \label{app:cut summary}
\begin{table}[htb]
  \centering
  \begin{tabular}{|l|l|l|}
   \hline
   Description &  Cut selection & Comment \\
   \hline
   acceptance              &  SDD-1/2 $\theta$=141.--299.\,mrad & \\
                           &   RICH-1$\;\;\;\theta$=141.--299.\,mrad & 2/3 ring accepantce\\
                           &   RICH-2$\;\;\;\theta$=141.--258.\,mrad & 2/3 ring accepantce\\
   \hline
   double dE/dx            &  SDD-1 dE/dx\,$<$\,1133 $\mathbf{OR}$&  5\,mrad   resummation\\
   rejection               &  SDD-2 dE/dx\,$<$\,1204     &  \\
   \hline
   SDD-1         &  SDD-2 dE/dx\,$<$\,1204  $\mathbf{OR}$& 8\,mrad resummation \\
   conversion                        &  RICH-1 sum amplitude\,$<$\,1550 & \\
   \hline
   track quality           &  SDD-1$-$2 $\Omega_{\rm match}$\,$<$\,1.3--1.5\,mrad& target disk dependence \\
                           &  SDD$-$RICH-1 $\Omega_{\rm match}$\,$<$\,$2\,\sigma_{\rm
                           match}(p)$ & $\sigma=\sqrt{1.7^2+1.0^2/p^2}$\\
                           &   RICH-1$-$2 $\theta_{\rm match}$\,$<$\,$2\,\sigma_{\rm match}(p)$ & $\sigma=\sqrt{1.5^2+1.2^2/p^2}$ \\
                           &   RICH-2 Hough-2 amplitude\,$>$\,360  & ring candidate quality\\
                           &   RICH-2 $\chi^2/(n_{\rm hits}-1)$\,$<$\,1.3 & ring fit quality \\
                           &   $\Delta\phi_{\rm
                           RICH-1-RICH-2}$\,$<$\,300\,mrad& max.
                           B-field deflection \\
                           &   $\Delta\phi_{\rm SDD-PD}\,>\,5\,$mrad& min.$\;\;$B-field deflection \\
  \hline
  high-$p_{\bot}$ pion     &  & pressure dependence \\
  rejection                &  \raisebox{1.3ex}[-1.3ex]{\mbox{$\displaystyle \frac{R_{\rm RICH-1}}{ 14.62}+\frac{R_{\rm RICH-2}}{16.20}\,>\,2.015$} $\mathbf{OR}$} & of nominal radius \\
                           &  $\Delta\phi_{\rm SDD-PD}\,>\,17\,$mrad & \\
  \hline
  final acceptance         &  $\eta$=2.1--2.65 (single track) & eq. $\theta$=141.--243.\,mrad\\
                           &
                           $p_{\bot}$\,$<$\,0.2(0.1)\,GeV/c (single track) & \\
                           &  $\alpha_{\rm ee}$\,$>$\,35\,mrad &  pair opening
                           angle \\
  \hline
  \end{tabular}
  \caption[Summary of rejection cuts]
  {Summary of rejection cuts. The rejection cuts were
  applied in sequence as presented in the table. The order chosen
  was motivated by the intention to apply the most powerful and reliable
  cuts first. Most cuts are strongly correlated and cannot be
  looked at individually. All individual rejection conditions must be
  fulfilled according to a logical $\mathbf{AND}$ operation unless otherwise stated.}
  \label{tab:cuts}
\end{table}

\end{appendix}

\cleardoublepage

\addcontentsline{toc}{chapter}{Bibliography} \cleardoublepage

\section*{Resume}

\begin{tabbing}
\hspace{3cm}\= \kill

{\bf Name:}\>Hering, Gunar\\[.2cm]

{\bf Date of Birth:} \>February 2, 1972 - Chemnitz,
Germany\\[.2cm]

{\bf Address:}\>Friedensstra{\ss}e 23\\
  \>D-69121 Heidelberg, Germany\\
  \>Tel.:  (49)6221-474904\\
  \>Email: Gunar.Hering@gsi.de\\[.3cm]

{\bf Education:}\>\\
 5/97 - present \> Technical University of Darmstadt\\
               \> $\cdot$ Ph.D. in Physics\\
               \> $\cdot$ Thesis on ``Dielectron production in heavy ion
               collisions\\
               \> $\;$ at 158\,GeV/c per nucleon'', adivsor: Prof. Peter Braun-Munzinger\\
               \> $\cdot$ Ph.D. scholarship of German Scholarship
               Foundation\\[.2cm]
3/98 - $\;\;$3/00    \> Karl Ruprechts University, Heidelberg\\
               \> $\cdot$ Major: Economics\\[.2cm]
2/99 - $\;\;$3/99    \> Universidad del la Frontera Temuco,
Chile\\
               \> $\cdot$ Internship at Department of Economics\\[.2cm]
9/91 - 12/96   \> Friedrich Schiller University, Jena\\
               \> $\cdot$ Major: Physics, Minor: Mathematics, Ecology\\
               \> $\cdot$ Diploma in Physics,  grade ``very good''\\
               \> $\cdot$ Intermediate examination, grade ``very good''\\
               \> $\cdot$ Scholarship of German Scholarship Foundation\\
               \> $\cdot$ Scholarship of The Melton Foundation\\[.2cm]
8/94 - $\;\;$8/95 \> State University of New York at Stony Brook
,USA\\
               \> $\cdot$ Masters of Arts, Major: Physics\\
               \> $\cdot$ Thesis on ``Hexadecapole deformations in actinide
               and\\
               \> $\;$ trans-actinide nuclei'', adivsor: Prof. Peter Paul\\
               \> $\cdot$ Fulbright scholarship\\[.2cm]
9/86 - $\;\;$7/90    \> Special school for natural sciences at
Chemnitz (senior high school)\\
               \> $\cdot$ Abitur, grade ``very good''\\
               \> $\cdot$ 2. Prize federal contest ``Jugend
       forscht'' for the\\
               \> $\;$ development of an electronic cardiological
               model\\[.2cm]

{\bf Work history:}\>\\

5/97 - present \> {\em Gesellschaft f\"ur Schwerionenforschung},
Darmstadt\\
               \> $\cdot$ Developed particle detector and analyzed experimental data\\[.2cm]
12/99          \> {\em Hanse Institute for Advanced Study},
Bremen\\
               \> $\cdot$ Organized conference on ``Problems of applied
               ethics''\\[1cm]
2/99 - 3/99    \>{\em Centro de Gestion Empresarial} - Project
manager, Temuco Chile\\
               \>$\cdot$ Conducted market study for start-up micro-credit\\
               \>$\;$ project ``Mujeres del sur del mundo''\\[.2cm]
1/97 - 5/97     \>{\em The Boston Consulting Group} - Visiting
Associate, Frankfurt/M.\\
               \> $\cdot$ Performed a strategy review for a software systems provider\\[.2cm]
9/92 - 7/94   \>{\em FSU Physics department} - Teaching assistant,
Jena\\
               \> $\cdot$ Trained students in physics labs\\[.2cm]
9/90 - 9/91 \> {\em Rettungsamt Chemnitz} - Paramedic (civil
service)\\[.2cm]
\end{tabbing}

\noindent Heidelberg, January 21, 2002\\

\addcontentsline{toc}{chapter}{Resume}

\end{document}